\documentclass[%
 reprint,
superscriptaddress,
 amsmath,amssymb,
 aps,
pra,
]{revtex4-2}
\usepackage[normalem]{ulem}
\usepackage{svg}
\usepackage{xr}
\usepackage{graphicx}%
\usepackage{dcolumn}%
\usepackage{tabularx}
\usepackage{xcolor}
\usepackage{bm}%
\usepackage{hyperref}
\hypersetup{breaklinks=true, colorlinks=true, linkcolor={blue!90!black}, citecolor={blue!90!black}, urlcolor={blue!90!black}}
\usepackage{braket}
\usepackage[utf8]{inputenc}
\usepackage{verbatim}
\usepackage{amsmath}
\usepackage{soul}
\usepackage{booktabs}
\usepackage{xr}
\usepackage{amssymb}
\usepackage{amsthm}
\usepackage{blindtext}
\usepackage{mathtools}
\usepackage{slashed}
\usepackage{comment}
\usepackage{newtxmath}
\usepackage{amsfonts}
\usepackage{makecell}
\usepackage{tabularx}
\usepackage{float}
\usepackage{color}
\usepackage{multirow}
\usepackage{physics}
\usepackage{empheq}
\usepackage{bbm}
\usepackage{bibunits}

\usepackage[nameinlink]{cleveref}

\usepackage[ruled,vlined,linesnumbered]{algorithm2e}

\SetKwInput{KwInput}{Input}
\SetKwInput{KwOutput}{Output}

\let\oldaddcontentsline\addcontentsline
\newcommand{\stoptocentries}{\renewcommand{\addcontentsline}[3]{}}
\newcommand{\starttocentries}{\let\addcontentsline\oldaddcontentsline}

\definecolor{cerise}{rgb}{0.87, 0.19, 0.39}

\newcommand{\TotalSamples}{23{,}651}
\newcommand{\VerifiedSamples}{11,961}

\newcommand{\AverageTimePerSample}{0.913}
\newcommand{\AverageTimeMeasurement}{30}
\newcommand{\PositionVerificationRadius}{4,500}

\newcommand{\XEBMeasured}{0.586}

\newtheorem{theorem}{Theorem}

\mathchardef\mhyphen="2D %

\newtheorem{assumption}[theorem]{Assumption}
\newtheorem{lemma}[theorem]{Lemma}
\newtheorem{corollary}[theorem]{Corollary}

\newtheorem{definition}[theorem]{Definition}

\newtheorem{remark}[theorem]{Remark}

\numberwithin{equation}{section}

\newcolumntype{Y}{>{\centering\arraybackslash}X}

\newcommand{\eq}[1]{(\ref{eq:#1})}

\renewcommand{\sec}[1]{\hyperref[sec:#1]{Section~\ref*{sec:#1}}}
\newcommand{\app}[1]{\hyperref[app:#1]{Appendix~\ref*{app:#1}}}
\newcommand{\thm}[1]{\hyperref[thm:#1]{Theorem~f\ref*{thm:#1}}}
\newcommand{\lem}[1]{\hyperref[lem:#1]{Lemma~\ref*{lem:#1}}}
\newcommand{\cor}[1]{\hyperref[cor:#1]{Corollary~\ref*{cor:#1}}}
\newcommand{\fgr}[1]{\hyperref[fgr:#1]{Figure~\ref*{fgr:#1}}}
\newcommand{\tab}[1]{\hyperref[tab:#1]{Table~\ref*{tab:#1}}}
\DeclareMathOperator*{\E}{\mathbb{E}}

\usepackage{subfiles}

\begin{document}
\begin{bibunit}[apsrev4-2]
\stoptocentries

\title{Certified randomness amplification \\ by dynamically probing remote random quantum states}
\author{Minzhao~Liu}
\email{minzhao.liu@jpmchase.com}
\affiliation{Global Technology Applied Research, JPMorganChase, New York, NY 10017, USA}
\author{Pradeep~Niroula}
\affiliation{Global Technology Applied Research, JPMorganChase, New York, NY 10017, USA}
\author{Matthew~DeCross} 
\affiliation{Quantinuum, Broomfield, CO 80021, USA}
\author{Cameron Foreman} 
\affiliation{Quantinuum, Partnership House, London SW1P 1BX, UK}
\author{Wen Yu Kon} %
\affiliation{Global Technology Applied Research, JPMorganChase, New York, NY 10017, USA}
\author{Ignatius William Primaatmaja}
\affiliation{Global Technology Applied Research, JPMorganChase, New York, NY 10017, USA}
\author{M.S. Allman} 
\affiliation{Quantinuum, Broomfield, CO 80021, USA}
\author{J.P. Campora III} 
\affiliation{Quantinuum, Broomfield, CO 80021, USA}
\author{Akhil Isanaka} 
\affiliation{Quantinuum, Broomfield, CO 80021, USA}
\author{Kartik Singhal} 
\affiliation{Quantinuum, Broomfield, CO 80021, USA}
\author{Omar Amer}
\affiliation{Global Technology Applied Research, JPMorganChase, New York, NY 10017, USA}
\author{Shouvanik Chakrabarti}
\affiliation{Global Technology Applied Research, JPMorganChase, New York, NY 10017, USA}
\author{Kaushik~Chakraborty} %
\affiliation{Global Technology Applied Research, JPMorganChase, New York, NY 10017, USA}
\author{Samuel F. Cooper}
\affiliation{Quantinuum, Broomfield, CO 80021, USA}
\author{Robert D. Delaney}
\affiliation{Quantinuum, Broomfield, CO 80021, USA}
\author{Joan M. Dreiling}
\affiliation{Quantinuum, Broomfield, CO 80021, USA}
\author{Brian Estey}
\affiliation{Quantinuum, Broomfield, CO 80021, USA}
\author{Caroline Figgatt}
\affiliation{Quantinuum, Broomfield, CO 80021, USA}
\author{Cameron Foltz} 
\affiliation{Quantinuum, Broomfield, CO 80021, USA}
\author{John P. Gaebler}
\affiliation{Quantinuum, Broomfield, CO 80021, USA}
\author{Alex Hall}
\affiliation{Quantinuum, Broomfield, CO 80021, USA}
\author{Zichang He}
\affiliation{Global Technology Applied Research, JPMorganChase, New York, NY 10017, USA}
\author{Craig A. Holliman}
\affiliation{Quantinuum K.K., Tokyo, Japan}
\author{Travis~S.~Humble} %
\affiliation{Quantum Science Center, Oak Ridge National Laboratory, Oak Ridge, TN 37831, USA}
\author{Shih-Han~Hung} %
\affiliation{Department of Electrical Engineering, National Taiwan University, Taipei City, 10617, ROC}
\author{Ali A. Husain}
\affiliation{Quantinuum, Brooklyn Park, MN 55422, USA}
\author{Yuwei Jin}
\affiliation{Global Technology Applied Research, JPMorganChase, New York, NY 10017, USA}
\author{Fatih Kaleoglu}
\affiliation{Global Technology Applied Research, JPMorganChase, New York, NY 10017, USA}
\author{Colin J. Kennedy}
\affiliation{Quantinuum, Broomfield, CO 80021, USA}
\author{Nikhil Kotibhaskar}
\affiliation{Quantinuum, Partnership House, London SW1P 1BX, UK}
\author{Nathan K. Lysne}
\affiliation{Quantinuum K.K., Tokyo, Japan}
\author{Ivaylo S. Madjarov}
\affiliation{Quantinuum, Broomfield, CO 80021, USA}
\author{Michael Mills}
\affiliation{Quantinuum, Broomfield, CO 80021, USA}
\author{Alistair R. Milne}
\affiliation{Quantinuum, Partnership House, London SW1P 1BX, UK}
\author{Kevin Milner} 
\affiliation{Quantinuum, Partnership House, London SW1P 1BX, UK}
\author{Louis Narmour}
\affiliation{Quantinuum, Broomfield, CO 80021, USA}
\author{Sivaprasad Omanakuttan}
\affiliation{Global Technology Applied Research, JPMorganChase, New York, NY 10017, USA}
\author{Annie J. Park}
\affiliation{Quantinuum, Broomfield, CO 80021, USA}
\author{Michael A. Perlin}
\affiliation{Global Technology Applied Research, JPMorganChase, New York, NY 10017, USA}
\author{Adam P. Reed}
\affiliation{Quantinuum, Broomfield, CO 80021, USA}
\author{Chris~N.~Self}
\affiliation{Quantinuum, Terrington House, Cambridge CB2 1NL, UK}
\author{Matthew Steinberg}
\affiliation{Global Technology Applied Research, JPMorganChase, New York, NY 10017, USA}
\author{David T. Stephen}
\affiliation{Quantinuum, Broomfield, CO 80021, USA}
\author{Joseph Sullivan}
\affiliation{Global Technology Applied Research, JPMorganChase, New York, NY 10017, USA}
\author{Alex Chernoguzov}
\affiliation{Quantinuum, Broomfield, CO 80021, USA}
\author{Florian J. Curchod} %
\affiliation{Quantinuum, Terrington House, Cambridge CB2 1NL, UK}
\author{Anthony Ransford}
\affiliation{Quantinuum, Broomfield, CO 80021, USA}
\author{Justin G. Bohnet}
\affiliation{Quantinuum, Broomfield, CO 80021, USA}
\author{Brian Neyenhuis}
\affiliation{Quantinuum, Broomfield, CO 80021, USA}
\author{Michael~Foss-Feig} %
\affiliation{Quantinuum, Broomfield, CO 80021, USA}
\author{Rob~Otter}
\affiliation{Global Technology Applied Research, JPMorganChase, New York, NY 10017, USA}
\author{Ruslan~Shaydulin}
\email{ruslan.shaydulin@jpmchase.com}
\affiliation{Global Technology Applied Research, JPMorganChase, New York, NY 10017, USA}

\date{\today}

\begin{abstract}
Cryptography depends on truly unpredictable numbers, but physical sources emit biased or correlated bits. 
Quantum mechanics enables the amplification of imperfect randomness into nearly perfect randomness, but prior demonstrations have required physically co-located, loophole-free Bell tests \cite{kulikov2024device}, constraining the feasibility of remote operation.
Here we realize certified randomness amplification across a network by dynamically probing large, entangled quantum states on Quantinuum's 98-qubit Helios trapped-ion quantum processor. Our protocol is secure even if the remote device acts maliciously or is compromised by an intercepting adversary, provided the samples are generated quickly enough to preclude classical simulation of the quantum circuits. We stream quantum gates in real time to the quantum processor, maintain quantum state coherence for $\approx 0.9$ seconds, and then reveal the measurement bases to the quantum processor only milliseconds before measurement. This limits the time for classical spoofing to 30 ms and constrains the location of hypothetical adversaries to a $4{,}500$ km radius. We achieve a fidelity of \XEBMeasured~on random circuits with 64 qubits and 276 two-qubit gates, enabling the amplification of realistic imperfect randomness with a low entropy rate into nearly perfect randomness.
\end{abstract}

\maketitle
\renewcommand{\theequation}{\arabic{equation}}

Unbiased and truly unpredictable numbers lie at the heart of many applications~\cite{certrand_apps,Acn2016}. Many cryptographic primitives are impossible unless the participants have access to nearly perfectly uniform randomness~\cite{dodis2004impossibility,Foreman2023practicalrandomness}. Yet, every physical source --- from thermal noise to atmospheric static --- inevitably generates bits that are biased or correlated. This problem is even more pronounced in distributed settings, including cryptographic protocols like blockchains and non-interactive zero-knowledge proofs, as well as societal functions such as auditing and lotteries \cite{certrand_apps}. In these settings, multiple parties who may not trust each other must collectively agree that a generated random number could not have been predicted or influenced by any participant.

{Typical methods of augmenting ‘weak’ randomness, i.e., randomness that may be biased or exhibit correlations, resort to deterministic  `whitening' or randomness extraction techniques and require strong assumptions on the weak source to guarantee the security of the output bits. In practice, this implies detailed modeling and characterization of the physical sources~\cite{Barker2015,Turan2018,Barker2025}. However, if the source has been tampered with, or if there is incomplete understanding of source imperfections and potential information-leakage channels --- possibilities that arise in both classical and quantum random number generators~\cite{zhou2005side,boneh1999importance,thewes2019eavesdropping} --- such deterministic post-processing cannot guarantee the unpredictability of the output. In fact, extracting nearly perfect randomness from most weak sources by classical means alone is impossible \cite{santha1986generating}.}

Fortunately, quantum mechanics enables the certifiable `amplification' of weak randomness into nearly perfect randomness without detailed source characterization. Remarkably, this can be achieved even when the outcomes of a quantum device used to enable the amplification process are public (i.e., unpredictable before generation, but fully revealed to a hypothetical adversary once produced) if the weak source of randomness remains private (i.e., hidden from the adversary)~\cite{Foreman2023practicalrandomness}. This process, known as randomness amplification, has been experimentally demonstrated using Bell inequality violation \cite{colbech2012free,kessler2020device,Foreman2023practicalrandomness,kulikov2024device}.

{However, Bell-test-based protocols require the enforcement of strict physical conditions, such as sufficient separation between experimental components. Such conditions cannot be enforced when quantum devices are only accessible remotely, as in cloud-based deployments and multi-party scenarios in which not all parties can have physical control of the hardware. A recent experiment based on random circuit sampling (RCS)~\cite{jpmc_cr}, where random quantum circuits are executed and measured to produce random bits, generated certifiable randomness with a remote quantum computer accessed over the internet. However, the experimental protocol required perfect randomness as input~\cite{aaronson2023certified,jpmc_cr} and therefore could not achieve randomness amplification.}

{In this work, we design and implement a protocol that simultaneously amplifies weak randomness and remains secure when using an untrusted remote quantum device. We first introduce a certified randomness protocol that outputs quantum randomness.} The protocol dynamically prepares and probes remote, computationally-complex quantum states; by delaying the delivery of measurement information until after the quantum state has been prepared, the protocol drastically reduces the time available for classical spoofing. This reduction ultimately improves operational characteristics such as cost and security. 
We then embed the certified randomness protocol in a larger randomness amplification scheme that uses quantum randomness sent over a public channel to amplify private weak randomness into nearly perfect private randomness.

Beyond its practical significance, this capability gives fundamental insight into quantum foundations. Randomness is not merely a feature but a signature of deeper physical principles. A long line of research has investigated the minimal assumptions under which randomness can arise in nature \cite{Acn2016}, culminating in the development of randomness amplification \cite{colbech2012free,brandao2016realistic,putz2016measurement,kessler2020device,foreman25,kulikov2024device}: if partially random events exist, they can be amplified into fully unpredictable ones. In other words, nature would have to be `superdeterministic' to preclude full randomness. {Although Bell-test protocols do not require detailed source characterization, they still require that each bit of input randomness contains some entropy---a weak form of device model assumption.} Our protocol relaxes this assumption to its minimal form, enabling the amplification of unstructured input randomness, at the cost of computational assumptions on the adversary that need only hold during the experiment.

\section*{Randomness with gate streaming}

We use the newly developed 98-qubit Quantinuum Helios trapped-ion quantum processor \cite{ransford_helios}, which is capable of generating and storing high-fidelity and computationally complex quantum states and allows remote users to manipulate and probe such states with instructions sent in real-time. Unlike conventional approaches that treat the quantum processor as a black box simply producing results after receiving instructions, we leverage the ability to manipulate a coherent quantum state dynamically.

The central idea behind our protocol is as follows. A randomness consumer (the `client') asks an untrusted remote quantum server to perform RCS, requesting bitstrings sampled from a random quantum circuit in a random measurement basis. 
We model the interaction as adversarial: the server may be dishonest, or an eavesdropper may mount a man-in-the-middle attack by impersonating the server.
Since RCS is classically intractable for large circuits~\cite{Arute2019,qntm_rcs,aaronson2016complexity,Zhu2022,Wu2021}, rapid completion of the task certifies that the returned samples originated from a quantum process and are therefore random, even if generated by an adversary.

\begin{figure}[!t]
    \centering
    \includegraphics[width=1\linewidth]{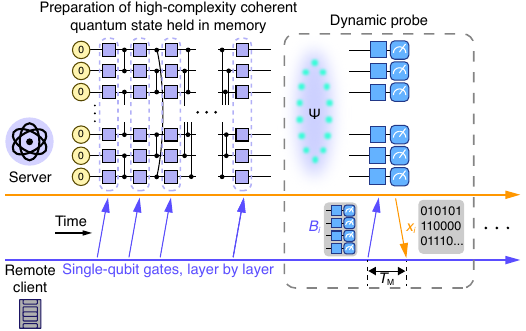}
    \caption{\textbf{Client-server interaction.} $B_i$ is the measurement basis sent by the client in the $i$th round, and $x_i$ is the corresponding bitstring returned by the server. $T_{\rm M}$ includes the time to apply single-qubit rotations (blue gates), as well as measurement readout and communication time. The full protocol repeats the interaction $L$ times and collects $L$ samples.}
    \label{fig:interactive_protocol}
\end{figure}

Our protocol consists of two steps. In the first step shown in Fig.~\ref{fig:interactive_protocol}, the client streams random quantum circuits to the server one layer at a time, recording a timestamp for each layer. All circuits have the same two-qubit gates and only the single-qubit gates differ. Therefore, the client transmits only the single-qubit gate information at each layer. After receiving the final layer of single-qubit gates specifying the measurement bases, the server must return a bitstring sampled from the circuit within a predetermined time window. If the server exceeds this deadline, the failure is recorded and the client proceeds to the next round. The timing requirement prevents the adversary from running long classical simulations that mimic the behavior of the quantum computer while not providing entropy.

In the second step, the client certifies that the bitstrings it receives correspond to the output distributions of the circuits it sent by computing the XEB score for a random subset of circuits and bitstrings:
\begin{align}\label{eqn:xeb_main}
\mathrm{XEB}=\frac{2^n}{L_{\rm val}}\sum_{i\in\mathcal{V}}P_{U_i}(x_i)-1,
\end{align}
where $n$ is the number of qubits for the circuits, $\mathcal{V}$ is a set validation set, $L_{\rm val}=\vert\mathcal{V}\vert$, $U_i$ is the $i$th sent random circuit, $x_i$ is the $i$th received bitstring, and $P_{U}(x)=\vert\langle x\vert U\vert 0\rangle\vert^2$ is the probability of measuring $x$ from the ideal quantum state after applying circuit $U$ to the zero state.

\begin{figure*}[!ht]
    \centering
    \includegraphics[width=1\textwidth]{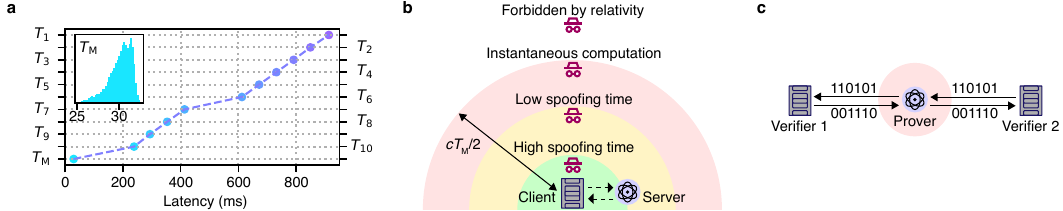}
    \caption{\textbf{Certified randomness with gate streaming.} \textbf{a}, Average latencies (measured by the client) between sending a layer of single qubit gates and receiving the bitstring. $T_j$ is the latency for the $j$th layer; $T_M$ is the latency for the final layer specifying the measurement basis. The inset shows the frequency distribution of measured $T_M$. Error bars are smaller than the marker size. \textbf{b}, As the adversary moves farther from the client, less time is available for spoofing. Adversaries located beyond $c\cdot T_{\rm M}/2$, where $c$ is the speed of light, are forbidden by relativity. \textbf{c}, Illustration of a position verification protocol. The verifiers and the prover exchange classical information. The red circle denotes the uncertainty of the certified position of the prover.}
    \label{fig:streaming}
\end{figure*}

Under suitable conditions, the XEB score coincides with the quantum device fidelity \cite{morvan2023phase}. If the bitstrings $x_i$ come from the ideal output distribution of $U_i$, the XEB score concentrates around 1. For $x_i$ generated independently from $U_i$, it concentrates around 0. The protocol aborts if the XEB score is below some threshold. For rounds that violate the timing requirement, we set $x_i$ to the zero bitstring and replace $P_{U_i}(x_i)$ in Eq.~\ref{eqn:xeb_main} with 0, thereby preventing an attacker from post-selecting samples.

An adversary may try to `spoof' the protocol by returning low-entropy samples that nonetheless achieve a high XEB score, e.g. by classically simulating sampling. Although the adversary's cost to produce classical samples is proportional to the client's cost to compute $P_{U_i}(x_i)$ during validation, the adversary must spoof within the short time window after receiving the circuit, whereas the client can compute the XEB score at leisure. {While the validation cost is exponential in $n$, the circuits need only be large enough to render classical spoofing intractable; absent unexpected breakthroughs in spoofing, $n$ will remain constant.}

{Because achieving a high XEB score classically is difficult, a high XEB score indicates that the samples could only have been produced by a quantum computer. It has also been shown that such samples from a quantum computer must have high entropy, i.e. an efficient deterministic quantum algorithm that does well on XEB does not exist \cite{aaronson2023certified,jpmc_cr}. Consequently, a high XEB score allows us to lower-bound the entropy of the bitstrings.}

Our dynamic approach--streaming layers of single-qubit gates and the measurement basis in real time--prevents classical spoofing from commencing until the basis is revealed.
The intuition is that the quantum device maintains a coherent state with high classical simulation complexity prior to measurement, and this state cannot be efficiently replaced by a classical representation that would allow a classical algorithm to perform well on the XEB test with randomized bases. The same intuition underpins the recent quantum information supremacy demonstration where, for Haar-random states, no polynomial-sized classical description of the state allows a classical algorithm to achieve high XEB \cite{kretschmer2025demonstrating}. While our setting is not equivalent to that of quantum information supremacy, we discuss their relationship in Supplementary Information Section \ref{sec:basis}.

Classical simulators could evolve the quantum state partially and cache it before the measurement basis is known. However, such approaches are infeasible for beyond-classical RCS experiments like ours. For example, matrix-product-state-based methods that compress the quantum state representation have been shown to require an impractical amount of memory to match experimental fidelity \cite{morvan2023phase,qntm_rcs}. We apply the same analysis to our circuits in Supplementary Information Section \ref{sec:mps}.

Similarly, without knowledge of the measurement basis, tensor network methods can only pre-compute partial tensor contractions before encountering an unspecified tensor due to the unknown measurement basis. This approach is not fruitful since tensor network methods need to exploit the knowledge of the output bits to dramatically reduce the computational cost, which is now obfuscated by the unknown measurement bases. Our numerical evidence in Supplementary Information Section \ref{sec:basis} shows that this strategy does not notably reduce the overall simulation cost for the circuits we use. Streaming gates one layer at a time provides an additional safeguard against useful pre-computation by further limiting the time available to perform pre-computation with each layer of the circuit.

We perform the experiment with the 98-qubit Quantinuum Helios trapped-ion quantum processor. We use random quantum circuits with 64 qubits, a two-qubit gate depth of 9, and arbitrary connectivity as detailed in Methods. The client runs on standard computer hardware connected to the control system of the quantum processor via Ethernet (see Ref.~\cite{pn_space_advantage} for implementation details). The control system initiates all requests for the gate layers and measurement basis, and the client's responses are precisely timed to arrive just as gates or measurements are ready to be executed, masking network latencies. A total of 23,651 circuits were requested, and the server only failed to return a bitstring within $40$ ms of receiving the measurement basis for 53 circuits. Fig.~\ref{fig:streaming}\textbf{a} shows the timing information of gate streaming, and more details are shown in Supplementary Information Fig.~\ref{fig:timing_results}. Finally, we validate a random subset of \VerifiedSamples\ circuit-sample pairs using the Aurora supercomputer at the Argonne Leadership Computing Facility~\cite{auroraoverview} and obtain an XEB score of \XEBMeasured, which also serves as an estimate of the quantum fidelity. 

The average time between the client sending the first layer of the quantum circuit and receiving the bitstring is $\langle T_1\rangle=\AverageTimePerSample$ s. However, the interval between the client sending the measurement basis and receiving the bitstring averages to $\langle T_M \rangle = \AverageTimeMeasurement$ ms. Since the cost of spoofing and verification are closely linked, smaller $T_M$ enables high security at a lower validation cost.

Furthermore, a short response time also allows us to use relativistic constraints to bound the adversary's power by their geographic distance. An adversary at distance $d$ from the client would require a round-trip signal delay of $2d/c$ where $c$ is the speed of light, leaving at most $T_{\rm M}-2d/c$ for classical simulation. If an adversary located adjacent to the client can achieve a simulation fidelity of $\Phi_{\rm C}$, an equally powerful but distant adversary can attain fidelity of at most $\Phi_{\rm C}\cdot[1-2d/(c\cdot T_{\rm M})]$ --- the further an adversary, the harder the task of spoofing. This effect is illustrated in Fig.~\ref{fig:streaming}\textbf{b}. For example, if the adversary is 3,000 km away from the client, the spoofing fidelity is reduced by a factor of $1-2d/(c\cdot T_{\rm M})\approx0.33$. 

The low response time also enables position verification of a remote party using classical communication \cite{kaleoglu2024equivalence}. Fig.~\ref{fig:streaming}\textbf{c} shows two verifiers that aim to confirm the position of a prover located far from both parties. This task is impossible with classical computation alone, and is useful as an additional layer of authentication \cite{chandran2014position}. With our 30 ms latency, the prover’s location could be certified within a \PositionVerificationRadius\ km radius, provided suitable classical position-verification infrastructure is implemented in the future. Further details on position verification are provided in Supplementary Information Section \ref{sec:position_verification}.

\section*{Randomness amplification}

\begin{figure*}[!ht]
    \centering
    \includegraphics[width=\textwidth]{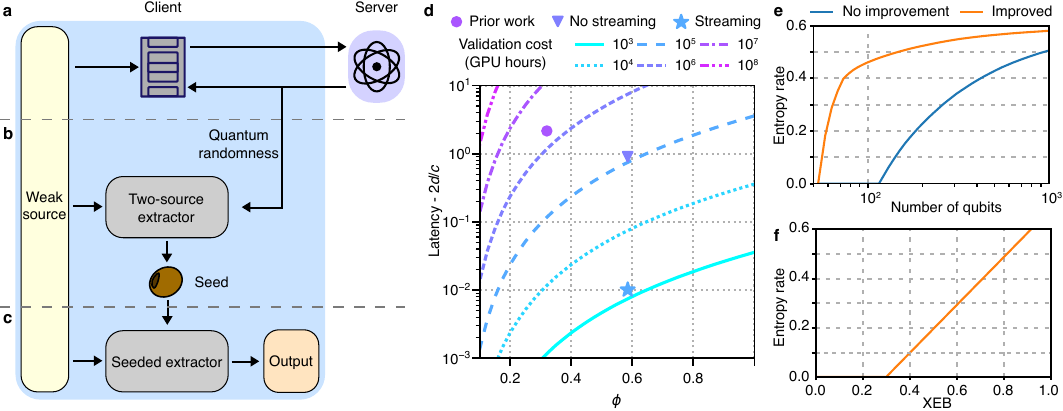}
    \caption{\textbf{Certified randomness amplification.} \textbf{a,b,c} The amplification protocol. \textbf{a}, The client takes weak randomness as input and generates challenge {quantum circuits} to send to the quantum randomness server. The server responds with randomness to the client, and the client verifies the response. \textbf{b}, After obtaining and verifying the quantum randomness, the client combines it with additional weak randomness and, using a two-source extractor, produces nearly perfect randomness termed the ‘seed’. \textbf{c}, After the seed is obtained, the client uses a seeded extractor to amplify weak randomness. {With the seed, the seeded extractor converts the weak source into nearly perfect randomness.} 
    This procedure can be repeated, amplifying successive outputs of the weak source with the same seed.
    \textbf{d}, Validation cost in GPU hours at different quantum device fidelities $\phi$ and sample latencies, assuming the restricted adversary with $10^5$ GPUs located at a distance $d\geq 3{,}000$ km. 
    `Prior work' uses the fidelity and latency achieved in Ref.~\cite{jpmc_cr}. `No streaming' uses the fidelity achieved in this work and latency of $T_1$, corresponding to the measurement bases not being streamed. `Streaming' uses the latency of $T_{\rm M}$. \textbf{e}, {Entropy rates of the quantum randomness in the oracle adversary model} for different numbers of qubits at $\mathrm{XEB}=0.586$, with and without the improvements described in the main text. \textbf{f}, {Entropy rates of the quantum randomness in the improved oracle adversary model} at $n=64$ at different achievable $\mathrm{XEB}$ scores.}
    \label{fig:amplification_main}
\end{figure*}

In randomness amplification, the input is a weak source and the output is nearly perfect randomness. The degree of randomness of a weak source is captured by the min-entropy rate $\alpha\in[0,1]$, defined as the ratio of min-entropy to the source length. A deterministic source has $\alpha=0$, while perfect randomness has $\alpha=1$. Min-entropy reflects the worst-case probability that an adversary can correctly guess the output. A weak source may be subject to additional assumptions. For example, in a Santha-Vazirani (SV) source, each new bit must retain at least $\alpha$ bits of min-entropy when conditioned on past outputs and side information. Other sources relax this demand, requiring min-entropy only over large blocks of bits. They do not restrict how the entropy is distributed within a block and are called block min-entropy sources.

Amplification of sources with $\alpha<0.98$ was previously out of reach \cite{kulikov2024device} unless one was willing to compromise on device independence, i.e., on the requirement that the security does not rely on the honesty of the devices involved \cite{Foreman2023practicalrandomness}. Prior experiments further assumed the weak source to be an SV source, excluding more realistic sources such as block min-entropy sources. Although proposals already exist for amplifying block min-entropy sources~\cite{bouda2014device,plesch2014device,chung2014physical}, they remain impractical and cannot be certified remotely.

In contrast, we demonstrate amplification of block min-entropy sources with entropy rate $\alpha<0.5$ under computational assumptions restricting the adversary's ability to analyze the weak source as detailed in Methods and Supplementary Information Section \ref{sec:amplify_private}. These computational assumptions need hold only during the protocol execution; even if they fail later, the amplified bits remain secure. This property is called everlasting security and is impossible with pseudo-randomly generated randomness.

Our randomness amplification protocol is illustrated in Fig.~\ref{fig:amplification_main}. We first obtain a nearly perfectly random seed, which we then use in a seeded extractor to repeatedly extract the weak source. 
We tune the protocol parameters to balance three objectives: security, the total number of extracted bits forming our protocol output, considering both the number of seeded-extraction rounds and the bits extracted per round, and the required assumptions on the entropy rate $\alpha$ of the weak source.
Additional details on parameterization of the amplification protocol are provided in Supplementary Information Section \ref{sec:amplification_concrete}.

We analyze the protocol's security under two adversarial models. First, we consider the model of Ref.~\cite{jpmc_cr}, which incorporates the best available spoofing techniques. In this model, each sample returned by the adversary is either sampled honestly from the quantum circuit generated via classically simulated. {Adversaries equipped with more GPUs can simulate the circuits at higher fidelities, making quantum and classical samples harder to distinguish. Consequently, maintaining the same security and entropy against stronger adversaries requires verifying more samples, thereby increasing validation cost.} 

Second, we consider an adversary capable of general quantum computation, encompassing a much broader class of potential future adversaries. We call this the oracle model, which we explain later. Our experiment shows that randomness amplification is possible even against this more general adversary.

\begin{table}
    \centering
        \begin{tabular}{p{1.7cm} p{1.4cm} p{0.8cm} p{2cm} p{2cm}}
        \hline
         Protocol & Remote & DI & Security & Source requirement\\
         \hline
         Ref. \cite{kulikov2024device} & No & Yes & IT & SV, $\alpha\geq0.98$ \\
         \hline
         Ref. \cite{Foreman2023practicalrandomness} & No & No & IT & SV, $\alpha\geq0.66$  \\
         \hline
          This work & Yes & Yes & C, restricted & BM, $\alpha\geq0.41$ \\
          \hline
          This work & Yes & Yes & C, oracle & BM, $\alpha\geq0.53$  \\
     \hline
\end{tabular}
    \caption{\textbf{Comparison of requirements on the weak source for experimentally demonstrated randomness amplification protocols under various security models}. DI stands for device independent. For security models, IT stands for information-theoretic, and C stands for computational assumptions during the protocol execution are required for everlasting security. BM stands for block min-entropy sources. See the main text for more details. }
    \label{tab:comparison}
\end{table}

For a comparison of experimentally demonstrated randomness amplification protocols, see Table \ref{tab:comparison}. Additional results for other security parameters and entropy rates are provided in Supplementary Information Table \ref{tab:amplification_results}. We note that Bell test-based protocols achieve information-theoretic (IT) security as the laws of physics forbid low-entropy spoofing of the protocol. On the other hand, quantum-computing-based protocols, including ours, provide everlasting security as long as the randomness-generating device satisfies reasonable computational restrictions during the protocol, denoted as C. In Table \ref{tab:comparison}, we assume an adversary that is five times more computationally powerful than Aurora supercomputer~\cite{auroraoverview} and set the soundness parameter to $\varepsilon_{\rm sou}=10^{-3}$. Here, $\varepsilon_{\rm sou}$ quantifies the permissible deviation from a perfectly secure protocol. A formal definition of security and soundness is provided in the Methods. 

Against the restricted adversary, we achieve a min-entropy rate of $\beta=0.53$, enabling amplification of block min-entropy sources with vanishing min-entropy rates. In particular, a two-source extractor succeeds only if at least one input (the quantum randomness or the weak source) has an entropy rate exceeding $0.5$. Because earlier demonstrations produced quantum randomness with very low entropy rates, they had to assume weak sources with $\alpha>0.5$; this assumption is unnecessary in our protocol. For a weak source with $\alpha=0.41$, we extract a seed of size $4{,}093$. This seed enables $10^8$ extraction rounds, each yielding $1{,}570$ bits from the weak source. We note that assuming $\alpha>0.5$ allows the protocol to maintain high security even against much stronger adversaries. For example, with $\alpha=0.54$ the protocol achieves a soundness of $\varepsilon_{\rm sou}=10^{-6}$ while remaining secure against an adversary 300 times more powerful than Aurora. Other potential trade-offs between the security parameters are summarized in Supplementary Information Table~\ref{tab:amplification_results}

We analyze the validation cost in the limit of infinitely many rounds. Fig.~\ref{fig:amplification_main}\textbf{d} shows the {total validation} cost required to achieve success probability $1-10^{-4}$, security parameter $\varepsilon_{\rm sou}=10^{-6}$, and min-entropy rate $\beta=0.04$ against an adversary equipped with $10^5$ GPUs, as a function of device fidelity and latency. High fidelity and low latency substantially reduce the validation cost. For details of the numerical analysis, see Supplementary Information Section \ref{sec:cost_analysis}.

We further analyze our protocol against adversaries in the oracle model, which is used in Refs.~\cite{kretschmer2021quantum,aaronson2023certified}. By oracle access, we mean the adversary can execute the challenge circuit as a whole, potentially as part of a bigger quantum circuit, but cannot exploit its internal structure. Each execution of the challenge circuit is referred to as a ‘query’. There are no additional restrictions on the computation the adversary may perform. Intuitively, if the circuit is sufficiently random and lacks meaningful structures, this is the adversary's optimal strategy. The restriction to oracle access is common in cryptography literature, typically applied to cryptographic hash functions that do not have obvious structures to exploit \cite{katz2014introduction}.

We derive entropy bounds against adversaries with oracle access to {Haar random challenge} quantum circuits, and plot the results in  Fig.~\ref{fig:amplification_main}\textbf{e}. 
We observe that when the number of qubits $n$ is small, no entropy is generated. At the same time, the verification cost becomes prohibitive as $n$ increases beyond 80 qubits, making the protocol impractical.
To address this challenge, we devise improvements that lower the minimum $n$ required to obtain non-zero entropy, as shown in Fig.~\ref{fig:amplification_main}\textbf{e}. The details of the improvements are presented in Supplementary Information Section \ref{sec:app-single-round}.

We also analyse how quantum device fidelity affects the attainable entropy. Fig.~\ref{fig:amplification_main}\textbf{f} shows the achievable entropy rate at different values of XEB score; the latter is expected to approximate the device fidelity. A key finding is that no entropy can be certified once the XEB score falls below a certain threshold, underscoring the need for high-fidelity trapped-ion devices.  For quantum computing platforms with lower fidelity --- even if they offer reduced latency --- the ability to perform beyond-classical sampling does not, by itself, enable certified randomness.

It is reasonable to assume that no existing device is capable of implementing more than one query to each circuit used in this experiment at the observed fidelity and within our latency constraints. Presently, only the Helios trapped-ion quantum processor used in this experiment is capable of executing RCS with arbitrary circuit geometries of this size while achieving high fidelity. Additionally, we understand the timing characteristics of the device well and attempted to reduce the circuit execution time as much as possible. We assume the oracle access adversary has one query to each challenge circuit at fidelity at most 0.65, leading the min-entropy rate of the quantum output to be $\beta=0.14$, enabling amplification of a weak source with entropy rate $\alpha=0.53$.

\section*{Outlook}
We experimentally demonstrate remotely certifiable randomness amplification of block min-entropy sources secure against general quantum adversaries in the oracle model and amplification of sources with vanishing entropy rates against restricted adversaries. Our dynamic state preparation and measurement strategy constrains the location of the adversary, improves the security, and reduces the validation cost, thus addressing the principal limitations of certified randomness for practical use. As an example, we discuss how multiparty protocols may use and benefit from certified randomness in Supplementary Information Section \ref{sec:joint}.

A crucial aspect of the protocol is the infeasibility of precomputation by classical adversaries before the measurement bases are available. Although we make connections with quantum information supremacy and existing techniques are ruled out in Supplementary Information Section \ref{sec:basis}, a complexity-theoretic proof remains an open challenge. It is possible that there may be an interesting interplay between quantum computational advantage and quantum information advantage.

\putbib[citations]

\clearpage

\section*{Methods}
\label{sec:methods}

\subsection{Experimental protocol}

The challenge random quantum circuits we implement in our experiment are closely related to the random circuits used in prior work \cite{jpmc_cr,qntm_rcs,morvan2023phase} and are constructed as follows. Starting from the $\ket{0}$ state, we apply a layer of random single-qubit gates drawn from the set of eight possible gates, namely $Z^p X^{1/2} Z^{-p}$ with $p\in\{-1,-3/4,-1/2,\dots,3/4\}$. The gate set can be implemented using the native gate
\begin{align}\nonumber
U_{1q}(\theta,\phi)&=e^{i(\cos\phi\hat{X}+\sin\phi\hat{Y})\frac{\theta}{2}}\\
&=\begin{pmatrix}
\cos{\frac{\theta}{2}}& -ie^{-i\phi}\sin{\frac{\theta}{2}} \\
-ie^{i\phi}\sin{\frac{\theta}{2}} & \cos{\frac{\theta}{2}}
\end{pmatrix}.
\end{align}
The gates are realized with $\theta=\pi/2$ and $\phi=p\pi$, modulo an overall phase $e^{-i\pi/4}$ that is common to all qubits and therefore omitted.
This is followed by a layer of $n/2$ native $RZZ(\pi/2)$ gates on distinct randomly chosen pairs of qubits. We repeat these single- and two-qubit gate layers several times with independent randomness, and end with a final layer of random single-qubit gates.

In our implementation, eight layers of 32 two-qubit gates are applied to random qubit pairs, followed by a final layer of 20 two-qubit gates acting on a randomly chosen set of 40 qubits. 
The two-qubit gates are identical across all circuits; only the single-qubit gates vary. 
One initial layer of 64 random single-qubit gates precedes everything, and an additional layer of 64 random single-qubit gates follows each of the eight two-qubit gate layers that have 32 two-qubit gates. After the final layer of 20 two-qubit gates, the 40 affected qubits are each rotated by a further random single-qubit gate. Each circuit therefore contains $32\times 8 + 20=276$ two-qubit gates and $64\times 9+40=616$ single-qubit gates. Finally, a random single-qubit measurement basis ($Z$ or $X$) is sent to the quantum processor control system. On receipt of the basis information, $H$ gates are applied to qubits that need to be measured in the $X$ basis, after which all qubits are measured.

We choose the architecture to achieve the desired validation hardness. Circuits that are too easy to validate are also easy to spoof, whereas overly hard circuits preclude validation of a sufficient number of samples.

We comment on the comparison between the circuit used in \cite{jpmc_cr} and this work. In Fig.~\ref{fig:amplification_main}\textbf{d}, for the purple circle, the circuit is assumed to have 64 qubits instead of 56 as in the experiment of Ref.~\cite{jpmc_cr}. Although the circuits used for our work and Ref.~\cite{jpmc_cr} differ in qubit count and depth, they contain approximately the same number of two-qubit gates. Therefore, the fidelities and latencies can be fairly compared. In particular, the circuit in \cite{jpmc_cr} has $56/2=28$ two-qubit gates per layer and 10 layers, resulting in a total of 280 two-qubit gates per circuit, which is very similar to the 276 two-qubit gates in this work. Furthermore, the circuits have similar computational hardness as shown in Supplementary Information Table~\ref{tab:prior_work}.

\subsection{Certified randomness amplification protocol}

We generate all the randomness used by the protocol beforehand. We use RDSEED~\cite{intel-rdseed} as a weak entropy source. We first generate $3\times616\times100{,}000=184{,}800{,}000$ bits of randomness. This is enough for 100,000 circuits, even though we only use $\TotalSamples$ of them. Because each circuit contains $616$ single-qubit gates, each selected from eight possibilities, three random bits are needed per gate. Using weak randomness for challenge generation is secure in our model, as is shown in Supplementary Information Section \ref{sec:app-pseudorandom-unitary-w-imperfect-rand}.

Additional randomness is required for selecting the validation set and for choosing the measurement bases. 
Because protocol security depends only on the adversary’s inability to predict the measurement bases and validation set, it is sufficient that this randomness be computationally indistinguishable from uniform.
For more details on how computational indistinguishability enters the proof of soundness, see Supplementary Information Section \ref{sec:amplify_private}.

{To amplify weak sources with entropy rate $\alpha>0.5$, we generate $2\times74{,}207{,}281$ random bits from RDSEED (denoted $W_1$), discard an equal-sized block to mitigate short-range correlations, obtain a further $74{,}207{,}281$ bits ($W_2$). We assume that $W_1$ and $W_2$ are computationally indistinguishable from independent. 
That is, no efficient computation can distinguish whether $W_1$ and $W_2$ originate from one weak source or from two independent sources.
This allows us to pass them through a two-source extractor to produce an extracted output $B$ of length $7{,}400{,}000$ that is computationally indistinguishable from uniformly random. By this, we mean that given $B$, the adversary cannot run some computation to efficiently determine whether it came from the uniform distribution or not.} One can also stretch the randomness using a cryptographically secure pseudo-random number generator. For the security analysis of the generation of computationally indistinguishable from uniformly random bits using two-source extractors and pseudo-random number generators, see Supplementary Information Section \ref{sec:computationally_indistinguishable_from_uniform}.

For sources with entropy rate $\alpha\leq0.5$, we instead draw a random  $B$ of length $7{,}400{,}000$ directly from the weak source.
Two-source extraction step is omitted because it requires at least one input with an entropy rate exceeding $0.5$. The protocol therefore assumes that the weak source is computationally indistinguishable from uniform.

In the prior demonstration \cite{jpmc_cr}, the client sends a batch of circuits each round, and checks for hardware failures during execution, such as ion loss, are performed at the end of the batch. If the batch fails the check, the samples are discarded from the protocol and do not penalize the server. In the our protocol, the client only sends the information for a single circuit per round and waits for the corresponding bitstring to be returned. If a check fails, the device is recalibrated or lost ions are reloaded; no sample post-selection is performed on the basis of these checks. In fact, there is no post-selection anywhere in the protocol: if the server fails to respond within the prescribed time window, the worst-case penalty is applied.

The collected samples are then validated using a supercomputer. For the adversarial model in \cite{jpmc_cr}, the analysis considers a protocol where $L_{\rm val}$ samples out of a total of $L$ samples are randomly chosen for validation. The XEB score is 
computed as
\begin{align}
\mathrm{XEB}=\frac{2^n}{L_{\rm val}}\sum_{i\in\mathcal{V}}P_{U_i}(x_i)-1\label{eqn:xeb}.
\end{align}
The contribution for each sample $P_{U_i}(x_i)$ is computed using exact tensor network contraction.

For the oracle access adaptive adversary model, for each sample, we generate a random Bernoulli variable $T_i$ with $\Pr[T_i=1]=\gamma$ and $\Pr[T_i=0]=1-\gamma$ for some $\gamma\in(0,1]$ chosen based on the validation budget available, and include the sample in the validation set if $T_i=1$. 
We adopt this random selection of samples to validate because the entropy accumulation theorem for multi-round analysis assumes precisely this validation structure.
Entropy accumulation is employed because it accommodates adversaries that adaptively change their behavior based on  the history of previous rounds.
The decision to validate a given sample is independent of all other samples. Consequently, the total number of validated samples is a binomial random variable, unlike in the previous protocol in which $L_{\rm val}$ was fixed a priori. In our experiment, we set $\gamma=0.59$.

Additionally, the score $s$ we use differs slightly from the XEB score in Eq.~\ref{eqn:xeb}. Specifically, each sample contributes to $s$ up to a truncated maximum value $p_{\rm max}$, i.e.,
\begin{align}
s=\frac{2^n}{L_{\rm val}}\sum_{i\in\mathcal{V}}\min(p_{\rm max, }P_{U_i}(x_i))-1.
\end{align}
This modification is again motivated by the entropy accumulation theorem, which requires the per-round test score to be bounded.

In our analysis we set $p_{\rm max}=2/2^n$. Although a large fraction of probability mass falls above $p_{\rm max}$, the retained information is sufficient to estimate the fidelity. Because $P_{U_i}(x_i)$ is close to the distribution produced by finite-fidelity sampling, the effect of the truncation can essentially be reversed (Supplementary Information Section \ref{sec:improve_variance}). Additionally, choosing a low value of $p_{\rm max}$ improves the entropy, as explained in Methods Section \ref{sec:methods_oracle_model_entropy}.

After the quantum randomness passes the test, we generate $2\times43{,}112{,}609$ random bits from RDSEED. These bits are combined with the quantum randomness by a two-source extractor to produce a uniformly random seed. Thereafter, any RDSEED output can be extracted with a seeded extractor using the seed.

\subsection{Entropy bound in the restricted model}

For a protocol to be secure, it must either output high entropy bitstrings or abort (denoted with event $\Omega^c$ as the complement of the success event $\Omega$) with probability close to 1, i.e. $\Pr[\Omega^c]=1-\varepsilon_{\rm accept}$ for some small $\varepsilon_{\rm accept}$. In particular, it is common to bound the smooth min-entropy of successful protocol execution. Min-entropy $H_{\rm min}$ is the negative logarithm of the highest probability guess, which quantifies the worst-case risk in the presence of a guessing adversary. Smooth min-entropy $H_{\rm min}^{\varepsilon_{\rm smooth}}$ with smoothing parameter $\varepsilon_{\rm smooth}$ is closely related to $H_{\rm min}$. Specifically, it is the highest min-entropy possible when we are allowed to optimize over $\varepsilon_{\rm smooth}$-perturbations of the original distribution. For formal definitions, see Supplementary Information Section \ref{sec:preliminaries}.

The restricted security model in \cite{jpmc_cr} considers the following adversary. For each sample, it either performs honest sampling on a quantum computer or classical simulation using a supercomputer. If a total of $Q$ samples are quantum samples, then the probability for the protocol to have an XEB score of at least $\chi$ is computable as a function $\varepsilon_{\rm adv}(Q,\chi)$. If we are willing to tolerate a malicious adversary passing the XEB test with probability at most $\varepsilon_{\rm accept}$, then we can safely assume that the adversary generates at least $Q_{\rm min}$ quantum samples to pass the test with greater probability. Specifically, we have
\begin{align}
Q_{\rm min}=\rm{arg}\:\min_{Q}\{\varepsilon_{\rm adv}(Q,\chi)\geq\varepsilon_{\rm accept}\}.
\end{align}
With this, the smooth min-entropy is at least
\begin{align}
H_{\min}^{\varepsilon_{\rm smooth}}\geq Q_{\rm min}(n-1) + \log \varepsilon_{\rm smooth}.
\end{align}

We use Theorem 1 in Methods of Ref.~\cite{jpmc_cr} to estimate the smooth min-entropy of the protocol at any given $\varepsilon_{\rm accept}$ and $\varepsilon_{\rm smooth}$. Since $\varepsilon_{\rm adv}$ should depend on the simulation fidelity, we need to specify assumptions on the adversary's simulation power. Assuming the adversary uses the same classical tensor network contraction technique that we use for validation, the computational cost of simulating a circuit with fidelity $f$ would be $f$ times our validation cost for a single circuit.

To estimate the classical simulation fidelity of the adversary, we perform benchmarks of tensor network contraction on the Intel Data Center GPU Max Series (codename Ponte Vecchio or PVC) devices available on the Aurora supercomputer. Aurora currently has 10,624 nodes, where each node on Aurora has 6 GPUs and each GPU has two tiles \cite{auroraoverview}. Denote the time for a single GPU or two tiles to compute a single amplitude or $P_{U_i}(x_i)$ as $T_{\rm PVC}$ seconds. Assuming the adversary has $T_{\rm M}$ seconds to simulate the circuit and is $X$ times as powerful as Aurora, the classical simulation fidelity is given by 
\begin{align}
\Phi_{\rm C}=X\times10{,}624\times 6\times T_{\rm M}/T_{\rm PVC}.\label{eqn:classical_fidelity}
\end{align}

One can estimate the validation cost required to validate a smaller number of circuits than in our experiment. This allows us to study the achievable security for various hypothetical adversaries and experimental parameters such as latency and fidelity. Supplementary Information Fig. \ref{fig:restricted_adversary_supplement} plots the results for adversaries with varying computational resources.

\subsection{Entropy bound in the oracle model}\label{sec:methods_oracle_model_entropy}
For the security analysis using the oracle model, we determine the classical simulation fidelity of the adversary using the same method as Eq.~\ref{eqn:classical_fidelity}. To quantitatively estimate the entropy, we use Supplementary Information Theorem \ref{thm:final_oracle_model_theorem}, which is the special case of one query access and no batching. We reproduce it below for an informal summary.

\begin{theorem}[Informal]\label{thm:query_adversary_security}
For an adversary with a quantum device with fidelity $\phi_{\rm adv}$ and a classical supercomputer whose output probability distribution is at most $d_{\rm C}$ from sampling a fidelity $\Phi_{\rm C}$ quantum computer, the $L$-round protocol using $n$-qubit random quantum circuits with a truncated XEB threshold $s^*$ either aborts with probability at least $1-\varepsilon_{\rm accept}$ or the output has conditional smooth min-entropy
\begin{equation}
H_{\rm min}^{\varepsilon_{\rm smooth}}\geq Lh(s^*)-\sqrt{L}c-c',
\end{equation}
where $c$ and $c'$ are some constants depending on the protocol,
\begin{equation}
h(s)=n\left[\left(t(s)-
    \Phi_{\rm C}\right)\cdot\left(1-\frac{(k+\ell)^2}{N}\right)-\phi_{\rm adv}C\right]-2,
\end{equation}
$t(s)$ is an affine function in $s$ and $C$ depends on $k$ and $\ell$ for any natural number $k,\ell\leq \sqrt{N}=2^{n/2}$.
\end{theorem}

One can see that $h(s^*)$ is an affine function in $s^*$, the XEB score threshold. Therefore, entropy accumulates linearly in the number of rounds $L$ with a square root penalty factor. Additionally, inspection of $h$ shows that the simulation fidelity $\Phi_{\rm C}$ is effectively a linear penalty on the XEB score as given by $t(s)-\Phi_{\rm C}$.

The entropy bound has the above form due to the application of the entropy accumulation theorem. It is commonly used to obtain multi-round entropy bounds for adaptive adversaries that could change their behavior depending on prior rounds \cite{dupuis2019entropy,dupuis2020entropy}. Such theorems state that the entropy of systems with suitable properties scales linearly in the number of rounds with a second order penalty term, which is precisely what we observe in the above theorem.

For protocols with testing, the constant in front of the linear contribution is the single-round entropy of system that passes the average testing outcome, and the second order constant depends on the verification rate $\gamma$ and the range of the possible test score in each round. In our case, the test score of each round is the contribution to the XEB score for each round, which is truncated to be in $[0,p_{\rm max}]$. In this case, the second order constant is $c\propto\sqrt{p_{\rm max}/\gamma}$. This is the reason why we instantiate our protocol with a low value of $p_{\rm max}$.

\subsection{Amplification protocol soundness}

Denote the output of the protocol as a bitstring on register $R$. If the protocol aborts, the state of the register is denoted as $\perp$. If $\Omega$ denotes the event that the protocol does not abort (and its complement is $\Omega^c$), then the quantum state should ideally be
\begin{align}
    \rho^{\rm ideal}=\tau_R\otimes\rho_{\wedge\Omega}+\vert\bot\rangle\langle\bot\vert_R\otimes\rho_{\wedge\Omega^c},
    \label{eq:ideal-state-security}
\end{align}
where $\tau_R$ is the maximally mixed state and $\rho_{\land\Omega}$
is the subnormalized state of the side information conditioned on the event $\Omega$. A protocol is $\varepsilon_{\rm sou}$-sound if for all input initial states, the output state $\rho$ satisfies
\begin{align}
    \|\rho-\rho^{\rm ideal}\|_1\leq\varepsilon_{\rm sou}.
\end{align}

We aim to show that our randomness amplification protocol is $\varepsilon_{\rm sou}$-sound. In the amplification protocol, we generate bits $B$ that are computationally indistinguishable from uniform up to a (very small) maximum distinguishing advantage $\varepsilon_B$, which are used as part of the input randomness for the certified randomness protocol. This is sufficient since the adversary is computationally bounded. As long as the assumption of $B$ holds during protocol execution, the generated randomness is genuinely random and the protocol output will remain secure even if the assumptions on the computational indistinguishability of $B$ no longer holds afterwards. This property is called everlasting security.

The certified randomness protocol itself can choose the level of security by making malicious adversaries abort with high probability $1-\varepsilon_{\rm accept}$ and making the smoothing parameter $\varepsilon_{\rm smooth}$ of the output smooth min-entropy small. We can choose the values of $\varepsilon_{\rm accept}$ and $\varepsilon_{\rm smooth}$ based on our desired security level and quantify the amount of entropy available.

After the certified randomness protocol generates raw quantum bits, the protocol uses a two-source extractor to extract a nearly perfectly random seed from the quantum randomness for later use. Typically, one considers two-source extractors as taking two independent min-entropy sources and extracting nearly perfect randomness. The extractor is parameterized by an extractor error $\varepsilon_{\rm ts}$, which bounds the distance of the output from uniform randomness when extracting two independent min-entropy sources. In our setting, however, one of the input randomness sources provides quantum randomness with bounded smooth min-entropy. This deviation from the two min-entropy source setting introduces an additional parameter $\varepsilon_2$, which contributes an extra additive term in the distance of the output from uniform randomness. Overall, if the two parameters are chosen to be very small for a higher level of security, the number of extractable bits decreases.%

Finally, a seeded extractor parameterized by extractor error $\varepsilon_{\rm seeded}$ is used to extract near perfect randomness from the weak source seeded by the aforementioned seed. This is repeated $M$ times, with the soundness guarantee degrading linearly in $M$.

The aforementioned security parameters $\varepsilon_{\rm accept},\varepsilon_{\rm smooth},\varepsilon_{\rm ts},\varepsilon_{2}$ and $M$ are chosen to be related to a single parameter $\varepsilon_{\rm sou}$. Specifically, we make the explicit choice $\varepsilon_2=\varepsilon_{\rm ts}=10^{-8},\varepsilon_{\rm seeded}=10^{-16},M=10^8$ and
\begin{align}
\varepsilon_{\rm accept}&=\varepsilon_{\rm sou}\\
\varepsilon_{\rm smooth}&=\frac{\varepsilon_{\rm sou}-2\varepsilon_2-2\varepsilon_{\rm ts}-M\varepsilon_{\rm seeded}}{6}.
\end{align}
With these choices, we prove that the randomness amplification protocol is $\varepsilon_{\rm sou}$-sound in Supplementary Information Section \ref{sec:amplify_private}.

\section*{Acknowledgments}
We thank Scott Aaronson, Harry Buhrman, Garnet Chan, Eli Chertkov, Davide DelVento, Srinivas Eswar, Johnnie C. Gray, Reza Haghshenas, Jeffrey M. Larson, and Valerii Dmytryk for the helpful discussions. We thank Anthony Armenakas and Soorya Rethinasamy for helpful feedback on the manuscript. We are grateful to Jamie Dimon, Lori Beer, and Scot Baldry for their executive support of JPMorganChase’s
Global Technology Applied Research Center and our work in quantum computing. We thank the technical staff at JPMorganChase’s Global Technology Applied Research for their support. 
T.H.~was supported by the U.S. Department of Energy, Office of Science, Advanced Scientific Computing Research program office under the quantum computing user program.
An award of computer time was provided by the INCITE program. This research used resources of the Argonne Leadership Computing Facility, which is a DOE Office of Science User Facility supported under Contract DE-AC02-06CH11357.

\section*{Author Contributions}
M.L. and R.S. led the overall project. M.D., M.F.-F., M.L., P.N., R.S., D.S. designed the random measurement bases protocol. M.L. derived the security guarantees of the certified randomness protocol, performed numerical analysis, implemented verification, designed the centralized verification protocol, analyzed position verification, and analyzed the frugal rejection sampling and top sampling adversaries. M.D., M.F.-F., C.S. investigated the classical simulation implications of the random measurement bases protocol. A.C. developed the gate-streaming concept. K.S., L.N., A.C. developed the software required for gate-streaming on the Quantinuum Helios trapped-ion processor. P.N. implemented the gate-streaming quantum program for executing the certified randomness protocol. M.D., M.F.-F., M.S.A, J.P.C. III, A.I. optimized the performance of the gate-streaming quantum program on the trapped-ion hardware. F.C., C.For., K.M. designed the randomness amplification protocol and proposed the use of computational sources, derived and implemented the randomness extraction and considered the constraints on the geographical location of the server. K.C., W.-Y.K., I.-W.P. revised the security guarantees of certified randomness. F.K. derived the message complexity of quantum information supremacy for random $Z$, $X$ basis. O.A., Sh.C. formalized jointly certifiable randomness expansion. Z.H., Y.J., M.L., S.O., M.P., M.S., R.S. performed some related exploratory work. F.K., revised the position verification analysis. S.H. derived the improvement in Supplementary Information Section \ref{sec:improve_quantum}. Sam.C., R.D., J.M.D., B.E., C.Fig., C.Fol., J.P.G., A.H., C.A.H., A.A.H, C.J.K, N.K., N.K.L., I.S.M., M.M., A.R.M., A.J.P., A.P.R., A.R. maintained, optimized, and/or operated the trapped-ion hardware. All authors contributed to technical discussions and the writing and editing of the manuscript and the supplementary information.

\section*{Additional Information}
Supplementary Information is available for this paper.

\section*{Data Availability}

The full data presented in this work is available at \url{https://doi.org/10.5281/zenodo.17527293}.

\section*{Code Availability}

The code required to verify and reproduce the results presented in this work is available at \url{https://github.com/jpmorganchase/global-technology-applied-research/tree/main/certified-randomness-amplification}.

\section*{Disclaimer}
This paper was prepared for informational purposes with contributions from the Global Technology Applied Research center of JPMorgan Chase \& Co. This paper is not a product of the Research Department of JPMorgan Chase \& Co. or its affiliates. Neither JPMorgan Chase \& Co. nor any of its affiliates makes any explicit or implied representation or warranty and none of them accept any liability in connection with this paper, including, without limitation, with respect to the completeness, accuracy, or reliability of the information contained herein and the potential legal, compliance, tax, or accounting effects thereof. This document is not intended as investment research or investment advice, or as a recommendation, offer, or solicitation for the purchase or sale of any security, financial instrument, financial product or service, or to be used in any way for evaluating the merits of participating in any transaction.  This manuscript has been authored in part by UT-Battelle, LLC, under contract DE-AC05-00OR22725 with the US Department of Energy (DOE). The US government retains and the publisher, by accepting the article for publication, acknowledges that the US government retains a nonexclusive, paid-up, irrevocable, worldwide license to publish or reproduce the published form of this
manuscript, or allow others to do so, for US government purposes. DOE will provide public access to these results of federally sponsored research in accordance with the DOE Public Access Plan (\url{http://energy.gov/downloads/doe-public-access-plan}).

\end{bibunit}
\clearpage
\newpage
\starttocentries

\begin{bibunit}[apsrev4-2]
\title{Supplementary Information for:\\ Certified randomness amplification by dynamically probing remote random quantum states}

\maketitle

\onecolumngrid

\tableofcontents
\newpage

\section{Summary of contributions}

We first provide numerical evidence and heuristic arguments why the latency $T_{\rm M}$ between sending the measurement basis and receiving the bitstring should be the available simulation time for a classical adversary. Specifically, we make connections with quantum information supremacy (Section \ref{sec:information_supremacy}) and also examine a variant of the tensor network contraction algorithm that performs as much computation as it can until it needs the measurement bases. We show that the adversary cannot perform a meaningful amount of computation before the bases become available (Section \ref{sec:tn_post_cost}).

For the security analysis, we provide security proofs in the single round for adversaries with oracle access to the challenge quantum circuits. Specifically, we generalize to the setting where a batch of circuits are sent at a time and show that the entropy scales linearly in the number of qubits (Section \ref{sec:app-single-round}) and the number of challenge circuits per round. To achieve better entropy guarantees, we provide several improvements to the original theory which are valid in different settings.

Notably, we show that one oracle access, i.e., calling the challenge circuit once in a potentially large quantum circuit, can be simulated by obtaining a sample from the challenge circuit and postprocessing it (Section \ref{sec:one_query_one_sample}). This hints at the interesting possibility that honest sampling may be the only relevant quantum algorithm for random circuit sampling.

We then discuss in Section \ref{sec:combine_classical}
how this kind of adversary can combine oracle access quantum algorithms and state-of-the-art classical simulations. This is important since the experiment operates on a finite-size regime, so taking into account the effect of large supercomputer simulation is crucial.

We then provide security analysis for a multi-round protocol against non-adaptive oracle access adversaries and use detailed statistical properties of the output of the adversary (Section \ref{sec:ad_hoc}). Intuitively, adaptive attacks cannot lead to meaningful gains especially when the fraction of verification circuits is small. We use detailed statistical properties of the output to obtain a good entropy bound.

We also analyze a classical-quantum hybrid oracle access adversary that can perform adaptive attacks (Section \ref{sec:app-multi-round}). We use the entropy accumulation theorem (EAT) to bound the entropy, making it robust against adaptive attacks. Specifically, the detailed statistical properties derived in the non-adaptive adversary analysis are used to significantly improve the entropy bound of EAT (Section \ref{sec:improve_variance}). We believe it would be interesting for future work to minimize the gap between the adaptive and non-adaptive adversary analyses.

We then formally prove the security of randomness amplification using our protocol, where the client has private randomness that is not uniformly random but wishes to obtain uniformly random bits through interacting with a quantum device (Section \ref{sec:amplification}). Previous results for generating certified randomness only consider randomness expansion, where the input randomness is uniformly random. In our scheme, imperfectly random bits are used to select challenge circuits, and we show that this still satisfies the requirement of RCS-based certified randomness (Section \ref{sec:app-pseudorandom-unitary-w-imperfect-rand}). We then show how to generate random validation sets that are computationally indistinguishable from random by using imperfect randomness that satisfy some reasonable assumptions (Section \ref{sec:computationally_indistinguishable_from_uniform}).

We then discuss the possibility of jointly certifiable protocols for randomness expansion and amplification and provide formal definitions and constructions (Section \ref{sec:joint}). However, each client must pay a high validation cost, which makes the practical adoption of joint protocols challenging. Another challenge in the protocol is the trust in the verifier. Clients may not have access to a trusted supercomputer for validation, and designated computation that is verifiable by a low-budget client is desirable. In the joint setting, multiple clients can now pool validation resources. We discuss the approach of verifying a random subset of slices of the verification results (sub-tasks of verification that can be computed independently) from an untrusted supercomputer and discuss security guarantees (Section \ref{sec:low-budget_client_verification}).

We characterize leading classical simulation techniques and show that they can be effectively analyzed and made compatible with our security analysis. In particular, we analyze the conventional approach of performing frugal rejection sampling (Section \ref{sec:frugal_rejection_sampling}), as well as the more recent postprocessing approach of choosing the bitstring with the maximum partially computed amplitude (Section \ref{sec:oversampling_classical_adversary}). We show that in suitable regimes, the statistical properties of these adversaries are comparable to honest finite fidelity sampling.

We finally analyze the affordability of certified randomness for various experimental parameters such as latency, fidelity, and quantum parallelization under the restricted adversary model (Section \ref{sec:cost_analysis}).

\section{Delayed measurement basis}\label{sec:basis}

The certification of randomness in the protocol implemented here is ultimately predicated on the inability of a classical adversary to faithfully sample the output of the quantum circuits submitted to the Helios quantum computer within the time ($T_{\rm M}$) between when measurement bases are specified by the client and when results are returned to the client.  Notably, parts of the circuit are revealed \emph{prior} to the specification of the measurement bases, and we should assume that some amount of precomputation may already have been performed (based on the revealed structure of the circuit) prior to learning the measurement bases, i.e. outside of the latency window $T_{\rm M}$. Therefore, we must assess how much work is involved in sampling from a circuit $C$ classically assuming some (potentially quite large) work has been done in the absence of knowledge about the measurement bases.

\subsection{Connection with quantum information supremacy}\label{sec:information_supremacy}

\begin{figure}[!t]
    \centering
    \includegraphics[width=1\columnwidth]{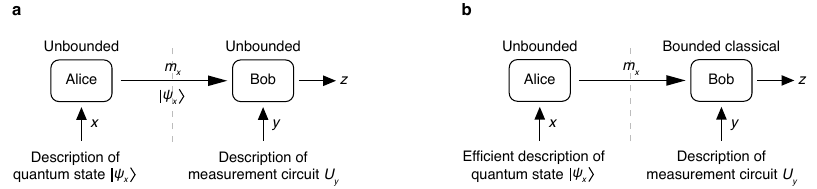}
    \caption{\textbf{Comparison between the information theoretic setting and the computational setting}. \textbf{a}, The setting of quantum information supremacy. The classical description $x$ is exponentially long since it describes a Haar random state. The message from Alice to Bob can be quantum or classical. \textbf{b}, The computational setting that models our delayed measurement basis classical adversary. Unlike \textbf{a}, the classical description $x$ and Bob are polynomially bounded.}
    \label{fig:information_supremacy}
\end{figure}

This is closely related to the work on \textit{quantum information supremacy}, where an unconditional separation between quantum and classical information resources is established. Specifically, there is a computational task that can be solved efficiently with a quantum device, but a classical device would require exponential memory \cite{kretschmer2025demonstrating}. The particular task considered is the Distributed Linear Cross-Entropy Heavy Output Generation (DXHOG) task that involves two parties Alice and Bob that are both unbounded. Alice is given the exponentially long classical description $x$ of a Haar random quantum state $\vert\psi_x\rangle=U\vert0\rangle$. Alice then performs some computation and sends a message, quantum or classical, to Bob. Bob is then provided a random measurement basis in the form of a circuit $U_y$ described by the classical string $y$ and asked to generate a bitstring $z$ that achieves a high value $\vert \langle z\vert U_y\vert \psi_x\rangle\vert^2=\vert\langle z\vert U_y U\vert 0\rangle\vert^2$. If Alice is allowed to send a quantum message, then Alice can simply send the quantum state itself and Bob can apply $U_y$ and measure it in the computational basis. On the other hand, if Alice is only allowed to send a classical message, then the length of the message must be exponential for Bob to do well on the DXHOG task. We call this the information theoretic setting and illustrate it in Fig. \ref{fig:information_supremacy}\textbf{a}.

The information theoretic setting is not completely the same as our setting of a classical adversary trying to spoof certified randomness. In the most general setting, we assume that precomputation is performed by an unbounded classical adversary Alice, which is the same as before, but Alice is now given an efficient classical representation of the circuit $U$. To model postprocessing after receiving the measurement basis, we model it with a polynomial time classical adversary Bob. Alice is allowed to share a polynomial-sized classical message with Bob, which models the result of precomputation. Alice and Bob aim to perform well on the DXHOG task like before. We illustrate this model in Fig. \ref{fig:information_supremacy}\textbf{b}, which we call the computational setting.

As discussed in the main text, the polynomial-size limitation on the classical message or memory is natural. This is because leading simulation algorithms such as tensor network contraction \cite{pan2022solving,huang2020classical} have polynomial space and exponential time. It is possible to run a simulation algorithm longer, but it is not possible to exponentially increase the RAM of a supercomputer, whose size is usually fixed ahead of time. This setting is exactly that of performing exact tensor network contraction. The memory complexity of tensor network contraction increases exponentially, but this can be mitigated by `slicing' indices of the tensor network. Slicing breaks the computational task into independent parts. In particular, slicing each index doubles the number of tasks, and the memory footprint reduces by at most a factor of two, and one can slice until the memory requirement for each task is satisfied by the available RAM. Other simulation techniques must similarly satisfy the fixed RAM requirement, hence the polynomial-size restriction on the classical message or memory is reasonable.

This result in the information theoretic setting confirms the intuition that there is no efficient classical representation of a generic quantum state that it is sufficient for the XHOG task. This provides some confidence to the conjecture that polynomial-sized messages do not allow Bob to do well on DXHOG in the computational setting as well in the sense that an approximate `state' does not exist. If Alice computes the full statevector and treats the observed statevector as if it is Haar random, Alice cannot find a polynomial message that allows Bob to do well on DXHOG.

Our protocol uses random $Z$ or $X$ basis single-qubit measurements, which is not a setting that was analyzed in Ref.~\cite{kretschmer2025demonstrating}. We provide an extended information supremacy result in this setting. Ref.~\cite{kretschmer2025demonstrating} gives a communication lower-bound for a classical protocol that passes $\varepsilon$-DXHOG when the measurement basis $U_y$ consists of random single-qubit Clifford gates. We observe that their proof also works for the Clifford subgroup of single-qubit gates that equal to either identity ($I$) or Hadamard ($H$), corresponding to single-qubit measurements in the $Z$ or $X$ basis.

To give a rigorous proof, we start by defining some useful notation. Let $U(2^n)$ be the set of unitaries over $n$ qubits that Bob receives. For a distribution $\mathcal{U}$ over $U(2^n)$ and a function $g : U(2^n) \to \{0,1\}^n$, define \begin{align*}
    M(g;\mathcal{U}) := \E_{U \sim \mathcal{U}} \left[ U^\dagger \ket{g(U)} \bra{g(U)} U \right].
\end{align*}

We list the main result and some lemmas from Ref.~\cite{kretschmer2025demonstrating} below. We say a protocol for $\varepsilon$-$\mathrm{DXHOG}$ is with respect to $\mathcal{S}$ and $\mathcal{U}$ if Alice's state is $\vert\psi_x\rangle\sim\mathcal{S}$ and Bob's unitary is $U_y\sim\mathcal{U}$.

\begin{theorem}[Theorem 3 of Ref. \cite{kretschmer2025demonstrating}] \label{thm:information_supremacy}
    Let $\mathcal{S}$ be the Haar measure over $n$-qubit states. Let $\mathcal{U}$ be a distribution over $U(2^n)$ such that for every $g : U(2^n) \to \{0,1\}^n$, \begin{align*}
        \|M(g;\mathcal{U})\|_{\mathrm{F}} \le A \quad \text{and} \quad \|M(g;\mathcal{U})\|_{\mathrm{op}} \le B.
    \end{align*}
    Then, any $m$-bit classical protocol for $\varepsilon$-$\mathrm{DXHOG}$ with respect to $\mathcal{S}$ and $\mathcal{U}$ satisfies \begin{align*}
        \varepsilon \le O\left(\max\{\sqrt{m}A, mB\}\right).
    \end{align*}
\end{theorem}

\begin{lemma}[Lemma 6 from Ref. \cite{kretschmer2025demonstrating}] \label{lem:X}
    Let $\mathcal{U}$ be a distribution over $U(2^n)$ and let $g : U(2^n) \to \{0,1\}^n$. Then, \begin{align*}
        \|M(g;\mathcal{U})\|_{\mathrm{F}}^2 \le \E_{U, U' \sim \mathcal{U}} \left[\max_{z_1, z_2 \in \{0,1\}^n} \abs{\braket{z_1}{U(U')^\dagger \big| z_2}}^2 \right].
    \end{align*}
\end{lemma}

\begin{lemma}[Adapted from Lemma 8 of Ref. \cite{kretschmer2025demonstrating}] \label{lem:norm_bound}
    Let $\mathcal{U}$ be the uniform distribution over $\left\{\bigotimes_{i=1}^n U_i : U_i \in \{I,H\} \text{ for } i \in [n]\right\}$ and $g : U(2^n) \to \{0,1\}^n$. Then, \begin{align*}
        \|M(g;\mathcal{U})\|_{\mathrm{F}} \le \left(\frac{3}{4}\right)^{n/2}.
    \end{align*}
\end{lemma}

\begin{proof}
    Let $U = \bigotimes_{i=1}^n U_i$. Then, \begin{align}
        \|M(g;\mathcal{U})\|_{\mathrm{F}}^2 &= \E_{U, U' \sim \mathcal{U}} \left[\max_{z_1, z_2 \in \{0,1\}^n} \abs{\braket{z_1}{U(U')^\dagger \big| z_2}}^2 \right] \label{eq:F1} \\
        &= \E_{U \sim \mathcal{U}} \left[\max_{z_1, z_2 \in \{0,1\}^n} \abs{\braket{z_1}{U \big| z_2}}^2 \right] \label{eq:F2} \\
        &= \E_{U \sim \mathcal{U}} \left[\max_{z \in \{0,1\}^n} \abs{\braket{z}{U \big| 0^n}}^2 \right] \label{eq:F3} \\
        &= \E_{U \sim \mathcal{U}} \left[\prod_{i=1}^n \max_{z_i \in \{0,1\}} \abs{\braket{z_i}{U_i \big| 0}}^2 \right] \label{eq:F4} \\
        &= \prod_{i=1}^n \E_{U \sim \mathcal{U}} \left[\max_{z_i \in \{0,1\}} \abs{\braket{z_i}{U_i \big| 0}}^2 \right] \label{eq:F5} \\
        &= \prod_{i=1}^n \frac{1}{2}\left( \abs{\braket{0}{0}}^2 + \abs{\braket{0}{+}}^2 \right) \label{eq:F6} \\
        &= \left(\frac{3}{4}\right)^n        .
    \end{align}
    Above, in Eq. (\ref{eq:F1}) we use Lemma \ref{lem:X}. In Eq. (\ref{eq:F2}) we use the group structure of $\{H,I\}$. In Eq. (\ref{eq:F3}) we use the fact that all non-zero entries of a Clifford unitary have the same magnitude. In Eq. (\ref{eq:F4}) we use the tensor product structure of $U$. In Eq. (\ref{eq:F5}) we use independence of terms in the product. The rest of the calculation is straightforward.
\end{proof}

Now we are ready to state the desired result:

\begin{corollary}
    Let $\mathcal{S}$ be the Haar measure over $n$-qubit states. Let $\mathcal{U}$ be the uniform distribution over tensor product of single-qubit identity/Hadamard gates. Then, any $m$-bit classical protocol for $\varepsilon$-DXHOG with respect to $\mathcal{S}$ and $\mathcal{U}$ satisfies \begin{align*}
        m \ge \Omega\left(\varepsilon \left(\frac{4}{3}\right)^{n/2}\right).
    \end{align*}
\end{corollary}
\begin{proof}
    By Lemma \ref{lem:norm_bound}, for any $g : U(2^n) \to \{0,1\}^n$ we have \begin{align*}
        \|M(g;\mathcal{U})\|_{\mathrm{op}} \le \|M(g;\mathcal{U})\|_{\mathrm{F}} \le \left(\frac{3}{4}\right)^{n/2}.
    \end{align*}
    Thus, the corollary follows from Theorem \ref{thm:information_supremacy}.
\end{proof}

However, in the computational setting, Alice does know the efficient description of the circuit $U$ that is capable of producing $\vert\psi_x\rangle=U\vert0\rangle$ and that the state is not Haar random. Due to this difference, one cannot simply use the information theoretic result to rule out the existence of a polynomial classical message that allows Bob to do well in the computational setting. If Alice has access to $U$, then the impossibility in the information theoretic setting simply does not hold. In fact, Alice can simply send $U$ to Bob, and Bob can execute an exponential time algorithm to do well on DXHOG.

Nevertheless, sending $U$ is not a viable strategy in the computational setting either since Bob is polynomial. On the one hand, if Alice is completely agnostic to the fact that the state can be prepared by an efficient circuit $U$, doing well on DXHOG is plausibly hard by the information theoretic setting results. On the other hand, if Alice only cares about using $U$, Bob must pay the full cost of classically simulating RCS. It is an open question whether other strategies that exploit the efficiency of $U$ and performs well on DXHOG exist.

\subsection{Analysis of tensor networks}\label{sec:tn_post_cost}

We analyze tensor network algorithms that try to perform as much precomputation as possible, which is an example of an algorithm that performs precomputation and exploits the structure of the quantum circuit $U$. We break the circuit up as $U=M\circ G$, where $G$ consists of all gates learned outside the latency window $T_{\rm M}$ and $M$ consists of the final layer of gates determining the measurement bases, and we assume that all of $G$ is known for a time $T_{\rm pre}$. A conservative approach would be to take $T_{\rm pre}=T_1$, though in reality much of the circuit is not learned this much in advance of the results being returned. We would like to ensure that even when $T_{\rm pre}\gg T_{\rm M}$, there is no feasible pre-computation that could be preformed during time $T_{\rm pre}$ that would substantially reduce the effort needed (within the time $T_{\rm M}$) to produce high-quality samples once the measurement bases are known.

One conceptually straightforward way to perform precomputation based on knowledge of $G$ is to directly produce the state vector $G\ket{0}$. Even given a large amount of time, such a strategy would be infeasible for several reasons, including that (1) storing a $64$-qubit state vector would be enormously challenging, (2) even if one could generate and store it, drawing samples from that statevector in the specified measurement basis will also be exponentially hard in the qubit number and must be performed in the time window $T_{\rm M}$. 

\begin{figure}[!t]
    \centering
    \includegraphics[width=0.8\linewidth]{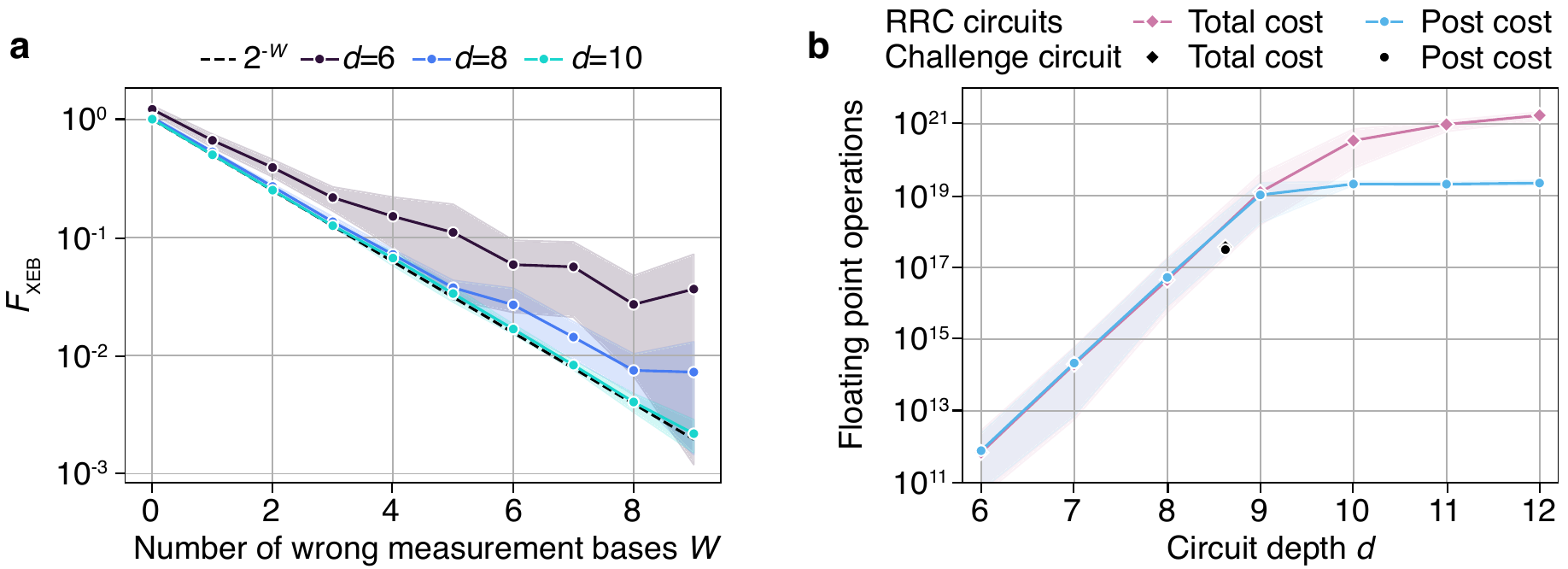}
    \caption{\textbf{Numerical evidence that no meaningful precomputation is possible}. \textbf{a}, XEB fidelity when sampling by contracting tensor networks corresponding to wrong measurement bases. \textbf{b}, Memory unconstrained post cost and total cost for tensor-network contraction of random-geometry circuits as a function of depth. Blue and pink plot markers indicate the mean over 10 random circuit geometries, with the associated shaded regions indicating the min and max cost obtained over those 10 circuits at each depth. Black diamond and circle indicate the optimized total cost and post cost of the challenge circuits used in this paper, respectively.}
    \label{fig:pt_results}
\end{figure}

In general, it is far more feasible to draw samples from the output of $U$ by tensor-network contraction than by producing (and sampling from) the output statevector. To produce a sample from a random quantum circuit by tensor-network contraction, one first computes the output probabilities of multiple random (though potentially correlated) bitstrings in the \emph{full} circuit $U$, and then selects amongst them according to their relative probabilities.
While this strategy cannot be directly employed absent knowledge of $M$, one could imagine pre-computing a valid sample for each of many possible measurement bases, and then deciding which sample to return once the measurement bases are learned (essentially generating an exponentially large look-up table of valid samples). The challenge circuits used in this work request measurement in either the computational ($Z$) or Hadamard ($X$) basis.  Given the binary choice of measurement basis, precomputation in the manner just described would require generating a sample for $2^N$ possible measurement bases, which is exponentially more costly than the (already quite costly) computational cost of computing a single bitstring amplitude (i.e. the per-circuit validation cost).  Alternatively, one could tabulate samples for the possible measurement bases of only a subset of qubits of size $M<N$, and simply guess the measurement bases of the remaining $N-M$ qubits, thereby reducing the multiplicative overhead of spoofing relative to verification from $2^N$ to $2^M$. However, one can show via simple analytical arguments, confirmed with numerical simulations in Fig.\,\ref{fig:pt_results}\textbf{a}, that the expected XEB score upon guessing $W$ of the measurement bases wrong decreases as $2^{-W}$. Given the high fidelity requirement for the generated samples, only a small number of measurement bases (likely just one or two) could be guessed while still passing the XEB test, implying no substantial reduction below the cost of tabulating samples across all possible measurement bases. We therefore view this precomputation strategy as deeply infeasible.

We believe the most viable precomputation method is to consider the output probability of an arbitrary bitstring as a tensor network in which only a subset of the tensors are known, and another subset (comprising the final layer of gates that determine the measurement bases) are unknown. We denote the set of known tensors as $\mathscr{T}_{\rm pre}$, and the set of unknown tensors as $\mathscr{T}_{\rm post}$, such that the complete set of tensors is $\mathscr{T}=\mathscr{T}_{\rm pre}\cup\mathscr{T}_{\rm post}$. The contraction of this tensor network can be partially carried out prior to learning the tensors in $\mathscr{T}_{\rm post}$ by iteratively merging any tensors in $\mathscr{T}_{\rm pre}$.  We care about the cost of this tensor network contraction that must be incurred during the time window $T_{\rm M}$, i.e.\ once the tensors in $\mathscr{T}_{\rm post}$ are known and become involved in the contraction, which we call the \emph{post-cost}. Therefore, we consider the problem of optimizing the contraction path of the complete tensor network while targeting only the cost of contractions that involve tensors in the set $\mathscr{T}_{\rm post}$.

Formally, the optimization problem is specified over the space of all possible contraction trees for the full tensor network.  A contraction tree is a rooted-binary tree in which the leaves are tensors in $\mathscr{T}$, the root is the (scalar) amplitude of a particular bitstring output $b$ ($a_b = \langle b\vert U\rangle$), and each node in the tree indicates the merger of two incoming tensors (children) into an output tensor (parent).

We attempt to minimize $\mathcal{C}_{\rm post}$ using an optimization algorithm based on the sub-tree reconfigurations described in \cite{Kalachev2021multi}. Instead of the simulated annealing employed in that work, we use parallel tempering \cite{katzgraber2011} with 56 replicas that are geometrically spaced in temperature. Each replica is seeded with a random balanced tree. We then attempt 3200 replica swaps, allowing each replica to evolve for 100 sweeps of reconfiguration attempts (from the root to leaves of the tree) between replica swaps. Fig. \ref{fig:pt_results}\textbf{b} shows both the optimized total cost of contracting $\mathscr{T}$ (red) and the optimized post-contraction cost (blue) over a range of circuit depths. We observe that there are two distinct regimes:
\begin{itemize}
\item Shallow circuits (depths $d\leq 9$). The optimitzed post-cost is nearly identical to the optimized total cost, implying that no appreciable savings from pre-contraction of tensors in $\mathscr{T}_{\rm pre}$ are possible. 
\item Deep circuits (depths $d\geq 10$). The contraction cost begins to saturate as the optimal contraction patterns look progressively more state-vector-like. In this regime, the post cost becomes lower than the total cost because much of the state vector evolution can be carried out without knowing the measurement bases.
\end{itemize}

The challenge circuits we use in this protocol have between eight and nine layers of gates (the ninth layer only has 20 two-qubit gates instead of $64/2=32$ two-qubit gates). Shown as black circle and diamond in Fig.\,\ref{fig:pt_results}\textbf{b}, they are within the shallow regime, where no substantial savings in post-cost are available. Note that while we call these circuits shallow, from the analysis in \cite{qntm_rcs} we expect them to be suitably deep for XEB to be well correlated with circuit fidelity.

\section{Single-round protocol security analysis}
\label{sec:app-single-round}

Work by Aaronson and Hung \cite{aaronson2023certified, aaronson2023certifiedArxiv} shows the existence of random circuit sampling (RCS) based protocols that results in $\Omega(n)$ bits of entropy and entropy accumulation over multiple rounds, where $n$ is the number of qubits of the challenge quantum circuits. Specifically, they show this for three types of adversaries: 1. A general polynomial time quantum device with no shared entanglement with the adversary, 2. A semi-honest adversary that must honestly execute the challenge circuits and measure the bitstring but share entanglement with the adversary, and 3. A general quantum device with shared entanglement with the adversary but only query access to the challenge circuits. The protocols discussed in \cite{aaronson2023certified, aaronson2023certifiedArxiv} require many samples to be returned by the quantum computer when the challenge circuit is received.

However, experimental demonstrations of certified randomness using RCS face unique challenges that make the idealized protocols infeasible \cite{jpmc_cr}. The first challenge is the near-constant time overhead (e.g. networking, compilation, etc.) that is significantly larger than the actual circuit execution time. This favors batched submission of circuits which amortizes this constant overhead. The second challenge is that obtaining many samples from the same circuit classically is much easier than obtaining one sample from one circuit for many different circuits. Since certified randomness requires that classical spoofing of the process is hard, this favors a protocol where many circuits are submitted and only one sample is expected for each circuit. Overall, this results in a protocol where $m$ circuits are submitted in each batch, only one sample is requested from each circuit, no circuits are reused, and this is repeated many times. The complexity theoretic arguments used by \cite{aaronson2023certifiedArxiv} for the first kind of adversary were corrected and adapted to show $\Omega(n)$ entropy per circuit for this batched multi-circuit protocol.

Despite the success of the complexity theoretic analysis, one major drawback is the seeming contradiction between the ability for a classical verifier to verify the quantum circuits and the inability for the adversary to classically simulate the circuits. Specifically, verification requires the computation of the bitstring probabilities $|\langle z\vert U\vert 0\rangle|^2$ for some samples, where $z$ is a measured bitstring and $U$ is the unitary corresponding to a challenge circuit. At the same time, current state-of-the-art methods for spoofing sampling classically aims to compute the same quantities approximately. One redeeming fact of the protocol is that only a small random fraction of quantum circuits need to be verified. If the fraction is sufficiently small, one can argue reasonably that verification is feasible but classical spoofing is not. However, this points to the critical flaw in the complexity theoretic analysis that simply assumes spoofing is impossible in the asymptotic limit. In currently feasible experiments, the fraction of verified circuits is not negligible, and the previous experimental demonstration verified $5\%$ of the total challenge circuits \cite{jpmc_cr}. Therefore, if verification can compute the probability amplitudes with perfect fidelity, classical spoofing must be able to simulate sampling with non-negligible fidelity. Although this fidelity is hopefully substantially smaller than the fidelity that one can obtain from honest quantum sampling, a theory that allows us to concretely understand the effect of finite-size circuits and adversary classical computational power on the entropy guarantees of the protocol is necessary.

As a result, \cite{jpmc_cr} considers a finite-size adversary that uses state-of-the-art classical simulation techniques and performs honest quantum sampling. Some samples are classically simulated and provide no entropy, while others are honestly executed on a quantum computer and provide almost $n$ bits of entropy. Since the probability distributions of $\vert\langle z\vert U\vert0\rangle\vert^2$ for $z$ obtained from classical spoofing and quantum sampling can be computed, one can analyze the security of the protocol and the entropy guarantees. Specifically, one can calculate the minimum number of quantum samples required for the adversary to pass the test with probability $\varepsilon$. Under this analytical framework, one can concretely understand the effect of execution time, fidelity, verification budget, adversary classical computational power, and many other factors.

A crucial drawback of the finite-size adversary analysis is that the action space of the adversary is fairly limited, particularly on the side of the quantum computer. As discussed earlier, the range of quantum adversaries considered by \cite{aaronson2023certifiedArxiv} is fairly general. It is, therefore, interesting to us to generalize the analysis of the classical-quantum mixture adversary to consider a more general class of quantum adversaries. However, we cannot readily apply the generalized complexity theoretic arguments for the polynomial time quantum adversary without entanglement with the adversary to this framework. Specifically, working in a finite-size regime, it is unclear whether we can ignore correction factors and only consider dominant terms in the security analysis, which reflects the paradoxical nature of combining finite-size and asymptotic analyses. Further, although $\Omega(n)$ entropy is guaranteed, there is an unknown proportionality constant factor that depends on the complexity theoretic assumption (i.e. the $B$ value of the Long List Hardness Assumption of Ref.~\cite{aaronson2023certified}, which is itself not proven for even Haar random circuits).

Fortunately, the analysis in \cite{aaronson2023certifiedArxiv} for the third type of adversary, which only has query access, is not just valid in the asymptotic limit. Although the results are presented in asymptotic notations, it turns out that all the constants can be propagated. This is indeed the analysis that we will build on top of.

We first summarize the high-level theoretical results in Section 7 of Ref.~\cite{aaronson2023certifiedArxiv} regarding this type of adversary. Informally, solving the so-called $\mathrm{XHOG}$ problem implies entropy. This calls for a definition of the $\mathrm{XHOG}$ problem. Below, we use $N=2^n$ throughout.
\begin{definition}[$b-\mathrm{XHOG}(\mathcal{D})$ \cite{kretschmer2021quantum}]
\label{def:xhog}
    A quantum algorithm $\mathcal{A}$ on input a quantum circuit $U\sim\mathcal{D}$ is said to solve $b$-$\mathrm{XHOG}$ if it outputs a bitstring $z$ such that
    \begin{equation}
    \underset{U\sim\mathcal{D}}{\mathbb{E}}\left[\underset{z\sim\mathcal{A}^{U}}{\mathbb{E}}\left[p_U(z)\right]\right]\geq\frac{b}{N},
    \end{equation}
    where $p_U(z)$ is the probability of sampling bitstring $z$ when executing circuit $U$ faithfully on an ideal quantum computer, given by $\vert\langle 0\vert U\vert z\rangle\vert^2$.
\end{definition}
Now, we restate the actual theorem of interest. The query model of the theorem is discussed in \cite{aaronson2023certifiedArxiv} and we will discuss this in more detail below, but we omit it for now.
\begin{theorem}[Theorem 7.15 of Ref.~\cite{aaronson2023certifiedArxiv}]\label{thm:original_theory}
For any algorithm $\mathcal{A}$ making $T$ queries to a Haar random unitary $U$ and $k\in[N]$, if $\mathcal{A}$ solves $(1+\delta)-\mathrm{XHOG}$, then there exists a state $\psi$ which is $O(Tk^{-1/2})$-close to $\mathcal{A}$'s output classical-quantum state such that
\begin{equation}
    H(Z\vert UE)_\psi\geq \left(\delta - O(k^3/N) - O(nTk^{-1/2}+nTN^{-1/2})\right)n - \log_2 n - 2\log_2 k - O(1).
\end{equation}
\end{theorem}

Although this only provides asymptotic entropy, the exact bound can be obtained by tracking all the constants carefully, which results in
\begin{equation}
    H(Z\vert UE)_\psi\geq \left(\delta - \frac{(k+1)k(k-1)}{N} - 2\varepsilon'(n\ln2+3)\right)n - \log_2 n - 2\log_2 k - O(1),
\end{equation}
where
\begin{equation}
    \varepsilon'=T\left(\frac{4}{2^{n/2}} + \sqrt{2\left(1-\left(\cos\frac{\pi}{k}\right)^{k/2}\right)}+2/\sqrt{k/2+1}\right).\label{eqn:varepsilon_prime}
\end{equation}
The parameter $\varepsilon'$ is the bound on the diamond distance between the actual algorithm $\mathcal{A}$ and an algorithm $\mathcal{F}$, where $\mathcal{F}$ only uses $k$ samples drawn from the bitstring probability distribution induced by the circuit as input. We reserve the notation $\varepsilon$ for later use as the smoothing parameter of the entropy. The second term of Eq. \ref{eqn:varepsilon_prime} in the parenthesis is due to Lemma 41, and the third term is due to Lemma 42 of Ref.~\cite{ambainis2014quantum}. They propagate to Theorem 3 of Ref.~\cite{ambainis2014quantum} and is paraphrased as Theorem 7.4 of Ref.~\cite{aaronson2023certifiedArxiv}. The factor of $2$ is due to setting $n$ of Lemma 41 and $m$ of Lemma 42 of Ref.~\cite{ambainis2014quantum} to $k/2$, so a total of $m+n=k$ copies of the state is needed. The cubic term in $k$ of the entropy is due to Theorem 7.10 of Ref.~\cite{aaronson2023certifiedArxiv}, which we can carefully calculate.

The above analysis is not sufficient. Specifically, protocols may submit $m$ circuits in a batch, which may be relevant for fast quantum devices whose classical communication latency dominates compared to the shot time, so the analysis for the query-access quantum adversary needs to be generalized. Another reason may be amortization of near-constant overhead per submission such as circuit compilation. We may also be interested in parallel execution of challenge circuits by the prover to reduce the average time per sample. Under our protocol where each round sends $m$ challenges circuits at once, the single-round entropy needs to scale with $m$, which is not obvious from the analysis of Ref.~\cite{aaronson2023certifiedArxiv}. We therefore provide a proof of this claim.

Since the original result is based on solving the $\mathrm{XHOG}$ problem, we similarly need to define a generalized problem for our analysis. Below, the notation $\vec{U}\sim\mathcal{D}^{\otimes m}$ means $\vec{U}=\{U^{(i)}\}_{i\in[m]}$ where $U^{(i)}\sim\mathcal{D}$ are i.i.d. random variables.
\begin{definition}\label{def:mxhog}
    A quantum algorithm $\mathcal{A}$ on input $\vec{U}\sim\mathcal{D}^{\otimes m}$ is said to solve $b$-$\mathrm{MXHOG}_m$ if it outputs $\vec{z}=\{z^{(i)}\}_{i\in[m]}$ such that
    \begin{equation}
    \underset{\vec{U}\sim\mathcal{D}^{\otimes m}}{\mathbb{E}}\left[\underset{\vec{z}\sim\mathcal{A}^{\vec{U}}}{\mathbb{E}}\left[\sum_i p_{U^{(i)}}(z^{(i)})\right]\right]\geq\frac{bm}{N}.
    \end{equation}
\end{definition}

Informally, the result we are after should be similar to Theorem \ref{thm:original_theory} with: explicit constants; $\varepsilon'$ and the overall entropy scaled by a factor of $m$; and improved correction terms. To proceed with this generalization, we follow the analysis of Section 7 of Ref.~\cite{aaronson2023certifiedArxiv} closely. Some key results use different techniques in order to obtain improved correction terms. The high level idea is as follows. We first show that any oracle access to a Haar random unitary can be simulated by an algorithm that is given $k$ bitstrings independently sampled from the measurement probability distribution corresponding to that unitary. Specifically, the quantum state from the oracle access algorithm is $O(1/\sqrt{k})$-close to the quantum state from the algorithm given the bitstrings. We then show that for any algorithm only given samples and no other information about the Haar random unitary, it must output one of the given bitstrings with sufficient probability to do well on $\mathrm{XHOG}$. Since the input samples have high entropy and the output is likely one of the samples, the output must have entropy. Finally, since the oracle access algorithm output quantum state is close to the sample access algorithm output quantum state, the quantum state of the oracle access algorithm must also have entropy.

Unfortunately, even then, the above theory does not result in a good entropy guarantee. Specifically, ignoring the $O(1)$ term, assuming $T=1$, $b=2$, and optimizing over $k$, the entropy guarantee is zero for all systems under 70 qubits, and the entropy rate is less than $0.32$ up to 100 qubits.

As a result, we seek to improve the entropy guarantee of the theory under the query-access model. Specifically, we can improve the correction term cubic in $k$ to quadratic, and improve the term $-2\log_2 k$ to $-\log_2 k$. We prove the improvement to $-\log_2 k$ against classical side information for all $m$. We also provide a proof against quantum side information for $m=1$, which we believe to be easily generalizable to all $m$ despite cumbersome notations.

Additionally, we prove an even more aggressive improvement from $-2\log_2 k$ to no $\log_2 k$ penalty in the special case of $m=1$ and $T=1$. Although much more specialized, we believe this is a reasonable adversary model. First, our experimentally implemented protocol is in the $m=1$ case. Additionally, only trapped ion devices are capable of executing the challenge circuits at such a high fidelity observed experimentally, and we know that no quantum device today is capable of executing the quantum circuit more than once within the latency window. We additionally believe that generalization to all $m$ and $T$ should be possible, otherwise an unnatural jump in entropy penalty is present when increasing $T$ from 1 to 2.

\subsection{Simulation of Haar random unitaries given sample access}

We first formally define what we mean by oracle access to a Haar random unitary. 
\begin{definition}
    For Haar random $U$, we say an efficient device that acts on system $D$ is given oracle access to $U$ if it can apply $U'$ to any subset of its system $D$ of the same size, where $U'=VUW$, $V$ is a random phase unitary $\sum_z e^{i\theta_z}\vert z\rangle\langle z\vert$, $\vert z\rangle$ are computational basis states, $\theta_z$ are independently sampled from $[0, 2\pi)$, and $W$ is a Haar random unitary on the subspace $\{\vert x\rangle:x\in\{0,1\}^n\setminus \{0^n\}\}$.
\end{definition}
This restriction is natural since the client only cares about the bitstring probability amplitudes corresponding to the state prepared acting on the $\vert0\rangle$ state, which is unchanged by applying $W$ and $V$. The client can `mask' the original circuit $U$ by applying $W$ and $V$ and the adversary would only have blackbox access to $U'$. We can generalize this notion to a batched setting.
\begin{definition}
    For $\vec{U}=\{U^{(i)}\}_{i\in[m]}$ such that $U^{(i)}$ are independent and Haar random, we say an efficient device is given oracle access to $\vec{U}$ if it is given oracle access to $U^{(i)}$ for all $i\in[m]$.
\end{definition}

Under this oracle access model, entropy conditioned on the challenge circuits is the same as entropy conditioned on the corresponding bitstring probability distributions due to the random phase symmetrization and the random unitary $W$ symmetrization of the oracle access model.
\begin{theorem}[Generalization of Theorem 7.1 of Ref.~\cite{aaronson2023certifiedArxiv}]\label{thm:entropy_conditioned_on_circuits_equal_distributions}
    For $\vec{U}=\{U^{(i)}\}_{i\in[m]}$ such that $U^{(i)}$ are independent and Haar random, any device $\mathcal{A}$ taking on input system $D$ of a bipartite state $\rho_{DE}$ and outputs a classical quantum state $\psi$ given oracle access to $\vec{U}$ must have $H(\vec{Z}\vert\vec{U}E)\psi=H(\vec{Z}\vert\vec{P}E)_\psi$, where $\vec{P}=\{P^{(i)}\}_{i\in[m]}$ and $P^{(i)}(z)=\vert\langle z\vert U^{(i)}\vert 0\rangle\vert^2$ is distributed according to the Dirichlet distribution $\mathrm{Dir}(1^N)$.
\end{theorem}
\begin{proof}
    Using the same arguments as in Theorem 7.1 of Ref.~\cite{aaronson2023certifiedArxiv}, for every distribution $P^{(i)}$, one can simulate a query to $U^{(i)}$ using one query to $U^{P^{(i)}}$, where $U^{P^{(i)}}$ is a `representative' in the set of all unitaries $U$ such that $\vert\langle z\vert U\vert 0^n\rangle\vert^2=P^{(i)}(z)$. Denote $\vec{U}^{\vec{P}}=\{U^{P^{(i)}}\}_{i\in[m]}$. Therefore, the quantum state given oracle access to $\vec{U}$ is the same as the quantum state given oracle access to $\vec{U}^{\vec{P}}$, meaning $\psi^{\vec{U}}=\psi^{\vec{U}^{\vec{P}}}$. Since $\vec{U}^{\vec{P}}$ only depends on $\vec{P}$, we can write $\psi^{\vec{U}^{\vec{P}}}=\sigma^{\vec{P}}$. Since the output state of $\mathcal{A}$ only depends on $\vec{P}$, the theorem follows.
\end{proof}

We now proceed with the main proof. For any $n$-qubit state $\vert\psi\rangle$, we define several objects in the same way as Section 7.1.1 of Ref.~\cite{aaronson2023certifiedArxiv}, which we reproduce for completeness. Let $C^\psi$ be a random unitary such that $C^\psi\vert 0 \rangle=\vert\psi\rangle$ and $C^\psi$ is Haar random for the subspace orthogonal to $\vert 0\rangle$. We would like to show that accessing such a unitary can be simulated by making queries to the so-called state preparation oracle. Unlike $C^\psi$, the state preparation oracle must not contain any additional information other than $\vert\psi\rangle$.

We first fix a canonical state $\vert\bot\rangle$ orthogonal to all $n$-qubit states, which one can obtain by extending the Hilbert space such as $\vert\bot\rangle=\vert0^n\rangle\vert0\rangle$ (see footnote 4 of Ref.~\cite{kretschmer2021quantum}). Define an oracle $\mathcal{O}^\psi$ that is a reflection about $\vert\psi_\bot\rangle=\frac{\vert\psi\rangle-\vert\bot\rangle}{\sqrt{2}}$, which means $\mathcal{O}^\psi\vert\psi\rangle=\vert\bot\rangle, \mathcal{O}^\psi\vert\bot\rangle=\vert\psi\rangle$, and $\mathcal{O}^\psi\vert\phi\rangle=\vert\phi\rangle$ for any $\vert\phi\rangle$ orthogonal to the span of $\left\{\vert\psi\rangle, \vert\bot\rangle\right\}$.

\begin{definition}
    A $(T,m)$-query algorithm $\mathcal{A}^{\vec{\mathcal{O}}}$, where $\vec{\mathcal{O}}=\{\mathcal{O}^{(i)}\}$, is an algorithm that makes at most $T$ queries to $\mathcal{O}^{(i)}$ for all $i\in[m]$.
\end{definition}
We now show that oracle access to the circuits can be approximated by querying the state preparation oracle.
\begin{lemma}[Generalization of Theorem 7.2 of Ref.~\cite{aaronson2023certifiedArxiv}]\label{lem:circuit_to_state_preparation_oracle}
    For every set of $m$ quantum states $\{\vert\psi^{(i)}\rangle\}_{i\in[m]}$, denote $\vec{C}=\{C^{\psi^{(i)}}\}_{i\in[m]}$ and $\vec{\mathcal{O}}=\{\mathcal{O}^{\psi^{(i)}}\}$. Every $(T,m)$-query algorithm $\mathcal{A}^{\vec{C}}$ can be approximated by a $(2T,m)$-query quantum algorithm $\mathcal{B}^{\vec{\mathcal{O}}}$ such that
    \begin{equation}
        \left\| \mathcal{B}^{\vec{\mathcal{O}}} - \underset{\vec{C}}{\mathbb{E}}\left[\mathcal{A}^{\vec{C}}\right] \right\|_\diamond\leq\frac{4Tm}{2^{n/2}}.
    \end{equation}
\end{lemma}
\begin{proof}
    Define $\mathcal{C}^\psi(\rho)=C^\psi\rho (C^\psi)^\dag$. As discussed in the proof of Theorem 7.2 of Ref.~\cite{aaronson2023certifiedArxiv}, for any state $\vert\psi^{\bot}\rangle$ orthogonal to $\vert\psi\rangle$, one can replace a call to a random $\mathcal{C}^\psi$ with $\Phi^\psi$ defined as (fixing a typo in Eq. 175 of Ref.~\cite{aaronson2023certifiedArxiv}),
    \begin{align}
    \Phi^{\psi}(\rho)=\underset{\vert\varphi\rangle\sim\mathrm{Haar(2^n)}}{\mathbb{E}}[\mathcal{O}^{\psi}\mathcal{O}^{\varphi}\mathcal{O}^{\psi}V^{\varphi}W\rho (\mathcal{O}^{\psi}\mathcal{O}^{\varphi}\mathcal{O}^{\psi}V^{\varphi}W)^{\dagger}],
    \end{align}
    satisfying $\|\Phi^\psi-\mathcal{C}^\psi\|_\diamond\leq\frac{4}{2^{n/2}}$ as stated in Eq. 178. Also, from the definition of $\Phi^\psi$ in Eq. 175 of Ref.~\cite{aaronson2023certifiedArxiv}, each $\mathcal{O}^{\psi^{(i)}}$ is queried $2T$ times. For $\vec{\Phi}=\{\Phi^{\psi^{(i)}}\}_{i\in[m]}$, we can choose the algorithm $\mathcal{B}^{\vec{\mathcal{O}}}=\mathcal{B}^{\vec{\Phi}}$. This is a $(2T,m)$-query algorithm. The total error bound of the lemma follows from the triangle inequality.
    
\end{proof}

We now show that the state preparation oracle can be approximated given some resource state that depends on the Haar random state. We reproduce the definition of the resource state considered in Theorem 7.4 of Ref.~\cite{aaronson2023certifiedArxiv} here:
\begin{equation}
    \vert R^\psi\rangle=\vert\psi\rangle^{\otimes\ell}\otimes\bigotimes_{i\in[k]}\left(\cos\left(\frac{i\pi}{2k}\right)\vert\psi\rangle+\sin\left(\frac{i\pi}{2k}\right)\vert\bot\rangle\right).\label{eqn:resource_state}
\end{equation}

\begin{lemma}[Generalization of Corollary 7.4 of Ref.~\cite{aaronson2023certifiedArxiv}]\label{lem:state_preparation_oracle_to_resource_state}
    For any state $\vert\psi^{(i)}\rangle$, define $\vert R^{\psi^{(i)}}\rangle$ as Eq. \ref{eqn:resource_state}, and define $\vert\vec{R}\rangle=\bigotimes_{i\in[m]}\vert R^{\psi^{(i)}}\rangle$. Let $\mathcal{O}$ be an oracle and $\rho$ be a quantum state. For every algorithm $\mathcal{B}$ that takes $\rho$ as input and makes $(T,m)$-query to $\vec{O}$, there exists an algorithm $\mathcal{G}$ that on input $\rho,\vert\vec{R}\rangle$ makes the same number of queries to $\mathcal{O}$ as $\mathcal{B}$ such that
    \begin{equation}
        \left\|\mathcal{B}^{\vec{\mathcal{O}},\mathcal{O}}(\rho)-\mathcal{G}^{\mathcal{O}}(\rho,\vert\vec{R}\rangle)\right\|_{1}\leq Tm\left(\sqrt{2\left(1-\left(\cos\frac{\pi}{2k}\right)^k\right)}+2/\sqrt{\ell+1}\right).
    \end{equation}
\end{lemma}
\begin{proof}
    Eq. 11 and 12 of Ref.~\cite{ambainis2014quantum} are true for each $\vert\psi^{(i)}\rangle$ (in the notations of Ref.~\cite{ambainis2014quantum}, $\vert\Psi\rangle$). For Lemma 41 of Ref.~\cite{ambainis2014quantum}, each invocation of $\mathcal{O}_\Psi$ is replaced with $U_\Psi$, which are $\varepsilon_n$ close to each other due to Eq. 11 of Ref.~\cite{ambainis2014quantum}. This allows us to write down a $(T,m)$-query version of Lemma 41 of Ref.~\cite{ambainis2014quantum} with trace distance $Tm\sqrt{2(1-(\cos\frac{\pi}{2k})^k)}$ (or $q_\Psi\sqrt{2(1-(\cos\frac{\pi}{2n})^n)}$ in the notations of Ref.~\cite{ambainis2014quantum}, where the expression is taken from the paragraph above Eq. 11 and $q_\Psi$ comes from the fact that it makes $q_\Psi$ queries).

    For Lemma 42 of Ref.~\cite{ambainis2014quantum}, exactly the same arguments apply from Eq. 12 of Ref.~\cite{ambainis2014quantum}. This allows us to write down a $(T,m)$-query version of Lemma 42 of Ref.~\cite{ambainis2014quantum} with trace distance $2Tm/\sqrt{\ell+1}$ (or $2q_{\rm Ref}/\sqrt{m+1}$ in the notations of Ref.~\cite{ambainis2014quantum}). Combining the $(T,m)$-query version of Lemma 41 and 42 concludes the proof.
\end{proof}

This now allows us to establish a generalization of Corollary 7.6 of Ref.~\cite{aaronson2023certifiedArxiv}, which states that oracle access to the Haar random circuit can be approximated by accessing the resource state. It turns out that the inclusion of the $\theta$ parameter throughout the proof of Ref.~\cite{aaronson2023certifiedArxiv} is unnecessary, so we omit it in further discussions.
\begin{corollary}[Generalization of Corollary 7.6 of Ref.~\cite{aaronson2023certifiedArxiv} but without $\theta$]\label{cor:circuit_to_resource_state}
    For quantum state $\{\vert\psi^{(i)}\rangle\}_{i\in[m]}$ and every quantum algorithm $\mathcal{A}$ making $(T,m)$-queries to $\vec{C}$, there exists a quantum algorithm $\mathcal{G}$ given access to $\vert\vec{R}\rangle$ such that
    \begin{equation}
        \left\| \underset{\vec{C}}{\mathbb{E}}\left[ \mathcal{A}^{\vec{C}} \right] - \mathcal{G}\left(\vert\vec{R}\rangle\langle\vec{R}\vert\right)\right\|_{\diamond}\leq Tm\left(\frac{4}{2^{n/2}} + \sqrt{2\left(1-\left(\cos\frac{\pi}{2k}\right)^k\right)}+2/\sqrt{\ell+1}\right).
    \end{equation}
\end{corollary}
\begin{proof}
    The proof follows the same idea as Corollary 7.6 of Ref.~\cite{aaronson2023certifiedArxiv} by combining Lemma \ref{lem:circuit_to_state_preparation_oracle} and Lemma \ref{lem:state_preparation_oracle_to_resource_state}.
\end{proof}

We now show that access to resource states can be simulated by access to samples drawn from $\vert\psi\rangle$.
We reproduce some definitions from \cite{aaronson2023certifiedArxiv}. Consider $P=(P_0,\dots,P_{N-1})\sim\mathrm{Dir}(1^N)$, a random state $\vert\psi\rangle=W\vert P\rangle$, where $\vert P\rangle=(\sqrt{P_0},\dots,\sqrt{P_{N-1}})$ and $W$ is a random diagonal phase matrix. We can now define the resource state
\begin{equation}
    \vert R^{P,W}\rangle=(W\vert P\rangle)^{\otimes\ell}\otimes\bigotimes_{i\in[k]}\left(\cos\left(\frac{i\pi}{2k}\right)W\vert P\rangle+\sin\left(\frac{i\pi}{2k}\right)\vert\bot\rangle\right).
\end{equation}

\begin{lemma}[Lemma 7.7 of Ref.~\cite{aaronson2023certifiedArxiv}, rephrased from Lemma 15 of Ref.~\cite{kretschmer2021quantum}]\label{lem:single_P_symmetrized_state}
    There is an algorithm which prepares
    \begin{equation}
        \sigma^P=\underset{W}{\mathbb{E}}\left[\vert R^{P,W}\rangle\langle R^{P,W}\vert\right]
    \end{equation}
    by measuring $(k+\ell)$ copies of $\vert\psi\rangle$ in the standard basis. Here, $\vert\psi\rangle$ is any state with probability distribution $P$.
\end{lemma}
We note that this Lemma follows when we consider the state $\vert R\rangle$ of Lemma 15 of Ref.~\cite{kretschmer2021quantum} with $j\in[k+\ell]$ and $\alpha_j=1,\beta_j=0$ for $j\in[\ell]$.

We now define $\vec{W}=\{W^{(i)}\}_{i\in[m]}$, where $W^{(i)}$ are independent random diagonal phase matrices. Similarly, we define $\vec{P}=\{P^{(i)}\}_{i\in[m]}$, and also use the notation $\vec{P}\sim \mathrm{Dir}(1^N)^{\otimes m}$ to denote $P^{(i)}\sim\mathrm{Dir}(1^N)$ for all $i\in[m]$. We now define the resource state
\begin{equation}
    \vert R^{\vec{P},\vec{W}}\rangle=\bigotimes_{i\in[m]}\vert R^{P^{(i)},W^{(i)}}\rangle.
\end{equation}

\begin{lemma}[Generalization of Lemma \ref{lem:single_P_symmetrized_state}]\label{lem:vector_P_symmetrized_state}
    There is an algorithm which prepares
    \begin{equation}
        \sigma^{\vec{P}}=\underset{\vec{W}}{\mathbb{E}}\left[\vert R^{\vec{P},\vec{W}}\rangle\langle R^{\vec{P},\vec{W}}\vert\right]
    \end{equation}
    by measuring $(k+\ell)$ copies of $\vert\psi^{(i)}\rangle$ for every $i\in[m]$. Here, $\vert\psi^{i}\rangle$ is any state with probability distribution $P^{(i)}$.
\end{lemma}
\begin{proof}
    It can be shown that $\sigma^{\vec{P}}=\bigotimes_{i\in[m]}\sigma^{P^{(i)}}$, hence the proof follows from Lemma \ref{lem:single_P_symmetrized_state}.
\end{proof}

We now show that any algorithm given oracle access to Haar random circuits can be simulated by an algorithm with sample access. We consider random circuits given distribution $P$ in the same way as \cite{aaronson2023certifiedArxiv}. For any distribution $P$, let $C^P=WVC'$, where $W$ is a random diagonal phase matrix, $V\vert 0\rangle=\vert P\rangle$, and $C'$ is Haar random on the subspace orthogonal to $\vert 0\rangle$. Similarly, $\vec{C}^{\vec{P}}=\{C^{P^{(i)}}\}_{i\in[m]}$, where each $C^{P^{(i)}}$ are independently and randomly sampled as above.

For $\vec{z}=\{z^{(i)}\}_{i\in[m]}$, we use the following notation $\vec{z}\sim\vec{P}$ to denote $z^{(i)}\sim P^{(i)}$ for all $i\in[m]$.
\begin{theorem}[Generalization of Theorem 7.8 of Ref.~\cite{aaronson2023certifiedArxiv}]\label{thm:gen_7.8}
    For any $\vec{P}$ and any algorithm $\mathcal{A}$ making $(T,m)$-query to $\vec{C}^{\vec{P}}$, there exists a quantum algorithm $\mathcal{F}$ given $(k+\ell)$ samples drawn from $P^{(i)}$ for all $i\in[m]$ such that
    \begin{equation}
    \|\bar{\mathcal{A}}^{\vec{P}}-\bar{\mathcal{F}}^{\vec{P}}\|_\diamond\leq Tm\left(\frac{4}{2^{n/2}} + \sqrt{2\left(1-\left(\cos\frac{\pi}{2k}\right)^k\right)}+2/\sqrt{\ell+1}\right),
    \end{equation}
    where $\bar{\mathcal{A}}^{\vec{P}}=\mathbb{E}_{\vec{C}^{\vec{P}}}[\mathcal{A}^{\vec{C}^{\vec{P}}}]$ and $\bar{\mathcal{F}}^{\vec{P}}(\rho)=\mathbb{E}_{\vec{z}_1,\dots,\vec{z}_k\sim \vec{P}}[\mathcal{F}(\rho, \vec{z}_1,\dots,\vec{z}_k)]$.
\end{theorem}
\begin{proof}
    We closely mirror the arguments presented in Theorem 7.8 of Ref.~\cite{aaronson2023certifiedArxiv}. Given $P$, sampling random $C^P$ is equivalent to sampling random diagonal phase matrix $W$ to produce $\vert\psi\rangle=W\vert P\rangle$, and then sample random circuit $C^\psi$ such that $C^\psi\vert 0\rangle=\vert\psi\rangle$ and $C^\psi$ is Haar random for the subspace orthogonal to $\vert 0\rangle$, as described after Theorem \ref{thm:entropy_conditioned_on_circuits_equal_distributions}. Similarly, given $m$ distributions $\vec{P}$, sampling random $\vec{C}^{\vec{P}}$ is equivalent to sampling $m$ random diagonal phase matrices $\vec{W}$, producing $m$ states $\{\vert\psi^{(i)}\rangle=W^{(i)}\vert P^{(i)}\rangle\}$, and then sample $m$ random circuits $\vec{C}=\{C^{\psi^{(i)}}\}$ in the same way as above. Therefore,
    \begin{equation}
    \underset{\vec{C}^{\vec{P}}}{\mathbb{E}}\left[\mathcal{A}^{\vec{C}^{\vec{P}}}\right]=\underset{\vec{W},\vec{C}}{\mathbb{E}}\left[\mathcal{A}^{\vec{C}}\right].
    \end{equation}
    Let the quantum process that takes samples to prepare $\sigma^{\vec{P}}$ in Lemma \ref{lem:vector_P_symmetrized_state} be $\Phi$. By the same argument as \cite{aaronson2023certifiedArxiv},
    \begin{equation}
        \underset{\vec{W}}{\mathbb{E}}\left[\mathcal{G}\left(\vert R^{\vec{P},\vec{W}}\rangle\langle R^{\vec{P},\vec{W}}\vert\right)\right]=\underset{\vec{z}_1,\dots,\vec{z}_k\sim \vec{P}}{\mathbb{E}}\left[\mathcal{G}\circ\Phi(\vec{z}_1,\dots,\vec{z}_k)\right].
    \end{equation}
    With $\mathcal{F}=\mathcal{G}\circ\Phi$, triangle inequality and Corollary \ref{cor:circuit_to_resource_state},
    \begin{align}\nonumber
    \left\|\underset{\vec{C}^{\vec{P}}}{\mathbb{E}}\left[\mathcal{A}^{\vec{C}^{\vec{P}}}\right]-\mathbb{E}_{\vec{z}_1,\dots,\vec{z}_k\sim \vec{P}}[\mathcal{F}(\vec{z}_1,\dots,\vec{z}_k)]\right\|_\diamond\leq & \underset{\vec{W}}{\mathbb{E}}\left\|\underset{\vec{C}}{\mathbb{E}}\left[\mathcal{A}^{\vec{C}}\right]-\mathcal{G}\left(\vert R^{\vec{P},\vec{W}}\rangle\langle R^{\vec{P},\vec{W}}\vert\right)\right\|_\diamond\\
        \leq & Tm\left(\frac{4}{2^{n/2}} + \sqrt{2\left(1-\left(\cos\frac{\pi}{2k}\right)^k\right)}+2/\sqrt{\ell+1}\right).
    \end{align}
\end{proof}

\subsection{A simplified device}

We now show that sample access algorithms must provide entropy if it passes solves $\mathrm{MXHOG}$. Below, we provide a generalization of Theorem 7.10 of Ref.~\cite{aaronson2023certifiedArxiv}, where we improve the correction from $k^3$ to $k^2$. This theorem shows that solving $\mathrm{MXHOG}$ implies some constant probability of returning one of the input samples.
\begin{theorem}[Generalization of Theorem 7.10 of Ref.~\cite{aaronson2023certifiedArxiv}]\label{thm:gen_7.10}
For a quantum algorithm $\mathcal{F}$ given $(k+\ell)$ samples draw from $P^{(i)}$ for all $i\in[m]$, if $n\geq50$, $k\leq N^{1/2}$, and
\begin{equation}
    \underset{\vec{P}\sim\mathrm{Dir}(1^N)^{\otimes m}}{\mathbb{E}}\underset{\vec{z}\sim\bar{\mathcal{F}}^{\vec{P}}(\rho)}{\mathbb{E}}\left[\sum_{i\in[m]} P^{(i)}(z^{(i)})\right]\geq m \frac{2-\varepsilon}{N+k},
\end{equation}
where $\vec{P}=\{P^{(i)}\}_{i\in[m]}$ and $\vec{z}=\{z^{(i)}\}_{i\in[m]}$, then with probability greater than $1-6m/N$,
\begin{equation}
    \frac{1}{m}\sum_{i\in[m]}\Pr\left[z^{(i)}\in\{z_j^{(i)}\}_{j\in[k]}\right]\geq 1-\varepsilon-1.001\times\frac{16k^2\ln^2 N}{N},
\end{equation}
where $\vec{z}_j=\{z_j^{(i)}\}_{i\in[m]}$, and the probability is over choices of $\vec{P}\sim\mathrm{Dir}(1^N)^{\otimes m},\vec{z}_1,\dots \vec{z}_k\sim \vec{P}$, and $\vec{z}=\{z^{(i)}\}_{i\in[m]}\sim\mathcal{F}(\rho, \vec{z}_1,\dots,\vec{z}_k)$.
\end{theorem}
\begin{proof}
For each set of samples $\{z_1^{(i)},\dots,z_k^{(i)}\}$ obtained from distribution $P^{(i)}$, let $f^{(i)}=\{f_0^{(i)},\dots,f_{N-1}^{(i)}\}$ be the frequency vector. Denote $\vec{f}=\{f^{(i)}\}_{i\in[m]}$, the output of $\mathcal{F}(\rho,\vec{z}_1,\dots,\vec{z}_k)$ can be described by a distribution $Q(\vec{f})$ that only depends on $\vec{f}$. Due to Lemma 7.9 of Ref.~\cite{aaronson2023certifiedArxiv}, $f^{(i)}$ is distributed according to the uniform distribution of all frequency vectors that has $N$ elements and sums to $k$, which we denote as $f^{(i)}\sim\Phi(N,k)$. We use the notation $\vec{f}\sim\Phi(N,k)^{\otimes m}$ to denote the fact that this holds for all $i\in[m]$.

Upon observing $f^{(i)}$, the posterior distribution of $P^{(i)}$ follows $\mathrm{Dir}(1^N + f^{(i)})$, as described by Eq. 200 of Ref.~\cite{aaronson2023certifiedArxiv}. For each $\vec{f}$ such that $\|f^{(i)}\|_\infty=j$ for some $i\in[m]$, the contribution of the $i$th circuit to the $\mathrm{MXHOG}$ score is given by
\begin{equation}
    \underset{\vec{z}\sim Q(\vec{f})}{\mathbb{E}}\underset{P^{(i)}\sim\mathrm{Dir}(1^N+f^{(i)})}{\mathbb{E}}\left[P^{(i)}(z^{(i)})\right]=\underset{\vec{z}\sim Q(\vec{f})}{\mathbb{E}}\left[\frac{f_{z^{(i)}}^{(i)}+1}{N+k}\right]\leq\frac{1}{N+k}\left(1+j\cdot\underset{\vec{z}\sim Q(\vec{f})}{\Pr}\left[f_{z^{(i)}}^{(i)}>0\right]\right)\label{eqn:score_to_contain_prob},
\end{equation}
where the first equality uses the property of the Dirichlet distribution in the same way as Eq. 200 of Ref.~\cite{aaronson2023certifiedArxiv}, and the inequality follows from $\|f^{(i)}\|_\infty=j\rightarrow \forall z, f_{z}^{(i)}\leq j$.

Denote the event that $\|f^{(i)}\|_\infty=j$ as $\Omega^{(i)}_j$, and the event $\|f^{(i)}\|_\infty\geq j$ as $\Omega^{(i)+}_j$. We now upper bound $\Pr\left[\Omega^{(i)+}_j\right]$, which is the probability that at least one of the $N$ possible bitstrings are sampled at least $j$ times. Given any bitstring $s$, the probability that a specific choice of a set of $j$ out of $k$ samples are $s$ (other samples not in the set are allowed to be $s$) is $P(s)^j\leq (\max_z P(z))^j$. Since there are $\binom{k}{j}$ choices of $j$ samples, by the union bound, the probability that any given bitstring is sampled at least $j$ times is upper bounded by $\binom{k}{j}\left(\max_z P(z)\right)^j$. By Lemma 3.14 of Ref.~\cite{aaronson2023certifiedArxiv}, with probability $1-6/N$ over $P\sim\mathrm{Dir}(1^N)$, we have $\max_z P(z)\leq 4\frac{\ln N}{N}=p_{\rm max}$. Therefore, with probability greater than $1-6m/N$, for all $i\in[m]$, the probability that any given bitstring is sampled at least $j$ times is at most $\binom{k}{j}\left(4\frac{\ln N}{N}\right)^j$. There are $N$ possible bitstrings, and by the union bound,
\begin{equation}
\Pr\left[\Omega^{(i)+}_j\right]\leq N\cdot\binom{k}{j}\cdot\left(4\frac{\ln N}{N}\right)^j\leq \frac{(4k\ln N)^j}{j}\cdot\frac{1}{N^{j-1}}\label{eqn:prob+bound},
\end{equation}
where we used the fact that
\begin{equation}
    \binom{k}{j}=\frac{k\times(k-1)\times\dots\times(k-j+1)}{j\times(j-1)\times\dots\times 1}\leq \frac{k^j}{j}.
\end{equation}

The score of $\mathcal{F}$ in expectation is
\begin{align}\nonumber
&\underset{\vec{f}\sim\Phi(N,k)^{\otimes m}}{\mathbb{E}}\underset{\vec{z}\sim Q(\vec{f})}{\mathbb{E}}\underset{\vec{P}\sim\mathrm{Dir}(1^N+\vec{f})}{\mathbb{E}}\left[\sum_{i\in[m]}P^{(i)}(z^{(i)})\right]\\\nonumber
=& \sum_{i\in[m]}\underset{\vec{f}\sim\Phi(N,k)^{\otimes m}}{\mathbb{E}}\underset{\vec{z}\sim Q(\vec{f})}{\mathbb{E}}\underset{\vec{P}\sim\mathrm{Dir}(1^N+\vec{f})}{\mathbb{E}}\left[P^{(i)}(z^{(i)})\right]\\\nonumber
\leq& \frac{1}{N+k}\sum_{i\in[m]}\sum_{j=1}^k\Pr\left[\Omega^{(i)}_j\right]\left(1+j\cdot\underset{\vec{f}\sim\Phi(N,k)^{\otimes m},\vec{z}\sim Q(\vec{f})}{\Pr}\left[f^{(i)}_{z^{(i)}}>0\Big\vert \Omega^{(i)}_j\right]\right)\\\nonumber
\leq& \frac{1}{N+k}\left\{m+\sum_{i\in[m]}\left(\underset{\vec{f}\sim\Phi(N,k)^{\otimes m},\vec{z}\sim Q(\vec{f})}{\Pr}\left[f^{(i)}_{z^{(i)}}>0\right]+\sum_{j=2}^k j\cdot\Pr\left[\Omega^{(i)+}_j\right]\right)\right\}\\\nonumber
\leq& \frac{1}{N+k}\left\{m+\frac{m16k^2\ln^2 N}{N}\sum_{j=0}^\infty \frac{(4k\ln N)^j}{N^j}+\sum_{i\in[m]}\left(\underset{\vec{f}\sim\Phi(N,k)^{\otimes m},\vec{z}\sim Q(\vec{f})}{\Pr}\left[f^{(i)}_{z^{(i)}}>0\right]\right)\right\}\\
\leq& \frac{1}{N+k}\left\{m+1.001\times\frac{m16k^2\ln^2 N}{N}+\sum_{i\in[m]}\left(\underset{\vec{f}\sim\Phi(N,k)^{\otimes m},\vec{z}\sim Q(\vec{f})}{\Pr}\left[f^{(i)}_{z^{(i)}}>0\right]\right)\right\},
\end{align}
where the first inequality uses Eq. \ref{eqn:score_to_contain_prob}, the second inequality uses the fact that
\begin{align}
\Pr\left[\Omega^{(i)}_1\right]\Pr\left[f^{(i)}_{z^{(i)}}>0\Big\vert\Omega^{(i)}_1\right]\leq &\Pr\left[f^{(i)}_{z^{(i)}}>0\right]\\
\Pr\left[\Omega^{(i)}_{j}\right]\Pr\left[f^{(i)}_{z^{(i)}}>0\Big\vert\Omega^{(i)}_j\right]\leq &\Pr\left[\Omega^{(i)}_j\right] \leq\Pr\left[\Omega^{(i)+}_j\right],
\end{align}
where we omit $\vec{f}\sim\Phi(N,k)^{\otimes m},\vec{z}\sim Q(\vec{f})$ in the notation $\Pr$, the third inequality uses Eq. \ref{eqn:prob+bound}, and the last inequality uses the fact that
\begin{equation}
    \sum_{j=0}^\infty \frac{(4k\ln N)^j}{N^j}=\frac{1}{1-(4k\ln N)/N}\leq 1.001
\end{equation}
for $n\geq 50$ and $k\leq N^{1/2}$.

Thus if the score is at least $m\frac{2-\varepsilon}{N+k}$,
\begin{equation}
    \frac{1}{m}\sum_{i\in[m]}\underset{\vec{f}\sim\Phi(N,k)^{\otimes m}\vec{z}\sim Q(\vec{f})}{\Pr}\left[f_{z^{(i)}}^{(i)}>0\right]\geq 1-\varepsilon-1.001\times\frac{16k^2\ln^2 N}{N}.
\end{equation}
\end{proof}

Below, we show that algorithms returning one of the input samples must have entropy by proving a generalization of Theorem 7.11 of Ref.~\cite{aaronson2023certifiedArxiv} to $\mathrm{MXHOG}$, where the proof techniques differ slightly to improve the $2\log_2 k$ factor to $\log_2 k$.

\begin{theorem}[Generalization of Theorem 7.11 and Corollary 7.12 of Ref.~\cite{aaronson2023certifiedArxiv}]\label{thm:gen_7.12}
For distributions $\vec{P}\sim\mathrm{Dir}(1^N)^{\otimes m}$, let $\mathcal{F}$ be any algorithm that produce an output register $\vec{Z}=Z^{(1)}\dots Z^{(m)}$ given access to $\vec{z}_1,\dots \vec{z}_k\sim \vec{P}$ over register $\vec{Z}_1\dots\vec{Z}_k$ and the first subsystem of any bipartition state $\rho_{DE}$. Denote $\bar{\mathcal{F}}^{\vec{P}}(\rho)=\mathbb{E}_{\vec{z}_1,\dots \vec{z}_k\sim \vec{P}}[\mathcal{F}(\rho, \vec{z}_1,\dots \vec{z}_k)]$. If $n\geq 50$, $k\leq N^{-1/2}$, $m,n\leq 10^6$, $\forall i\in[m],\max_z P^{(i)}(z)\leq p_{\rm max}=4\ln N/N$, and
\begin{equation}
    \frac{1}{m}\sum_{i\in[m]}\underset{\vec{z}_1,\dots \vec{z}_k\sim \vec{P},\vec{z}\sim\mathcal{F}(\rho, \vec{z}_1,\dots,\vec{z}_k)}{\Pr}\left[z^{(i)}\in\{z_j^{(i)}\}_{j\in[k]}\right]\geq1-\delta,\label{eqn:gen_7.11}
\end{equation}
then, conditioned on classical side information $E$,
\begin{equation}
    H(\vec{Z}\vert \vec{P}E)_{\bar{\mathcal{F}}^{\vec{P}}(\rho)}\geq m\left[n(1-\delta)-\log_2n-\log_2k-3\right].
\end{equation}
\end{theorem}

\begin{proof}
First, consider a fixed set of distributions $\vec{P}$ and evaluate the entropy conditioned on $E$. Define the classical-quantum state
\begin{equation}
    \psi_{\vec{Z}_1\dots\vec{Z}_k\vec{Z}E}=\sum_{\vec{z}_1,\dots,\vec{z}_k}\vec{P}(\vec{z}_1)\dots\vec{P}(\vec{z}_k)\vert \vec{z}_1,\dots,\vec{z}_k\rangle\langle\vec{z}_1,\dots,\vec{z}_k\vert_{\vec{Z}_1\dots\vec{Z}_k}\otimes\mathcal{F}(\rho,\vec{z}_1,\dots,\vec{z}_k)_{\vec{Z}E},
\end{equation}
where $\vec{P}(\vec{z}_j)=\prod_{i\in[m]}P^{(i)}(z^{(i)}_j)$.
For the time being, let us ignore $E$, and the state can be stated in terms of some conditional distribution $Q(\cdot\vert \vec{z}_1,\dots,\vec{z}_k)$,
\begin{equation}
\psi_{\vec{Z}_1\dots\vec{Z}_k\vec{Z}}=\sum_{\vec{z}_1,\dots,\vec{z}_k}\vec{P}(\vec{z}_1)\dots\vec{P}(\vec{z}_k)\vert \vec{z}_1,\dots,\vec{z}_k\rangle\langle\vec{z}_1,\dots,\vec{z}_k\vert_{\vec{Z}_1\dots\vec{Z}_k}\otimes\sum_{\vec{z}}Q(\vec{z}\vert \vec{z}_1,\dots,\vec{z}_k)\vert\vec{z}\rangle\langle\vec{z}\vert.
\end{equation}
We decompose each distribution
\begin{equation}
    Q(\vec{z}\vert \vec{z}_1,\dots,\vec{z}_k)=\sum_{\vec{s}\in\{0,1\}^m}Q_{\vec{s}}(\vec{z}\vert\vec{z}_1,\dots,\vec{z}_k),
\end{equation}
where for $\vec{s}=\{s_i\}_{i\in[m]}$ and $\vec{z}=\{z^{(i)}\}_{i\in[m]}\in\mathrm{support}(Q_{\vec{s}}(\cdot\vert \vec{z}_1,\dots,\vec{z}_k))$, we have $z^{(i)}\in \{z_1^{(i)},\dots,z_k^{(i)}\}$ for all $i\in[m]$ such that $s_i=1$ and $z^{(i)}\notin \{z_1^{(i)},\dots,z_k^{(i)}\}$ for all $i\in[m]$ such that $s_i=0$. Let us formally denote this condition as
\begin{align}
    \vec{z}\stackrel{\vec{s}}{\in}\{\vec{z}_1,\dots,\vec{z}_k\}&\Leftrightarrow\forall i\in[m],(s_i=1\wedge z^{(i)}\in \{z_j^{(i)}\}_{j\in[k]})\vee (s_i=0\wedge z^{(i)}\notin \{z_j^{(i)}\}_{j\in[k]}).
\end{align}
Define the following probabilities:
\begin{align}
p_{\vec{s},\vec{z}_1,\dots,\vec{z}_k}&=\sum_{\vec{z}}Q_{\vec{s}}(\vec{z}\vert \vec{z}_1,\dots,\vec{z}_k)\\
p_{\vec{s}}&=\sum_{\vec{z}_1,\dots,\vec{z}_k}p_{\vec{s},\vec{z}_1,\dots,\vec{z}_k}\vec{P}(\vec{z}_1)\dots\vec{P}(\vec{z}_k)\\
\bar{Q}_{\vec{s}}(\vec{z}\vert \vec{z}_1,\dots,\vec{z}_k)&=Q_{\vec{s}}(\vec{z}\vert \vec{z}_1,\dots,\vec{z}_k)/p_{\vec{s},\vec{z}_1,\dots,\vec{z}_k},
\end{align}
we can define normalized states
\begin{equation}
    \psi_{\vec{s}}=\frac{1}{p_{\vec{s}}}\sum_{\vec{z}_1,\dots,\vec{z}_k}p_{\vec{s},\vec{z}_1,\dots,\vec{z}_k}\vec{P}(\vec{z}_1)\dots\vec{P}(\vec{z}_k)\vert \vec{z}_1,\dots,\vec{z}_k\rangle\langle\vec{z}_1,\dots,\vec{z}_k\vert_{\vec{Z}_1\dots\vec{Z}_k}\otimes\sum_{\vec{z}}\bar{Q}_{\vec{s}}(\vec{z}\vert \vec{z}_1,\dots,\vec{z}_k)\vert\vec{z}\rangle\langle\vec{z}\vert,
\end{equation}
and therefore $\psi=\sum_{\vec{s}\in\{0,1\}^m}p_{\vec{s}}\psi_{\vec{s}}$.

Consider $\sigma_{\vec{s}}=\mathrm{tr}_{\vec{Z}_1\dots\vec{Z}_k}(\psi_{\vec{s}})$. The operator $\sigma_{\vec{s}}$ requires that $\vec{z}\stackrel{\vec{s}}{\in}\{\vec{z}_1,\dots,\vec{z}_k\}$ as described earlier.
\begin{align}\nonumber
\langle\vec{z}\vert\sigma_{\vec{s}}\vert\vec{z}\rangle&=\frac{1}{p_{\vec{s}}}\sum_{\vec{z}_1,\dots,\vec{z}_k}p_{\vec{s},\vec{z}_1,\dots,\vec{z}_k}\vec{P}(\vec{z}_1)\dots\vec{P}(\vec{z}_k)\bar{Q}_{\vec{s}}(\vec{z}\vert \vec{z}_1,\dots,\vec{z}_k)\\\nonumber
&\leq\frac{1}{p_{\vec{s}}}\sum_{\vec{z}_1,\dots,\vec{z}_k}p_{\vec{s},\vec{z}_1,\dots,\vec{z}_k}\vec{P}(\vec{z}_1)\dots\vec{P}(\vec{z}_k)\mathbb{I}[\vec{z}\stackrel{\vec{s}}{\in}\{\vec{z}_1,\dots,\vec{z}_k\}]\\\nonumber
&\leq\frac{1}{p_{\vec{s}}}\sum_{\vec{z}_1,\dots,\vec{z}_k}p_{\vec{s},\vec{z}_1,\dots,\vec{z}_k}\vec{P}(\vec{z}_1)\dots\vec{P}(\vec{z}_k)\prod_{i\in[m]:s_i=1}\sum_{j=1}^k\mathbb{I}[z^{(i)}=z^{(i)}_j]\\\nonumber
&=\frac{1}{p_{\vec{s}}}\prod_{i\in[m]:s_i=1}\sum_{\vec{z}_1,\dots,\vec{z}_k}p_{\vec{s},\vec{z}_1,\dots,\vec{z}_k}\vec{P}(\vec{z}_1)\dots\vec{P}(\vec{z}_k)\sum_{j=1}^k\mathbb{I}[z^{(i)}=z^{(i)}_j]\\
&=\frac{1}{p_{\vec{s}}}\prod_{i\in[m]:s_i=1}\sum_{j\in[k]}p^{(ij)}_{z^{(i)}}P^{(i)}(z^{(i)}),
\end{align}
where
\begin{align}
p^{(ij)}_{z^{(i)}}\equiv\underset{\vec{z}_1,\dots,\vec{z}_{j-1},\vec{z}_{j+1},\dots,\vec{z}_k}{\mathbb{E}}\left[\underset{z^{(1)}_j,\dots,z^{(m-1)}_j,z^{(m+1)}_j,\dots,z^{(m)}_j}{\mathbb{E}}\left[p_{\vec{s},\vec{z_1},\dots,\vec{z}_j=\{z^{(1)}_j,\dots,z^{(i-1)}_j,z^{(i)},z^{(i+1)}_j,\dots,z^{(m)}_j\},\dots,\vec{z}_k}\right]\right].
\end{align}
Further, by definition, $p^{(ij)}_{z^{(i)}}\in[0,1]$, and thus 
\begin{align}
\langle\vec{z}\vert\sigma_{\vec{s}}\vert\vec{z}\rangle&\leq\frac{1}{p_{\vec{s}}}\prod_{i\in[m]:s_i=1}k\max_zP^{(i)}(z)\leq\frac{1}{p_{\vec{s}}}(kp_{\rm max})^{w(\vec{s})}\\
H(\vec{Z})_{\sigma_{\vec{s}}}&\geq H_{\rm min}(\vec{Z})_{\sigma_{\vec{s}}}\geq w(\vec{s})\left(\log_2 \frac{1}{p_{\rm max}}-\log_2 k\right)-\log_2\frac{1}{p_{\rm s}}\geq  w(\vec{s})\left(n - \log_2 n - 2-\log_2 k\right)-\log_2\frac{1}{p_{\vec{s}}},
\end{align}
where $w(\vec{s})$ is the Hamming weight of $\vec{s}$, and we used the fact that $\log_2 \frac{1}{p_{\rm min}}\geq -\log_2\frac{4\ln N}{N}=n-\log_2 n - \log_2\ln2 - 2\geq n - \log_2 n - 2$.

Back to the general state unconditioned on any $\vec{s}$,
\begin{align}\nonumber
    H(\vec{Z})_{\psi}&\geq H(\vec{Z}|\vec{s})_{\psi}=\sum_{\vec{s}\in\{0,1\}^m}p_{\vec{s}}H(\vec{Z})_{\sigma_{\vec{s}}}\\\nonumber
    &=(n-\log_2n-2-\log_2k)\sum_{\vec{s}\in\{0,1\}^m}p_{\vec{s}}w(\vec{s})+\sum_{\vec{s}\in\{0,1\}^m}p_{\vec{s}}\log_2p_{\vec{s}}\\\nonumber
    &=(n-\log_2n-2-\log_2k)\sum_{i\in[m]}\underset{\vec{z}_1,\dots \vec{z}_k\sim \vec{P},\vec{z}\sim\mathcal{F}(\rho, \vec{z}_1,\dots,\vec{z}_k)}{\Pr}\left[z^{(i)}\in\{z_j^{(i)}\}_{j\in[k]}\right]-H(\vec{s})_{\psi}\\
    &\geq m\left[n(1-\delta)-\log_2n-\log_2k-3\right],
\end{align}
where the last line uses the fact that the sum in the previous line is $1-\delta$. However, the theorem demands entropy conditioned on $E$. For classical side information, we are done since any classical side information can be absorbed into the strategy of the adversary to minimize entropy. Further, since the proof is for fixed $\vec{P}$, conditioning on $\vec{P}$ gives the same bound.
\end{proof}

We note that the above theorem with the improvement is conditioned on classical side information. For randomness amplification, since the quantum device is assumed to be the adversary, the quantum bits themselves do not remain private, only unpredictable upon initial snapshot of information of the adversary. In this case, entropy conditioned on quantum side information is unnecessary. This is similarly the case for other use cases where the quantum randomness will be public such as multi-party protocols.

We can obtain a version with quantum side information that also improves the entropy penalty from $2\log_2 k$ to $\log_2 k$. We do not generalize to the multiple input setting due to cumbersome notation. We defer the proof to Subsection \ref{sec:improve_quantum}

Combining Theorem \ref{thm:gen_7.10} and Theorem \ref{thm:gen_7.12} yields the following corollary, which states that sample access algorithms solving $\mathrm{MXHOG}$ must have entropy.
\begin{corollary}[Generalization of Corollary 7.13 of Ref.~\cite{aaronson2023certifiedArxiv}]\label{cor:gen_7.13}
Let $\mathcal{F}$ be any device given access to $k$ samples from each distribution $P^{(i)}=P_{C^{(i)}}$ in $\vec{P}$ for $C^{(i)}\sim\mathrm{Haar}(N)$. If $n\geq 50$, $k\leq N^{-1/2}$, $m,n\leq 10^6$, and $\mathcal{F}$ solves $(1+\delta)-\mathrm{MXHOG}_m$, then
\begin{equation}
    H(\vec{Z}\vert \vec{P}E)_\Psi\geq m\left[\left(\delta-1.001\times\frac{16k^2\ln^2 N}{N}\right)n-\log_2 n -\log_2 k - 3\right].
\end{equation}
\end{corollary}
\begin{proof}
    We first use Theorem \ref{thm:gen_7.10}. Solving $(1+\delta)-\mathrm{MXHOG}_m$ requires $\frac{1+\delta}{N}=\frac{2-\varepsilon}{N+k}\rightarrow\varepsilon\leq1-\delta+k/N$, which implies the average probability that the output sample is in the set of $k$ samples is at least $\delta-k/N-1.001\times16k^2\ln^2 N/N$. For the parameters of interest, $k/N$ can be safely absorbed into $1.001\times16k^2\ln^2 N/N$ since the constant $1.001$ is loose. Combining with Theorem \ref{thm:gen_7.12} completes the proof. The $1-6m/N$ probability in Theorem \ref{thm:gen_7.10} results in a small proportional decrease in the entropy. For $m<10^6$ and $n>50$, this results in a correction smaller than $10^{-8} m$ and can be absorbed into constants of correction terms since these corrections are loose.
\end{proof}

\subsection{The analysis of a general device}

We now show that algorithms with oracle access to Haar random circuits and solve $\mathrm{MXHOG}$ must have entropy. We first provide a generalization of Lemma 3.13 of Ref.~\cite{aaronson2023certifiedArxiv} while fixing some technical errors.
\begin{lemma}[Generalization of Lemma 3.13 of Ref.~\cite{aaronson2023certifiedArxiv}]\label{lem:gen_3.13}
\begin{equation}\underset{\vec{P}\sim\mathrm{Dir}(1^N)^{\otimes m}}{\mathbb{E}}\left[\frac{1}{m}\max_{\vec{z}}\sum_{i\in[m]}P^{(i)}(z^{(i)})\right]\leq \frac{2}{N}(n\ln 2 + 3)\label{eqn:3.13_second},
\end{equation}
where $P^{(i)}(z)$ is the probability of sampling $z$ from $P^{(i)}$.
\end{lemma}
\begin{proof}
\begin{align}\nonumber
    & \underset{P\sim\mathrm{Dir}(1^N)}{\mathbb{E}}\left[\max_z P(z)\right]\\\nonumber
    = & \underset{Q\sim\Gamma(1,1)^N}{\mathbb{E}}\left[\frac{\max Q}{\bar{Q}}\right]\\\nonumber
    = & \underset{Q\sim\Gamma(1,1)^N}{\mathbb{E}}\left[\frac{\max Q}{\bar{Q}}\Big\vert\bar{Q} \geq N/2\right]\Pr[\bar{Q} \geq N/2]+\underset{Q\sim\Gamma(1,1)^N}{\mathbb{E}}\left[\frac{\max Q}{\bar{Q}}\Big\vert \bar{Q} < N/2\right]\Pr[\bar{Q} < N/2]\\\nonumber
    \leq & \frac{2}{N}\underset{Q\sim\Gamma(1,1)^N}{\mathbb{E}}\left[\max Q\Big\vert\bar{Q} \geq N/2\right]\Pr[\bar{Q} \geq N/2]+\Pr[\bar{Q} < N/2]\\\nonumber
    \leq &  \frac{2}{N}\underset{Q\sim\Gamma(1,1)^N}{\mathbb{E}}\left[\max Q\right]+\Pr[\bar{Q} < N/2]\\
    \leq & \frac{2}{N}(\ln N + 1) + \frac{4}{N}= \frac{2}{N}(n\ln 2+3)\label{eqn:max_prob_out_of_m_distributions},
\end{align}
where $Q$ is a list of $N$ independent random variables drawn from $\Gamma(1,1)$, $\bar{Q}$ is the sum of all values in $Q$,
$\mathbb{E}[\max(Q_1,\dots,Q_m)]$ is from Eq. 31 of Ref.~\cite{aaronson2023certifiedArxiv}, $\Pr[\bar{Q} < N/2]$ is from Eq. 32 of Ref.~\cite{aaronson2023certifiedArxiv}, 
and second last line uses the fact that
\begin{equation}
    \mathbb{E}[X\vert A]\Pr[A]=\mathbb{E}[X]-\mathbb{E}[X\vert \neg A](1-\Pr[A])\rightarrow \mathbb{E}[X\vert A]\Pr[A] \leq \mathbb{E}[X].
\end{equation}
Using the fact that
\begin{equation}
    \frac{1}{m}\max_{\vec{z}}\sum_{i\in[m]}P^{(i)}(z^{(i)})=\frac{1}{m}\sum_{i\in[m]}\max_{z^{(i)}}P^{(i)}(z^{(i)}),
\end{equation}
we have
\begin{equation}
    \underset{\vec{P}\sim\mathrm{Dir}(1^N)^{\otimes m}}{\mathbb{E}}\left[\max_{\vec{z}}\sum_{i\in[m]}P^{(i)}(z^{(i)})\right]=\underset{\vec{P}\sim\mathrm{Dir}(1^N)^{\otimes m}}{\mathbb{E}}\left[\sum_{i\in[m]}\max_z P^{(i)}(z)\right]=\sum_{i\in[m]}\underset{P^{(i)}\sim\mathrm{Dir}(1^N)}{\mathbb{E}}\left[\max_z P^{(i)}(z)\right]\leq \frac{2m}{N}(n\ln 2+3),
\end{equation}
where the last inequality uses Eq. \ref{eqn:max_prob_out_of_m_distributions}
\end{proof}

\begin{theorem}[Generalization of Theorem 7.14 of Ref.~\cite{aaronson2023certifiedArxiv}]\label{thm:gen_7.14}
For every $(T,m)$-query device $\mathcal{A}$ that solves $b-\mathrm{MXHOG}_m$, there exists a device $\mathcal{F}$ given access to $(k+\ell)$ samples for each of the $m$ circuits such that $\mathcal{F}$ solves $b'-\mathrm{MXHOG}_m$ for
\begin{equation}
    b'=b-2m\varepsilon'(n\ln 2 + 3)
\end{equation}
and $\varepsilon'=T\left(\frac{4}{2^{n/2}} + \sqrt{2\left(1-\left(\cos\frac{\pi}{2k}\right)^k\right)}+2/\sqrt{\ell+1}\right)$. Similarly, for every $\mathcal{A}$ that cannot solve $b-\mathrm{MXHOG}_m$, there exists a device $\mathcal{F}$ that cannot solve $b'-\mathrm{MXHOG}_m$ for
\begin{equation}
    b'=b+2m\varepsilon'(n\ln 2 + 3).
\end{equation}
\end{theorem}
\begin{proof}
The argument mirrors that of Theorem 7.14 of Ref.~\cite{aaronson2023certifiedArxiv}. For every $\mathcal{A}^{\vec{P}}$, by Theorem \ref{thm:gen_7.8}, there exists $\mathcal{F}^{\vec{P}}$ such that
\begin{align}\nonumber
\left\vert\underset{\vec{z}\sim\bar{\mathcal{A}}^{\vec{P}}(\rho)}{\mathbb{E}}\left[\frac{1}{m}\sum_{i\in[m]}P^{(i)}(z^{(i)})\right]-\underset{\vec{z}\sim\bar{\mathcal{F}}^{\vec{P}}(\rho)}{\mathbb{E}}\left[\frac{1}{m}\sum_{i\in[m]}P^{(i)}(z^{(i)})\right]\right\vert
&=\left\vert\frac{1}{m}\sum_{\vec{z}}\left(p_{\mathcal{A}}(\vec{z})-p_{\mathcal{F}}(\vec{z})\right)\cdot\sum_{i\in[m]}P^{(i)}(z^{(i)})\right\vert\\\nonumber
&\leq \frac{1}{m}\left(\max_{\vec{z}}\sum_{i\in[m]}P^{(i)}(z^{(i)})\right)\sum_{\vec{z}}\left\vert p_{\mathcal{A}}(\vec{z})-p_{\mathcal{F}}(\vec{z})\right\vert\\\nonumber
&\leq \frac{1}{m}\left(\max_{\vec{z}}\sum_{i\in[m]}P^{(i)}(z^{(i)})\right)\|\bar{\mathcal{A}}^{\vec{P}}(\rho)-\bar{\mathcal{F}}^{\vec{P}}(\rho)\|_{\mathrm{tr}}\\
&\leq m\varepsilon'\frac{1}{m}\max_{\vec{z}}\sum_{i\in[m]}P^{(i)}(z^{(i)}),
\end{align}
where $\vec{P}(\vec{z})$ is the probability of sampling $\vec{z}$ from $\vec{P}$. Taking the expectation over $\vec{P}\sim\mathrm{Dir}(1^N)^{\otimes m}$, the upper bound is at most $m\varepsilon'\frac{2}{N}(n\ln 2 + 3)$ by Lemma \ref{lem:gen_3.13}.
\end{proof}

\begin{lemma}[Lemma 2 of Ref.~\cite{winter2016tight}, rephrased]\label{lem:continuity_of_entropy}
For states $\rho$ and $\sigma$ on Hilbert space $A\otimes B$, if $\frac{1}{2}\|\rho-\sigma\|_{1}\leq\varepsilon\leq 1$ and $A$ is classical, then
\begin{equation}
    \vert S(A\vert B)_\rho - S(A\vert B)_\sigma \vert \leq \varepsilon\log_2\vert A\vert + (1+\varepsilon)h\left(\frac{\varepsilon}{1+\varepsilon}\right).
\end{equation}
\end{lemma}

\begin{theorem}[Generalization of Theorem 7.15 of Ref.~\cite{aaronson2023certifiedArxiv}]\label{thm:gen_7.15}
Let $\mathcal{A}$ be any algorithm making $(T,m)$-queries to $\vec{U}\sim\mathrm{Haar}^{\otimes m}$. If $n\geq 50$, $k\leq N^{-1/2}$, $m,n\leq 10^6$, and $\mathcal{A}$ solves $(1+\delta)-\mathrm{MXHOG}_m$, then
\begin{equation}
    H(\vec{Z}\vert\vec{U}E)_{\mathcal{A}^{\vec{U}}}\geq
    m\left[\left(\delta-2m\varepsilon'(n\ln 2 + 3.5)-1.001\times\frac{16(k+\ell)^2\ln^2 N}{N}\right)n-\log_2 n -\log_2(k+\ell) - 3\right]-2
\end{equation}
and $\varepsilon'=T\left(\frac{4}{2^{n/2}} + \sqrt{2\left(1-\left(\cos\frac{\pi}{2k}\right)^k\right)}+2/\sqrt{\ell+1}\right)$.
\end{theorem}
\begin{proof}
We follow similar arguments in \cite{aaronson2023certifiedArxiv}. By Theorem \ref{thm:gen_7.8}, for every $(T,m)$-query $\mathcal{A}$, there exists $\mathcal{F}$ given access to $(k+\ell)$ samples for each circuit (drawn according the probability distribution induced by the circuit) such that $\|\bar{\mathcal{A}}^{\vec{P}}-\bar{\mathcal{F}}^{\vec{P}}\|_\diamond\leq m\varepsilon'$. Denote the output state of $\bar{\mathcal{F}}$ as $\Psi$. Also, by Theorem \ref{thm:gen_7.14}, $\mathcal{F}$ solves $b-\mathrm{MXHOG}_m$ for $b=1+\delta-2m\varepsilon'(n\ln 2 + 3)$. Combining Theorem \ref{thm:gen_7.14}, Corollary \ref{cor:gen_7.13} and Theorem \ref{thm:entropy_conditioned_on_circuits_equal_distributions}, this yields
\begin{equation}
H(\vec{Z}\vert\vec{U}E)_\Psi\geq
m\left[\left(\delta-2m\varepsilon'(n\ln 2 + 3)-1.001\times\frac{16(k+\ell)^2\ln^2 N}{N}\right)n-\log_2 n -\log_2(k+\ell) - 3\right].
\end{equation}
This provides a bound on a state $\Psi$ close to the actual state of the algorithm $\mathcal{A}^{\vec{U}}$. We can now use the continuity bound of entropy in Lemma \ref{lem:continuity_of_entropy} to bound the entropy of the actual state. The dimension of the first system is $2^{mn}$, so entropy is reduced by $m\varepsilon'mn+(1+m\varepsilon')h\left(\frac{m\varepsilon'}{1+\varepsilon'}\right)\leq m\varepsilon'mn+2$, which completes the proof.
\end{proof}

\subsection{Further improvement in the $\log (k+\ell)$ entropy penalty}\label{sec:one_query_one_sample}

The $\log (k+\ell)$ penalty in the entropy is unnatural. Specifically, the oracle is only capable of outputting a single $n$-qubit state, and the $k+\ell$ samples are not utilized in the simulation of the oracles in a very exploitative way (e.g. only having a cyclic shift applied to it). It is not actually possible for the adversary to read the bitstrings and select a specific sample to minimize the entropy.

To see this, we sketch the following intuitive argument. Having oracle access to a unitary $V$ is equivalent to having access to $fU$, where $f$ is a random one-to-one function that maps all bitstrings to all bitstrings and $U=f^{-1}V$. Further, having access to $U$ is equivalent to having access to $k+\ell$ samples from $p_U$, which is equivalent to having access to $k+\ell$ samples of $f^{-1}(z_i)$ for $z_i\sim p_V$. The random function means that the sample access algorithm receives uniformly random bitstrings which will be inversely mapped back to $z_i$. The algorithm cannot select the bitstring to output to minimize the entropy, since the unknown function $f$ performs the inverse mapping.

Note that in the above argument, it is not the case that the client chooses some $f$ and hides from the adversary and provide it with the blackbox access. We are saying that the oracle access to $V$ is equivalent to having oracle access to random $f$ and use $f$ and oracle access to $U$ in this specific form.

To sketch out this formally, we consider the special case of $m=1$ and $T=1$, which is analytically more tractable. We believe the argument is also applicable to general cases, although the proof would be more involved due to the complexities introduced by correlated random phases applied to the unitaries as well as the fact that the oracles may not be implemented in parallel.

\begin{theorem}\label{thm:oracle_close_to_1_access}
For any $P$ and any algorithm $\mathcal{A}$ making $(1,1)$-query to $C^{P}$, there exists a quantum algorithm $\mathcal{B}$ given one sample drawn from $P$ such that
\begin{equation}
\|\bar{\mathcal{A}}^{P}-\bar{\mathcal{B}}^{P}\|_\diamond\leq \frac{4}{2^{n/2}}+\sqrt{2(1-(\cos\frac{\pi}{2k})^k)}+2/\sqrt{\ell+1}+\frac{2(k+\ell)(k+\ell-1)}{N}+\frac{2k}{N(1-k/N)},
\end{equation}
where $\bar{\mathcal{A}}^{P}=\mathbb{E}_{C^{P}}[\mathcal{A}^{C^{P}}]$ and $\bar{\mathcal{B}}=\underset{z\sim P}{\mathbb{E}}[\mathcal{B}(z)]$.
\end{theorem}

\begin{proof}
For $V$ such that $V\vert 0\rangle=\vert P\rangle$,
\begin{align}\nonumber
\bar{\mathcal{A}}^{P}&=\mathbb{E}_{C^{P}}[\mathcal{A}^{C^{P}}]=\underset{W,C'}{\mathbb{E}}\left[\mathcal{A}^{WVC'}\right]=\underset{W_1,W_2,C',f}{\mathbb{E}}\left[\mathcal{A}^{W_1ff^{-1}W_2VC'}\right]=\underset{W_1,W_2,C',f}{\mathbb{E}}\left[\mathcal{A}^{W_1f(f^{-1}W_2f)f^{-1}VC'}\right]\\
&=\underset{W_1,W_2,C',f}{\mathbb{E}}\left[\mathcal{A}^{W_1fW_2(f^{-1}V)C'}\right]=\underset{W_1,W_2,C',f}{\mathbb{E}}\left[\mathcal{A}^{W_1f(W_2U_fC')}\right],
\end{align}
where the first equality is the definition, the second equality uses the definition of how $C^P$ is sampled as described above Theorem \ref{thm:gen_7.8} where $W$ is the random phase and $C'$ is a random unitary for which $\vert 0\rangle$ is stationary, the third equality averages over a uniformly random function $f$ that permutes all bitstrings and holds due to the fact that applying one random phase is equivalent to applying two random phases, the fifth equality uses the fact that the measure of random phases does not change when conjugated by a random permutation $f$, and the last equality uses newly defined unitaries $U_f=f^{-1}V$.

Since $W_2U_fC'$ is a trivial quantum algorithm calling $W_2U_fC'$, by Theorem \ref{thm:gen_7.8}, there exists a quantum channel $\Xi$ given $(k+\ell)$ samples drawn from the bitstring probability distributions $P_{U_f}$ of $U_f\vert 0\rangle$ such that
\begin{align}
\left\|\underset{W_2,C'}{\mathbb{E}}[W_2U_fC']-\bar{\Xi}^{P_{U_f}}\right\|_\diamond\leq \frac{4}{2^{n/2}} + \sqrt{2\left(1-\left(\cos\frac{\pi}{2k}\right)^k\right)}+2/\sqrt{\ell+1},\label{eqn:distance_oracle_sim_Xi}
\end{align}
where $\bar{\Xi}^{P_{U_f}}=\underset{z_1,\dots,z_k\sim p_U}{\mathbb{E}}[\Xi(z_1,\dots,z_k)]$.
To clarify what kind of mathematical object $\Xi$ is, since $W_2 U_f C':\mathcal{L}(\mathcal{H})\rightarrow\mathcal{L}(\mathcal{H})$ is a quantum channel on $n$-qubits, where $\mathcal{H}$ is the $n$-qubit Hilbert space and $\mathcal{L}$ represents the set of bounded linear operators, we know $\bar{\Xi}^{P_{U_f}}:\mathcal{L}(\mathcal{H})\rightarrow\mathcal{L}(\mathcal{H})$ and $\Xi(z_1,\dots,z_k):\mathcal{L}(\mathcal{H})\rightarrow\mathcal{L}(\mathcal{H})$. More explicitly, $\bar{\Xi}^{P_{U_f}}$ internally sample $z_1,\dots,z_k\sim p_{U_f}$, which controls the action of $\Xi$.

Having access to samples from $P_{U_f}$ is equivalent to having access to samples from $P_V$ followed by permutation $f^{-1}$. Therefore, $\bar{\Xi}$ can be modeled by a quantum channel
\begin{align}
\bar{\Xi}^{P_{U_f}}=\underset{z_1,\dots,z_k\sim P_V}{\mathbb{E}}\left[\Xi(f^{-1}(z_1),\dots,f^{-1}(z_k))\right].
\end{align}
Therefore,
\begin{align}
\underset{W_1,f}{\mathbb{E}}\left[W_1\circ f\circ\bar{\Xi}^{P_{U_f}}\right]=\underset{z_1,\dots,z_k\sim P_V}{\mathbb{E}}\left[\underset{W_1,f}{\mathbb{E}}\left[W_1\circ f\circ\Xi(f^{-1}(z_1),\dots,f^{-1}(z_k))\right]\right],
\end{align}
where $f$ now also means the unitary or quantum channel on $n$-qubit quantum states that perform the remapping $f$ of the bitstrings basis states. The above expression is a quantum channel for $n$-qubit states. Denote $\bar{\Xi}^{P_{U_f}}$ conditioned on the $k$ samples not repeating (which we call event $R^c$ where $R$ is the event of repeating and the superscript $c$ indicates the complement of the set) as $\bar{\Xi}^{P_{U_f}}_{R^c}$, we have
\begin{align}
\left\|\underset{W_1,f}{\mathbb{E}}\left[W_1\circ f\circ\bar{\Xi}^{P_{U_f}}\right]-\underset{W_1,f}{\mathbb{E}}\left[W_1\circ f\circ\bar{\Xi}^{P_{U_f}}_{R^c}\right]\right\|_\diamond\leq \Pr[R].\label{eqn:distance_repeat_and_no_restriction}
\end{align}

We now analyze the effect of the channel $\underset{f}{\mathbb{E}}\left[f\circ\bar{\Xi}^{P_{U_f}}_{R^c}\right]$. To do this, we focus on the special case where the input quantum state to the channel is an $n$-qubit state not entangled with any other quantum systems. In this case, the output is an $n$-qubit state which depends on the input state. We denote the output of $\underset{f}{\mathbb{E}}\left[f\circ\bar{\Xi}^{P_{U_f}}\right]$ as $\rho$ and that of $\underset{f}{\mathbb{E}}\left[f\circ\bar{\Xi}^{P_{U_f}}_{R^c}\right]$ as $\rho_{R^c}$. We have
\begin{align}\nonumber
\rho&=\underset{z_1,\dots,z_k}{\mathbb{E}}[\rho_{z_1,\dots,z_k}]\\
\rho&=\Pr[R^c]\cdot\rho_{R^c}+\Pr[R]\cdot\rho_R\\
\rho_{R^c}&\equiv\underset{(z_1,\dots,z_k)\in B}{\mathbb{E}}\left[\rho_{(z_1,\dots,z_k)}\right]\\
\rho_R&\equiv\underset{\{z_1,\dots,z_k\}\in C}{\mathbb{E}}\left[\rho_{z_1,\dots,z_k}\right]\\
\rho_{(z_1,\dots,z_k)}&\equiv\frac{1}{k!}\sum_{\{s_i\}_{i\in[k]}\in S((z_1,\dots,z_k))}\rho_{s_1,\dots,s_k}\\
&=\frac{1}{k!}\sum_{\pi\in\Pi(z_1,\dots,z_k)}\sum_f\Pr[f]\cdot f(\Xi(\{f^{-1}(\pi(z_i))\}_{i\in[k]})),
\end{align}
where $B$ is the set of all unordered sets of $k$ unique bitstrings, $C$ is the set of all ordered lists of $k$ bitstrings with repeated entries, $S((z_1,\dots,z_k))$ is the set of all possible reorderings of the $k$ non-repeating bitstrings in $(z_1,\dots,z_k)$, and $\Pi(z_1,\dots,z_k)$ is the set of all functions that map the set of bitstrings in $(z_1,\dots,z_k)$ onto itself and act trivially on all other bitstrings. We use $(\cdots)$ to denote an unordered set and $\{\cdots\}$ to denote an ordered list.

For each $\pi$ and $f$, we can define $f'$ such that $f^{-1}\circ\pi=f'^{-1}$, which implies $f=\pi\circ f'$.
\begin{align}
\rho_{(z_1,\dots,z_k)}&=\frac{1}{k!}\sum_{\pi\in\Pi(z_1,\dots,z_k)}\sum_{f'}\Pr[f']\cdot \pi (f'(\Xi(\{f'^{-1}(z_i)\}_{i\in[k]})))\\
&=\frac{1}{k!}\sum_{f}\Pr[f]\sum_{\pi\in\Pi(z_1,\dots,z_k)}\pi (f(\Xi(\{f^{-1}(z_i)\}_{i\in[k]}))).\label{eqn:symmetry_under_pi}
\end{align}
For a generic state
\begin{align}
\Xi(\{s_i\}_{i\in[k]})&=\sum_{\alpha,\beta}\vert \alpha\rangle\langle \beta\vert\xi_{\{s_i\}_{i\in[k]}}(\alpha,\beta),
\end{align}
we can write
\begin{align}
\langle x\vert \rho_{(z_1,\dots,z_k)}\vert y\rangle&=\frac{1}{k!}\underset{(s_1,\dots,s_k)}{\mathbb{E}}\left[\underset{f:\{f^{-1}(z_i)\}_{i\in[k]}=\{s_i\}_{i\in[k]}}{\mathbb{E}}\left[\sum_{\pi\in\Pi(z_1,\dots,z_k)}\sum_{\alpha,\beta}\xi_{\{s_i\}_{i\in[k]}}(\alpha,\beta)\langle x\vert \pi f(\alpha)\rangle\langle \pi f(\beta)\vert y\rangle\right]\right],\label{eqn:symmetry_under_f}
\end{align}
where $(s_1,\dots,s_k)$ are uniformly sampled sets of non-repeating $k$ bitstrings. We can write the sum this way because enumeration over all $f$ means that all possible sets of non-repeating samples $(s_1,\dots,s_k)$ appear. Further,
\begin{align}\nonumber
&\sum_{\pi\in\Pi(z_1,\dots,z_k)}\sum_{\alpha,\beta}\xi_{\{s_i\}_{i\in[k]}}(\alpha,\beta)\langle x\vert \pi f(\alpha)\rangle\langle \pi f(\beta)\vert y\rangle\\
=&\sum_{\pi\in\Pi(z_1,\dots,z_k)}\xi_{\{s_i\}_{i\in[k]}}(f^{-1}\pi^{-1}(x),f^{-1}\pi^{-1}(y)).
\end{align}
In the case where $x,y\in\{z_i\}_{i\in[k]}$ and $x=y$, we have $\pi^{-1}(x)\in\{z_i\}_{i\in[k]}$ and $f^{-1}\pi^{-1}(x)\in\{s_i\}_{i\in[k]}$. Summing over all $\pi$ means each $\{s_i\}_{i\in[k]}$ appears the same number of times. Therefore,
\begin{align}
\sum_{\pi\in\Pi(z_1,\dots,z_k)}\xi_{\{s_i\}_{i\in[k]}}(f^{-1}\pi^{-1}(x),f^{-1}\pi^{-1}(y))=\frac{k!}{k}\sum_{i\in[k]}\xi_{\{s_i\}_{i\in[k]}}(s_i, s_i).
\end{align}
For $x\neq y$, summing over all $\pi$ means every possible pair by choosing from $\{s_i\}_{i\in[k]}$ appears the same number of times, except those equal pairs. Therefore,
\begin{align}
\sum_{\pi\in\Pi(z_1,\dots,z_k)}\xi_{\{s_i\}_{i\in[k]}}(f^{-1}\pi^{-1}(x),f^{-1}\pi^{-1}(y))=\frac{k!}{k(k-1)}\sum_{i,j\in[k], i\neq j}\xi_{\{s_i\}_{i\in[k]}}(s_i, s_j).
\end{align}
In the case where $x,y\notin\{z_i\}_{i\in[k]}$ and $x=y$, we have $\pi^{-1}(x)=x$ and $f^{-1}\pi^{-1}(x)\notin\{s_i\}_{i\in[k]}$. Averaging over all $f:\{f^{-1}(z_i)\}_{i\in[k]}=\{s_i\}_{i\in[k]}$, all samples not in $\{s_i\}_{i\in[k]}$ appears with equal probability. Therefore,
\begin{align}
\underset{f:\{f^{-1}(z_i)\}_{i\in[k]}=\{s_i\}_{i\in[k]}}{\mathbb{E}}\left[\sum_{\pi\in\Pi(z_1,\dots,z_k)}\xi_{\{s_i\}_{i\in[k]}}(f^{-1}\pi^{-1}(x),f^{-1}\pi^{-1}(y))\right]=\frac{k!}{N-k}\sum_{s\notin\{s_i\}_{i\in[k]}}\xi_{\{s_i\}_{i\in[k]}}(s, s).
\end{align}
For $x\neq y$, we have
\begin{align}
\underset{f:\{f^{-1}(z_i)\}_{i\in[k]}=\{s_i\}_{i\in[k]}}{\mathbb{E}}\left[\sum_{\pi\in\Pi(z_1,\dots,z_k)}\xi_{\{s_i\}_{i\in[k]}}(f^{-1}\pi^{-1}(x),f^{-1}\pi^{-1}(y))\right]=\frac{k!}{(N-k)(N-k-1)}\sum_{s_1,s_2\notin\{s_i\}_{i\in[k]},s_1\neq s_2}\xi_{\{s_i\}_{i\in[k]}}(s_1, s_2).
\end{align}
For $x\in\{z_i\}_{i\in[k]}$ and $y\notin\{z_i\}_{i\in[k]}$,
\begin{align}
\underset{f:\{f^{-1}(z_i)\}_{i\in[k]}=\{s_i\}_{i\in[k]}}{\mathbb{E}}\left[\sum_{\pi\in\Pi(z_1,\dots,z_k)}\xi_{\{s_i\}_{i\in[k]}}(f^{-1}\pi^{-1}(x),f^{-1}\pi^{-1}(y))\right]=\frac{k!}{k(N-k)}\sum_{s_1\in\{s_i\}_{i\in[k]},s_2\notin\{s_i\}_{i\in[k]}}\xi_{\{s_i\}_{i\in[k]}}(s_1, s_2),
\end{align}
and for $x\in\{z_i\}_{i\in[k]}$ and $y\notin\{z_i\}_{i\in[k]}$, the value is the complex conjugate.

Collecting all cases,
\begin{align}
\langle x\vert \rho_{(z_1,\dots,z_k)}\vert y\rangle=
\begin{cases}
\frac{1}{k}\underset{(s_1,\dots,s_k)}{\mathbb{E}}\left[\sum_{i\in[k]}\xi_{\{s_i\}_{i\in[k]}}(s_i, s_i)\right]&~\text{if}~x=y\wedge x\in\{z_i\}_{i\in[k]}\\
\frac{1}{k(k-1)}\underset{(s_1,\dots,s_k)}{\mathbb{E}}\left[\sum_{i,j\in[k], i\neq j}\xi_{\{s_i\}_{i\in[k]}}(s_i, s_j)\right]&~\text{if}~x\neq y\wedge x,y\in\{z_i\}_{i\in[k]}\\
\frac{1}{N-k}\underset{(s_1,\dots,s_k)}{\mathbb{E}}\left[\sum_{s\notin\{s_i\}_{i\in[k]}}\xi_{\{s_i\}_{i\in[k]}}(s, s)\right]&~\text{if}~x=y\wedge x,y\notin\{z_i\}_{i\in[k]}\\
\frac{1}{(N-k)(N-k-1)}\underset{(s_1,\dots,s_k)}{\mathbb{E}}\left[\sum_{s_1,s_2\notin\{s_i\}_{i\in[k]},s_1\neq s_2}\xi_{\{s_i\}_{i\in[k]}}(s_1, s_2)\right]&~\text{if}~x\neq y\wedge x,y\notin\{z_i\}_{i\in[k]}\\
\frac{1}{k(N-k)}\sum_{s_1\in\{s_i\}_{i\in[k]},s_2\notin\{s_i\}_{i\in[k]}}\xi_{\{s_i\}_{i\in[k]}}(s_1, s_2)&~\text{if}~x\in\{z_i\}_{i\in[k]},y\notin\{z_i\}_{i\in[k]}\\
\frac{1}{k(N-k)}\sum_{s_1\notin\{s_i\}_{i\in[k]},s_2\in\{s_i\}_{i\in[k]}}\xi_{\{s_i\}_{i\in[k]}}(s_1, s_2)&~\text{otherwise.}\\
\end{cases}
\end{align}
The special case of the classical state discussed before corresponds to the special case where the non-diagonal terms are zero. Additionally, the values of the matrix elements are independent of the samples $z_1,\dots,z_k$ although the positions of these terms obviously depend on the samples.

Then, averaging over all $(z_1,\dots,z_k)$, we have
\begin{align}
\rho_{R^c}=\underset{(z_1,\dots,z_k)}{\mathbb{E}}\left[\rho_{(z_1,\dots,z_k)}\right],
\end{align}
whose diagonal elements are
\begin{align}
\langle x\vert \rho_{R^c}\vert x\rangle=
\frac{u}{k}\cdot p^{R^c}_V(x)+
\frac{1-u}{N-k}\cdot\left(1-p^{R^c}_V(x)\right),
\end{align}
where we used the definitions
\begin{align}
p^{R_c}_V(x)&=\Pr_{(z_1,\dots,z_k)}[x\in(z_1,\dots,z_k)]\\
u=\underset{(s_1,\dots,s_k)}{\mathbb{E}}\left[\sum_{i\in[k]}\xi_{\{s_i\}_{i\in[k]}}(s_i, s_i)\right]&=1-\underset{(s_1,\dots,s_k)}{\mathbb{E}}\left[\sum_{s\notin\{s_i\}_{i\in[k]}}\xi_{\{s_i\}_{i\in[k]}}(s, s)\right].
\end{align}
Finally, averaging over all phases, the state $\underset{W_1}{\mathbb{E}}\left[W_1\rho_{R^c}W_1^\dagger\right]$ becomes classical and all non-diagonal elements vanish. Therefore, to prepare the state, we can simply perform the following sampling process: first, sample a random set of non-repeating $k$ bitstrings $(z_1,\dots,z_k)$ with probability proportional to $\prod_{i=1}^k p_V(z_i)$; then with probability $u$, sample a random bitstring from the set. Otherwise, sample a random bitstring not in the set. The first case is equivalent to drawing a random $z$ from the set conditioned on $R^c$, whose probability we denote as $\Pr[z\vert R^c]$. We denote the sampling processing corresponding to $\underset{W_1}{\mathbb{E}}\left[W_1\rho_{R^c}W_1^\dagger\right]$ as $\mathrm{Sim}_{R^c}$.

We now show that the probability distribution of $z$ for $\underset{W_1}{\mathbb{E}}\left[W_1\rho_{R^c}W_1^\dagger\right]$ is close to a mixture of one random sample from $p_V$ with probability $u$ and one random sample from the uniform distribution with probability $1-u$. Consider another sampling process $A$: sample $z_1,\dots,z_k\sim p_V$. With probability $u$, choose a random $i\in[k]$ and output $z_i$. Otherwise, output a random bitstring not in $\{z_1,\dots,z_k\}$. The probability of outputting $z$ is
\begin{align}
p_A(z)=\Pr[R^c]\cdot\Pr[z\vert R^c]+\Pr[R]\cdot\Pr[z\vert R],
\end{align}
so the total variation distance (TVD) between $A$ and $\mathrm{Sim}_{R^c}$ is at most $\Pr[R]$. Now consider another sampling process $B$: with probability $u$, output a random sample from $p_V$; otherwise, output a random bitstring from the uniform distribution. An equivalent process of $B$ is as follows: sample $z_1,\dots,z_k\sim p_V$. With probability $u$, choose a random $i\in[k]$ and output $z_i$. Otherwise, output a random bitstring from the uniform distribution. Only the second case differs from the sampling process $A$. Denote the second case as $S$, the probability distributions for this case is
\begin{align}
p_A(z\vert S)&=\frac{1}{N-k}\Pr_{z_1,\dots,z_k}[\{z_1,\dots,z_k\}\not\ni z]\\
p_B(z\vert S)&=\frac{1}{N}=\frac{1}{N}\Pr_{z_1,\dots,z_k}[\{z_1,\dots,z_k\}\not\ni z]+\frac{1}{N}\Pr_{z_1,\dots,z_k}[\{z_1,\dots,z_k\}\ni z]\\\nonumber
\big\vert p_B(z\vert S)-p_A(z\vert S)\big\vert&\leq \frac{1}{N}\left(\frac{1}{1-k/N}-1\right)\Pr_{z_1,\dots,z_k}[\{z_1,\dots,z_k\}\not\ni z]+\frac{1}{N}\Pr_{z_1,\dots,z_k}[\{z_1,\dots,z_k\}\ni z]\\
&\leq\frac{1}{N}\left(\frac{1}{1-k/N}-1\right)+\frac{1}{N}\cdot k\cdot p_V(z).
\end{align}
Therefore,
\begin{align}\nonumber
\mathrm{TVD}(A,B) &= (1-u)\sum_z\left[\big\vert p_B(z\vert S)-p_A(z\vert S)\big\vert\right]\\\nonumber
&\leq (1-u)\sum_z\left[\frac{1}{N}\left(\frac{1}{1-k/N}-1\right)+\frac{1}{N}\cdot k\cdot p_V(z)\right]\\\nonumber
&=(1-u)\left[\left(\frac{1}{1-k/N}-1\right)+k/N\right]\\
&\leq\frac{2k}{N(1-k/N)}.
\end{align}
Therefore, the total TVD between $B$ and $\mathrm{Sim}_{R^c}$ is
\begin{align}
\mathrm{TVD}(\mathrm{Sim}_{R^c},B)\leq \Pr[R]+\frac{2k}{N(1-k/N)}.
\end{align}

Although we only analyzed the special case where the input to the channel is an $n$-qubit state unentangled with any other quantum systems, the result shows that the action of the channel is close to sampling $z$ from $B$, where the probability $u$ depends on the input quantum state. The result can be immediately extended to the case where the input to the channel is entangled to a quantum register held by an adversary. Notice that while the channel $\bar{\Xi}^{P_{U_f}}$ and the channel $f$ acts coherently, the channel $W_1$ will apply a random phase on the state, thus breaking any entanglement between the oracle's output and the adversary's quantum register. In other words, the state shared between the oracle's output and the adversary's side information can only be correlated classically, which implies that it is optimal to employ shared randomness to prepare this state. Therefore, our earlier claim also covers the case where the adversary prepares quantum states that are initially entangled. Denoting the quantum channel that corresponds to the sampling process $B$ which has access to one random sample from $p_V$ as $\bar{\mathcal{B}}=\underset{z\sim p_V}{\mathbb{E}}[\mathcal{B}(z)]$,
\begin{align}
\left\|\underset{W_1,f}{\mathbb{E}}\left[W_1\circ f\circ\bar{\Xi}^{P_{U_f}}_{R^c}\right]-\bar{\mathcal{B}}\right\|_\diamond\leq\Pr[R]+\frac{2k}{N(1-k/N)}.
\end{align}
Using Eq. \ref{eqn:distance_repeat_and_no_restriction},
\begin{align}
\left\|\underset{W_1,f}{\mathbb{E}}\left[W_1\circ f\circ\bar{\Xi}^{P_{U_f}}\right]-\bar{\mathcal{B}}\right\|_\diamond\leq2\Pr[R]+\frac{2k}{N(1-k/N)}.
\end{align}

Finally, using Eq. \ref{eqn:distance_oracle_sim_Xi},
\begin{align}
\left\|\underset{W_1,W_2,C',f}{\mathbb{E}}\left[W_1 f W_2 U_f C'\right]-\bar{\mathcal{B}}\right\|_\diamond\leq\frac{4}{2^{n/2}}+\sqrt{2(1-(\cos\frac{\pi}{2k})^k)}+2/\sqrt{\ell+1}+\frac{2(k+\ell)(k+\ell-1)}{N}+\frac{2k}{N(1-k/N)},
\end{align}
where we used the fact that
\begin{align}
\Pr[R]=1-\Pr_{m\sim\Phi(N,k+\ell)}[\|m\|_\infty=1]\leq\frac{(k+\ell)(k+\ell-1)}{N}.
\end{align}
Specifically, $\Pr_{m\sim\Phi(N,k+\ell)}[\|m\|_\infty=1]$ is the probability that a random frequency vector sampled uniformly from all possible frequency vectors has infinite norm $1$ as defined in the proof of Theorem \ref{thm:gen_7.10} and substituting $k\rightarrow k+\ell$, which is the same as the probability of obtaining non-repeating samples averaged over samples and distributions $P$ from the Dirichlet distribution.

\end{proof}

As a corollary, we can obtain an improved entropy bound for the general adversary for the special case of $m=1$ and $T=1$.
\begin{corollary}\label{cor:improved_general_adv_m1_T1}
Let $\mathcal{A}$ be any algorithm making $(1,1)$-queries to $U\sim\mathrm{Haar}(2^n)$. If $n\geq 50$, $k\leq N^{-1/2}$, $n\leq 10^6$, and $\mathcal{A}$ solves $(1+\delta)-\mathrm{XHOG}_m$, then
\begin{equation}
    H(Z\vert UE)_{\mathcal{A}^{U}}\geq
    \left(\delta-2m\varepsilon''(n\ln 2 + 3.5)-1.001\times\frac{16^2\ln^2 N}{N}\right)n-\log_2 n - 5
\end{equation}
and $\varepsilon''=\frac{4}{2^{n/2}} + \sqrt{2\left(1-\left(\cos\frac{\pi}{2k}\right)^k\right)}+2/\sqrt{\ell+1}+\frac{2(k+\ell)(k+\ell-1)}{N}+\frac{2k}{N(1-k/N)}$.
\end{corollary}
\begin{proof}
We follow similar arguments in \cite{aaronson2023certifiedArxiv}. By Theorem \ref{thm:oracle_close_to_1_access}, for every $(1,1)$-query $\mathcal{A}$, there exists $\mathcal{B}$ given access to one sample the circuit (drawn according the probability distribution induced by the circuit) such that $\|\bar{\mathcal{A}}^{P}-\bar{\mathcal{B}}^P\|_\diamond\leq \varepsilon''$. Denote the output state of $\bar{\mathcal{B}}$ as $\Psi$. Also, by similar arguments as Theorem \ref{thm:gen_7.14}, we can show that $\mathcal{B}$ solves $b-\mathrm{XHOG}_m$ for $b=1+\delta-2\varepsilon''(n\ln 2 + 3)$. Combining the aforementioned result, Corollary \ref{cor:gen_7.13} and Theorem \ref{thm:entropy_conditioned_on_circuits_equal_distributions}, this yields
\begin{equation}
H(Z\vert UE)_\Psi\geq
\left(\delta-2m\varepsilon''(n\ln 2 + 3)-1.001\times\frac{16\ln^2 N}{N}\right)n-\log_2 n - 3.
\end{equation}
This provides a bound on a state $\Psi$ close to the actual state of the algorithm $\mathcal{A}^{U}$. We can now use the continuity bound of entropy in Lemma \ref{lem:continuity_of_entropy} to bound the entropy of the actual state. The dimension of the first system is $2^{mn}$, so entropy is reduced by $\varepsilon''n+(1+\varepsilon'')h\left(\frac{\varepsilon''}{1+\varepsilon''}\right)\leq \varepsilon''n+2$, which completes the proof.
\end{proof}

\subsection{Summary of improvements}

Overall we make improvements in the correction terms of the single-round entropy. The improvements valid for general $T$ query adversaries are reducing $(k+\ell)^3\rightarrow (k+\ell)^2$ and $2\log_2(k+\ell)\rightarrow\log_2 (k+\ell)$. The improvements valid only for one-access adversaries are reducing $(k+\ell)^3\rightarrow (k+\ell)^2$ and $2\log_2(k+\ell)\rightarrow 0$. For the main text figure, the baseline we compare our one-access adversaries to is the theory of Ref.~\cite{aaronson2023certifiedArxiv} with explicit constants as shown in Theorem \ref{thm:original_theory}.

\subsection{Effect of classical simulation}\label{sec:combine_classical}

The above analysis shows the entropy guarantee of a quantum device with query access to the challenge circuits. However, since we are working with finite-size challenge circuits, the adversary may be able to simulate the circuits with a non-negligible fidelity, and we need to consider the classical simulation capabilities. Consider the case where the adversary simulates $m'$ out of $m$ circuits classically, and each simulated circuit is expected to solve $(1+\delta_j)-\mathrm{XHOG}$ over the Haar measure, meaning the $m'$ simulated circuits are expected to solve $(1+\delta')-\mathrm{MXHOG}_{m'}$, where $\delta'=\frac{1}{m'}\sum_{j=1}^{m'}\delta_j$. Note that this situation allows simulations for different circuits to be dependent on each other, although there are no known ways for this to be meaningfully helpful.

If all $m$ circuits, quantum and classical, solve $(1+\delta)-\mathrm{MXHOG}_m$ for $\delta > \delta'$, then the remaining $m''=m-m'$ quantum samples must solve $(1+\delta'')-\mathrm{MXHOG}_{m''}$ with
\begin{equation}
    \delta''=\frac{1}{1-m'/m}\left(\delta-\frac{m'}{m}\delta'\right)=\frac{\delta-F_{\rm C}}{1-m'/m},\label{eqn:xhog_score_of_quantum_part_of_hybrid}
\end{equation}
and $F_{\rm C}=\delta' m' / m$ can be interpreted as the average classical simulation fidelity per circuit. The entropy of the $m''=m-m'=m(1-m'/m)$ quantum samples, and therefore the entropy of all $m$ samples since classical samples contribute no entropy, is
\begin{align}\nonumber
    H(\vec{Z}\vert \vec{C}E)&\geq m''\left[\left(\delta''-2m\varepsilon'(n\ln 2 + 3.5)-1.001\frac{16(k+\ell)^2\ln^2N}{N}\right)n-\log_2 n -\log_2(k+\ell) - 3\right]-2\\\nonumber
    &=m(1-m'/m)\left[\left(\frac{\delta-F_{\rm C}}{1-m'/m}-2m\varepsilon'(n\ln 2 + 3.5)-1.001\frac{16(k+\ell)^2\ln^2N}{N}\right)n-\log_2 n -\log_2(k+\ell) - 3\right]-2\\
    &\geq m\left[\left(\delta-F_{C}-2m\varepsilon'(n\ln 2 + 3.5)-1.001\frac{16(k+\ell)^2\ln^2N}{N}\right)n-\log_2 n -\log_2(k+\ell) - 3\right]-2.\label{eqn:entropy_with_classical}
\end{align}

It is then sufficient for the client to upper bound $F_{\rm C}$. For example, for frugal rejection sampling based on contracting a fraction of slices of the tensor network corresponding to the probability amplitude, which is the finite-size adversary analyzed in \cite{jpmc_cr}, the average simulation fidelity is $F_{\rm C}=\frac{C_{\rm eff}T_{\rm batch}\mathcal{A}}{m\mathcal{B}}$, where $C_{\rm eff}$ is the numerical efficiency of the adversary, $T_{\rm batch}$ is the allowed time for the adversary to return all samples since receiving the batch of circuits, $\mathcal{A}$ is the adversary's classical computational power, and $\mathcal{B}$ is the cost of exactly simulating one circuit. For an adversary that computes $2^k$ amplitudes by contraction a fraction $f$ of slices, which we discuss in more detail in a later section, the fidelity is $F_{\rm C}=f(k-1)$. Similarly, given the total time per batch and the assumed adversarial computational power, one can in principle upper bound $F_{\rm C}$ for any classical simulation algorithm.

\section{Multi-round security under non-adaptive hybrid adversaries}\label{sec:ad_hoc}

In this section, we analyze an adversary whose output across rounds are independent, though not necessarily identical. For convenience, we only consider classical side information for both the multi-round output and single-round outputs. This is justified because, in settings where the randomness is used publicly, certified randomness is concerned only with entropy conditioned on an initial snapshot of classical information. This is the case for multi-party protocols, but also for randomness amplification where the quantum randomness is not assumed to be private. Conversely, in settings where the certified randomness needs to remain private, the client must have physical control over the quantum device, and any shared initial entanglement can be destroyed by simple measures such as powering down the quantum device.

To motivate the non-adaptive adversary, we observe that past output bitstrings do not aid future adversarial interactions since future challenge circuits are generated independently. The exception to this would be if the adversary computes the potential contribution to the XEB score, which may allow the adversary to use lower entropy samples if it knows the past samples are likely to do well on XEB. In the limit where the number of rounds is large, the fraction of verified circuits is small, and the classical simulation fidelity the adversary can achieve in the limited time is small, computing the effect on the XEB score has little effect. If the history is short, the probability that these samples will be selected for validation is small, which means the adversary should not change the strategy very much depending on the history. If the history is long, then with high probability the history should look `typical'. If the classical simulation fidelity is small, then the contribution to the XEB score cannot be accurately characterized.

\subsection{Preliminaries}\label{sec:preliminaries}
We first define some notations before proceeding with the formal analysis. A quantum state $\rho$ is a positive-semidefinite operator with $\Tr(\rho) = 1$. A multipartite state $\rho$ with a classical subsystem over register $X$ and a quantum subsystem over register $A$ is called a classical-quantum state, and has the form
\begin{equation}
\rho_{XA}=\sum_{x}p(x)\dyad{x}\otimes\rho_A^x,
\end{equation}
where $\sum_x p(x)=1$. One may also be interested in analyzing subnormalized classical-quantum states, which may arise when we condition a normalized state on some event $\Omega$ on the classical register $X$:
\begin{equation}
\rho_{XA\wedge\Omega}=\sum_{x\in\Omega}p(x)\dyad{x}\otimes\rho_A^x,
\end{equation}
where the probability of event $\Omega$ is $\Pr[\Omega]=\sum_{x\in\Omega}p(x)$. Once normalized, the above state is given by \begin{equation}
\rho_{XA|\Omega}=\Pr[\Omega]^{-1}\rho_{XA\wedge\Omega}.
\end{equation}

For cryptographic analysis, it is typical to consider the worst-case probability of an adversary guessing the outcome of a source of randomness. For a classical system, this corresponds to the probability of the mostly likely outcome, the negative logarithm of which is called the min-entropy. 

However, this definition is not applicable to quantum systems and requires generalization. For certified randomness, since the random outputs are always classical but the adversary side information might be quantum, we are interested in the min-entropy for a classical-quantum state which is discussed in Section 6.14 of Ref.~\cite{tomamichel2015quantum}.

\begin{definition}
Given a (potentially subnormalized) classical-quantum state $\rho_{XA}=\sum_{x}p(x)\dyad{x}\otimes\rho_A^x$, the conditional min-entropy $H_{\min}(X|A)_\rho$ is given by
\begin{equation}
    H_{\min}(X|A)_\rho=-\log_2 p_{\rm guess}(X|A)_\rho,
\end{equation}
where $p_{\rm guess}(X|A)_\rho:=\sup_{\{M_x\}_x}\sum_x p(x)\lVert\rho_A^xM_x\rVert_{1},$ with supremum over POVMs on register $A$, is known as the guessing probability.
\end{definition}

The min-entropy can be generalized by including a smoothing parameter.
\begin{definition}
    Given a classical-quantum state $\rho_{XA}$ and a smoothing parameter $\varepsilon\in(0,\sqrt{\lVert\rho_{XA}\rVert_{1}})$, the conditional $\varepsilon$-min-entropy is given by
\begin{equation}
     H_{\min}^{\varepsilon}(X|A)_{\rho}:=\sup_\sigma H_{\min}(X|A)_{\sigma}\,,
\end{equation}
where the supremum is over states $\sigma_{XA}$ in the $\varepsilon$-ball in purified distance centered around $\rho$ \cite{tomamichel2015quantum}.
\end{definition}

\subsection{Single-round statistical properties}

For the classical part of the adversary, we consider a specific class of adversaries whose probability distribution of the output bitstring probability amplitudes can be explicitly modeled. This allows us to compute the probability distribution of the XEB score of a protocol in Fig. \ref{fig:protocol_ad_hoc}.

We first obtain the following result.
\begin{figure*}
    \hrule
    \vspace{.5em}
    \begin{flushleft}\underline{Protocol Arguments:}\end{flushleft}
 \vspace{-1.5em}
 \begin{align*}
  n \in \mathbb{N}~&:~\text{Number of qubits}\\
  m \in \mathbb{N}~&:~\text{Number of circuits per round (batch size)}\\
  L \in \mathbb{N}~&:~\text{Number of rounds}\\
  T_{\rm batch}~&:~\text{Threshold on the time allowed for the samples to return}\\
  L_{\rm val} \in [L]~&~\text{Number of validation circuits}\\
  p_{\rm max}\in (1/2^n, 1]~&:~\text{Maximum contribution probability}\\
  s^*~&:~\text{Threshold for test score}\\
  \end{align*}
  \vspace{-2em}
  \begin{flushleft}\underline{The Protocol:}\end{flushleft}
  \begin{enumerate}
  \item Set $s=0$.
  \item For $i\in[L]$, do:
  \begin{enumerate}
      \item Sample $m$ Haar random unitaries.
      \item Send them to the server and start the timer.
      \item If does not receive the bitstrings in time $T_{\rm batch}$, abort. Otherwise, store the bitstrings $\vec{Z}_i$.
  \end{enumerate}
  \item Sample validation sets of batch indices $i$ uniformly from all possible size $L_{\rm val}$ subsets of $[L]$.
  \item For each $i\in\mathcal{V}$, uniformly sample one of the $m$ circuit and bitstring. Compute the bitstring probability $p$ for the chosen validation circuit of that batch, and add $\frac{N\min(p, p_{\rm max})}{L_{\rm val}}$ to $s$. Finally, abort if $s<s^*$.
  \end{enumerate}
  \hrule\vspace{1em}
  \caption{Description of the protocol for the ad hoc analysis.}
  \label{fig:protocol_ad_hoc}
  \end{figure*}

\begin{lemma}\label{lem:low_h_implies_low_delta}
    Let $\mathcal{A}$ be any algorithm making $(T,m)$-queries to $\vec{U}\sim\mathrm{Haar}^{\otimes m}$. If $n\geq 50, k\leq N^{-1/2}, m,n\leq 10^6$, and
    \begin{equation}
        H(\vec{Z}\vert\vec{U}E)_\Psi\leq h,
    \end{equation}
    then there exists a quantum algorithm $\mathcal{F}$ which, given $(k+\ell)$ samples drawn from $P^{(i)}$ for all $i\in[m]$, outputs a state $m\varepsilon'$-close to the output of $\mathcal{A}$ and cannot solve $(1+\delta)-\mathrm{MXHOG}_m$, where
    \begin{equation}
        \delta=\frac{1}{n}\left(\frac{h+2}{m}+\log_2 n+\log_2(k+\ell) + 3\right)+4m\varepsilon'(n\ln2+3.5)+1.001\times\frac{16(k+\ell)^2\ln^2N}{N},
    \end{equation}
    and $\varepsilon'$ is as defined in Theorem \ref{thm:gen_7.15}.
\end{lemma}
\begin{proof}
    Theorem \ref{thm:gen_7.15} implies that if entropy is at most $h$, then $\mathcal{A}$ cannot solve $(1+\delta')-\mathrm{MXHOG}$ with
    \begin{equation}
        \delta'=\frac{1}{n}\left(\frac{h+2}{m}+\log_2 n+\log_2(k+\ell) + 3\right)+2m\varepsilon'(n\ln2+3.5)+1.001\times\frac{16(k+\ell)^2\ln^2N}{N}.
    \end{equation}
Combining this with Theorem \ref{thm:gen_7.14} completes the proof.
\end{proof}

We also have a lemma similar to Theorem \ref{thm:gen_7.10}.
\begin{lemma}\label{lem:low_delta_implies_low_contain_prob}
For a quantum algorithm $\mathcal{F}$ given $k$ samples drawn from $P^{(i)}$ for all $i\in[m]$, if
\begin{equation}
    \underset{\vec{P}\sim\mathrm{Dir}(1^N)^{\otimes m}}{\mathbb{E}}\underset{\vec{z}\sim\bar{\mathcal{F}}^{\vec{P}}(\rho)}{\mathbb{E}}\left[\sum_{i\in[m]} P^{(i)}(z^{(i)})\right]\leq m \frac{2-\varepsilon}{N+k},
\end{equation}
then
\begin{equation}
    \frac{1}{m}\sum_{i\in[m]}\Pr\left[z^{(i)}\in\{z_j^{(i)}\}_{j\in[k]}\right]\leq 1-\varepsilon.
\end{equation}
\end{lemma}
\begin{proof}
    Similar to Theorem \ref{thm:gen_7.10},
    \begin{equation}
        \underset{\vec{z}\sim Q(\vec{f})}{\mathbb{E}}\underset{\vec{P}\sim\mathrm{Dir}(1^N+f^{(i)})}{\mathbb{E}}\left[P^{(i)}(z^{(i)})\right]=\underset{\vec{z}\sim Q(\vec{f})}{\mathbb{E}}\left[\frac{f_{z^{(i)}}^{(i)}+1}{N+k}\right]\geq \frac{1}{N+k}\left(1+\underset{\vec{z}\sim Q(\vec{f})}{\Pr}\left[f_{z^{(i)}}^{(i)}>0\right]\right).
    \end{equation}
    The score of $\mathcal{F}$ in expectation is
\begin{align}
    &\underset{\vec{f}\sim\Phi(N,k)^{\otimes m}}{\mathbb{E}}\underset{\vec{z}\sim Q(\vec{f})}{\mathbb{E}}\underset{\vec{P}\sim\mathrm{Dir}(1^N+\vec{f})}{\mathbb{E}}\left[\sum_{i\in[m]}P^{(i)}(z^{(i)})\right]\\
    =& \sum_{i\in[m]}\underset{\vec{f}\sim\Phi(N,k)^{\otimes m}}{\mathbb{E}}\underset{\vec{z}\sim Q(\vec{f})}{\mathbb{E}}\underset{\vec{P}\sim\mathrm{Dir}(1^N+\vec{f})}{\mathbb{E}}\left[P^{(i)}(z^{(i)})\right]\\
    \geq& \sum_{i\in[m]}\underset{\vec{f}\sim\Phi(N,k)^{\otimes m}}{\mathbb{E}} \frac{1}{N+k}\left(1+\underset{\vec{z}\sim Q(\vec{f})}{\Pr}\left[f_{z^{(i)}}^{(i)}>0\right]\right)\\
    =&\frac{1}{N+k}\left(m+\sum_{i\in[m]}\left(\underset{\vec{f}\sim\Phi(N,k)^{\otimes m},\vec{z}\sim Q(\vec{f})}{\Pr}\left[f^{(i)}_{z^{(i)}}>0\right]\right)\right).
\end{align}
Thus if the score is at most $m\frac{2-\varepsilon}{N+k}$,
\begin{equation}
    \frac{1}{m}\sum_{i\in[m]}\underset{\vec{f}\sim\Phi(N,k)^{\otimes m}\vec{z}\sim Q(\vec{f})}{\Pr}\left[f_{z^{(i)}}^{(i)}>0\right]\leq 1-\varepsilon.
\end{equation}
\end{proof}

Lemma \ref{lem:low_h_implies_low_delta} and \ref{lem:low_delta_implies_low_contain_prob} implies the following lemma.
\begin{lemma}\label{lem:low_h_implies_low_contain_p_prob}
    Let $\mathcal{A}$ be any algorithm making $(T,m)$-queries to $\vec{U}\sim\mathrm{Haar}^{\otimes m}$. If $n\geq 50, k\leq N^{-1/2}, m,n\leq 10^6$, and
    \begin{equation}
        H(\vec{Z}\vert\vec{U}E)_\Psi\leq h,
    \end{equation}
    then there exists a quantum algorithm $\mathcal{F}$ given $(k+\ell)$ samples drawn from $P^{(i)}$ for all $i\in[m]$ such that the output state is $m\varepsilon'$-close to the output of $\mathcal{A}$ and
    \begin{equation}
        \frac{1}{m}\sum_{i\in[m]}\Pr\left[z^{(i)}\in\{z_j^{(i)}\}_{j\in[k+\ell]}\right]\leq \delta,
    \end{equation}
    where
    \begin{align}
        \delta&=\frac{h+2}{mn}+C,\\
        C&=\frac{1}{n}\left(\log_2 n + \log_2(k+\ell) + 3\right)+4m\varepsilon'(n\ln2+3.5)+1.001\times\frac{16(k+\ell)^2\ln^2N}{N}+0.001.
    \end{align}
\end{lemma}
\begin{proof}
    Lemma \ref{lem:low_h_implies_low_delta} implies that $\mathcal{F}$ cannot solve $(1+\delta)-\mathrm{MXHOG}_m$, then
    \begin{equation}
        \underset{\vec{P}\sim\mathrm{Dir}(1^N)^{\otimes m}}{\mathbb{E}}\underset{\vec{z}\sim\bar{\mathcal{F}}^{\vec{P}}(\rho)}{\mathbb{E}}\left[\sum_{i\in[m]} P^{(i)}(z^{(i)})\right]\leq m \frac{2-\varepsilon}{N+k}=m\frac{1+\delta}{N}\rightarrow \varepsilon=2-\frac{N+k}{N}(1+\delta).
    \end{equation}
    By Lemma \ref{lem:low_delta_implies_low_contain_prob},
    \begin{equation}
        \frac{1}{m}\sum_{i\in[m]}\Pr\left[z^{(i)}\in\{z_j^{(i)}\}_{j\in[k+\ell]}\right]\leq 1-\varepsilon=\delta+\frac{k}{N}(1+\delta)\leq \delta+10^{-22}.
    \end{equation}
    The negligible constant is due to the fact that $n\geq 50$ and $k\leq N^{-1/2}$, and can be safely absorbed into the constant $0.001$ in $\delta$.
\end{proof}

Let us consider the simple case where the probability amplitudes for all rounds are independent. Consider the case where the bitstring probability distributions for all rounds are fixed except the $i$th round. Recall that to the adversary, higher XHOG score is better. If the $i$th round has two possible bitstring probability distributions, it is always more favorable for the adversary to choose the \textit{first-order stochastically dominant} probability distribution.

\begin{definition}
A random variable $A$ on $\mathbb{R}$ has \textit{first-order stochastic dominance} over a random variable $B$ on $\mathbb{R}$ if $F_B(x)\geq F_A(x)$ for all $x$ and $F_B(x) > F_A(x)$ for some $x$, where $F_A(x)=\Pr[A\leq x]$ and $F_B(x)=\Pr[B\leq x]$ are the cumulative density functions of $A$ and $B$. For probability distributions $\rho_A$ of $A$ and $\rho_B$ of $B$, the dominance of random variable $A$ over random variable $B$ is denoted as $\rho_A\succeq\rho_B$.
\end{definition}

The following lemma will be useful to analyze the cumulative density functions of the bitstring probabilities of various processes we will discuss below.
\begin{lemma}\label{lem:cdf_of_mixture}
    The cumulative density function of the following process:
    \begin{enumerate}
        \item With probability $1-q$, sample $Q\sim\Gamma(1, 1)$.
        \item Otherwise, sample $Q\sim\Gamma(2,1)$,
    \end{enumerate}
    has the following cumulative density function:
    \begin{equation}
        F(x) = 1 - (1 + qx) e^{-x}.
    \end{equation}
\end{lemma}

Now we analyze the single-round output of a sample access algorithm $\mathcal{F}$.
\begin{lemma}\label{lem:dist_F}
Consider a quantum algorithm $\mathcal{F}$ given $k$ samples drawn from $P^{(i)}$ for all $i\in[m]$. Randomly select one of the $m$ outputs of the algorithm and compute the bitstring probability $p$ and cap the value at $p_{\rm max}$. Then, the probability distribution $\rho_p'$ with the following cumulative density function dominates $\rho_p$:
\begin{equation}
    F(p)=
    \begin{cases}
    0 & \mathrm{if~} 0\leq p \leq p_{\rm min}\\
    (1-(1+qN(1-N^{-1/3})p)e^{-N(1-N^{-1/3})p})-(mN^{-1/3}+mk^2/N) & \mathrm{if~} p_{\rm min} < p < p_{\rm max}\\
    1 & \mathrm{otherwise},
    \end{cases}
\end{equation}
where
\begin{align}
    q&=\frac{1}{1-k^2/N}\frac{1}{m}\sum_{b\in[m]}\Pr[z^{(b)}\in\{z^{(b)}_{c}\}_{c\in[k]}]\\
    0&=(1-(1+qNp_{\rm min})e^{-Np_{\rm min}})-(mN^{-1/3}+mk^2/N).
\end{align}
\end{lemma}
\begin{proof}

We use the same notation as Theorem \ref{thm:gen_7.10} for the frequency vector $f^{(i)}$ of the samples. Conditioned on the observed frequency vector, the posterior distribution is $P^{(i)}\sim\mathrm{Dir}(1^N+f^{(i)})$. Sampling $P^{(i)}$ can be done using the procedure discussed in Section 3.4.1 of Ref.~\cite{aaronson2023certifiedArxiv}, which we paraphrase here. We first sample $Q_j\sim\Gamma(1+f_j^{(i)}, 1)$ independently for $j\in[N]$, where $\Gamma(\alpha, \beta)$ is the Gamma distribution with parameters $\alpha,\beta$. Then, we set $P^{(i)}(\mathrm{str}(j))=Q_j/\bar{Q}$, where $\mathrm{str}(j)$ is the $j$th bitstring in $\{0,1\}^N$ and $\bar{Q}=\sum_{j=1}^N Q_j$. As shown in Eq. 199 of Ref.~\cite{aaronson2023certifiedArxiv}, $\Pr[\|f^{(i)}\|_\infty=1]\geq 1-\frac{k^2}{N}$ over choices of frequency vectors.

Let us focus on attention on the special where there is only one distribution $P^{(i)}$. Denote the $k$ samples the from $P^{(i)}$ given to algorithm $\mathcal{F}$ as $\{z_j^{(i)}\}_{j\in[k]}$. There are two possibilities for the $k$ samples. On possibility is that the samples do not repeat, which means $\|f^{(i)}\|_\infty=1$ since $f^{(i)}$ is the frequency vector of the samples. The other possibility is $\|f^{(i)}\|_\infty>1$, which we ignore for the time being since it is very unlikely. For the output $z^{(i)}$ of algorithm $\mathcal{F}$, there are also two possibilities, namely it is or is not from the set of input samples $\{z_j^{(i)}\}_{j\in[k]}$.\newline

\noindent \textbf{The $z^{(i)}\in\{z_j^{(i)}\}_{j\in[k]}$ and $\|f^{(i)}\|_\infty=1$ case.}
The bitstring probability $p$ can be sampled by sampling $Q\sim\Gamma(2,1)$, sample $Q_j\sim\Gamma(1,1)$ for $j\in[N-k]$, and sample $Q_\ell\sim\Gamma(2,1)$ for $\ell\in[k-1]$, and return
\begin{equation}
p=Q\cdot\left(Q+\sum_{j\in[N-k]}Q_j+\sum_{\ell\in[k-1]}Q_\ell\right)^{-1}\leq Q\cdot\left(\sum_{j\in[N-k]}Q_j+\sum_{\ell\in[k-1]}Q_\ell\right)^{-1}.
\end{equation}
We denote the corresponding probability density function (PDF) of this distribution as $f(p; N, k)$. The sum in parenthesis concentrates to $N+k$, which means
\begin{equation}
    f(p; N,k)\approx N^2 xe^{-Np}.
\end{equation}
This is equivalent to saying that $f$ is very close to the PDF of the bitstring probability of sampling honestly from a perfect quantum state \cite{brandao2020notes, morvan2023phase, jpmc_cr}. Moreover, with integer shape parameters $\alpha$, Gamma distributions are Erlang distributions. By the property of sum of independent Erlang distributed variables (PDF of the sum is Erlang distributed with the same rate $\beta$ and the sum of the shape parameters as the shape parameter),
\begin{equation}
    \sum_{j\in[N-k]}Q_j+\sum_{\ell\in[k-1]}Q_\ell\sim \Gamma(N+k-2, 1).
\end{equation}
Further, $\Gamma(N+k-2, 1)$ dominates $\Gamma(N, 1)$. Therefore, if we sample $p'$ by sampling $Q\sim\Gamma(2,1)$ and $Q'\sim\Gamma(N,1)$ and setting $p'=Q/Q'$, then $p'$ should dominate $p$. We therefore only consider the sampling process for $p'$.

Since the mean and variance of $\Gamma(N,1)$ are both $N$, the Chebyshev's inequality implies
\begin{equation}
    \Pr[\left\vert Q'-N\geq c\sqrt{N}\right\vert ]\leq\frac{1}{c^2}.
\end{equation}
Setting $c=N^{1/6}$ yields
\begin{equation}
    \Pr[Q'/N\leq 1-N^{-1/3}]\leq N^{-1/3}.
\end{equation}
Therefore, in the case of $Q'/N>1-N^{-1/3}$, we have $p'<\frac{Q}{N(1-N^{-1/3})}$.

The following sampling process, which assumes $Q'/N>1-N^{-1/3}$, is at most $N^{-1/3}$ away from the actual sampling process of $p'$ which dominates $p$: sample $Q\sim\Gamma(2,1)$ and set $p'=\frac{Q}{N(1-N^{-1/3})}$.\newline

\noindent \textbf{The $z^{(i)}\notin\{z_j^{(i)}\}_{j\in[k]}$ and $\|f^{(i)}\|_\infty=1$ case.} 
We can sample $p$ by sampling $Q$ and $Q_j$ from $\Gamma(1,1)$ for $j\in[N-1]$ and set
\begin{equation}
    p=Q\cdot\left(Q+\sum_{j\in[N-1]}Q_j\right)^{-1}.
\end{equation}
The term in the parenthesis should concentrate to $N$, which means \begin{equation}
    f(p; N,k)\approx N e^{-Np}.
\end{equation}
This is equivalent to saying that $f$ is very close to the PDF of the bitstring probability of sampling honestly from the uniform distribution \cite{brandao2020notes, morvan2023phase, jpmc_cr}. By similar arguments as in the previous case, the following sampling process, which assumes $Q'/N>1-N^{-1/3}$, is at most $N^{-1/3}$ away from the sampling process of another random variable $p'$ which dominates $p$: set $p'=\frac{Q}{N(1-N^{-1/3})}$, where $Q\sim\Gamma(1,1)$.\newline

\noindent \textbf{The general case for $m$ samples when $\|f^{(i)}\|_\infty=1$ for all $i\in[m]$.}
Above, we only examined the sampling of one bitstring probability from a single round of $\mathcal{A}$ with $m$ circuits. In general, however, the outputs for different circuits can be correlated. 

For a binary bitstring $\vec{s}\in\{0,1\}^m=\{s^{(1)}, s^{(2)}, \dots, s^{(m)}\}$, denote as $\Omega_{\vec{s}}$ the event where $z^{(i)}\in\{z_j^{(i)}\}_{j\in[k]}$ if $s^{(i)}=1$. Additionally, require that $Q'/N>1-N^{-1/3}$ for all circuits, which introduces a TVD of $mN^{-1/3}$ from the actual distribution. Therefore, the sampling process is as follows: with probability $\Pr[\Omega_{\vec{s}}\big\vert \forall i\in[m],\|f^{(i)}\|_\infty=1]$, set $p^{(i)}=\frac{Q}{N(1-N^{-1/3})}$ with $Q\sim\Gamma(1+s^{(i)},1)$.\newline

\noindent \textbf{Bitstring probability of a randomly chosen sample}

Since $p=\frac{Q}{N(1-N^{-1/3})}$ either has $Q\sim\Gamma(1,1)$ (shape 1 samples) or $Q\sim\Gamma(2,1)$ (shape 2 samples), we need the probability that a randomly selected sample is a shape 1 sample or a shape 2 sample.
\begin{align}
    q\equiv\Pr[Q\sim\Gamma(2,1)]&= \frac{1}{m}\sum_{\vec{s}\in\{0,1\}^m}w(\vec{s})\Pr[\Omega_{\vec{s}}\big\vert \forall i\in[m],\|f^{(i)}\|_\infty=1]\\
    &\leq \frac{1}{m}\sum_{\vec{s}\in\{0,1\}^m}w(\vec{s})\frac{\Pr[\Omega_{\vec{s}}]}{1-mk^2/N}\\
    &= \frac{1}{1-k^2/N}\frac{1}{m}\sum_{b\in[m]}\Pr[z^{(b)}\in\{z^{(b)}_{c}\}_{c\in[k]}],
\end{align}
where $w(\vec{s})$ denotes the Hamming weight of the bitstring. The first inequality uses the fact that 
\begin{equation}
    \Pr[\Omega_{\vec{s}}\big\vert \forall i\in[m],\|f^{(i)}\|_\infty=1]\leq \Pr[\Omega_{\vec{s}}]/\Pr[\forall i\in[m],\|f^{(i)}\|_\infty=1]\leq \Pr[\Omega_{\vec{s}}]/(1-mk^2/N).
\end{equation}

For the final output, we need to incorporate the TVD distance of the distribution as well as the cap to $p_{\rm max}$. The cutoff is obviously accomplished by setting the CDF to 1 for $p>p_{\rm max}$. For the most dominant distribution with TVD equal to $mN^{-1/3}+mk^2/N$ (the $mk^2/N$ contribution comes from the upper bound on the probability that $\|f^{(i)}\|_\infty\neq1$ for some $i$), any probability mass under $p_{\rm min}$ are moved to $p_{\rm max}$. Here, $p_{\rm min}$ is chosen such that the original probability of $p<p_{\rm min}$ is $mN^{-1/3}+mk^2/N$. This has the effect of setting the CDF to zero for $p<p_{\rm min}$ and shift the CDF down by $mN^{-1/3}+mk^2/N$ between $p_{\rm min}$ and $p_{\rm max}$. Combined with Lemma \ref{lem:cdf_of_mixture}, this completes the proof.

\end{proof}

Now that we have a dominant distribution of the output of $\mathcal{F}$ and we know the distance between $\mathcal{A}$ and $\mathcal{F}$, we can write down a dominant distribution for $\mathcal{A}$. Combining Lemma \ref{lem:low_h_implies_low_contain_p_prob} and Lemma \ref{lem:dist_F} and moving the additional bottom probability mass of $m\varepsilon'$ to $p_{\rm max}$ in a similar way as Lemma \ref{lem:dist_F}, we have the following lemma.
\begin{lemma}\label{lem:dist_A}
Let $\mathcal{A}$ be any algorithm making $(T,m)$-queries to $\vec{U}\sim\mathrm{Haar}^{\otimes m}$. Let $p$ be the bitstring probability of a randomly chosen bitstring. If $n\geq 50, k\leq N^{-1/2}, m,n\leq 10^6$, and
\begin{equation}
    H(\vec{Z}\vert\vec{U}E)_\Psi\leq h,
\end{equation}
then $p$ is dominated by a random variable with the following cumulative density function:
\begin{equation}
    F(p)=
    \begin{cases}
    0 & \mathrm{if~} 0\leq p \leq p_{\rm min}\\
    (1-(1+qN(1-N^{-1/3})p)e^{-N(1-N^{-1/3})p})-md_{\rm Q} & \mathrm{if~} p_{\rm min} < p < p_{\rm max}\\
    1 & \mathrm{otherwise},
    \end{cases}
\end{equation}
where
\begin{align}
    d_{\rm Q}&=N^{-1/3}+(k+\ell)^2/N+\varepsilon'\\
    q&=\frac{1}{1-(k+\ell)^2/N}\left(\frac{h+2}{mn}+C\right)\\
    0&=(1-(1+qNp_{\rm min})e^{-Np_{\rm min}})-m d_{\rm Q}.
\end{align}
\end{lemma}

State of the art classical simulation algorithms that spoofs the cross entropy benchmarking (XEB) test for RCS rely on tensor networks \cite{pan2022solving, kalachev2021classical, Kalachev2021multi, zhao2024leapfrogging, fu2024achieving, morvan2023phase, jpmc_cr, qntm_rcs}. In the case of frugal rejection sampling \cite{markov2018quantum}, the output bitstring probability $p$ behaves very similarly to honest finite fidelity quantum sampling \cite{jpmc_cr}, which we provide more analysis in Section \ref{sec:frugal_rejection_sampling}. For the more recent algorithm that uses post-processing by choosing the largest probability bitstring out of many bitstrings \cite{zhao2024leapfrogging, fu2024achieving}, we later show in Section \ref{sec:oversampling_classical_adversary} that when the effective fidelity is low, $p$ also behaves very similarly to honest finite fidelity quantum sampling. Overall, this motivates us to approximate the behavior of classical algorithms with that of honest finite fidelity quantum sampling from a depolarized state. Concretely, one can characterize the `effective fidelity' of the algorithm as well as the TVD between the actual distribution and that of honest finite fidelity quantum sampling, either theoretically or empirically.

\begin{lemma}\label{lem:dist_classical}
Consider any classical sampling process whose bitstring probability distribution for the $i$th circuit conditioned on arbitrary side information, for all $i\in[m]$, is at most $d_{\rm C}$ from the distribution with the following cumulative density function:
\begin{equation}
    F(x)=1-(1+\phi_i N(1-N^{-1/3})x)e^{-N(1-N^{-1/3})x}\label{eqn:cdf_classical_ideal}.
\end{equation}
If $\sum_{i\in[m]}[\phi_i]=\Phi_{\rm C}$ and $p$ is the bitstring probability of a randomly selected output string, then $p$ is dominated by a random variable with the following cumulative density function:
\begin{equation}
    F(p)=
    \begin{cases}
    0 & \mathrm{if~} 0\leq p \leq p_{\rm min}\\
    1-(1+\frac{\Phi_{\rm C}}{m} N(1-N^{-1/3})p)e^{-N(1-N^{-1/3})p}-md_{\rm C} & \mathrm{if~} p_{\rm min} < p < p_{\rm max}\\
    1 & \mathrm{otherwise},
    \end{cases}
\end{equation}
where
\begin{align}
    0=1-(1+\frac{\Phi_{\rm C}}{m} Np_{\rm min})e^{-Np_{\rm min}}-md_{\rm C}.
\end{align}
\end{lemma}
\begin{proof}
    Consider the distribution of $p'$ which is the bitstring probability of a randomly selected output string out of $m$ circuits, where each circuit samples from Eq. \ref{eqn:cdf_classical_ideal}. The overall CDF is a weighted sum of Eq. \ref{eqn:cdf_classical_ideal}, given by
    \begin{equation}
        F'(p')=1-(1+\frac{\Phi_{\rm C}}{m} N(1-N^{-1/3})p')e^{-N(1-N^{-1/3})p'}\label{eqn:cdf_classical_ideal_sum}.
    \end{equation}
    However, $p$ is sampled from bitstring probabilities with distance $d_{\rm C}$ from Eq. \ref{eqn:cdf_classical_ideal}, so the total distance is bounded by $md_{\rm C}$. The most dominant distribution that is $md_{\rm C}$ from Eq. \ref{eqn:cdf_classical_ideal_sum} shifts the CDF down by $md_{\rm C}$ within $[p_{\rm min}, p_{\rm max}]$, which completes the proof.
\end{proof}

We are now finally ready to consider a mixture of classical and quantum oracle access adversary.
\begin{lemma}\label{lem:dist_mixture}
Let $\mathcal{A}$ be any algorithm making $(T,m_{\mathrm{Q}})$-queries to $m_{\mathrm{Q}}$ Haar random independent unitaries producing $m_{\mathrm{Q}}$ bitstrings. Let the rest $m_{\mathrm{C}}=m-m_{\mathrm{Q}}$ bitstrings be produced by a classical simulation algorithm satisfying the conditions in Lemma \ref{lem:dist_classical} and that $\sum_{i\in[m_{\mathrm{C}}]}\phi_i=\Phi_{\rm C}$. If $n\geq 50, k\leq N^{-1/2}, m,n\leq 10^6$, and
\begin{equation}
    H(\vec{Z}\vert\vec{U}E)_\Psi\leq h,
\end{equation}
then $p$ is dominated by a random variable with the following cumulative density function:
\begin{equation}
    F(p)=
    \begin{cases}
    0 & \mathrm{if~} 0\leq p \leq p_{\rm min}\\
    \left[1-\left(1+\Phi N(1-N^{-1/3})p\right)e^{-N(1-N^{-1/3})p}\right]-m\max(d_{\rm Q}, d_{\rm C}) & \mathrm{if~} p_{\rm min} < p < p_{\rm max}\\
    1 & \mathrm{otherwise},
    \end{cases}
\end{equation}
where
\begin{align}
    d_{\rm Q}&=N^{-1/3}+(k+\ell)^2/N+\varepsilon'\\
    \Phi&=\frac{1}{1-(k+\ell)^2/N}\left(\frac{h+2}{mn}+C\right)+\frac{\Phi_{\rm C}}{m}\\
    0&=(1-(1+\Phi Np_{\rm min})e^{-Np_{\rm min}})-m\max(d_{\rm Q}, d_{\rm C}).
\end{align}
\end{lemma}
\begin{proof}
We know the $m_{\mathrm{Q}}$ quantum samples have entropy at most $h$. The resulting cumulative density function of the mixture is a weighted sum of the cumulative density functions given by Lemma \ref{lem:dist_A} and Lemma \ref{lem:dist_classical}, with $m$ set to $m_{\mathrm{Q}}$ and $m_{\mathrm{C}}$, respectively. The `middle' part of the cumulative density function is given by
\begin{align}
    F(q)=&\left[1-\left(1+\frac{m_{\mathrm{Q}}q+m_{\mathrm{C}}\phi}{m}N(1-N^{-1/3})p\right)e^{-N(1-N^{-1/3})p}\right]-m_{\mathrm{Q}}d_{\rm Q}+m_{\mathrm{C}}d_{\rm C}\\
    \geq&\left[1-\left(1+\frac{m_{\mathrm{Q}}q+m_{\mathrm{C}}\phi}{m}N(1-N^{-1/3})p\right)e^{-N(1-N^{-1/3})p}\right]-m\max(d_{\rm Q}, d_{\rm C}),
\end{align}
where
\begin{align}
    q&=\frac{1}{1-(k+\ell)^2/N}\left(\frac{h+2}{m_{\mathrm{Q}}n}+C\right)\\
    \phi&=\frac{1}{m_{\mathrm{C}}}\sum_{i\in[m_{\mathrm{C}}]}\phi_i=\frac{1}{m_{\mathrm{C}}}\Phi_{\rm C}.
\end{align}
Since
\begin{equation}
    m_{\mathrm{Q}}q+m_{\mathrm{C}}\phi=\frac{1}{1-(k+\ell)^2/N}\left(\frac{h+2}{n}+m_{\mathrm{Q}}C\right)+\Phi_{\rm C}\leq \frac{1}{1-(k+\ell)^2/N}\left(\frac{h+2}{n}+mC\right)+\Phi_{\rm C},
\end{equation}
we have
\begin{equation}
    F(q)\geq\left[1-\left(1+\Phi N(1-N^{-1/3})p\right)e^{-N(1-N^{-1/3})p}\right]-m\max(d_{\rm Q},d_{\rm C}).
\end{equation}
With this middle part, we then need to appropriately adjust $p_{\rm min}$ as required.
\end{proof}

We can additionally show that samples with the above CDF allows for an interpretation of sampling shape 2 samples with probability $\Phi$ and shape 1 samples with probability $1-\Phi$, modulo a small shift.
\begin{lemma}\label{lem:cdf_to_effective_honest_sampling}
A random variable with the following cumulative density function:
\begin{equation}
    F(x)=
    \begin{cases}
    0 & \mathrm{if~} 0\leq x \leq x_{\rm min}\\
    \left[1-\left(1+\Phi x\right)e^{-x}\right]-d & \mathrm{if~} x_{\rm min} < x < x_{\rm max}\\
    1 & \mathrm{otherwise},
    \end{cases}
\end{equation}
where
\begin{align}
    0=\left[1-\left(1+\Phi x_{\rm min}\right)e^{-x_{\rm min}}\right]-d,
\end{align}
is dominated by a random variable with the following cumulative density function:
\begin{equation}
F(x)=
\begin{cases}
0 & \mathrm{if~} 0\leq x \leq x_{\rm min}\\
(1-d)\left\{1-e^{-(x-x_{\rm min})}\left[1+\Phi(x-x_{\rm min})\right]\right\} & \mathrm{if~} x_{\rm min} < x < x_{\rm max}\\
(1-d)\left\{1-e^{-(x-x_{\rm min})}\left[1+\Phi(x-x_{\rm min})\right]\right\}+d & \mathrm{otherwise}.
\end{cases}
\end{equation}
\end{lemma}
\begin{proof}
For $x_{\rm min} < x < x_{\rm max}$,
\begin{align}
    F(x)&=1-e^{-x}(1+\Phi x)-d\\
    &=(1-d)\left[1-\frac{1}{1-d}e^{-x}(1+\Phi x)\right]\\
    &=(1-d)\left[1-\frac{e^{x_{\rm min}}}{1+\Phi x_{\rm min}}e^{-x}(1+\Phi x)\right]\\
    &=(1-d)\left\{1-\frac{1}{1+\Phi x_{\rm min}}e^{-(x-x_{\rm min})}\left[1+\Phi (x-x_{\rm min})+\Phi x_{\rm min}\right]\right\}\\
    &=(1-d)\left\{1-e^{-(x-x_{\rm min})}\left[1+\frac{\Phi }{1+\Phi x_{\rm min}}(x-x_{\rm min})\right]\right\}\\
    &\geq (1-d)\left\{1-e^{-(x-x_{\rm min})}\left[1+\Phi (x-x_{\rm min})\right]\right\},
\end{align}
where the third equality uses the definition of $x_{\rm min}$ which requires that $1-e^{-x_{\rm min}}(1+\Phi)=d$ and $\frac{1}{1-d}=\frac{e^{x_{\rm min}}}{1+\Phi x_{\rm min}}$. For $x\geq x_{\rm max}$, the new cumulative density function is also always smaller than 1.
\end{proof}

\subsection{Multi-round results}

With the single round statistical properties fully characterized, we can obtain a bound on the multi-round entropy. In this analysis, we will still use the entropy accumulation theorem, although statistical testing is handled separately. We first reproduce a version of entropy accumulation without statistical testing in the notation of Ref.~\cite{dupuis2020entropy}. We high light the following notation. Consider a register for the $i$th round $A_i$, denote the collection of registers $A_i A_{i+1}\dots A_j=A_i^j$ and $A_1^j=A^j$.

\begin{theorem}[Corollary 4.9 of Ref.~\cite{dupuis2020entropy}]\label{thm:eta_no_test}
    Let $\mathcal{M}_1,\dots,\mathcal{M}_n$ and $\rho_{A_1^n B_1^n E}$ be such that Eq. 26 and 27 of Ref.~\cite{dupuis2020entropy} hold. Then
    \begin{equation}
        H_{\rm min}^\varepsilon(A_1^n\vert B_1^n E)_\rho > \sum_{i=1}^n\underset{\omega_{R_{i-1}R}}{\inf}H(A_i\vert B_i R)_{(\mathcal{M}_i\otimes \mathcal{I}_R)(\omega_{R_{i-1}R})}-c\sqrt{n},
    \end{equation}
    where $c=3\log_2(1+2d_A)\sqrt{1-2\log_2\varepsilon}$.
\end{theorem}

With this theorem, if the entropy of the multi-round protocol is less than some value, then we can upper bound the sum of single round entropy. Since query access quantum devices are close to sample access quantum devices, an upper bounded entropy allows us to limit the possible probability distribution of the output bitstring probability amplitudes. We note that the above theorem results in an estimate of the smooth min-entropy of the unconditional state $\rho$. Techniques in \cite{dupuis2019entropy,dupuis2020entropy} can be used to convert the smooth min-entropy below for $\rho$ into the smooth min-entropy for $\rho\vert\Omega$.

\begin{theorem}\label{thm:multi_round_entropy_ad_hoc}
Consider an adversary with access to a quantum computer given at most $m$ query access to each challenge circuit and a classical computer satisfying Lemma \ref{lem:dist_classical}. If $n\geq 50, k\leq N^{-1/2}, m,n\leq 10^6$, then for any $k, \ell, \epsilon_1$, and $\epsilon_2$, either the smooth min-entropy is bounded by
\begin{equation}
    H_{\rm min}^\varepsilon(\vec{Z}^L\vert \vec{U}^L E)_\rho>H,
\end{equation}
or the protocol of Fig. \ref{fig:protocol_ad_hoc} aborts with probability at least
\begin{equation}
    1=\Pr[\Omega]\geq 1-(\epsilon_1+\epsilon_2+\epsilon_3),
\end{equation}
where
\begin{gather}
    \epsilon_3 = 1 - \sum_{i=0}^{L_{\rm val}-n_{\rm max, max}}\Tilde{\Gamma}(L_{\rm val}-n_{\rm max, max} + i, \chi)\cdot\Pr[n_2=i\vert n_{\rm max}=n_{\rm max, max}, N_2=N_{2, \max}],\\
    \Pr[n_2=i\vert n_{\rm max, max}=n_{\rm max, max}, N_2=N_{2, \max}]=\left[\binom{L-n_{\rm max, max}}{L_{\rm val}-n_{\rm max, max}}^{-1} {\binom{N_{2, \rm max}}{i}\binom{L-n_{\rm max, max}-N_{2,\rm max}}{L_{\rm val}-n_{\rm max, max}-i}}\right],\\
    n_{\rm max, max}=\min\left(L_{\rm val}, \left(1+\sqrt{\frac{3}{L_{\rm val}d}\ln\frac{1}{\epsilon_1}}\right)L_{\rm val}d\right),\\
    N_{2, \rm max}=\min\left(L-n_{\rm max, max}, \left(1+\sqrt{\frac{3}{L\Phi}\ln\frac{1}{\epsilon_2}}\right)L\Phi\right),\\
    \chi=(L_{\rm val}-n_{\rm max,max})\left(s^*-\frac{N}{L_{\rm val}}(n_{\rm max, max}p_{\rm max}+(L_{\rm val}-n_{\rm max,max})p_{\rm min})\right)(1-N^{-1/3}),\\
    d=m\max(d_{\rm Q},d_{\rm C}),\\
    0=1-e^{-x_{\rm min}}(1+x_{\rm min})-d,\\
    p_{\rm min}=\frac{x_{\rm min}}{N(1-N^{-1/3})},\\
    d_{\rm Q}=N^{-1/3}+(k+\ell)^2/N+\varepsilon',\\
    \varepsilon'=T\left(\frac{4}{2^{n/2}} + \sqrt{2\left(1-\left(\cos\frac{\pi}{2k}\right)^k\right)}+2/\sqrt{\ell+1}\right),\\
    \Phi=\frac{\Phi_{\rm C}}{m} + \frac{1}{1-(k+\ell)^2 / N} \left(\frac{2}{mn}+\phi C+\frac{1}{Lmn} (H+c\sqrt{L})\right),\\
    C=\frac{1}{n}\left(\log_2 n + \log_2(k+\ell) + 3\right)+4m\varepsilon'(n\ln2+3.5)+1.001\times\frac{16(k+\ell)^2\ln^2N}{N}+0.001,\\
    c=3\log_2(1+2\times 2^{mn})\sqrt{1-2\log_2\varepsilon}.
\end{gather}
\end{theorem}
\begin{proof}
For a single round, we consider the cumulative density function given by Lemma \ref{lem:dist_mixture}, relabeling $\Phi,h$ as $\Phi_a,h_a$ for the $a$th round. Applying Lemma \ref{lem:cdf_to_effective_honest_sampling} with rescaling of $x\rightarrow N(1-N^{-1/3})p$ and using the observation from Lemma \ref{lem:cdf_of_mixture}, the dominant cumulative density function can be described by the following sampling process of $p$:
\begin{enumerate}
    \item With probability $d=m\max(d_{\rm Q},d_{\rm C})$, set $p=p_{\rm max}$.
    \item With probability $(1-d)(1-\Phi_a)$, set $p=\frac{Q}{N(1-N^{-1/3})}+p_{\rm min}$ with $Q\sim\Gamma(1,1)$.
    \item Otherwise, set $p=\frac{Q}{N(1-N^{-1/3})}+p_{\rm min}$ with $Q\sim\Gamma(2,1)$.
\end{enumerate}
Here, all rounds are considered to have the same $p_{\rm min}$, which is taken as the maximum possible value attainable when $\Phi_a=1$, which is given by the conditions of this theorem.

We obtain an upper bound on the sum of $\Phi_a$, given by
\begin{align}
    \frac{1}{L} \sum_a \Phi_a &\leq \left[\frac{\Phi_{\rm C}}{m} + \frac{1}{1-(k+\ell)^2/N}\left(\frac{h_a+2}{mn}+C\right)\right]\\
    & = \frac{\Phi_{\rm C}}{m} + \frac{1}{1-(k+\ell)^2 / N} \left(\frac{2}{mn}+C+\frac{1}{Lmn} \sum_a h_a\right)
\end{align}
If $H_{\rm min}^\varepsilon(\vec{Z}^L\vert \vec{C}^L E)_\rho<H$, Theorem \ref{thm:eta_no_test} implies
\begin{equation}
\sum_{a\in[L]}h_a=\sum_{a\in[L]}\underset{\omega_{R_{a-1}R}}{\inf}H(\vec{Z}_a\vert \vec{C}_a R)_{(\mathcal{M}_i\otimes \mathcal{I}_R)(\omega_{R_{a-1}R})}<H+c\sqrt{L},
\end{equation}
and therefore
\begin{equation}
    \frac{1}{L} \sum_a \Phi_a < \frac{\Phi_{\rm C}}{m} + \frac{1}{1-(k+\ell)^2 / N} \left(\frac{2}{mn}+C+\frac{1}{Lmn} (H+c\sqrt{L})\right)\equiv\Phi.
\end{equation}

For validation set $\mathcal{V}$ of size $L_{\rm val}$, we can upper bound the probability of obtaining $n_{\rm max}$ validation samples of $p_{\rm max}$. Since every sample has probability $d$ of being $p_{\rm max}$ and are independent, $n_{\rm max}$ is given by the binomial distribution, and the tail bound can be obtained using the known Chernoff bound for sum of Poisson trials (Theorem 4.4 of Ref.~\cite{Mitzenmacher_Upfal_2005}):
\begin{align}
    \Pr[n_{\rm max}\geq (1+\delta)L_{\rm val}d] &< e^{-\frac{\delta^2 L_{\rm val}d}{3}}\leq \epsilon_1\\
    \Pr[n_{\rm max}\geq n_{\rm max, max}] &< \epsilon_1, \quad
    n_{\rm max, max}=\left(1+\sqrt{\frac{3}{L_{\rm val}d}\ln\frac{1}{\epsilon_1}}\right)L_{\rm val}d.
\end{align} 
Similarly, we can bound the number $n_2$ of shape $2$ samples in the validation set. We can even consider the worse case where each sample is shape 2 with probability $\Phi_a$ instead of $(1-d)\Phi_a$. We can first bound the number $N_2$ of shape 2 samples in total, which is given by the same Chernoff bound above:
\begin{align}
    \Pr[N_2\geq N_{2, \max}]&<\epsilon_2,\quad
    N_{2, \max}=\left(1+\sqrt{\frac{3}{\sum_{a\in[L]}\Phi_a}\ln\frac{1}{\epsilon_2}}\right)\sum_{a\in[L]}\Phi_a=\left(1+\sqrt{\frac{3}{L\Phi}\ln\frac{1}{\epsilon_2}}\right)L\Phi.
\end{align}

Since the client performs verification only on a much smaller set of circuits $\mathcal{V}$, the distribution of the score over the verification set depends on the number $n_2$ of shape 2 samples in the verification set, not the number $N_2$ of shape 2 samples in total. Given $n_{\rm max}$ samples are set to $p_{\rm max}$, each of the remaining $L_{\rm val}-n_{\rm max}$ verification samples is drawn from a total of $L-n_{\rm max}$ samples without replacement. If $N_2$ out of $M$ samples are shape 2, then the probability that $n_2$ out of $L_{\rm val}-n_{\rm max}$ of the remaining validation samples are shape 2 is given precisely by the hypergeometric distribution $\text{Hypergeometric}(L-n_{\rm max}, N_{2}, L_{\rm val}-n_{\rm max})$. \cite{rice2007mathematical}.

Given $n_2$ shape 2 samples and $n_{\rm max}$ samples set to $p_{\rm max}$ in the validation set, there should be $n_1=L_{\rm val}-n_2-n_{\rm max}$ shape 1 samples in the validation set. The sum of all the $Q$'s is distributed according to $\Gamma(n_1+2n_2, 1)$. Further, the contribution to the final score due to shape 1 and shape 2 samples is scaled by $(1-N^{-1/3})^{-1}$. The $n_{\rm max}$ samples set to $p_{\rm max}$ increases the score by $\frac{n_{\rm max}Np_{\rm max}}{L_{\rm val}}$. The offset in $Q$ by $x_{\rm min}$ for all shape 1 and shape 2 samples contributes $\frac{(n_1+n_2)Np_{\rm min}}{L_{\rm val}}$ to the score. Therefore,
\begin{align}
    &\Pr[s\leq s^*\Big\vert n_{\rm max}=n_{\rm max, max}\land N_2=N_{2, \max}]\\
    =&\sum_{i=0}^{L_{\rm val}-n_{\rm max, max}}\Pr[\text{Sum of all $Q$'s}\leq\chi\vert n_{\rm max}=n_{\rm max, max}, n_2=i]\cdot\Pr[n_2=i\vert n_{\rm max}=n_{\rm max, max}, N_2=N_{2, \max}]\\
    \leq& \sum_{i=0}^{L_{\rm val}-n_{\rm max, max}}\Tilde{\Gamma}(L_{\rm val}-n_{\rm max, max} + i, \chi)\cdot\Pr[n_2=i\vert n_{\rm max}=n_{\rm max, max}, N_2=N_{2, \max}]=1-\epsilon_3,
\end{align}
where
\begin{gather}
    s^*=\frac{1}{L_{\rm val}-n_{\rm max,max}}\chi(1-N^{-1/3})^{-1}+\frac{N}{L_{\rm val}}(n_{\rm max, max}p_{\rm max}+(L_{\rm val}-n_{\rm max,max})p_{\rm min}),\\
    \chi=(L_{\rm val}-n_{\rm max,max})\left(s^*-\frac{N}{L_{\rm val}}(n_{\rm max, max}p_{\rm max}+(L_{\rm val}-n_{\rm max,max})p_{\rm min})\right)(1-N^{-1/3}),\\
    \Pr[n_2=i\vert n_{\rm max, max}=n_{\rm max, max}, N_2=N_{2, \max}]=\left[\binom{L-n_{\rm max, max}}{L_{\rm val}-n_{\rm max, max}}^{-1} {\binom{N_{2, \rm max}}{i}\binom{L-n_{\rm max, max}-N_{2,\rm max}}{L_{\rm val}-n_{\rm max, max}-i}}\right].
\end{gather}
Putting it all together, the probability that the adversary exceeds threshold $\chi$ is
\begin{align}
    1=\Pr[\Omega]=&\Pr[s\geq s^*]\\
    \leq& \Pr[n_{\rm max}>n_{\rm max, max} \lor N_2>N_{2,\max}]\\
    &+\Pr[n_{\rm max}\leq n_{\rm max, max} \land N_2\leq N_{2,\max}]\Pr[s\geq\chi\vert n_{\rm max}\leq n_{\rm max, max} \land N_2\leq N_{2,\max}]\\
    \leq& \Pr[n_{\rm max}>n_{\rm max, max} \lor N_2>N_{2,\max}] + \Pr[s\geq\chi\vert n_{\rm max}=n_{\rm max, max} \land N_2=N_{2,\max}]\\
    \leq& \epsilon_1 + \epsilon_2 + \epsilon_3.
\end{align}

This concludes the analysis when we have quantum device fidelity $\phi=1$.
\end{proof}

\begin{remark}\label{rem:finite_fidelity}
    We can also consider noisy adversaries. Now, consider a noisy quantum device whose output $\vec{Z}=\vec{z}$ follows the probability distribution $\rho_{\rm noisy}(\vec{z})=\phi\rho_{\rm ideal}(\vec{z})+(1-\phi)\sigma(\vec{z})$, where $\rho_{\rm ideal}$ is the probability distribution of an ideal quantum device and $\sigma$ is some other normalized probability distribution. Further, assume that $\sigma$ is not usefully dependent on the input quantum circuit. If the noisy adversary solves $(1+\delta)-\mathrm{MXHOG}_m$, then the ideal algorithm should be able to solve $(1+\delta/\phi)-\mathrm{MXHOG}_m$, which means the ideal algorithm would have entropy $h(\delta/\phi)$. The noisy algorithm entropy can be lower bounded by $\phi h(\delta/\phi)$. This allows us to restate a noisy version of Lemma \ref{lem:low_h_implies_low_contain_p_prob} and Theorem \ref{thm:multi_round_entropy_ad_hoc} with
\begin{equation}
C\rightarrow\phi C.
\end{equation}

To see why, we first claim that $C\rightarrow\phi C$ of Lemma \ref{lem:low_h_implies_low_contain_p_prob} has the above form. If for the noisy algorithm, $H(\vec{Z}\vert\vec{U}E)_\rho\leq h$, then the worse case entropy bound for the ideal algorithm is $H(\vec{Z}\vert\vec{U}E)_\Psi\leq h/\phi$. Now, for the ideal algorithm we have $\delta=(h/\phi+2)/(mn)+C$, and the noisy algorithm multiplies $\delta$ by $\phi$, yielding $\delta=(h+2\phi)/mn+\phi C\leq(h+2)/mn+\phi C$.
\end{remark}

\begin{remark}
Here, we suffer from the $\sqrt{L}$ penalty in $\Phi$, which is a consequence of bounding the multi-round smooth min-entropy using EAT. However, one can obtain a better bound using Eq. 5 of Ref.~\cite{Tomamichel09_aep}. This would require obtaining single-round collision entropy bound for our certified randomness protocol. Additionally, if we only assume an independent but not i.i.d. adversary, then the asymptotic equipartition theorem requires generalization.

One can also potentially resolve the non-i.i.d. issue by ``postselecting'' on having $Q$ total samples returned from one of the $k$ samples provided to the simulation algorithm $\mathcal{F}$. This would allow a similar analysis to \cite{jpmc_cr} to be conducted.
\end{remark}

We can also devise a further improved entropy estimate using the improved single-round entropy bound without the $\log_2(k+\ell)$ penalty that is valid in the $T=1$, $m=1$ setting.
\begin{theorem}\label{thm:nonadaptive_special}
For the special case of $T=1$ and $m=1$, Theorem \ref{thm:multi_round_entropy_ad_hoc} holds with modified parameters
\begin{align}
C = \frac{1}{n}\left(\log_2 n+3\right)+4\varepsilon''(n\ln2+3.5)+1.001\times\frac{16\ln^2N}{N}+0.001,
\end{align}
where $\varepsilon''$ is as defined in Corollary \ref{cor:improved_general_adv_m1_T1}.
\end{theorem}
\begin{proof}
The proof is identical to Theorem \ref{thm:multi_round_entropy_ad_hoc}, except we change the proof of Lemma \ref{lem:low_h_implies_low_delta}. Instead of using Theorem \ref{thm:vanilla_EAT}, we use Theorem \ref{thm:vanilla_EAT_special} with the improve entropy. This improves the bound on $\delta$ for Lemma \ref{lem:low_h_implies_low_delta}.
\end{proof}

\section{Multi-round security under adaptive hybrid adversaries}\label{sec:multiround}
\label{sec:app-multi-round}

\subsection{Direct application of entropy accumulation}
To be secure against the most general adaptive adversaries, analyses are typically carried out under the framework of entropy accumulation, and each round is tested independently with some fixed probability. We define a multi-round protocol suitable for entropy accumulation analysis in Figure \ref{fig:protocol}. Consider the situation where each batch has validation probability $\gamma$, and to validate a batch means choosing a random circuit in the batch to validate. To analyze the protocol, we use an entropy accumulation theorem, originally proposed by \cite{dupuis2020entropy}. Specifically, we consider the improved version described in \cite{dupuis2019entropy}, and further improve the constant correction term, which we describe below.

\begin{figure*}
    \hrule
    \vspace{.5em}
    \begin{flushleft}\underline{Protocol Arguments:}\end{flushleft}
 \vspace{-1.5em}
 \begin{align*}
  n \in \mathbb{N}~&:~\text{Number of qubits}\\
  m \in \mathbb{N}~&:~\text{Number of circuits per round (batch size)}\\
  L \in \mathbb{N}~&:~\text{Number of rounds}\\
  T_{\rm batch}~&:~\text{Threshold on the time allowed for the samples to return}\\
  p_{\rm max}\in (1/2^n, 1]~&:~\text{Maximum contribution probability}\\
  s^*~&:~\text{Threshold for test score}\\
  \gamma~&:~\text{Probability of verifying a round}\\
  \end{align*}
  \vspace{-2em}
  \begin{flushleft}\underline{The Protocol:}\end{flushleft}
  \begin{enumerate}
  \item Set $s=0$.
  \item For $i\in[L]$, do:
  \begin{enumerate}
      \item Sample $m$ Haar random unitaries.
      \item Send them to the server and start the timer.
      \item If does not receive the bitstrings in time $T_{\rm batch}$, abort. Otherwise, store the bitstrings $\vec{Z}_i$.
      \item Sample $T_{i}\sim\mathsf{Bernoulli}(\gamma)$.
      \item If $T_i=1$, sample a circuit uniform from the batch of $m$ circuits for the $i$th round and add it to the validation set.
  \end{enumerate}
  \item For each $T_i=1$, compute the bitstring probability $p$ for the chosen validation circuit of that batch and add $\frac{N\min(p, p_{\rm max})}{\gamma L}$ to $s$. Finally, abort if $s<s^*$.
  \end{enumerate}
  \hrule\vspace{1em}
  \caption{Description of the certified randomness protocol.}
  \label{fig:protocol}
  \end{figure*}

In this portion of the text, we use the notation of Section V of Ref.~\cite{dupuis2019entropy}. We first provide an improved version of the entropy accumulation theorem.

\begin{theorem}[Theorem V.2 of Ref.~\cite{dupuis2019entropy}, improved]\label{thm:gen_v2}
    For $w=2\log_2 d_A+\max(f)-\min_\Sigma(f)$, if $w\geq3$ and $n\geq\frac{8\ln2\log_2\frac{2}{\varepsilon^2\Pr[\Omega]^2}}{V^2}$, then the smooth min-entropy conditioned on the event of not aborting ($\Omega$) is
    \begin{equation}
        H_{\rm min}^\varepsilon(A^n_1\vert B^n_1E)_{\rho\vert\Omega}>nh-c\sqrt{n}-c',
    \end{equation}
    where
    \begin{align}
        c&=V\sqrt{(2\ln2)\log_2\frac{2}{\varepsilon^2\Pr[\Omega]^2}}\\
        c'&=\frac{2\log_2\frac{2}{\varepsilon^2\Pr[\Omega]^2}}{V^2\ln2}K\\
        K&=4\times(w+1)^3 \times2^{\frac{w\sqrt{2\log_2\frac{2}{\varepsilon^2\Pr[\Omega]^2}}}{V\sqrt{n\ln2}}},
    \end{align}
    and
    \begin{equation}
        V=\sqrt{\mathrm{var}(f)+2}+\log_2(2d_A^2+1)
    \end{equation}
    if $A$ is quantum, or
    \begin{equation}
        V=\sqrt{\mathrm{var}(f)+2}+\log_2(2d_A+1)
    \end{equation}
    if $A$ is classical.
\end{theorem}
\begin{proof}
    Eq. 29 of Ref.~\cite{dupuis2019entropy} is
    \begin{equation}
        H_{\rm min}^\varepsilon(A^n_1\vert B^n_1E)_{\rho\vert\Omega}\geq nh-n\frac{(\alpha-1)\ln2}{2}V^2-\frac{1}{\alpha-1}\log_2\frac{2}{\varepsilon^2\Pr[\Omega]^2}-n(\alpha-1)^2K_\alpha,\label{eqn:entropy_w_alpha}
    \end{equation}
    where $K_\alpha$ is as in Eq. 31 of Ref.~\cite{dupuis2019entropy}. Note that we do not bound $K_\alpha$ with $K$ as in \cite{dupuis2019entropy}, since the effect of alpha optimization will be important here. We also note that the value of $V$ here arises due to the invocation of Corollary IV.2 in Eq. 35 of Ref.~\cite{dupuis2019entropy}, and the discrepancy between quantum and classical $A$ can be traced back to Eq. 37 and understood with Corollary III.5 of Ref.~\cite{dupuis2019entropy}.
    Now, we optimize $\alpha$. Choose
    \begin{equation}
        \alpha=1+\frac{\sqrt{2\log_2\frac{2}{\varepsilon^2\Pr[\Omega]^2}}}{V\sqrt{n\ln2}}.
    \end{equation}
    The assumptions on $n$ and $w$ of the theorems means that this choice of $\alpha$ satisfies $\alpha\leq1+\frac{1}{2\ln2}$, and
    \begin{equation}
        K_\alpha=\frac{1}{6\cdot(2-\alpha)^3\ln2}2^{\frac{w\sqrt{2\log_2\frac{2}{\varepsilon^2\Pr[\Omega]^2}}}{V\sqrt{n\ln2}}}\ln^3(2^w+e^2).
    \end{equation}
    Further, with the aforementioned lower bound on $n$, the upper bound on $\alpha$, and the lower bound on $w$, we have
    \begin{equation}
        K_\alpha\leq \frac{1}{6\cdot(1-\frac{1}{2\ln2})^3\ln2}\ln^3(2^{w+1})2^{\frac{w\sqrt{2\log_2\frac{2}{\varepsilon^2\Pr[\Omega]^2}}}{V\sqrt{n\ln2}}}\leq 4\times(w+1)^3 \times2^{\frac{w\sqrt{2\log_2\frac{2}{\varepsilon^2\Pr[\Omega]^2}}}{V\sqrt{n\ln2}}}.
    \end{equation}
    Finally, substituting $\alpha$ into the expression of the smooth min-entropy completes the proof.
\end{proof}

This results in the following theorem.
\begin{theorem}\label{thm:vanilla_EAT}
    Let $\mathcal{A}_1,\dots,\mathcal{A}_L$ be $(T,m)$-query sequential processes given access to the first system of a bipartite state $\rho_{DE}$ and outputting $\vec{z}_1,\dots,\vec{z}_L$. If $n\geq 50$, $k\leq N^{-1/2}$, $m,n\leq 10^6$, and the protocol in Figure \ref{fig:protocol} does not abort, then
    \begin{equation}
        H_{\rm min}^\varepsilon(\vec{Z}^L\vert \vec{U}^L E)_{\mathcal{M}_L\circ\cdots\circ\mathcal{M}_1(\rho)\vert\Omega}\geq Lh(s^*)-\sqrt{L}c-c',
    \end{equation}
    where $c,c',V$ follow the definitions in Theorem \ref{thm:gen_v2} in the case of classical $A$,
    \begin{align}
        h(s)&=m\left[\left(s - 1 - F_{C} - 2m\varepsilon'(n\ln 2 + 3.5)-1.001\frac{16(k+\ell)^2\ln^2N}{N}-0.001\right)n-\log_2 n -\log_2(k+\ell) - 5\right]-2\\
        w &= 2mn+h(Np_{\rm max})-h(0)\\
        \mathrm{var}(f)&\leq \frac{1}{\gamma}(h(Np_{\rm max})-h(0))^2,
    \end{align}
    and $\varepsilon'$ is as defined in Theorem \ref{thm:gen_7.15}.
\end{theorem}
\begin{proof}
Consider a quantum channel that takes as input the bitstring for the $i$th round on register $\vec{Z}_i$ and outputs a test output on register $X_i$. Specifically, the domain of $X_i$ is $\mathcal{X}=\mathcal{X}'\cup \{\perp\}$ and $\mathcal{X}'=[0, Np_{\rm max}]$. With probability $\gamma$, the channel output a test score $X_i\in\mathcal{X}'$; with probability $1-\gamma$, the channel does not test and outputs $X_i=\perp$. For each test round, say the $i$th round, where the single validated circuit has bitstring probability $p$, the test output is the score $N\min(p, p_{\rm max})$.

For entropy accumulation, we use Theorem \ref{thm:gen_v2}, and define the min-tradeoff function as Lemma V.5 of Ref.~\cite{dupuis2019entropy}. We first define $g:\mathbb{P}(\mathcal{X}')\rightarrow\mathbb{R}$, where $\mathbb{P}(\mathcal{X}')$ is the set of all probability distributions over $\mathcal{X}'$.
\begin{equation}
    g(q')=m\left[\left(s(q') - 1 -F_{C}-2m\varepsilon'(n\ln 2 + 3.5)-1.001\frac{16(k+\ell)^2\ln^2N}{N}-0.001\right)n-\log_2 n -\log_2(k+\ell) - 5\right]-2,
\end{equation}
where
\begin{equation}
    s(q')=\int_{\mathcal{X}'}xq'(x)dx
\end{equation}
is the average score of the distribution $q'$, and $q'$ is the probability distribution of the test output in the case of a test round. Since having an average score of $s$ implies solving $(1+s)-\mathrm{MXHOG}$, the definition of $g(q')$ immediately follows from Eq. 42 of Ref.~\cite{dupuis2019entropy} and Eq. \ref{eqn:entropy_with_classical}.

Then, by Lemma V.5 of Ref.~\cite{dupuis2019entropy}, the following function $f(q):\mathbb{P}(\mathcal{X})\rightarrow\mathbb{R}$ is a min-tradeoff function:
\begin{align}
    f(\delta_x) &=\max(g)+\frac{1}{\gamma}(g(\delta_x)-\max(g)) \forall x\in\mathcal{X}'\\
    f(\delta_\perp)&=\max(g).
\end{align}

By the same lemma,
\begin{align}
    \max(f)=&\max(g)\\
    \min_\Sigma(f)\geq&\min(g)\\
    \mathrm{var}(f)\leq & \frac{1}{\gamma}(\max(g)-\min(g))^2.
\end{align}
Since $\max(g)=h(Np_{\rm max})$ and $\min(g)=h(0)$, plugging everything into Theorem \ref{thm:gen_v2} completes the proof.
\end{proof}

We can evaluate the entropy guarantee of the protocol using Theorem \ref{thm:gen_v2}. Alternatively, we can use Eq. \ref{eqn:entropy_w_alpha}, substitute the value of $V$ as in Theorem \ref{thm:gen_v2} but keep the $\alpha$ dependence of $K_\alpha$, and optimize over $\alpha$. Since the result of Theorem \ref{thm:gen_v2} is with a specific non-optimal value of $\alpha$, the entropy guarantee obtained by the second approach is strictly better. However, in the regime of high entropy rate, the two approaches yield similar results as the choice of $\alpha$ made by Theorem \ref{thm:gen_v2} is nearly optimal.

Finally, we consider the improved single-round entropy without the $\log_2(k+\ell)$ penalty for the special case of $m=1$ and $T=1$. We additionally assume that the adversary's quantum computer has fidelity $\phi_{\rm adv}$. We assume that with probability $1-\phi$, at least a single error occurs when executing the oracle of the quantum circuit. We can assume that a single error results in the output state of a completely random Haar random quantum circuit. This motivates us to define the following notion of fidelity.

\begin{definition}\label{def:operational_fidelity_definition}
The quantum device output is $\rho_{\rm noisy}(\vec{z})=\phi\rho_{\rm ideal}(\vec{z})+(1-\phi)\sigma(\vec{z})$, where $\rho_{\rm ideal}$ is the probability distribution of an ideal quantum device and $\sigma$ is some other normalized probability distribution independent of the quantum circuit.
\end{definition}
Now, we can improve the entropy bound considering a finite fidelity quantum adversary.
\begin{theorem}\label{thm:vanilla_EAT_special}
Let $\mathcal{A}_1,\dots,\mathcal{A}_L$ be a fidelity $\phi_{\rm adv}$ (as defined in Definition \ref{def:operational_fidelity_definition}) $(1,1)$-query sequential processes with given access to the first system of a bipartite state $\rho_{DE}$ and outputting $\vec{z}_1,\dots,\vec{z}_L$. If $n\geq 50$, $k\leq N^{-1/2}$, $n\leq 10^6$, and the protocol in Figure \ref{fig:protocol} does not abort, then
\begin{equation}
    H_{\rm min}^\varepsilon(Z^L\vert U^L E)_{\mathcal{M}_L\circ\cdots\circ\mathcal{M}_1(\rho)\vert\Omega}\geq Lh(s^*)-\sqrt{L}c-c',
\end{equation}
where $c,c',V$ follow the definitions in Theorem \ref{thm:gen_v2} in the case of classical $A$, and
\begin{align}
    h(s)&=\phi_{\rm adv}\left[\left(\frac{s - 1 - F_{C}}{\phi_{\rm adv}} - 2\varepsilon''(n\ln 2 + 3.5)-1.001\frac{16\ln^2N}{N}-0.001\right)n-\log_2 n - 7\right]\\
    w &= 2n+h(Np_{\rm max})-h(0)\\
    \mathrm{var}(f)&\leq \frac{1}{\gamma}(h(Np_{\rm max})-h(0))^2,
\end{align}
where $\varepsilon''$ is as defined in Corollary \ref{cor:improved_general_adv_m1_T1}.
\end{theorem}
\begin{proof}
The proof follows the same structure as Theorem \ref{thm:vanilla_EAT} with the appropriate new single-round bound due to Corollary \ref{cor:improved_general_adv_m1_T1}. The effect of finite fidelity is to rescale the score of $\rho_{\rm ideal}$ since the ideal state must do better on the test score for the average score to be $s$. For a mixture of $m'$ classical samples per round and $m''=m-m'$ quantum samples where the quantum computer is ideal, the quantum algorithm must solve $(1+\delta'')-\mathrm{MXHOG}_{m''}$ with
\begin{equation}
\delta''=\frac{1}{1-m'/m}\left(\delta-\frac{m'}{m}\delta'\right)=\frac{\delta-F_{\rm C}}{1-m'/m}
\end{equation}
as discussed in Eq. \ref{eqn:xhog_score_of_quantum_part_of_hybrid}. For a quantum device with fidelity $\phi_{\rm adv}$, we have $\delta''\rightarrow \delta''/\phi_{\rm adv}$.
The entropy is scaled by $\phi_{\rm adv}$ since we assume $\sigma$ may not contribute any entropy. Propagating the new $\delta''$ and rescaling entropy by $\phi_{\rm adv}$ completes the proof.
\end{proof}

\subsection{Improving the variance}\label{sec:improve_variance}
However, both approaches still do not provide very good entropy estimates. Although the dependence on the testing probability $\gamma$ is improved from \cite{dupuis2020entropy} to \cite{dupuis2019entropy}, the dependence on the system size is worse. Further, the test bitstring probability range for each round is $[0, p_{\rm max}]$, while the mean value of the is much lower than $p_{\rm max}$ (e.g. one can reasonably choose $p_{\rm}=10/2^n$ to not lose too much of the probability mass, but the mean score is at most $2/2^n$). This means that the variance $\mathrm{var}(f)$ is large, which results in a significant penalty.

One approach to improve the variance is to reduce $p_{\rm max}$. Naively speaking, reducing $p_{\rm max}$ would cause a reduction in the XEB score in the honest case. For the malicious case, an adversary could output samples completely within $[0, p_{\rm max}]$ and the XEB score may not be affected. Fortunately, this is not the case, and reducing $p_{\rm max}$ must reduce the adversary's XEB score as well. This is because the distribution of the bitstring probability $p$ must be close to that of finite-fidelity honest sampling from a depolarized state as given by Lemma \ref{lem:dist_mixture}. Overall, this allows us to reduce $p_{\rm max}$. Although the honest case XEB score would reduce, the entropy guarantee should not reduce significantly due to the inability for any adversary to do well.

\begin{theorem}\label{thm:truncated_XEB_implies_entropy}
Let $\mathcal{A}$ be any algorithm making $(T,m_{\mathrm{Q}})$-queries to $m_{\mathrm{Q}}$ Haar random independent unitaries producing $m_{\mathrm{Q}}$ bitstrings. Let the rest $m_{\mathrm{Q}}=m-m_{\mathrm{Q}}$ bitstrings be produced by a classical simulation algorithm satisfying the conditions in Lemma \ref{lem:dist_classical} and that $\sum_{i\in[m_{\mathrm{Q}}]}\phi_i=\Phi_{\rm C}$. If $n\geq 50$, $k\leq N^{-1/2}$, $m,n\leq 10^6$, and $\mathcal{A}$ solves $(1+\delta)-\mathrm{MXHOG}_m$ when the bitstring probability contributions are capped at $p_{\rm max}$, then
\begin{align}
H(\vec{Z}\vert\vec{U}E)_{\mathcal{A}^{\vec{C}}}&\geq mn\left\{\left[\frac{\delta\cdot(1-N^{-1/3}) - 1 - dx_{\rm max} - x_{\rm min}+e^{-(x_{\rm max}-x_{\rm min})}}{1-e^{-(x_{\rm max}-x_{\rm min})}(1+x_{\rm max}-x_{\rm min})}-\frac{\Phi_{\rm C}}{m}\right]\left[1-\frac{(k+\ell)^2}{N}\right]-C\right\}-2\\
0&=(1-(1+Np_{\rm min})e^{-Np_{\rm min}})-m\max(d_{\rm Q}, d_{\rm C}),
\end{align}
where $x_{\rm min}=N(1-N^{-1/3})p_{\rm min}, x_{\rm max}=N(1-N^{-1/3})p_{\rm max}$, and $d_{\rm Q},d_{\rm C},\Phi_{\rm C},C$ be defined as in Lemma \ref{lem:dist_mixture}. Note that $C$ of Lemma \ref{lem:dist_mixture} is defined in Lemma \ref{lem:low_h_implies_low_contain_p_prob}.
\end{theorem}

\begin{proof}
    By the same arguments as Theorem \ref{thm:multi_round_entropy_ad_hoc}, the dominant cumulative density function of the truncated bitstring probability can be described by the following sampling process of $p$:
\begin{enumerate}
    \item With probability $d=m\max(d_{\rm Q},d_{\rm C})$, set $p=p_{\rm max}$.
    \item With probability $(1-d)(1-\Phi)$, set $p=\frac{Q}{N(1-N^{-1/3})}+p_{\rm min}$ with $Q\sim\Gamma(1,1)$.
    \item Otherwise, set $p=\frac{Q}{N(1-N^{-1/3})}+p_{\rm min}$ with $Q\sim\Gamma(2,1)$.
    \item After this, if $p>p_{\rm max}$, set $p$ to $p_{\rm max}$.
\end{enumerate}
Here, all rounds are considered to have the same $p_{\rm min}$, which is taken as the maximum possible value attainable when $\Phi_a=1$, which is given by the conditions of Theorem \ref{thm:multi_round_entropy_ad_hoc}. We have added the final truncation here. This is in contrast to Theorem \ref{thm:multi_round_entropy_ad_hoc}, where not truncating after leads to easier analysis.

The expected value of $x=N(1-N^{-1/3})p$ is therefore a weighted sum of the expected value of $x$ in three cases: setting $p=p_{\rm max}$ directly, setting $p=\min(p_{\rm max}, \frac{Q}{N(1-N^{-1/3})}+p_{\rm min})$ with $Q\sim\Gamma(1,1)$, and setting $p=\min(p_{\rm max}, \frac{Q}{N(1-N^{-1/3})}+p_{\rm min})$ with $Q\sim\Gamma(2,1)$.

In the second case, the expected value is
\begin{align}
    \langle x\rangle_2&=\int_{0}^{x_{\rm max}-x_{\rm min}}(Q+x_{\rm min})\mathrm{PDF}(Q)dQ+\Pr[Q>x_{\rm max}-x_{\rm min}]\cdot x_{\rm max}\\
    &=\int_{0}^{x_{\rm max}-x_{\rm min}}(Q+x_{\rm min})e^{-Q}dQ+x_{\rm max}\left(1-\int_{0}^{x_{\rm max}-x_{\rm min}}e^{-Q}dQ\right)\\
    &=x_{\rm max} - (x_{\rm max} - x_{\rm min})[1-e^{-(x_{\rm max}-x_{\rm min})}]+[1-(1+x_{\rm max}-x_{\rm min})e^{-(x_{\rm max}-x_{\rm min})}]\\
    &=1+x_{\rm min}-e^{-(x_{\rm max}-x_{\rm min})}.
\end{align}
In the third case, the expected value is
\begin{align}
    \langle x\rangle_3&=\int_{0}^{x_{\rm max}-x_{\rm min}}(Q+x_{\rm min})\mathrm{PDF}(Q)dQ+\Pr[Q>x_{\rm max}-x_{\rm min}]\cdot x_{\rm max}\\
    &=\int_{0}^{x_{\rm max}-x_{\rm min}}(Q+x_{\rm min})Qe^{-Q}dQ+x_{\rm max}\left(1-\int_{0}^{x_{\rm max}-x_{\rm min}}Qe^{-Q}dQ\right)\\
    &=x_{\rm max} - (x_{\rm max} - x_{\rm min})[1-(1+x_{\rm max}-x_{\rm min})e^{-(x_{\rm max}-x_{\rm min})}]+2\cdot\left[1-\left(1+x_{\rm max}-x_{\rm min}+\frac{1}{2}(x_{\rm max}-x_{\rm min})^2\right)e^{-(x_{\rm max}-x_{\rm min})}\right]\\
    &=2+x_{\rm min}-e^{-(x_{\rm max}-x_{\rm min})}[2+x_{\rm max}-x_{\rm min}].
\end{align}
Therefore, the overall mean value of $x$ is
\begin{align}
    \langle x\rangle&\leq dx_{\rm max} + (1-d)(1-\Phi)\langle x\rangle_2 + \Phi(1-d)\langle x\rangle_3\\
    &\leq dx_{\rm max} + (1-\Phi)\langle x\rangle_2 + \Phi\langle x\rangle_3\\
    &=dx_{\rm max}+\langle x\rangle_2+\Phi\cdot(\langle x\rangle_3-\langle x\rangle_2)\\
    &=dx_{\rm max}+1+x_{\rm min}-e^{-(x_{\rm max}-x_{\rm min})}+\Phi\cdot\left[1-e^{-(x_{\rm max}-x_{\rm min})}(1+x_{\rm max}-x_{\rm min})\right].
\end{align}
We then have 
\begin{equation}
    \delta=\frac{\langle x\rangle}{1-N^{-1/3}}-1\Rightarrow\langle x\rangle=(\delta+1)(1-N^{-1/3}),
\end{equation}
which implies
\begin{align}
    \Phi&\geq\frac{(\delta+1)\cdot(1-N^{-1/3}) - 1 - dx_{\rm max} - x_{\rm min}+e^{-(x_{\rm max}-x_{\rm min})}}{1-e^{-(x_{\rm max}-x_{\rm min})}(1+x_{\rm max}-x_{\rm min})}.
\end{align}
Using the definition of $\Phi$ in Lemma \ref{lem:dist_mixture} allows us to lower bound $h$ and complete the proof.
\end{proof}

\begin{remark}
When $p_{\rm max}$ is too small, more of the tail of the distribution for large $p$ is ignored, which hinders the ability to discriminate between different fidelities.
\end{remark}

Another approach to improve $\mathrm{var}(f)$ is to use the equality in Lemma V.5 of Ref.~\cite{dupuis2019entropy} instead of the upper bound.
\begin{lemma}\label{lem:improved_var}
The variance $\mathrm{var}(f)$ obeys
\begin{align}
\mathrm{var}(f)&\leq\frac{1}{\gamma}\int_{0}^{x_{\rm max}}q'(x)(\max(g)-g(\delta_x))^2dx-(\max(g)-g(q'))^2\\
g(q')&=\int_{0}^{x_{\rm max}}q'(x)g(\delta_x)dx\\
q'(x)&=m\max(d_{\rm Q}, d_{\rm C})\cdot \delta(x)+e^{-x}\cdot I_{(0,x_{\rm max})}(x)+\left[e^{-x_{\rm max}}-m\max(d_{\rm Q}, d_{\rm C})\right]\cdot\delta(x-x_{\rm max})\\
I_{(0, a)}(x) &= 
\begin{cases}
1~\mathrm{if}~0<x<a\\
0~\mathrm{otherwise},
\end{cases}\\
d_{\rm Q}&=N^{-1/3}+(k+\ell)^2/N+\varepsilon',\\
    \varepsilon'&=T\left(\frac{4}{2^{n/2}} + \sqrt{2\left(1-\left(\cos\frac{\pi}{2k}\right)^k\right)}+2/\sqrt{\ell+1}\right).
\end{align}
\end{lemma}

\begin{proof}
As shown in Lemma V.5 of Ref.~\cite{dupuis2019entropy},
\begin{align}
\sum_{x\in\mathcal{X}}q(x)\left(f(\delta_x)-\sum_{x\in\mathcal{X}}q(x)f(\delta_x)\right)^2=\frac{1}{\gamma}\sum_{x\in\mathcal{X}'}q'(x)((\max(g)-g(\delta_x))-\gamma(\max(g)-g(q')))^2+(1-\gamma)(\max(g)-g(q'))^2,
\end{align}
and the first term is expanded as
\begin{align}
\sum_{x\in\mathcal{X}'}\frac{q'(x)}{\gamma}(\max(g)-g(\delta_x))^2-2(\max(g)-g(q'))^2+\gamma(\max(g)-g(q'))^2.
\end{align}
Combining these existing results, we have
\begin{align}
\sum_{x\in\mathcal{X}}q(x)\left(f(\delta_x)-\sum_{x\in\mathcal{X}}q(x)f(\delta_x)\right)^2=\sum_{x\in\mathcal{X}'}\frac{q'(x)}{\gamma}(\max(g)-g(\delta_x))^2-(\max(g)-g(q'))^2.\label{eqn:dupuis_var_f}
\end{align}
Further, by Definition V.1. of Ref.~\cite{dupuis2019entropy},
\begin{equation}
    \mathrm{var}(f)=\max_{q:\Sigma_i(q)\neq\emptyset}\sum_{x\in\mathcal{X}}q(x)\left(f(\delta_x)-\sum_{x\in\mathcal{X}}q(x)f(\delta_x)\right)^2,
\end{equation}
we must maximize Eq. \ref{eqn:dupuis_var_f} over all valid $q'$. Fortunately, we know the test output probability distribution must be close to that of finite fidelity sampling. For example, see Lemma \ref{lem:dist_mixture}. By examining Eq. \ref{eqn:dupuis_var_f}, we see that pushing as much probability mass to $x=0$ maximizes the variance. However, the CDF in Lemma \ref{lem:dist_mixture} pushes as much probability mass to $p_{\rm max}$ as possible. Following the same steps that lead to Lemma \ref{lem:dist_mixture} except pushing probability mass to low values, we get the CDF
\begin{align}
Q'(x)=
\begin{cases}
 1-e^{-x}+m\max(d_{\rm Q}, d_{\rm C}) & \mathrm{if~} 0 < x < x_{\rm max}\\
1 & \mathrm{otherwise},
\end{cases}
\end{align}
which corresponds to the PDF $q'$ as stated in the Lemma. Substituting $q'$ into Eq. \ref{eqn:dupuis_var_f} completes the proof.
\end{proof}

With the aforementioned two improvements, applying the entropy accumulation theorem results in an improved multi-round entropy guarantee.
\begin{corollary}\label{cor:limit_pmax}
    If the protocol in Figure \ref{fig:protocol} does not abort, then
    \begin{equation}
        H_{\rm min}^\varepsilon(\vec{Z}^L\vert \vec{U}^L E)_{\mathcal{M}_L\circ\cdots\circ\mathcal{M}_1(\rho)\vert\Omega}\geq Lh(s^*)-\sqrt{L}c-c',
    \end{equation}
    where $c,c',V$ follow the definitions in Theorem \ref{thm:gen_v2} in the case of classical $A$,
    \begin{align}
        h(s)&=mn\left\{\left[\frac{s\cdot(1-N^{-1/3}) - 1 - dx_{\rm max} - x_{\rm min}+e^{-(x_{\rm max}-x_{\rm min})}}{1-e^{-(x_{\rm max}-x_{\rm min})}(1+x_{\rm max}-x_{\rm min})}-\frac{\Phi_{\rm C}}{m}\right]\left[1-\frac{(k+\ell)^2}{N}\right]-C\right\}-2\\
        w &= 2mn+h(Np_{\rm max})-h(0)\\
        \mathrm{var}(f)&\leq\frac{1}{\gamma}\int_{0}^{x_{\rm max}}q'(x)(h(Np_{\rm max})-h(x))^2dx-(h(Np_{\rm max})-h(q'))^2\\
        h(q')&=\int_{0}^{x_{\rm max}}q'(x)h(x)dx,
    \end{align}
    $x_{\rm min},x_{\rm max},\Phi_{\rm C}$ are as defined in \ref{thm:truncated_XEB_implies_entropy}, $q'$ is as defined in Lemma \ref{lem:improved_var}, and $C$ is as defined in Lemma \ref{lem:dist_mixture}.
\end{corollary}
\begin{proof}
The proof follows the same steps as Theorem \ref{thm:vanilla_EAT} using the single round entropy bound in Theorem \ref{thm:truncated_XEB_implies_entropy} as well as $\mathrm{var}(f)$ in Lemma \ref{lem:improved_var}.
\end{proof}

For the special case of $T=1,m=1$, we can use the improved single-round entropy bound.
\begin{corollary}\label{cor:limit_pmax_special}
For $T=1,m=1$, Corollary \ref{cor:limit_pmax} holds with $C,\varepsilon''$ as defined in Theorem \ref{thm:nonadaptive_special} and $\varepsilon'\rightarrow\varepsilon''$.
\end{corollary}

Additionally, we can put heuristic adjustments assuming the adversary's quantum computer has limited fidelity like before. The output probability distribution of the quantum algorithm is modified from Lemma \ref{lem:dist_A}.
\begin{corollary}\label{cor:finite_fidelity_truncated_eat_entropy}
    Let $\mathcal{A}_1,\dots,\mathcal{A}_L$ be a fidelity $\phi_{\rm adv}$ (as defined in Definition \ref{def:operational_fidelity_definition}) $(T,m)$-query sequential processes given access to the first system of a bipartite state $\rho_{DE}$ and outputting $\vec{z}_1,\dots,\vec{z}_L$. If $n\geq 50$, $k\leq N^{-1/2}$, $m,n\leq 10^6$, and the protocol in Figure \ref{fig:protocol} does not abort, then the output has the same entropy as Corollary \ref{cor:limit_pmax} except $C\rightarrow\phi C$. For $T=1,m=1$, the output has the same entropy as Corollary \ref{cor:limit_pmax_special} except $C\rightarrow\phi C$.
\end{corollary}
\begin{proof}
    We start with the difference in Lemma \ref{lem:dist_A} which specifies the distribution of the quantum algorithm given an upper bound of the entropy. For the sample of $\mathcal{A}'$ to still have entropy $h$, the entropy of $\mathcal{A}$ must be $h/\phi_{\rm adv}$. The cumulative density function should be a weighted sum of the original function with $h\rightarrow h/\phi_{\rm adv}$ and $1-e^{-Np}$ (corresponding to the bitstring probability of a fixed bitstring, which is the same as that of uniformly random bitstrings since it is a probability over Haar random unitaries) with weight $\phi_{\rm adv}$ and $1-\phi_{\rm adv}$. The additional discrepancy is acceptable since we want a dominant function.
    
    This difference from Lemma \ref{lem:dist_A} can be propagated to Lemma \ref{lem:dist_mixture}, resulting in
    \begin{equation}
        \Phi\rightarrow \frac{1}{1-(k+\ell)^2/N}\left(\frac{h+2\phi_{\rm adv}}{mn}+\phi_{\rm adv}C\right)+\frac{\Phi_{\rm C}}{m}.
    \end{equation}
    Finally, propagating this into Theorem \ref{thm:truncated_XEB_implies_entropy} and recognizing that $-2$ is worse than $-2\phi_{\rm adv}$ completes the proof.
\end{proof}

Incorporating all improvements, we have the following theorem which explicitly tracks all expressions.

\begin{theorem}\label{thm:final_oracle_model_theorem}
Let $\mathcal{A}_1,\dots,\mathcal{A}_L$ be a fidelity $\phi_{\rm adv}$ (as defined in Definition \ref{def:operational_fidelity_definition}) $(1,1)$-query sequential processes with given access to the first system of a bipartite state $\rho_{DE}$ and outputting $\vec{z}_1,\dots,\vec{z}_L$. Denote the event that the protocol in Figure \ref{fig:protocol} succeeds as $\Omega$. If $n\geq 50$, $n\leq 10^6$, then the protocol either aborts with probability at least $1-\varepsilon_{\rm accept}$, or
\begin{equation}
H_{\rm min}^{\varepsilon_{\rm smooth}}(\vec{Z}^L\vert \vec{U}^L E)_{\mathcal{M}_L\circ\cdots\circ\mathcal{M}_1(\rho)\vert\Omega}\geq Lh(s^*)-\sqrt{L}c-c',
\end{equation}
where
\begin{align}\nonumber
    h(s)=&n\left\{\left(t(s)-
    \Phi_{\rm C}\right)\cdot\left[1-\frac{(k+\ell)^2}{N}\right]-\phi_{\rm adv}C\right\}-2\\\nonumber
    t(s)=&\frac{s\cdot(1-\frac{1}{N^{\frac{1}{3}}}) - 1 - dx_{\rm max} - x_{\rm min}+e^{x_{\rm min}-x_{\rm max}}}{1-e^{x_{\rm min}-x_{\rm max}}(1+x_{\rm max}-x_{\rm min})}\\\nonumber
    C =& \frac{1}{n}\left(\log_2 n+3\right)+4\varepsilon''(n\ln2+3.5)+1.001\times\frac{16\ln^2N}{N}+0.001\\\nonumber
    x_{\rm max}=&N(1-N^{-1/3})p_{\rm max}\\\nonumber
    x_{\rm min}=&N(1-N^{-1/3})p_{\rm min}\\\nonumber
    0=&(1-(1+Np_{\rm min})e^{-Np_{\rm min}})-d\\\nonumber
    d=&\max(d_{\rm Q}, d_{\rm C})\\\nonumber
    d_{\rm Q}=&N^{-1/3}+(k+\ell)^2/N+\varepsilon''\\\nonumber
    \varepsilon''=&\frac{4}{2^{n/2}} + \sqrt{2\left(1-\left(\cos\frac{\pi}{2k}\right)^k\right)}+2/\sqrt{\ell+1}+\frac{2(k+\ell)(k+\ell-1)}{N}+\frac{2k}{N(1-k/N)}\\\nonumber
    c=&V\sqrt{(2\ln2)\log_2\frac{2}{\varepsilon_{\rm smooth}^2\varepsilon_{\rm accept}^2}}\\\nonumber
    c'=&\frac{2\log_2\frac{2}{\varepsilon_{\rm smooth}^2\varepsilon_{\rm accept}^2}}{V^2\ln2}K\\\nonumber
    K=&4\times(w+1)^3 \times2^{\frac{w\sqrt{2\log_2\frac{2}{\varepsilon_{\rm smooth}^2\varepsilon_{\rm accept}^2}}}{V\sqrt{n\ln2}}}\\\nonumber
    V=&\sqrt{\mathrm{var}+2}+\log_2(2\times2^n+1)\\\nonumber
    w =& 2n+h(Np_{\rm max})-h(0)\\\nonumber
    \mathrm{var}\leq&\frac{1}{\gamma}\int_{0}^{x_{\rm max}}q'(x)(h(Np_{\rm max})-h(x))^2dx-(h(Np_{\rm max})-h(q'))^2\\\nonumber
    h(q')=&\int_{0}^{x_{\rm max}}q'(x)h(x)dx\\\nonumber
    q'(x)=&\max(d_{\rm Q}, d_{\rm C})\cdot \delta(x)+e^{-x}\cdot I(x)+\left[e^{-x_{\rm max}}-\max(d_{\rm Q}, d_{\rm C})\right]\cdot\delta(x-x_{\rm max})\\\nonumber
    I(x) =& 
    \begin{cases}
    1~\mathrm{if}~0<x<x_{\rm max}\\\nonumber
    0~\mathrm{otherwise},
    \end{cases}
\end{align}
any natural number $k,\ell\leq \sqrt{N}=2^{n/2}$.
\end{theorem}

\begin{figure}
    \centering
    \includegraphics[width=\linewidth]{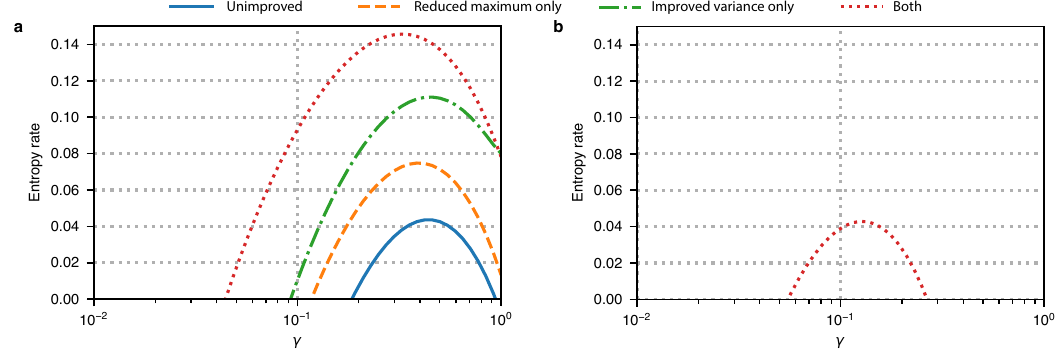}
    \caption{Entropy rate at different values of validation probability $\gamma$ for various entropy bounds. \textbf{a}, $n=64,m=1,L=100,000,\phi=0.6,\phi_{\rm adv}=0.65,\varepsilon_{\rm accept}=d_{\rm C}=\varepsilon=10^{-4},T_{\rm QC}=50$ ms, adversary's classical simulation algorithm is 4 times as powerful as Aurora and the quantum device has $T=1$ oracle access, the total validation budget is 300k Aurora node hours. \textbf{b}, same as \textbf{a} but for 15 times Aurora adversary. The unimproved version uses Theorem \ref{thm:vanilla_EAT_special}. The reduced maximum version only improves with the single round bound from Theorem \ref{thm:truncated_XEB_implies_entropy}. The improved variance only version only improves with the reduction in variance from Lemma \ref{lem:improved_var}. The version with both improvements uses Corollary \ref{thm:final_oracle_model_theorem}. In all cases, for each $\gamma$, the hardness of the challenge circuits is set to be the total validation budget divided by $100,000\times\gamma$, the average number of validated circuits.}\label{fig:multiround_improvement}
\end{figure}

We can observe the effect of the above improvements in Fig. \ref{fig:multiround_improvement}. Being able to obtain entropy at small $\gamma$ means that the protocol is more robust against stronger adversaries. For example, when the adversary power is increased to 10 times Aurora, only incorporating both improvements achieves non-zero entropy.

\section{Randomness amplification}\label{sec:amplification}

The protocol described in Secs.~\ref{sec:app-single-round} and~\ref{sec:app-multi-round} is for randomness \emph{expansion}: it takes near-perfect randomness (randomness that is negligibly close to uniform in statistical distance) as input and produces more of it as output. This leads to a circularity and, in practice, we often only have \emph{weak sources} that produce imperfect or biased randomness. Randomness \emph{amplification} converts such imperfect or biased randomness into near-perfect randomness. While this is impossible classically without strong assumptions on the source~\cite{santha1986generating}, it becomes possible with quantum methods. In this section, we extend our protocol to achieve certified randomness amplification.

To perform randomness amplification rather than expansion, we must address two challenges arising from the lack of near-perfect randomness as a resource: executing the protocol itself and extracting randomness from its outputs.
\begin{itemize}
    \item One does not have near-perfect randomness for generating circuits to challenge the quantum computer, nor to select the samples to verify. In Sec.~\ref{sec:app-pseudorandom-unitary-w-imperfect-rand}, we show that imperfect randomness suffices for challenge circuit generation. In Sec.~\ref{sec:computationally_indistinguishable_from_uniform}, we propose methods for generating computationally indistinguishable randomness and show it is sufficient for both measurement-basis and verification-sample selection.
    \item One does not have access to near-perfect randomness to perform randomness extraction. This necessitates the use of a \emph{two-source extractor}. However, this is challenging because quantum computer responses often have very low entropy rates (for instance, in~\cite{jpmc_cr} the certified entropy rate is roughly $0.04$). Typical two-source extractors require $\tfrac{k_1}{n_1}+\tfrac{k_2}{n_2}>1$, where $\tfrac{k_i}{n_i}$ is the min-entropy rate of the $i$th source, which would imply an unrealistically strong assumption ($k_1 \approx n_1$) on the weak source if relying on prior extractor constructions. Even when the entropy rate of the quantum responses become higher (say $\frac{k_1}{n_1}>0.5$), this would still require assuming $\frac{k_2}{n_2}>1-\frac{k_1}{n_1}$ on the weak source. To address this, we use the recently developed efficient implementation of Raz’s extractor~\cite{foreman25} with improved parameters.
\end{itemize}

We assume access to a weak source that is a $(n,k)$ block min-entropy source, producing $n$-bit blocks with min-entropy $k$, conditioned on the adversary’s side information $E$ and on all previously generated blocks. Such source is sufficient to perform randomness extraction and to select the unitaries, as we describe later. For the measurement basis and for the selection of samples for verification, however, we only require pseudo randomness. The idea is that the randomness used for the measurement bases and select verification samples only needs to be unpredictable \textit{during the protocol execution only} against the adversaries we consider. Overall, in addition to being a min-entropy source, we also assume that the weak source of randomness is computationally indistinguishable from a source that can be extracted from, a condition that only needs to hold temporarily. The protocol is schematized in Fig.\ref{fig:amplification_diagram}.%

\subsection{Pseudorandom unitary with imperfectly random keys}\label{sec:app-pseudorandom-unitary-w-imperfect-rand}
Unfortunately, the current theory does not readily permit the use of pseudorandom unitaries generated from imperfectly random keys, as the existing security analyses assumes ideal randomness as input. For the ensemble of challenge circuits to be suitable for certified randomness, \cite{aaronson2023certifiedArxiv} argues that it is sufficient for the ensemble to be quantum statistical zero-knowledge (QSZK) indistinguishable from the Haar measure. We prove that with imperfect randomness as input, the assumption that there exists pseudorandom unitaries QSZK-indistinguishable from the Haar measure holds relative to a random oracle. Specifically, we generalize Assumption 8.8 of Ref.~\cite{aaronson2023certifiedArxiv} by allowing the input $k$ to come from some probability distribution $\mathcal{D}$ of $k$ that has min-entropy $O(\ell)$ as opposed to the uniform distribution used in Ref.~\cite{aaronson2023certifiedArxiv}.

\begin{assumption}[Generalization of Assumption 8.8 of Ref.~\cite{aaronson2023certifiedArxiv}, pseudorandom unitary assumption]\label{assum:gen_8.8}
Let $\kappa\in\mathbb{N}$ be the security parameter, $\ell,m$ be polynomially bounded functions. There exists a family of keyed unitaries $\{U_k\in\mathbb{U}(2^{m(\kappa)}):k\in\{0,1\}^{\ell(\kappa)}\}$ such that the following conditions hold:
\begin{enumerate}
    \item There exists a polynomial-time quantum algorithm $G$ that implements $U_k$ on input $k\in\{0,1\}^{\ell(\kappa)}$, i.e., on input $k$ and quantum state $\vert\psi\rangle\in\mathbb{S}(2^{m(\kappa)}), G(k,\vert\psi\rangle)=U_k(\psi)$.
    \item For every pair of quantum algorithm $\mathcal{A},\mathcal{B}$ that makes $\mathrm{poly}(\kappa)$ queries and probability distribution $\mathcal{D}$ of $k$ that has min-entropy $O(\ell)$, it holds that
    \begin{equation}
        \bigg\vert\underset{k\sim D}{\mathbb{E}}\|\mathcal{A}^{U_k}-\mathcal{B}^{U_k}\|_{1}-\underset{U\sim\mathrm{Haar}(2^{m(\kappa)})}{\mathbb{E}}\|\mathcal{A}^{U}-\mathcal{B}^{U}\|_{1}\bigg\vert\leq\mathrm{negl}(\kappa).
    \end{equation}
\end{enumerate}
Here, $\mathbb{U}(d)$ is the set of all unitaries on a Hilbert space with dimension $d$ and $\mathbb{S}(d)$ is the set of all statevectors with dimension $d$.
\end{assumption}
We use the same construction relative to a oracle as in \cite{aaronson2023certifiedArxiv, kretschmer:LIPIcs.TQC.2021.2}. Namely, $\mathcal{O}=\sum_{k\in\{0,1\}^\ell}\vert k\rangle\langle k\vert\otimes\mathcal{O}_k$ where $\mathcal{O}_k\in\mathrm{Haar}(N)$ is a Haar random unitary for every $k\in\{0,1\}^\ell$. Let $\mu=\mathrm{Haar}(N)^{2^\ell}$ denote the measure of $\mathcal{O}$. To proceed, we modify the definition of the advantage we aim to upper bound in light of this new distribution of the keys $k$, i.e.,
\begin{align}\nonumber
\mathrm{adv}(\mathcal{O})&\equiv\underset{k\sim\mathcal{D},U\sim\mathrm{Haar}(N)}{\mathbb{E}}f(\mathcal{O},k,U)\\
&=\underset{k\sim\mathcal{D}}{\mathbb{E}}\|\mathcal{A}^{\mathcal{O},\mathcal{O}_k}-\mathcal{B}^{\mathcal{O},\mathcal{O}_k}\|_{1}-\underset{U\sim\mathrm{Haar}(N)}{\mathbb{E}}\|\mathcal{A}^{\mathcal{O},U}-\mathcal{B}^{\mathcal{O},U}\|_{1},
\end{align}
which is also $8T$-Lipschitz due to Lemma 8.10 of Ref.~\cite{aaronson2023certifiedArxiv}. Now we aim to upper bound this advantage by modifying subsequent lemmas. Specifically, $F$ is as defined in \cite{aaronson2023certifiedArxiv}.

\begin{lemma}[Generalization of Lemma 8.11 and 8.12 of Ref.~\cite{aaronson2023certifiedArxiv}]
Let $\mathcal{A},\mathcal{B}$ be a $T$-query algorithm, $\mathcal{D}$ be a distribution of $x\in\{0,1\}^{\ell(\kappa)}$ with min-entropy greater than $\alpha\ell$, and $F_0$ be the zero function. Then
\begin{equation}
    \underset{x\sim\mathcal{D}}{\mathbb{E}}\|\mathcal{A}^{F_x}-\mathcal{A}^{F_0}\|_{1}\leq 4T2^{-\alpha\ell/2},
\end{equation}
and
\begin{equation}
    \bigg\vert \underset{x\sim\mathcal{D}}{\mathbb{E}}\|\mathcal{A}^{F_x}-\mathcal{B}^{F_x}\|_{1} - \underset{x\sim\mathcal{D}}{\mathbb{E}}\|\mathcal{A}^{F_0}-\mathcal{B}^{F_0}\|_{1} \bigg\vert \leq 8T2^{-\alpha\ell/2}.
\end{equation}
\end{lemma}
\begin{proof}
    Everything is identical to the proof of Lemma 8.11 and 8.12 of Ref.~\cite{aaronson2023certifiedArxiv} except Eq.~263. Instead, we have
        \begin{align}
        \underset{x\sim\mathcal{D}}{\mathbb{E}}\|\mathcal{A}^{F_x}-\mathcal{A}^{F_0}\|_{1} &\leq 4\underset{x\sim\mathcal{D}}{\mathbb{E}}\left[\sum_{t=0}^{T-1}\|P_x\vert\psi_t^0\rangle\|\right]=4\sum_{t=0}^{T-1}\underset{x\sim\mathcal{D}}{\mathbb{E}}\left[\sqrt{\langle\psi_t^0\vert P_x\vert\psi_t^0\rangle}\right]=4\sum_{t=0}^{T-1}\underset{x\sim\mathcal{D}}{\mathbb{E}}\left[\sqrt{p_t^x}\right]=4\sum_{t=0}^{T-1}\sum_x\mathcal{D}(x)\sqrt{p_t^x},\label{eqn:imperfect_entropy_overlap}
    \end{align}
    where $\mathcal{D}(x)$ is the probability of obtaining $x$ from distribution $\mathcal{D}$ and $p_t^x\equiv\langle\psi_t^0\vert P_x\vert\psi_t^0\rangle$. By Cauchy-Schwarz inequality,
    \begin{equation}
        \left(\sum_x\mathcal{D}(x)\sqrt{p_t^x}\right)^2 \leq \left(\sum_x\mathcal{D}(x)^2\right)\cdot\left(\sum_x p_t^x\right)=\sum_x\mathcal{D}(x)^2,
    \end{equation}
    where we used the fact that $\sum_x p_t^x=\sum_x \langle\psi_t^0\vert x\rangle\langle x\vert\psi_t^0\rangle=1$. Since $\mathcal{D}(x)\leq 2^{-\alpha\ell}$ for all $x$ due to the min-entropy bound on $\mathcal{D}$, $\sum_x\mathcal{D}(x)^2$ is maximized when $\mathcal{D}(x)=2^{-\alpha\ell}$ for as many $x$ as possible. The number of such $x$ is at most $2^{\alpha\ell}$, and $\sum_x\mathcal{D}(x)^2\leq 2^{\alpha\ell} \left(2^{-\alpha\ell}\right)^2=2^{-\alpha\ell}$, which implies
    \begin{equation}
        \underset{x\sim\mathcal{D}}{\mathbb{E}}\|\mathcal{A}^{F_x}-\mathcal{A}^{F_0}\|_{1}\leq 4\sum_{t=0}^{T-1}\sum_x\mathcal{D}(x)\sqrt{p_t^x}=4T\sqrt{\sum_x\mathcal{D}(x)^2}\leq 4T2^{-\alpha\ell/2}.
    \end{equation}
    This results in the first result. Following the same proof as Lemma 8.12 of Ref.~\cite{aaronson2023certifiedArxiv} yields the second result.
\end{proof}

\begin{lemma}[Generalization of Lemma 8.13 of Ref.~\cite{aaronson2023certifiedArxiv}]\label{lem:gen_8.13}
    For every pair of $T$-query algorithms $\mathcal{A},\mathcal{B}$, $\vert\mathbb{E}_{\mathcal{O}\sim\mu}\mathrm{adv}(\mathcal{O})\vert\leq8T2^{-\alpha\ell/2}$.
\end{lemma}
\begin{proof}
    The proof is identical to  Lemma 8.13 of Ref.~\cite{aaronson2023certifiedArxiv}.
\end{proof}

\begin{theorem}[Generalization of Theorem 8.15 of Ref.~\cite{aaronson2023certifiedArxiv}]
    For $\alpha\ell=\lceil n-\log_2 n\rceil$ and $L=2^\ell$, $\Pr_{\mathcal{O}\sim\mu}[\vert\mathrm{adv}(\mathcal{O})\vert\geq16TL^{-\alpha}]\leq2^{-\Omega(n)}$.
\end{theorem}
\begin{proof}
    The proof is identical to Theorem 8.15 of Ref.~\cite{aaronson2023certifiedArxiv}. Note that there is a typo in Eq.~271 of Ref.~\cite{aaronson2023certifiedArxiv} when applying the constants.
\end{proof}

One undesirable feature of the above theorem is that it requires the key length to be precisely $\lceil n-\log_2 n\rceil$. Intuitively, increasing the key length should only improve the indistinguishability. We choose a slightly different parameterization for a more desirable result.
\begin{theorem}\label{thm:imperfect_challenge_ok}
    For $\alpha\ell\geq n$, $\Pr_{\mathcal{O}\sim\mu}[\vert\mathrm{adv}(\mathcal{O})\vert\geq2^{-\Omega(n)}]\leq2^{-\Omega(n)}$.
\end{theorem}
\begin{proof}
By the same argument as Theorem 8.15 of Ref.~\cite{aaronson2023certifiedArxiv}, Theorem 8.14 of Ref.~\cite{aaronson2023certifiedArxiv} implies
\begin{equation}
\Pr_{\mathcal{O}\sim\mu}\left[\mathrm{adv}(\mathcal{O})\geq\underset{\mathcal{P}\sim\mu}{\mathbb{E}}[\mathrm{adv}(\mathcal{P})]+t\right]\leq\exp\left(-\frac{(N-2)t^2}{24K^2}\right).
\end{equation}
Choosing $t=n^{1/2}2^{-n/2}$, by Lemma~\ref{lem:gen_8.13} and the aforementioned fact that $\mathrm{adv}(\mathcal{O})$ is $8T$-Lipschitz,
\begin{equation}
\Pr_{\mathcal{O}\sim\mu}\left[\mathrm{adv}(\mathcal{O})\geq 8T2^{-\alpha\ell/2}+n^{1/2}2^{-n/2}\right]\leq\exp\left(-\frac{(N-2)n 2^{-n}}{1536T^2}\right).
\end{equation}
\end{proof}

The following corollary similarly holds.
\begin{corollary}[Generalization of Corollary 8.16 of Ref.~\cite{aaronson2023certifiedArxiv}]
    There exists an oracle relative to which Assumption \ref{assum:gen_8.8} holds with probability $1$.
\end{corollary}

\subsection{Uplifting imperfect randomness to computationally indistinguishable from uniform randomness}\label{sec:computationally_indistinguishable_from_uniform}

We now describe how we obtain computational randomness suitable for generating random measurement bases in our protocol, as well as for selecting the samples to be verified. At a high level, this involves deriving from the weak source a seed that is computationally indistinguishable from uniform, and then optionally expanding it using a cryptographically secure pseudo-random number generator (CS-PRNG). We note that the PRNG is not strictly needed, as we can use a sufficiently long seed directly (which is what we do in the experiment). To formalize this, we introduce the notion of \textit{pseudosources}, inspired by the pseudoentropy framework of Ref.~\cite{DBLP:journals/siamcomp/HastadILL99}, which imposes requirements only against computationally bounded distinguishers. We begin by defining computational indistinguishability, pseudosources, and CS-PRNGs.
\begin{definition}[Computational distinguishability]
Let $\mathcal{C}$ be a class of (possibly randomized) functions representing the admissible distinguishers. The computational distinguishability between two random variables $X \in \{0,1\}^\ell$ and $Y\in \{0,1\}^\ell$, with respect to $\mathcal{C}$, is defined as
\begin{align}
    \mathrm{comp\mhyphen dist}_{\mathcal{C}}(X, Y) :=& \sup_{D \in \mathcal{C}} \left| \Pr[D(X) = 1] - \Pr[D(Y) = 1] \right|
\end{align}
where $\Pr[D(\cdot) = 1]$ denotes the probability that the function $D$ outputs $1$ on input drawn from $\cdot$.
\end{definition}

In our case, we assume access to additional (classical) side information, the random variable $E$, to help with this distinguishing.  
\begin{definition}[Conditional computational distinguishability]
Let $\mathcal{C}$ be a class of (possibly randomized) functions available to an adversary, and let $E$ be a random variable representing the adversary's side information. Let $X, Y \in \{0,1\}^\ell$ be random variables that may be correlated with $E$. The computational distinguishability of $X$ and $Y$ conditioned on $E$ with respect to $\mathcal{C}$ is defined as
\begin{align}
    \mathrm{comp\mhyphen dist}_{\mathcal{C}}(X, Y |E)
    := \sup_{D \in \mathcal{C}} \left| \Pr[D(X, E) = 1] - \Pr[D(Y, E) = 1] \right|,
\end{align}
where the probabilities are taken over the joint distributions of $(X, E)$ and $(Y, E)$ respectively, as well as the internal randomness of $D$ (if randomized).
\end{definition}

Using this, we can define a pseudosource. 
\begin{definition}[Pseudosource]
    A random variable $X$ is a $(\mathcal{C}, \varepsilon_{\rm source})$-pseudosource for a random variable $Y$ if, for all distinguishers $D \in \mathcal{C}$ and side information $E$, the advantage of $D$ to distinguishing between $X$ and $Y$ is at most $\varepsilon_{\rm source}$, i.e.  $\mathrm{comp\mhyphen dist}_{\mathcal{C}}(X, Y |E) \leq \varepsilon_{\rm source}$.
\end{definition}

While CS-PRNGs are typically defined without side information, their security naturally extends to the case with side information independent the seed, which we adopt below.

\begin{definition}[CS-PRNG with side information]
Let $\mathcal{C}$ be the class of functions computable in polynomial time. A deterministic function $\mathrm{PRF} : \{0,1\}^\ell \to \{0,1\}^{n_{\rm PRF}}$ is a CS-PRNG if $n_{\rm PRF} > \ell$ and for any random variable $E$ independent of $U_s$,
\begin{align}
    \mathrm{comp\mhyphen dist}_{\mathcal{C}}(\mathrm{PRF}(U_\ell), U_{n_{\rm PRF}} |  E) \leq \varepsilon_{\mathrm{PRF}},
\end{align}
where $U_\ell$ and $U_{n_{\rm PRF}}$ are uniform distributions over $ \{0,1\}^\ell $ and $ \{0,1\}^{n_{\rm PRF}}$, respectively, and $\varepsilon_{\mathrm{PRF}}$ is negligible in $n_{\rm PRF}$.
\end{definition}

To use a CS-PRNG, we require a seed consisting of uniform and independent bits. However, in randomness amplification, we only have access to a weak source. To obtain randomness that is computationally indistinguishable from uniform, we must first perform randomness extraction. Although there are several choices, we make one of the weakest possible assumptions, namely that our $(n,k)$ block min-entropy source is a $[(n_1, k_1){,}(n_2,k_2)]$ Markov pseudosource. In the main text we referred to the weak source being computationally indistinguishable from two independent sources, we now relax this condition to the weaker Markov independence. In words, Markov independence allows the two sources to be correlated through a common cause (the adversary's side-information in our case) and the independence condition now only needs to hold when conditioned on this common cause.

Formally, Markov pseudosource are defined as follows: 
\begin{definition}[Markov source]\label{def:markov_source}
    The random variables $(Y_1, Y_2)$ is a $[(n_1, k_1){,}(n_2, k_2)]$ Markov source, conditioned on side information $E$, if $Y_1\in\{0,1\}^{n_1}$, $Y_2\in\{0,1\}^{n_2}$, and
    \begin{align}\nonumber
        H_{\rm min}(Y_1|E)&=k_1\\\nonumber
        H_{\rm min}(Y_2|E)&=k_2\\\nonumber
        I(Y_1{:}Y_2 | E)&=0,
    \end{align}
    where $H_{\min}(Y_1 | E)$ denotes the min-entropy of $Y_1$ given $E$, and $I(Y_1 {:} Y_2 | E)$ the mutual information between $Y_1$ and $Y_2$ conditioned on $E$.
\end{definition}

\begin{definition}[Markov pseudosource]
The random variables $(X_1, X_2)$ are a $[(n_1, k_1){,}(n_2, k_2)]$ Markov $(\mathcal{C}, \varepsilon_{\rm source})$-pseudosource if there exists an $[(n_1, k_1){,}(n_2, k_2)]$ Markov source $(Y_1,Y_2)$ conditioned on $E$ such that
\begin{align}
\mathrm{comp\mhyphen dist}_{\mathcal{C}}((X_1, X_2), (Y_1, Y_2)|E) \leq \varepsilon_{\rm source}.
\end{align}
\end{definition}

Markov sources can be extracted by two-source extractors secure in the Markov model; 
\begin{definition}[Definition 5 of Ref.~\cite{arnonfriedman2016quantum}]\label{def:markov}
A function $\mathrm{Ext} : \{0,1\}^{n_1} \times \{0,1\}^{n_2} \to \{0,1\}^m$ is a classical-proof $(n_1, k_1, n_2, k_2, m, \varepsilon)$ two-source extractor secure in the Markov model if the following holds: For any $[(n_1, k_1){,}(n_2, k_2)]$ Markov source $(Y_1, Y_2)$ conditioned on classical side information $E$, we have
\begin{align}
    \Delta((\mathrm{Ext}(Y_1, Y_2), E), (U_m, E)) \leq \varepsilon,
\end{align}
where $U_m$ denotes the uniform distribution over $\{0,1\}^m$ and $\Delta(\cdot,\cdot)$ denotes the statistical distance. We say that $\mathrm{Ext}$ is strong in the $i$th input (for $i \in \{1,2\}$) if
\begin{align}
    \Delta(\mathrm{Ext}(Y_1, Y_2), Y_i, E),(U_m, Y_i, E)) \leq \varepsilon.
\end{align}
\end{definition}

As one might expect, any two-source extractor remains secure against computationally bounded distinguishers, as shown in the following lemma.
\begin{lemma}[Two-source extractors are secure against computational distinguishers] \label{lem: exts comp secure}
Let $\mathrm{Ext} : \{0,1\}^{n_1} \times \{0,1\}^{n_2} \to \{0,1\}^m$ be a classical-proof $(n_1, k_1, n_2, k_2, m, \varepsilon_{\mathrm{ext}})$ two-source extractor secure in the Markov model. Then for every $[(n_1, k_1){,}(n_2,k_2)]$ Markov source, conditioned on $E$, we have that 
\begin{align}
     \mathrm{comp\mhyphen dist}_{\mathcal{C}}(\mathrm{Ext}(Y_1, Y_2), U_m| E) \leq\varepsilon_{\mathrm{ext}}.
\end{align}
\end{lemma}

\begin{proof}
We have
\begin{align}
&\mathrm{comp\mhyphen dist}_{\mathcal{C}}(\mathrm{Ext}(Y_1, Y_2), U_m | E)\\ 
&= \sup_{D \in \mathcal{C}} \left|
\Pr[D(\mathrm{Ext}(Y_1, Y_2), E) = 1]
- \Pr[D(U_m, E) = 1]
\right| \\
&= \sup_{D \in \mathcal{C}} \left|
\sum_{z, e}  \delta_{D(z, e)}^{1} \Pr[\mathrm{Ext}(Y_1, Y_2) = z, E = e]
- \sum_{z, e}  \delta_{D(z, e)}^{1} \Pr[U_m = z, E = e]
\right| \\
&= \sup_{D \in \mathcal{C}} \left|
\sum_{z, e} \delta_{D(z, e)}^{1} \left( \Pr[\mathrm{Ext}(Y_1, Y_2) = z, E = e]
- \Pr[U_m = z, E = e] \right)
\right| \\
&\leq \left| \sum_{z,e} \max(\Pr[\mathrm{Ext}(Y_1,Y_2)=z, E=e] - \Pr[U_m=z, E=e], 0) 
\right| \\
&= \Delta((\mathrm{Ext}(Y_1, Y_2), E), (U_m, E)) \leq \varepsilon_{\mathrm{ext}}
\end{align}
where $\delta_{x}^{y}$ denotes the Kronecker delta which is $1$ if $x=y$ and $0$ otherwise, $\Delta$ denotes the statistical distance, and the final inequality follows from the security guarantee of two-source extractors in the Markov model.
\end{proof}

We will also use the fact that applying any function contained in $\mathcal{C}$ does not increase the computational distinguishability. This is shown in the following lemma.

\begin{lemma}
\label{lem:comp-dist-ext}
For any function $\mathrm{f}$ computable in $\mathcal{C}$, if 
\begin{align}
    \mathrm{comp\mhyphen dist}_{\mathcal{C}}(X, Y | E) \leq \varepsilon,
\end{align}
then 
\begin{align}
    \mathrm{comp\mhyphen dist}_{\mathcal{C}}(\mathrm{f}(X), \mathrm{f}(Y) | E) \leq \varepsilon.
\end{align}
\end{lemma}

\begin{proof}
Suppose for contradiction that there exists a distinguishing algorithm $D' \in \mathcal{C}$ such that
\begin{align}
    \left\vert  \Pr[D'(\mathrm{f}(X), E) = 1] - \Pr[D'(\mathrm{f}(Y), E) = 1] \right\vert  > \varepsilon.
\end{align}
Define a distinguishing algorithm $D \in \mathcal{C}$ by $D(X, E) := D'(\mathrm{f}(X), E)$.
Then
\begin{align}
    \left\vert  \Pr[D(X, E) = 1] - \Pr[D(Y, E) = 1] \right\vert  = \left\vert  \Pr[D'(\mathrm{f}(X), E) = 1] - \Pr[D'(\mathrm{f}(Y), E) = 1] \right\vert  > \varepsilon,
\end{align}
contradicting $\mathrm{comp\mhyphen dist}_{\mathcal{C}}(X,Y|E) \leq \varepsilon$.
\end{proof}

Together, these results can be used to obtain the following theorem.
\begin{theorem} \label{thm:secure-ext-then-prf}
Let $\mathrm{Ext}:\{0,1\}^{n_1} \times \{0,1\}^{n_2} \to \{0,1\}^m$ be any classical-proof two-source $(n_1, k_1, n_2, k_2, m, \varepsilon_{\mathrm{ext}})$-extractor secure in the Markov model and computable in $\mathcal{C}$, 
$\mathrm{PRF}:\{0,1\}^m \to \{0,1\}^{n}$ a CS-PRNG secure against $\mathcal{C}$, computable in $\mathcal{C}$, with distinguishing advantage $\varepsilon_{\mathrm{PRF}}$, and $(X_1, X_2)$ any $[(n_1, k_1){,}(n_2, k_2)]$ Markov $(\mathcal{C}, \varepsilon_{\rm source})$-pseudosource. Then,
\begin{align}
    \text{comp-dist}_{\mathcal{C}} (\mathrm{PRF}(\mathrm{Ext}(X_1, X_2)), U_n | E) \leq \varepsilon_{\mathrm{PRF}} + \varepsilon_{\mathrm{ext}} + \varepsilon_{\rm source}.
\end{align}
\end{theorem}

\begin{proof}
Using the triangle inequality, 
\begin{align}
&\text{comp-dist}_{\mathcal{C}} (\mathrm{PRF}(\mathrm{Ext}(X_1, X_2)), U_n | E) \\
&=\sup_{D \in \mathcal{C}}\left| \Pr[D(\mathrm{PRF}(\mathrm{Ext}(X_1, X_2)), E) = 1] - \Pr[D(U_{n}, E) = 1] \right| \\
&\leq \sup_{D \in \mathcal{C}}\left| \Pr[D(\mathrm{PRF}(\mathrm{Ext}(X_1, X_2)), E) = 1] - \Pr[D(\mathrm{PRF}(U_m),E) = 1] \right| \\ 
&\quad + \sup_{D \in \mathcal{C}} \left|\Pr[D(\mathrm{PRF}(U_m),E) = 1] - \Pr[D(U_{n}, E) = 1] \right| \\
&\leq \sup_{D \in \mathcal{C}}\left| \Pr[D(\mathrm{PRF}(\mathrm{Ext}(X_1, X_2)), E) = 1] - \Pr[D(\mathrm{PRF}(U_m), E) = 1] \right| + \varepsilon_{\text{PRF}}.
\end{align}
Now it remains to bound the term $\sup_{D \in \mathcal{C}}\left| \Pr[D(\mathrm{PRF}(\mathrm{Ext}(X_1, X_2)), E) = 1] - \Pr[D(\mathrm{PRF}(U_m),E) = 1] \right|$. Using Lemma~\ref{lem:comp-dist-ext}, since $\text{PRF}$ can be computed within $\mathcal{C}$, we have that
\begin{align}
    &\sup_{D \in \mathcal{C}}\left| \Pr[D(\mathrm{PRF}(\mathrm{Ext}(X_1, X_2)), E) = 1] - \Pr[D(\mathrm{PRF}(U_m), E)= 1] \right| \\
    &\leq \sup_{D \in \mathcal{C}}\left| \Pr[D(\mathrm{Ext}(X_1, X_2), E) = 1] - \Pr[D(U_m, E) = 1] \right|,
\end{align}
and using the triangle inequality, the fact that $(X_1, X_2)$ is an $[(n_1, k_1){,}(n_2, k_2)]$ Markov $(\mathcal{C},\varepsilon_{\rm source})$-pseudosource for some $[(n_1, k_1){,}(n_2, k_2)]$ Markov source $(Y_1, Y_2)$ and that Ext can also be computed within $\mathcal{C}$ (hence Lemma~\ref{lem:comp-dist-ext} again), we obtain 
\begin{align}
    &\sup_{D \in \mathcal{C}}\left| \Pr[D(\mathrm{Ext}(X_1, X_2), E) = 1] - \Pr[D(U_m, E) = 1] \right| \\
    &\leq \sup_{D \in \mathcal{C}}\left| \Pr[D(\mathrm{Ext}(X_1, X_2), E) = 1] - \Pr[D(\mathrm{Ext}(Y_1, Y_2), E) = 1] \right| \\
    &\quad + \sup_{D \in \mathcal{C}}\left| \Pr[D(\mathrm{Ext}(Y_1, Y_2), E) = 1] -  \Pr[D(U_m, E) = 1] \right| \\
    &\leq \sup_{D \in \mathcal{C}}\left| \Pr[D(X_1, X_2, E) = 1] - \Pr[D(Y_1, Y_2, E) = 1] \right| \\
    &\quad + \sup_{D \in \mathcal{C}}\left| \Pr[D(\mathrm{Ext}(Y_1, Y_2), E) = 1] -  \Pr[D(U_m, E) = 1] \right| \\
    &\leq \varepsilon_{\rm source} + \varepsilon_{\mathrm{ext}}.
\end{align}
Putting this all together completes the proof. 
\end{proof}

\noindent \textbf{Comparison with existing results.}
Several works have considered computational notions of randomness, i.e., randomness in the presence of a computationally restricted adversary. One approach, due to Yao~\cite{yao1982theory}, defines computational entropy through efficient compression, extending Shannon’s classical view of entropy as the minimal number of bits needed to describe a typical outcome. Another approach, which we refer to as the indistinguishability-based definition~\cite{DBLP:journals/siamcomp/HastadILL99}, characterizes the entropy through indistinguishability from distributions with high entropy. A third, more geometric perspective defines computational entropy as closeness (under computationally restricted metrics) to the set of high-entropy distributions. Computational entropy has also been studied in the quantum setting; see, for example,~\cite{chen2017computational, avidan2025fully, avidan2025quantum}. 

In this work, we focus on the indistinguishability-based notion of computational entropy and extend it to capture properties beyond entropy itself. To the best of our knowledge, this has not been previously explored. The indistinguishability-based framework naturally supports such a generalization, whereas it is less clear how to formulate analogous notions in other frameworks.\newline

\noindent \textbf{Some potential classifications for $\mathcal{C}$.}
The class $\mathcal{C}$ denotes the set of all algorithms (classical and quantum) that an adversary may use to distinguish the input random variables, using their side information $E$. So far, we intentionally leave $\mathcal{C}$ unspecified to keep the framework general. However, $\mathcal{C}$ can be instantiated in various ways, for example by restricting circuit size (e.g., to polynomial-time algorithms), imposing specific resource bounds (such as those based on multiples of Frontier, as used in \cite{jpmc_cr}), or choosing to restrict/allow quantum operations. In~\cite{avidan2025quantum}, there is some particularly nice discussion surrounding the types of restrictions that can be imposed on realistic computationally-restricted quantum adversaries.

We also note that although we have only considered classical side-information, the definitions generalize to adversaries holding quantum side-information. In this case, the class of adversaries is the one additionally acting on a quantum memory using their (quantum) algorithms. All extractors that we use in our work have quantum-proof extensions.\newline

\noindent \textbf{Alternative approaches do not need computational indistinguishability.}
It is also possible to perform certified randomness amplification without the need to generate randomness that is computationally indistinguishable from uniform. For validation set selection, one could validate all circuits. 
Alternatively, one can break the rounds into blocks of $B$ circuits, where one random circuit is verified in each block. This reduces the total number of rounds and will result in worse second order correction terms due to the entropy accumulation theorem. Nevertheless, consider the case where $B=2^b$ for some positive integer $b$. If the randomness used to choose the validation circuit out of $B$ circuits has min-entropy rate $\alpha$, the worst case distribution looks like as follows: for $2^{\alpha b}$ circuits, each will be chosen for validation with probability $2^{-\alpha b}$, and all other circuits will never be chosen. As a result, the adversary can concentrate all the XEB budget onto the $2^{\alpha b}$ circuits, leading to an apparent $2^b/2^{\alpha b}=2^{(1-\alpha)b}$ fold increase in the expected XEB score, which may be acceptable for large $\alpha$ and small $b$. As for the selection of random measurement basis, using imperfect randomness would still be very challenging for the adversary to guess. Although the basis for a larger fraction of qubits may be guessed correctly, performing tensor network contraction before the information is available is still prohibited since a large number of basis will still be guessed incorrectly.

\subsection{Amplification of private weak randomness}\label{sec:amplify_private}

To amplify a private weak source to uniform private randomness, a remote quantum device is used. Existing proposals are based on Bell tests \cite{colbech2012free,kessler2020device,Foreman2023practicalrandomness}. For a user in possession of the physical devices and capable of verifying that Bell tests are faithfully executed, this approach is suitable. However, for a user with only cloud access to the results of the quantum experiments, an adversarial quantum device may simulate the behavior of a quantum device. Certified randomness with RCS offers a natural alternative to Bell test-based approaches, where inspecting the device is not possible.

In our proposal, we first obtain random bits $B$ that are computationally indistinguishable from uniform. This can be accomplished in the aforementioned protocol that uses a two-source extractor applied to two outputs of the weak source that is assumed to be a $[(n_1, k_1){,}(n_2, k_2)]$ Markov $(\mathcal{C},\varepsilon_{\rm source})$-pseudosource, followed by expansion by a PRNG. We choose this assumption because a Markov two-source is arguably one of the weakest possible assumptions to allow randomness extraction. For the sake of simplicity, in the main text we called this independence only. %
Using the extracted randomness to select the validation set and random measurement bases, along with challenge circuits generated by the weak source, the output of the quantum computer in the certified randomness protocol achieves everlasting security. This means that although we start with computational assumptions, the protocol's output is ``upgraded'' to information-theoretic thereafter. Combining the quantum outputs with the weak source and extracting through a two-source extractor yields a nearly perfectly random seed. This seed can then be used by a seeded extractor to extract randomness from the weak source many times. At a high level, our protocol manages to amplify a weak source that is merely a block min-entropy source by additionally assuming it is a pseudosource. We also give a formal algorithm description of the protocol in Fig.~\ref{protocol:amplification}.\\

\begin{figure*}
    \hrule
    \vspace{.5em}
    \begin{flushleft}\underline{Protocol Arguments:}\end{flushleft}
 \vspace{-1.5em}
 \begin{align*}
  n \in \mathbb{N}~&:~\text{Number of qubits}\\
  L \in \mathbb{N}~&:~\text{Number of rounds}\\
  T_{\rm batch}~&:~\text{Threshold on the time allowed for the samples to return}\\
  p_{\rm max}\in (1/2^n, 1]~&:~\text{Maximum contribution probability}\\
  s^*~&:~\text{Threshold for test score}\\
  \alpha~&:~\text{Min-entropy rate of the private weak source}\\
  k~&:~\text{Size of key to the pseudorandom unitary, which depends on $\alpha$}\\
  \beta~&:~\text{$\varepsilon_{\rm smooth}$-smooth min-entropy rate of the certified randomness protocol if $\Pr[\Omega]\geq\varepsilon_{\rm accept}$}\\
  \mathrm{Ext}_{\rm t}~&:~\text{A strong quantum-proof $(n_{\rm weak}, \alpha n_{\rm weak}, Ln, \beta Ln-\log_2(1/\varepsilon_2), n_{\rm seed}, \varepsilon_{\rm ts})$ two-source extractor}\\
  \mathrm{Ext}_{\rm s}~&:~\text{A strong quantum-proof $(m, \alpha m, n_{\rm seed}, o, \varepsilon_{\rm seeded})$ seeded extractor}\\
  B~&:~\text{Private bits that are computationally indistinguishable from uniform, with distinguishing advantage at most $\varepsilon_B$.}
  \end{align*}
  \begin{flushleft}\underline{The Protocol:}\end{flushleft}
  \begin{enumerate}
  \item For $i\in[L]$, do:
  \begin{enumerate}
      \item Obtain $k$ bits from the private weak source.
      \item Sample a pseudorandom unitary using those $k$ bits as the key.
      \item Send it to the quantum device.
      \item After a suitable delay, generate pseudorandom measurement bases with an unused portion of $B$ and send them to the quantum device. Start a timer.
      \item At time $T_{\rm batch}$, proceed to the next round; if the quantum device has already returned a bitstring, record it as $Z_i$. Else set $Z_i$ to the zero bitstring and record a flag indicating the failure.
      \end{enumerate}
  \item For each round, include the index of the round in the validation set $\mathcal{V}$ with probability $\gamma$ pseudorandomly using an unused part of $B$.
  \item For each circuit–bitstring pair in the validation set, compute the bitstring probability $p_i$. If the sample failed to be returned in the time required, set $p_i=0$.
  \item If $s=\sum_{i\in \mathcal{V}}\frac{ 2^n\min(p_i, p_{\rm max})}{\vert\mathcal{V}\vert} <s^*$ abort. Else:
  \begin{enumerate}
  \item Obtain $n_{\rm weak}$ bits $X$ from the private weak source.
  \item Let $Z$ denote the collection of all quantum outputs $\{Z_i\}$. Apply the extractor $\mathrm{Ext}_{\rm t}$ (strong in $X$) to $(X,Z)$ to obtain output $S$.
  \item For $j\in[M]$, do:
  \begin{enumerate}
      \item From the private weak source, obtain $m$ bits $Y_j$.
      \item Apply the extractor $\mathrm{Ext}_{\rm s}$ (strong in the seed $S$) to $(S,Y_j)$ and denote the output randomness by $O_j$.
  \end{enumerate}
  \item Output $O_1, \ldots, O_M$.
  \end{enumerate}
  \end{enumerate}
  \hrule\vspace{1em}
  \caption{The protocol for the amplification of private weak randomness.}
  \label{protocol:amplification}
  \end{figure*}

To prove the security properties of the protocol, we begin with definitions of quantum-proof seeded and two-source extractors. These are the information-theoretic (and quantum-proof) version of the computational extractors used to generate the pseudorandomness $B$.

\begin{definition}[Definition 7 of Ref.~\cite{Foreman2023practicalrandomness}]\label{def:seeded}
    A function $\mathrm{Ext}:\{0,1\}^{n}\times\{0,1\}^{d}\rightarrow\{0,1\}^{m}$ is called a strong quantum-proof $(n,k,d,m,\varepsilon_{\rm seeded})$ seeded extractor if for sources $\rho_{XQ}$ with $H_{\rm min}(X\vert Q)_\rho\geq k$, we have
    \begin{equation}
        \|\rho_{\mathrm{Ext}(X,\tau_D)DQ}-\tau_M\otimes\tau_D\otimes\rho_Q\|_1\leq \varepsilon_{\rm seeded},
    \end{equation} where $\| \cdot \|_1$ denotes the trace distance.
\end{definition}

For the definition of a two-source extractor, see Definition 3 in~\cite{Foreman2023practicalrandomness}, which introduces a quantum-proof two-source extractor secure in the Markov model. Such extractors apply to pairs of sources with min-entropy. In our setting, however, we are interested in extracting from one source with min-entropy and another with smooth min-entropy. The following lemma addresses the case of two smooth min-entropy sources.
\begin{lemma}[Lemma 17 of Ref.~\cite{arnonfriedman2016quantum}]\label{lem:two_source_min_entropy}
    Let $\mathrm{Ext}:\{0,1\}^{n_1}\times\{0,1\}^{n_2}\rightarrow\{0,1\}^{m}$ be a quantum-proof $(n_1, k_1, n_2, k_2, m, \varepsilon_{\rm ts})$ two-source extractor strong in the source $X_i$. Then, for $\kappa_1,\kappa_2\in(0,1], \varepsilon_1,\varepsilon_2\in(0,1)$ and any Markov source $\rho_{X_1X_2Q}$ with
    \begin{align}
        H_{\rm min}^{\kappa_i}(X_i\vert Q)_\rho&\geq k_i+\log_2(1/\varepsilon_i),
    \end{align}
    for $i\in\{1,2\}$, we have that
    \begin{equation}
        \|\rho_{\mathrm{Ext}(X_1X_2)X_iQ}-\tau_M\otimes\rho_{X_iQ}\|_1\leq6\kappa_1+6\kappa_2+2\varepsilon_1+2\varepsilon_2+2\varepsilon_{\rm ts},
    \end{equation}
    where $\tau$ is the maximally mixed state.
\end{lemma}
It is straightforward to adapt the proof of the above lemma to our setting, where one is a min-entropy source and the other is a smooth min-entropy source.
\begin{lemma}\label{lem:two_source}
    Let $\mathrm{Ext}:\{0,1\}^{n_1}\times\{0,1\}^{n_2}\rightarrow\{0,1\}^{m}$ be a quantum-proof $(n_1, k_1, n_2, k_2, m, \varepsilon_{\rm ts})$ two-source extractor strong in the source $X_i$. Then, for $\varepsilon_{\rm smooth}\in(0,1], \varepsilon_2\in(0,1)$ and any Markov source $\rho_{X_1X_2Q}$ with
    \begin{align}
        H_{\rm min}(X_1\vert Q)_\rho&\geq k_1\\
        H_{\rm min}^{\varepsilon_{\rm smooth}}(X_2\vert Q)_\rho&\geq k_2+\log_2(1/\varepsilon_2),
    \end{align}
    we have that
    \begin{equation}
        \|\rho_{\mathrm{Ext}(X_1X_2)X_iQ}-\tau_M\otimes\rho_{X_iQ}\|_1\leq6\varepsilon_{\rm smooth}+2\varepsilon_{\rm ts}+2\varepsilon_2,
    \end{equation}
    where $\tau$ is the maximally mixed state.
\end{lemma}
\begin{proof}
    The proof for one smooth min-entropy source follows directly from that of Lemma 17 of Ref.~\cite{arnonfriedman2016quantum}, considering smoothing on just one source, i.e. obtaining a subnormalized state analogous to Lemma 18 using only $\hat{\sigma}_{X_2C_2Z}$, then using Lemma 37 for extraction with subnormalized states.
\end{proof}

We now state the security guarantee when the protocol does not abort, and then derive the protocol \emph{soundness}.
\begin{theorem}\label{thm:amplification}
    Under assumption \ref{assum:gen_8.8}, if the protocol of Fig. \ref{protocol:amplification} does not abort (an event denoted $\Omega$) the final state $\rho_{SZQ\vert\Omega}$ of the protocol satisfies
    \begin{equation}
        \|\rho_{SZQ\vert\Omega}-\tau_S\otimes\rho_{ZQ\vert\Omega}\|_1\leq 6\varepsilon_{\rm smooth}+2\varepsilon_2+2\varepsilon_{\rm ts},\label{eqn:two_source}
    \end{equation}
    and
    \begin{equation}
        \bigg\|\rho_{O_1\dots O_MSZQ\vert\Omega}-\bigotimes_{j\in[M]}\tau_{O_j}\otimes\tau_S\otimes \rho_{ZQ\vert\Omega}\bigg\|_1\leq 6\varepsilon_{\rm smooth}+2\varepsilon_2+2\varepsilon_{\rm ts} + M\varepsilon_{\rm seeded},\label{eqn:seeded}
    \end{equation}
    where $\tau$ denotes the maximally mixed state and we denote any side information as $Q$, which includes the classical side information that is a snapshot of the beginning of the protocol, all prior history of the weak randomness, and all the keys used to generate the pseudorandom challenge unitaries, the measurement bases, and the validation set.
\end{theorem}

\begin{proof}
    The argument follows from the use of the previous lemmas and definitions for extraction, from using assumption \ref{assum:gen_8.8} in the same way as in \cite{aaronson2023certifiedArxiv}, and finally from showing that computational randomness for measurement-bases and verification-sample selection suffices. We first show that $\rho_{XZQ}$ is a Markov source that can be extracted from. $Z$ only depends on the keys obtained from the weak source (denote these bits as $K$) used to generate the challenge circuits, the pseudorandomly generated measurement bases and validation set which are determined by $B$ and the side information before the protocol starts. $X$ only depends on the prior history of the weak source, which includes $K$ and $B$ (we note the implicit assumption that the weak source of randomness does not update its state depending on the quantum output $Z$, which is standard in amplification protocols \cite{kessler2020device,Foreman2023practicalrandomness}). All side information is included in $Q$, meaning that the mutual information between $Z$ and $X$ conditioned on $Q$ is zero. Additionally, since both $Z$ and $X$ have conditional (on $Q$) min-entropy, one can extract on them using a two-source extractor as in Lemma~\ref{lem:two_source}.

    By Theorem~\ref{thm:vanilla_EAT_special}, in the case where the keys for generating the challenge circuits, the measurement bases, and the validation sets are drawn from the uniform distribution, if the certified randomness protocol that generates $Z$ does not abort with probability greater than $1-\varepsilon_{\rm accept}$, then the $\varepsilon_{\rm smooth}$-smooth min-entropy of $Z$ conditioned on $Q$ is $H_{\rm min}^{\varepsilon_{\rm smooth}}(Z\vert Q)\geq\beta Ln$. Although the actual protocol in Fig. \ref{protocol:amplification} uses weak randomness for challenge circuit generation, Theorem \ref{thm:imperfect_challenge_ok} states that the advantage to distinguish is $\mathrm{negl}(n)$. The $\mathrm{negl}$ term can then be added directly to the final state distance by the same contrapose argument as in \cite{aaronson2023certifiedArxiv}: if the device, given a pseudorandom circuit like we generate instead of a random one, outputs a quantum state that changes the state by a non-negligible amount, then such a device implies a protocol that distinguishes a pseudorandom circuit from a random one, violating assumption \ref{assum:gen_8.8}.

    As for the randomness used for the measurement bases and the validation set, we only need them to be \emph{computationally} unpredictable to the adversary. Unlike unitaries, they do not need to be \emph{statistically} indistinguishable from uniform. For the quantum part of the adversary (with oracle access to the challenge circuit), the measurement bases can even be fixed and the entropy analysis still goes through by noting that the unitary before the measurements is sufficient for the claims. Therefore, the measurement bases randomness only matters to the classical part of the adversary. For the classical part of the adversary, it does not provide entropy and the only aspect that matters is the simulation fidelity. Having computationally indistinguishable from uniformly random measurement bases does not change the simulation fidelity of the classical adversary compared to having truly uniformly random bases, except by a negligible amount proportional to $\varepsilon_B$. Nevertheless, it forces the classical adversary to wait for the measurement bases to start simulating.

    For the validation set, let us first consider the adversary's perspective before validation happens. The adversary can either provide entropy greater than $\beta Ln$ or output less than that. In the latter case, if the validation set was uniformly random, then the protocol aborts with probability at least $1-\varepsilon_{\rm accept}$. In reality, the validation set is computationally indistinguishable from uniformly random (up to a negligible amount proportional to $\varepsilon_B$), which is to say from the perspective of the adversary, the abort probability remains at least $1-\varepsilon_{\rm acc ept}$.
    
    Since $\mathrm{negl}(n)$ and $\varepsilon_B$ can be made arbitrarily small, we ignore them. Therefore, either the abort probability is greater than $1-\varepsilon_{\rm accept}$, or the $\varepsilon_{\rm smooth}$-smooth min entropy is at least $\beta Ln$.
    
    Additionally, the imperfect private randomness output $X$ has min-entropy $\alpha n_{\rm weak}$ conditioned on $Q$. By Lemma \ref{lem:two_source}, Eq.~\ref{eqn:two_source} holds since the extractor can be made strong in $X$. It is necessary that the inputs $Y_i$ and $S$ of the seeded extractor are not correlated. We know $Y_i$ is correlated to $XQ$, but $S$ is uncorrelated to $XQ$ up to distance given by Eq.~\ref{eqn:two_source}, and $Y_i$ and $S$ are uncorrelated. Finally, one obtains an additive $M\varepsilon_{\rm seeded}$ term by the repeated use of the seeded extractor, applying the same argument as Appendix A of Ref.~\cite{frauchiger2013true}. Eq.~\ref{eqn:two_source} holds directly without having to use the seeded extractor, the only difference would be to set the two-source extractor strong in $Z$.
\end{proof}

We additionally show that our protocol satisfies the definition of soundness. An ideal protocol would either abort (denoted as $\Omega^c$) or output perfectly uniform randomness on register $R$ relative to some side information $Q$. We can write the ideal state as
\begin{align}
\rho^{\mathrm{ideal}}_{RQ}=\tau_R\otimes\rho_{Q\wedge\Omega}+\vert\bot\rangle\langle\bot\vert_R\otimes\rho_{Q\wedge\Omega^c},
\end{align}
where $\tau_R$ denotes the maximally mixed state on the bitstrings, $\vert\bot\rangle$ in the event that the protocol aborts and we use subnormalised states $\rho_{Q\wedge\Omega}$ and $\rho_{Q\wedge\Omega^c}$ with traces equal to $\Pr[\Omega]$ and $\Pr[\Omega^c]$, respectively. We say the protocol is $\varepsilon_{\rm sou}$-sound if for all initial states over the input to the protocol as well as side information $Q$, the (real) output state $\rho_{RQ}$ is $\varepsilon_{\rm sou}$-close to $\rho_{RQ}^{\rm ideal}$ in trace distance.

\begin{definition}[Soundness]\label{def:soundness}
    For $\varepsilon_{\rm sou}\in(0,1]$, 
    a protocol is $\varepsilon_{\rm sou}$-sound if for an honest classical client and any server, for all possible input states over the input of the protocol and side information $Q$, the output state $\rho_{RQ}$ satisfies
    \begin{align}
    \norm{\rho_{RQ\wedge\Omega}-\tau_R\otimes\rho_{Q\wedge \Omega}}_1\leq \varepsilon_{\rm sou},
    \end{align}
\end{definition}

We now prove soundness under this definition.
\begin{corollary}
The protocol of Fig. \ref{protocol:amplification} with
\begin{align}
\varepsilon_{\rm accept}&=\varepsilon_{\rm sou}\\
\varepsilon_{\rm smooth}&=\frac{\varepsilon_{\rm sou}-2\varepsilon_2-2\varepsilon_{\rm ts}-M\varepsilon_{\rm seeded}}{6}
\end{align}
is $\varepsilon_{\rm sou}$-sound.
\end{corollary}
\begin{proof}
Theorem \ref{thm:amplification} says that the protocol either aborts with probability at least
\begin{align}
1-\varepsilon_{\rm accept}=1-\varepsilon_{\rm sou},
\end{align}
or the trace distance from the ideal state is $\varepsilon_{\rm sou}$. In the first case, $\Pr[\Omega]\leq\varepsilon_{\rm sou}$, so $\norm{\rho_{RQ\wedge\Omega}-\tau_K\otimes\rho_{Q\wedge \Omega}}_{1}\leq \varepsilon_{\rm sou}$ is trivially satisfied. In the second case, Eq.~\ref{eqn:seeded} shows that the trace distance of the normalized conditional states is less than
\begin{align}
&6\varepsilon_{\rm smooth}+2\varepsilon_2+2\varepsilon_{\rm ts} + M\varepsilon_{\rm seeded}\\
=&[\varepsilon_{\rm sou}-2\varepsilon_2-2\varepsilon_{\rm ts}-M\varepsilon_{\rm seeded}]+2\varepsilon_2+2\varepsilon_{\rm ts} + M\varepsilon_{\rm seeded}=\varepsilon_{\rm sou}
\end{align}
so that of the subnormalized conditional state is also less than $\varepsilon_{\rm sou}$.
\end{proof}

\subsubsection{Concrete instantiation and analysis}\label{sec:amplification_concrete}
\begin{figure}[H]
    \centering
    \includegraphics[width=1\linewidth]{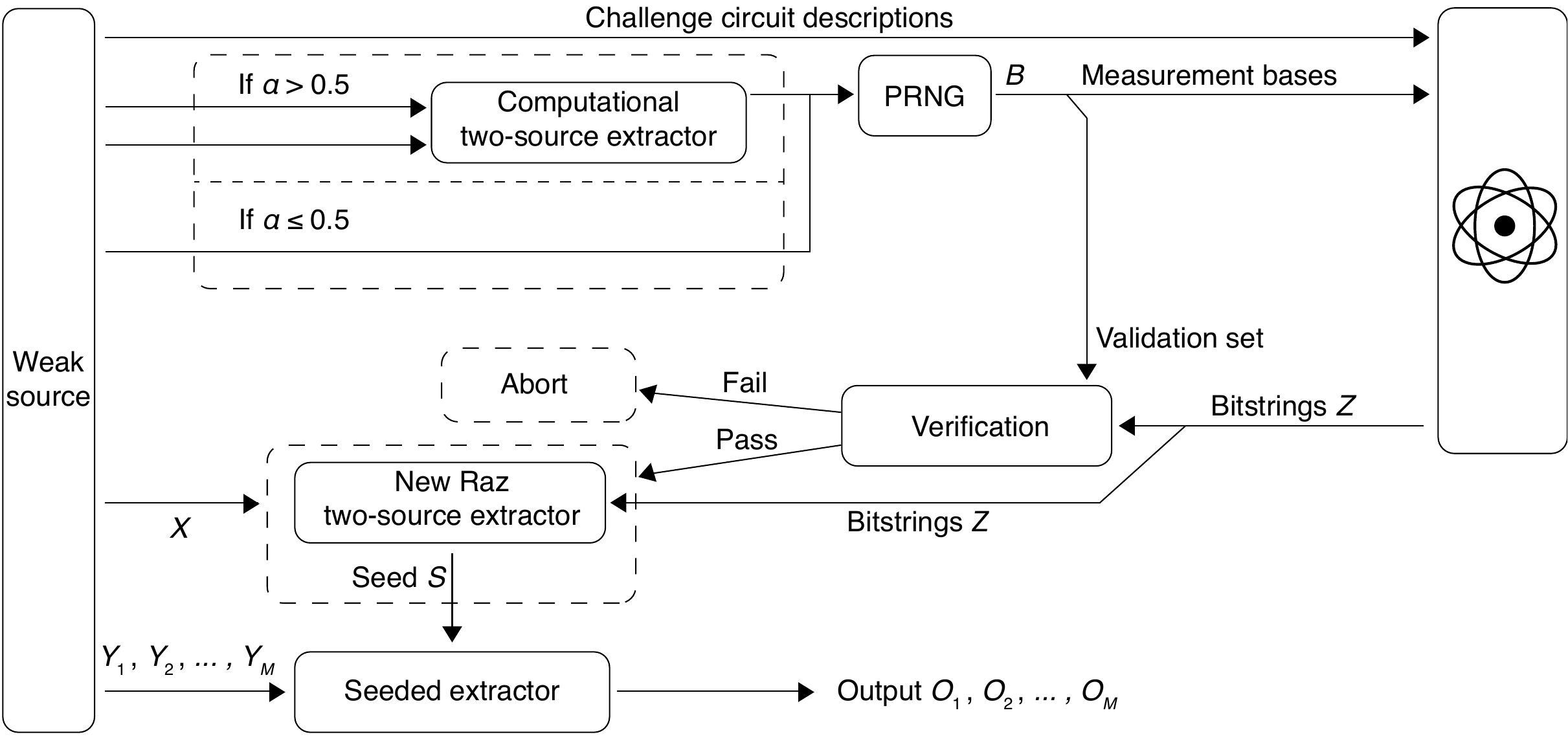}
    \caption{A schematics of the amplification protocol. A ``weak'' source here means that it has min-entropy. We note that the measurement bases and the validation set are generated using different parts of $B$.}\label{fig:amplification_diagram}
\end{figure}

Overall, the protocol in Fig.~\ref{protocol:amplification} enables a user with remote access to an untrusted quantum computer, equipped with a private weak source of randomness, to generate certifiably near-perfect random bits. For a concrete instantiation, we use Intel's RDSEED as our weak source~\cite{intel-rdseed}, the improved Raz extractor~\cite{raz2005extractors, foreman25} as our two-source extractor and the Circulant extractor~\cite{foreman2024cryptomite} as our seeded extractor. This scheme is illustrated in Fig. \ref{fig:amplification_diagram} with a PRNG, although we did not use one. In the case of amplifying a weak source with entropy rate $\alpha>0.5$, the weak source is assumed to be computationally indistinguishable from two independent weak sources in the Markov model and a computational two-source extractor extracts an output from two blocks of the weak source. In the case of amplifying a weak source with entropy rate $\alpha\leq 0.5$, the weak source is assumed to be computationally indistinguishable from the uniform distribution and no two-source extraction takes place.

We now provide more detail on this set up. Lemma~\ref{lem:raz} formalizes improved Raz extractor's parameters as a strong quantum-proof two-source extractor.
\begin{lemma}[Improved Raz extractor, Corollary 2 of Ref.~\cite{foreman25}] \label{lem:raz}
    For any $k_1,k_2$,
    the function $\mathrm{Ext}$ in Corollary 2 of Ref.~\cite{foreman25} is a strong (in either input) quantum-proof $(n_{1},k_{1}',n_{2},k_{2}',m, \varepsilon_{\rm ts} = 2^{3m/4} \, \sqrt{3\gamma/2})$ two-source extractor in the Markov model, for any $\gamma \geq 2^{(n_{1}-k_{1})/p} \cdot \big[ 2^{(l-n_1/2 + 1)/p} + p \cdot 2^{-k_{2}/2}\big]$, $k_{1}' \geq k_{1} + 1 + 2\log_2(1/\gamma)$ and $k_{2}' \geq k_{2} + 1 + 2\log_2(1/\gamma)$, any integers $n_2 \leq n_1/2$, $m \leq n_{1}/2$ and $l \leq n_{2} + \log_2(n_{1}/2)$, and any even integers $n_1$ and $p \leq 2^l/m$, implementable in $O(n_1 \log^2n_1)$ computation time if an irreducible polynomial exists over the field $\mathbb{GF}(2^{n_1/2})$.
\end{lemma}

\begin{remark}
The extractor in Lemma \ref{lem:raz} can be applied to sources with min-entropy $k_1'$ and $k_2'$ or sources with smooth min-entropy greater than $k_1'$ and $k_2'$.
\end{remark}

\noindent At a high level, the improved Raz extractor~\cite{foreman25} extracts under the following conditions:
\begin{itemize}
    \item The first input, of length $n_1$, has min-entropy at least $k_1 \geq \tfrac{n_1}{2} + \delta$ for some small $\delta$.
    \item The second input, of length $n_2 \leq n_1/2$, has min-entropy at least $k_2 \geq \log_2(n_2)$.
    \item An irreducible polynomial exists over the field $\mathbb{GF}(2^{n_1/2})$.
\end{itemize}
This implementation runs in time $O(n_1 \log^2 n_1)$, a significant improvement over the original $O(n_1^4)$, which is impractical for the input lengths required by our protocol. In practice, irreducible polynomials over large fields are hard to find and are only known for certain $n_{1}$ values; in this work, we use those identified by the Great Trinomial Hunt~\cite{greattrinomialhunt}.

For the strong quantum-proof seeded extractor, we turn to the Circulant seeded extractor~\cite{foreman2024cryptomite}.
\begin{lemma}[Circulant extractor, Theorem 9 of Ref.~\cite{foreman2024cryptomite}] \label{lem:circulant}
    The function $\mathrm{Ext}$ in Theorem 9 of Ref.~\cite{foreman2024cryptomite} is a strong quantum-proof $(n, k, d, m, \varepsilon_{\rm seeded})$ seeded extractor, for any $d = n + 1$ that is prime with primitive root 2 and $m \leq k - 2\log_2(1/\varepsilon_{\rm seeded})$, implementable in $O(n \log(n))$ computation time. 
\end{lemma}

\noindent \textbf{Generating the computationally secure randomness.}
We generate a bitstring $B \in \{0,1\}^{n_B}$ of length $n_B = 7{,}400{,}000$ that is computationally indistinguishable from uniform. 
The weak source $(W_1,W_2)$ is modeled as a Markov $[(n_1 = 2 \times 74{,}207{,}281,\; k_1 = \alpha_{\rm source} n_1){,}(n_2 = 74{,}207{,}281,\; k_2 = \alpha_{\rm source} n_2)]$ $(\mathcal{C}, \varepsilon_{\rm source})$-pseudosource with parameter $\alpha_{\rm source}$, and we apply the improved classical-proof Raz two-source extractor secure in the Markov model~\cite{foreman25,arnonfriedman2016quantum}. 
By Theorem~\ref{thm:secure-ext-then-prf}, the output error is bounded by $\varepsilon_B = \varepsilon_{\rm source} + \varepsilon_{\rm ext}$, where $\varepsilon_{\rm source}$ is the computational distance of $(W_1,W_2)$ from a Markov source, and $\varepsilon_{\rm ext}$ is the extractor error.  

To obtain $B$ in practice, we sample $n_1 = 2 \times 74{,}207{,}281$ bits from Intel’s RDSEED as $W_1$, discard the next $74{,}207{,}281$ bits to mitigate short-range correlations, and then sample $n_2 = 74{,}207{,}281$ bits as $W_2$. 
The choices of $n_1$ and $n_2$ correspond to the maximum input lengths supported by the improved Raz extractor implementation of Ref.~\cite{foreman25}. 
We set $\varepsilon_{\rm source} + \varepsilon_{\rm ext} = 10^{-8}$.\newline

\noindent \textbf{Minimum entropy rate analysis of the weak source.}
We determine the required min-entropy rate $\alpha$ of the weak source and the smooth min-entropy rate $\beta$ of the quantum randomness needed to extract $n_{\rm seed}=4{,}093$ bits. This value of $n_{\rm seed}$ corresponds to the maximum seed length supported by the Circulant extractor with 16-bit coefficients in a Number Theoretic Transform based implementation~\cite{foreman2024cryptomite}. Setting the seed length to $4{,}093$ means that the length of the weak source input is $4{,}092$ bits. We set $\epsilon_2 = \epsilon_{\rm ts} = 10^{-8}$, $\epsilon_{\rm seeded} = 10^{-16}$ and $M = 10^8$. Under these parameters, each of the $M$ seeded extractions yields an output length of $\left\lfloor 4{,}092 \cdot \alpha + 2\log_2(\epsilon_{\rm seeded}) \right\rfloor \geq \left\lfloor 4{,}092 \cdot \alpha - 107 \right\rfloor$.

We also set $\alpha_{\rm source}$ (the parameter associated to the quality of the pseudosource) to the min-entropy rate of the weak source, i.e. $\alpha_{\rm source}=\alpha$. In some cases, this value of $\alpha_{\rm source}$ is insufficient to generate $B$ of length $n_{B}$ using the method described above. For simplicity, when this occurs, we assume that the weak source is computationally indistinguishable from uniform. 

\begin{remark}
    Whilst we set $\alpha_{\rm source}=\alpha$, $\alpha_{\rm source}$ could, in principle, be taken (much) larger: the min-entropy rate $\alpha$ is an information-theoretic parameter, whereas $\alpha_{\rm source}$ characterizes the class of sources that a computationally bounded distinguisher cannot distinguish from the true source.
\end{remark}

\begin{figure}
    \centering
    \includegraphics[width=\linewidth]{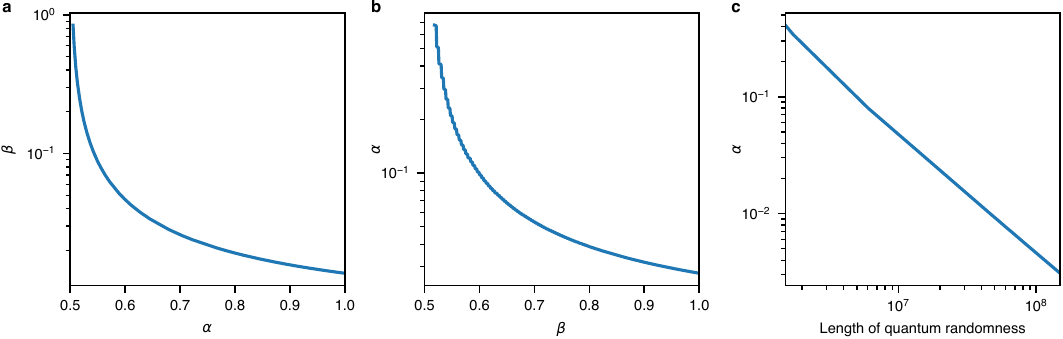}\caption{The required entropy rate on the quantum randomness and weak source to enable the extraction of $4{,}093$ bits. The quantum randomness is assumed to be $23{,}651\times 64$ bits. \textbf{a}, When the quantum randomness has entropy rate less than 0.5 and weak source with entropy rate greater than 0.5. The weak source randomness length is assumed to be $74{,}207{,}281$, the maximum support length. \textbf{b}, When the quantum randomness has entropy rate greater than 0.5 and weak source with entropy rate less than 0.5. The weak source length is $756{,}839$. \textbf{c}, Extractable weak source entropy rate for different lengths of quantum randomness. The smooth-min entropy rate of the quantum randomness is assumed to be $\beta=0.528$, which is the experimentally obtained by our protocol against an adversary five times as powerful as Aurora, as discussed in the main text. }\label{fig:extractor_entropy_rate_requirement}
\end{figure}

\noindent \textbf{(1) When the quantum output has smooth min-entropy rate less than $0.5$:}
In this case, the weak source must have min-entropy rate $\alpha>0.5$ to allow extraction using the improved Raz extractor. As long as the weak source is a pseudosource of two sources with $\alpha>0.5$, computational two-source extraction allows the extraction of bits that are computationally indistinguishable from uniformly random. To analyze the minimum min-entropy rate $\alpha$ of the weak source, we calculate if a quantum-proof $(n_{\rm weak}, \alpha n_{\rm weak}, Ln, \beta Ln-\log_2(1/\epsilon_2), n_{\rm seed}, \epsilon_{\rm ts})$ strong two-source extractor exists by Lemma \ref{lem:raz}, numerically optimizing over $l$ and $p$ of the Lemma. We consider $L=23{,}651$ rounds, the two-source Raz extractor has $n_1=n_{\rm weak}=2\times 43,112,609=86,225,218$ (which we choose, since it is the maximum supported length given the RAM constraint of our computer, and so reduces the entropy rate requirement on the quantum output), $n_2=Ln=1{,}513{,}678=23{,}651\times 64$, and $m=n_{\rm seed}=4093$. Fig.~\ref{fig:extractor_entropy_rate_requirement}\textbf{a} presents the results of this analysis. \newline

\noindent \textbf{(2) When the quantum output has smooth min-entropy rate greater than $0.5$:}
In this case, the min-entropy rate $\alpha$ of the weak source could (in-principle) be arbitrarily low. Unlike the previous case, one cannot perform computational two-source extraction since the entropy rate of the extractor inputs would be below 0.5. Therefore, we assume the weak source is already computationally indistinguishable from uniformly random. Note that the efficient implementation of the improved Raz' extractor extractor only supports certain input lengths for the first source (the source with min-entropy rate rate greater than $0.5$). When the quantum output is the first source, since we cannot in general make it have exactly the same length as one of the supported input lengths, we have to pad zeros to the quantum randomness until the total length is one of the supported lengths. The min-entropy of the padded randomness remains unchanged by padding, so the entropy rate goes down. It is therefore best to choose quantum randomness lengths such that minimum padding is required to minimize the reduction in the input entropy rate. For $n=64$, a total length of $23{,}651\times 64$ bits requires only $14$ padded bits, so we choose to analyze a protocol with $23{,}651$ rounds. Therefore, the extractor has $n_1=1{,}513{,}678=2\times 756{,}839>23{,}651\times 64=Ln$, $n_2=n_{\rm weak}=756{,}839$, and $m=n_{\rm seed}=4{,}093$. To study the entropy required, we see if a quantum-proof $(n_1, \beta Ln-\log_2(1/\epsilon_2), n_{\rm weak}, \alpha n_{\rm weak}, n_{\rm seed}, \epsilon_{\rm ts})$ strong two-source extractor exists. The results are shown in Fig.~\ref{fig:extractor_entropy_rate_requirement}\textbf{b}. Similarly, Fig.~\ref{fig:extractor_entropy_rate_requirement}\textbf{c} shows the extractable entropy rate of the weak source if the length of the quantum randomness was greater. This confirms that as the input randomness length increases, weak sources with very low entropy rates can be extracted.

We show the experimental result of the amplification protocol in Table \ref{tab:amplification_results}.

\begin{table}
\centering
\begin{tabular}{p{2.5cm} p{3cm} p{2.5cm} p{2.5cm} p{2.5cm} p{2.5cm}}
\hline
Adversary & Multiples of Aurora & $\varepsilon_{\rm sou}$ & $\beta$ & $\alpha$ \\
\hline
Restricted, T & 300 & $10^{-3}$ & 0.134 & 0.532 \\
\hline
Restricted, T & 40 & $10^{-3}$ & 0.479 & 0.509 \\
\hline
Restricted, U & 5 & $10^{-3}$ & 0.528 & 0.409\\
\hline
Restricted, T & 300 & $10^{-6}$ & 0.108 & 0.540 \\
\hline
Restricted, T & 40 & $10^{-6}$ & 0.458 & 0.510 \\
\hline
Restricted, T & 5 & $10^{-6}$ & 0.509 & 0.509 \\
\hline
Oracle, T & 300 & $10^{-3}$ & 0 & N/A \\
\hline
Oracle, T & 40 & $10^{-3}$ & 0.092 & 0.548 \\
\hline
Oracle, T & 5 & $10^{-3}$ & 0.137 & 0.532 \\
\hline
Oracle, T & 300 & $10^{-6}$ & 0 & N/A \\
\hline
Oracle, T & 40 & $10^{-6}$ & 0.025 & 0.710 \\
\hline
Oracle, T & 5 & $10^{-6}$ & 0.071 & 0.563 \\
\hline
\end{tabular}
    \caption{\textbf{Randomness amplification results}. The adversaries analyzed in this work are assumed to be at least $3{,}000$ kilometers away. T stands for the case where the adversary cannot efficiently computationally distinguish two blocks from the weak source from two blocks from two independent weak sources. U stands for the case where the adversary cannot efficiently computationally distinguish a block from the weak source from a block from the uniform distribution. The adversary classical compute power is given in multiples of Aurora. $\varepsilon_{\rm sou}$ is the soundness security parameter formally defined in Methods, which captures the deviation from a perfectly secure protocol. $\beta$ is the smooth min-entropy rate of the quantum randomness. $\alpha$ is the min-entropy rate of the weak source that can be amplified. }
    \label{tab:amplification_results}
\end{table}

\newpage
\section{Jointly certifiable randomness}\label{sec:joint}

\subsection{Jointly certifiable randomness expansion}

A desirable property of a certified randomness protocol is support for the joint verification of randomness. Specifically, one or more parties may want to jointly certify the generated randomness such that each party is independently assured of the randomness of the output even if a subset of the verifiers were to collude with the generator (to trick honest parties into thinking that the generated string has more entropy than it does in reality). Such joint verification can have several applications: for instance, there are cryptographic protocols such as non-interactive zero-knowledge proofs (NIZK) that require parties to have access to a common reference string (CRS) where the certified randomness of the common string may be crucial for the security of such protocols. The existing realizations of NIZK require either a single trusted authority or multiple authorities generating a single CRS with the assumption of an honest trusted majority among the authorities \cite{groth2014cryptography}. It may thus be practically desirable to reduce the trust assumption among the authorities to enable realization of NIZK protocols for large proof sizes \cite{certrand_apps}. Another possible application is in scenarios where the generated randomness is used as seeds for the implementation of algorithms whose fairness, privacy, etc. may be compromised if the seed is predictable to the party implementing the algorithm. Examples of such scenarios include the processing of private or sensitive data and randomized auditing of procedures. It then may be practically desirable for a third party to independently certify the randomness used for the implementation of the algorithm. 

We first define the desired notions abstractly. We define two versions of the protocol: one that generates a jointly certified string with high entropy~(Definition~\ref{defn:joint-cert-entropy-protocol}), and one that generates a jointly certified string that is near-perfectly random~(Definition~\ref{defn:joint-cert-rand-protocol}). Below, the definitions of conditional states and smooth min-entropy are as given in Section \ref{sec:preliminaries}.

\begin{definition}[Jointly Certifiable Entropy Generation Protocol]
\label{defn:joint-cert-entropy-protocol}
    Let $n,m,m'$ be integers, $\varepsilon, \delta_{\rm c}, \delta_{\rm s} \in [0,1]$ be real parameters, and $p_I,p_O,p_H \in O(\mathrm{poly}(n))$ be integer-valued functions with $p_I \in o(p_O(n))$. We denote by $\mathsf{JCertE}^{m,m'}_{n,p_I,p_O,p_H,\varepsilon,\delta_{\rm c}, \delta_{\rm s}}$ an interactive protocol between $m$ classical verifiers and a quantum polynomial time prover $(\mathsf{Ver}_1,\ldots, \mathsf{Ver}_m, \mathsf{Gen})$ where each $\mathsf{Ver}_i$ takes $p_I(n)$ bits of randomness as input and outputs ($O_i$, $\mathrm{acc}_i$), where $O_i$ is a register of size $p_O(n)$ bits and $\mathrm{acc}_i$ is a flag that indicates whether the transcript is accepted or rejected. $\mathsf{JCertE}^{m,m'}_{n,p_I,p_O,p_H,\varepsilon, \delta_{\rm c},\delta_{\rm s}}$ is a valid \textit{jointly certifiable entropy generation protocol} if it satisfies the following conditions:
    \begin{enumerate}
    \item \textbf{Completeness}: There exists an honest generator $\mathsf{Gen}$ that runs in quantum polynomial time and a set of $m-m'+1$ honest verifiers such that, when interacting with $\mathsf{Gen}$ and any set of fewer than $m'$ malicious verifiers, all honest verifiers accept the transcript with probability at least $1-\delta_{\rm c}$ over the $(m-m'+1)*p_I(n)$ bits used by the verifiers.

    \item \textbf{Soundness:} Let $\Omega_i$ denote the event that verifier $\mathcal{C}_i$ accepts the transcript, C be all messages sent by the verifiers, and E a register with side information correlated with the initial state of $\mathsf{Gen}'$ and any malicious verifiers. Then, for any malicious $\mathsf{Gen}'$ colluding with any set of malicious verifiers $\{\mathsf{Ver}_i'\}_{i\in I}$, indexed by $I \subset [m]$ with $|I| < m'$, it holds that for all $\mathsf{Ver}_i$ for $i \not\in I$, either $\mathsf{Ver}_i$ rejects the transcript with probability at least $1-\delta_{\rm s}$ over the $p_I(n)$ bits of randomness used by it, or it contains a register $O_i$ such that 
   \begin{align}
        H_{\min}^{\varepsilon}(O_i|CE)_{\rho_{OCE\wedge\Omega}} \ge p_H(n),
    \end{align}
    and the contents of all classical registers $O_{i}$ for $i \in [m]\setminus I$ where $\mathcal{C}_i$ accepts the transcript are identical.
\end{enumerate}
\end{definition}

\begin{definition}[Jointly Certifiable Randomness Expansion Protocol]
\label{defn:joint-cert-rand-protocol}
    Let $n,m,m'$ be integers, $\varepsilon, \delta_{\rm c}, \delta_{\rm s} \in [0,1]$ be real parameters, and $p_I,p_O \in O(\mathrm{poly}(n))$ be integer-valued functions with $p_I \in o(p_O(n))$. We denote by $\mathsf{JCertR}^{m,m'}_{n,p_I,p_O,\varepsilon,\delta_{\rm c}, \delta_{\rm s}}$ an interactive protocol between $m$ classical verifiers and a quantum polynomial time prover $(\mathsf{Ver}_1,\ldots, \mathsf{Ver}_m, \mathsf{Gen})$ where each $\mathsf{Ver}_i$ takes $p_I(n)$ bits of randomness as input and outputs ($O_i$, $\mathrm{acc}_i$), where $O_i$ is a register of size $p_O(n)$ bits and $\mathrm{acc}_i$ is a flag that indicates whether the transcript is accepted or rejected by $\mathsf{Ver}_i$. $\mathsf{JCertR}^{m,m'}_{n,p_I,p_O,\varepsilon, \delta_{\rm c},\delta_{\rm s}}$ is a valid \textit{jointly certifiable entropy generation protocol} if it satisfies the following conditions:

    \begin{enumerate}
    \item \textbf{Completeness}: There exists an honest generator $\mathsf{Gen}$ that runs in quantum polynomial time and a set of $m-m'+1$ honest verifiers such that, when interacting with $\mathsf{Gen}$ and any set of fewer than $m'$ malicious verifiers, all honest verifiers accept the transcript with probability at least $1-\delta_{\rm c}$ over the $(m-m'+1) \cdot p_I(n)$ bits used by the verifiers. 

    \item \textbf{Soundness:} Let $\Omega_i$ denote the event that verifier $\mathcal{C}_i$ accepts the transcript, C be all messages sent by the verifiers, and E a register with side information correlated with the initial state of $\mathsf{Gen}'$ and any malicious verifiers. Then, for any malicious $\mathsf{Gen}'$ colluding with any set of malicious verifiers $\{\mathsf{Ver}_i'\}_{i\in I}$, indexed by $I \subset [m]$ with $|I| < m'$, it holds that for all $\mathsf{Ver}_i$ for $i \not\in I$, either $\mathsf{Ver}_i$ rejects the transcript with probability at least $1-\delta_{\rm s}$ over the $p_I(n)$ bits of randomness used by it, or it contains a register $O_i$ such that 
   \begin{align}
        \lVert \rho_{O_i CE\wedge\Omega_i} - \tau_{p_O(n)} \otimes \rho_{CE\wedge\Omega_i} \rVert_{\mathrm{Tr}} \le \varepsilon,
    \end{align}
    and the contents of all classical registers $O_{i}$ for $i \in [m]\setminus I$ where $\mathcal{C}_i$ accepts the transcript are identical.

\end{enumerate}
\end{definition}

We now describe how Jointly Certifiable Randomness Expansion can be constructed from an underlying \textit{keyless} Certified Entropy Generation protocol, which we define below in Definition~\ref{defn:cert-rand-generation}. Protocols that satisfy Definition~\ref{defn:cert-rand-generation} guarantee that accepted responses from the generator are truly random by providing a lower bound on their conditional entropy given the messages of the verifier.
\begin{definition}[Certified Entropy Generation]
\label{defn:cert-rand-generation}
Let $n$ be an integer, $\varepsilon, \delta_{\rm c}, \delta_{\rm s} \in [0,1]$ be real parameters, and $p_I,p_O,p_H$ be integer-valued functions of $n$ such that $p_I(n),p_O(n),p_H(n) \in O(\mathrm{poly}(n))$, $p_H(n) \le p_O(n)$ for all $n$, and $p_I(n) \in o(p_H(n))$. We denote by $\mathsf{CertE}_{n,p_I,p_O,p_H\varepsilon,\delta_{\rm c},\delta_{\rm s}}$ an interactive protocol between a classical verifier $\mathsf{Ver}$ and a quantum polynomial time prover $\mathsf{Gen}$ such that the protocol uses $p_I(n)$ input bits of randomness, and the output of $\mathsf{Ver}$ contains a register $O$ of size $p_O(n)$ bits and a flag $\mathrm{acc}$ that indicates whether the transcript is accepted or rejected. $\mathsf{CertE}_{n,p_I,p_O,p_H,\varepsilon,\delta_{\rm c},\delta_{\rm s}}$ is a valid \emph{certified entropy generation} protocol if it satisfies the following conditions:
\begin{itemize}
    \item \textbf{Completeness}: There exists an honest prover $\mathsf{Gen}$ such that after engaging in the protocol \linebreak $\mathsf{CertE}_{n,p_I,p_O,p_H,\varepsilon,\delta_{\rm c}, \delta_{\rm s}}$ with $\mathsf{Gen}$, the verifier $\mathsf{Ver}$ accepts the transcript with probability at least $1 - \delta_{\rm c}$ over the $p_I(n)$ bits of randomness used by the protocol.
    \item \textbf{Soundness:} For any potentially malicious quantum polynomial time prover $\mathsf{Gen}'$, let $\Omega$ denote the event that the transcript is accepted by the verifier, $C$ a classical register that contains the messages sent by the verifier $\mathsf{Ver}$ to the prover $\mathsf{Gen}'$, and $E$ a register with side information correlated with the initial state of $\mathsf{Gen}'$. Then, it holds that either $\mathsf{Ver}$ rejects the transcript with probability at least $1-\delta_{\rm s}$ over the $p_I(n)$ bits of randomness used by the protocol, or it contains a register $O$ such that
    \begin{align}
        H_{\min}^{\varepsilon}(O|CE)_{\rho_{OCE\wedge\Omega}} \ge p_H(n).
    \end{align}
\end{itemize}
Of the $p_I(n)$ bits of randomness used by the protocol, $p_I^{\mathrm{external}}(n)$ bits are used to compute the messages corresponding to the register $C$ sent by the verifier, and the remaining $p_I^{\mathrm{internal}}(n) = p_I(n) - p_I^{\mathrm{external}}(n)$ are kept independent.
\end{definition}

For multiple verifiers to obtain the same random string, they must present the same challenge to the generator in some way. However, if one ore more verifiers is untrusted, they may attempt to disproportionately bias the challenge generation process towards those challenges that are easier for the generator (for example the generator can pre-compute deterministic responses to a subset of possible challenges). This bias invalidates the central assumptions that are needed for the security of the Certified Entropy Generation protocol. From the perspective of Definition~\ref{defn:cert-rand-generation}, the parties must share the $p_I^{\mathrm{external}}$ bits used for computing the challenges and the guarantees require these shared bits to be near uniformly random.

The restriction to keyless protocols here is to avoid the possibility of a simple attack resulting from collusion between some verifiers and the generator. If the protocol is keyed, the key determines the challenge and thus must be jointly generated by all parties. However, a colluding verifier may disclose the key to the generator after it is jointly generated. This invalidates the security assumptions for the certified randomness generation protocol. Our discussion below only applies to keyless certified randomness protocols such as those of Aaraonson and Hung~\cite{aaronson2023certified}, and Yamakawa and Zhandry~\cite{yamakawa2022verifiable}. Extending joint verification to keyed protocols like those of Refs.~\cite{brakerski2021cryptographic,mahadev2022efficient} requires modifications to the protocol, which we defer to future work.

In order to jointly construct challenges, we will rely on distributed protocols that allow parties to jointly generate random strings where the parties have access to a source of private randomness. We remark that these classical protocols differ from the quantum protocols that are the main subject of this paper, as they do not offer \emph{randomness expansion}. The randomness generated jointly is asymptotically no larger than the inputs of the parties. These have been extensively studied in the cryptography literature, beginning with the seminal work of Blum~\cite{blum1983coin} and are often referred to as protocols for \emph{Coin Flipping}. There have been many protocols proposed for coin flipping. They can be broadly classified by whether the protocols use quantum resources, the network model, whether they offer computational or information-theoretic security, the tradeoff between communication cost and the bias in the final random string generated, and the assumptions required for security. There have also been results that place restrictions on the bias achievable with a fixed number of rounds for \emph{any} protocol (quantum or classical). In particular~\cite{moran2009optimally}, for any two round protocol where both parties are allowed to abort the protocol at any time, the bias achievable in $r$ rounds of communication is $\Omega\left(1/r\right)$. This is relevant to our setting, as many protocols for certified randomness only realize expansion and not amplification, and thus can only tolerate bias in the initial seed that is negligible in the primary security parameters. 
To achieve this in the two party setting requires an exponential number of communication rounds, and this translates to an exponentially large requirement on input randomness, making randomness expansion impossible with polynomial length certified randomness protocols. We thus focus on multi-party protocols, or on security models where no aborts are allowed. For simplicity we focus on a concrete variant of the multi-party case and leave the incorporation of other settings to future work.

We first define an abstract notion of a coin flipping protocol in the multi-party setting. In this setting, information-theoretic security can be achieved when the number of malicious parties is bounded and so we restrict our definition to this notion of security.
\begin{definition}
\label{defn:coin-flip}
    Let $\mathcal{C}_1,\dots,\mathcal{C}_m$ for $m \in \mathbb{N}$ denote a set of (quantum)-polynomial time parties that wish to jointly generate a random string in $\{0,1\}^n$. Let $\mathsf{CoinFlip}_{m,n,\mathrm{cost}}$ be an interactive protocol requiring the exchange of $\mathrm{cost}$ uniformly random bits, at the end of which each party $i$ obtains an output $\mathsf{out}_i \in \{0,1\}^n$ in a classical $n$-bit register $K_i$ respectively. 

     We say the protocol $\mathsf{CoinFlip}_{m,n,\mathrm{cost}}$ is secure with parameters $(m',\varepsilon)$ if and only if the following condition is met. Denote the set of malicious parties fixed before the execution of the protocol as $I\subset[m]$. If $\|I\|\leq m'$, then after the protocol, the output register for every honest party $\{\mathcal{C}\}_{i \in [m]\setminus I}$ holds an identical $n$-bit classical string that is $\varepsilon$-close in total variation distance to a uniform string uncorrelated with $E$, where $E$ is a register holding a snapshot of the state of all malicious parties before the start of the protocol. In the density matrix formulation,
    \begin{align}
        \lVert \rho_{\{K_i\}_{i \in [m]\setminus I},E} - \chi_{\{K_i\}_{i \in [m]\setminus I}}\otimes \rho_E \rVert_{1} \le \varepsilon,
    \end{align}
    where $\chi_{\{K_i\}_{i \in [m]\setminus I}}$ is the maximally entangled state on registers $\{K_i\}_{i \in [m]\setminus I}$. 
\end{definition}

A concrete coin flipping protocol that satisfies the above definition can be obtained using the secure multi party computation protocols of Ben-Or, Goldwasser, and Wigderson~\cite{wigderson1988completeness}. We state a simple consequence of their main result that we use in the subsequent protocol.

\begin{lemma}[Existence of Efficient Multi-Party Coin Flipping (Consequence of~{\cite[Theorem 1, Theorem 6.6]{asharov2017full}}]
\label{lem:supermajority-coin-flip}
    Let $\mathcal{C}_1,\dots,\mathcal{C}_m$ for $m \in \mathbb{N}$ denote a set of (quantum)-polynomial time parties, with access to peer-to-peer private channels, as well as an authenticated broadcast channel. There exists a protocol $\mathsf{CoinFlip}_{m,n,\mathrm{cost}}$ with $\mathrm{cost} = n\cdot\mathrm{poly}(m)$ that is $(m',0)$ secure as per Definition~\ref{defn:coin-flip}, for any $m' \le m/3$. 
\end{lemma}

\begin{figure}
\hrule
\vspace{.5em}
\begin{flushleft}\underline{Protocol Arguments:}\end{flushleft}
\vspace{-1.5em}

\begin{itemize}

    \item Let $(\mathcal{C}_1, \ldots \mathcal{C}_m)$ be some set of classical verifiers and $Q$ be some quantum party. We allow $Q$ and some collection of $m' < \lfloor\frac{m}{2}\rfloor$verifiers to collude with $Q$ and act maliciously. We assume an authenticated broadcast channel connecting all parties. 
    
    \item Let $\mathsf{CertE}_{n,p_I,p_O,p_H,\varepsilon_{\rm cert},\delta_c, \delta_{\rm s}} = (\mathsf{Ver}, \mathsf{Gen})$ be a description of a certified entropy generation protocol between a classical verifier and a generator as defined in Definition \ref{defn:cert-rand-generation} with associated parameters.

    \item Let $\mathrm{Ext}\colon \{0,1\}^{p_O(n)} \times \{0,1\}^{n_{\rm seed}} \to \{0,1\}^{p'_O(n)}$ be a quantum-proof $(p_H(n),\varepsilon_{\rm ext})$-strong extractor with uniform seed.

    \item Let $\mathsf{CoinFlip}^{\rm (chall)}_{m,p_I^{\mathrm{external}}(n),\mathrm{cost}^{\rm (chall)}}$ and $\mathsf{CoinFlip}^{\rm (ext)}_{m,n_{\rm seed},\mathrm{cost}^{\rm (ext)}}$  be two ($m',\varepsilon_{\rm cf})$-secure coin flipping protocol as defined in Definition \ref{defn:coin-flip}. 
   
\end{itemize}
\vspace{.5em}
\begin{flushleft}\underline{The Protocol:}\end{flushleft}
\vspace{-1.5em}
\begin{enumerate}

    \item $\mathcal{C}_1, \ldots, \mathcal{C}_m$ sample uniformly random bit strings $r_1, \ldots, r_m$ bit strings respectively, with $|r_i| = \mathrm{cost}^{\rm (chall)}$.
        
    \item Each $\mathcal{C}_i$ takes part in $\mathsf{CoinFlip}^{\rm (chall)}$ with $r_i$ as their randomness, each outputting string $R_i$ with $|R_i|=p^{\mathrm{external}}_I(n)$ %

    \item Each $\mathcal{C}_i$ executes $\mathsf{Ver}$ using randomness $R_i$ as the external randomness and using its own privately sampled $p_I^{\mathrm{internal}}$ bits of randomness. Each verifier interacts with $Q$, which executes potentially multiple instances of $\mathsf{Gen}$. All messages, including the final output/acceptance flag,  are sent over the broadcast channel. $Q$ avoids doing additional work by responding identically to identical sequences of messages received from different $\mathcal{C}_i$. If $\mathsf{Ver}_i$ rejects the transcript, $\mathcal{C}_i$ rejects.
    
    \item Each $\mathcal{C}_i$ checks that there are at least $m - m'+1$ transcripts that are identical up to their final message containing the acceptance flag, and compatible with $R_i$. If the check fails, $\mathcal{C}_i$ rejects, otherwise it sets $O_i$ with $|O_i| = p_O(n)$ as the entropy output by $\mathsf{CertE}$

   \item $\mathcal{C}_1, \ldots, \mathcal{C}_m$ sample uniformly random bit strings $r'_1, \ldots, r'_m$ bit strings respectively, with $|r'_i| = \mathrm{cost}^{\rm (seed)}$.

    \item Each $\mathcal{C}_i$ takes part in $\mathsf{CoinFlip}^{\rm (ext)}$ with $r'_i$ as their randomness, each outputting string $R'_i$ with $|R'_i|=n_{\rm seed}$.

    \item Each $\mathcal{C}_i$ sets their output register $\mathsf{out}_i$ to $\mathrm{Ext}(O_i, R'_i)$ and accepts. 
\end{enumerate}
\hrule
\vspace{1em}
\caption{The protocol for jointly certified randomness expansion.}
\label{prot:joint_cr}
\end{figure}

Protocol in Fig.~\ref{prot:joint_cr} describes the steps required to obtain jointly certifiable randomness expansion. We formalize this in the following lemma. We note that while Protocol in Fig.~\ref{prot:joint_cr} satisfies Definition \ref{defn:joint-cert-rand-protocol}, if we instead terminate it after Step 4, a protocol which accepts at that point and outputs $O_i$ satisfies Definition \ref{defn:joint-cert-entropy-protocol}.

\begin{lemma}[Existence of Jointly Certifiable Randomness Expansion]
\label{thm:JCRE-exist}
Assume the existence of 
\begin{itemize}   
    \item A certified entropy generation protocol between a classical verifier and a generator \\$\mathsf{CertE}_{n,p_I,p_O,p_H,\varepsilon_{\rm cert},\delta_{\rm c}, \delta_{\rm s}} = (\mathsf{Ver},\mathsf{Gen})$ as defined in Definition \ref{defn:cert-rand-generation} with associated parameters, such that $\varepsilon_{\rm cert} \ge \frac{1}{2^{p_I(n)}}$.
    \item Two ($m',\varepsilon_{\rm cf})$-secure coin flipping protocols $\mathsf{CoinFlip}^{\rm (chall)}_{m,p_I^{\mathrm{external}}(n),\mathrm{cost}^{\rm (chall)}}$ and $\mathsf{CoinFlip}^{\rm (ext)}_{m,n_{\rm seed},\mathrm{cost}^{\rm (ext)}}$ as defined in Definition \ref{defn:coin-flip} where $n_{\rm seed} = \big\lceil 2\log\left(\frac{2p_O(n) p_H(n)}{\varepsilon_{\rm cert}}\right)\big\rceil$, such that $\mathrm{cost}^{\rm (chall)} = O(p_I(n))$ and $\mathrm{cost}^{\rm (ext)} = O(n_{\rm seed})$.
\end{itemize}
Then there exists a jointly certifiable randomness expansion scheme, as defined in Definition \ref{defn:joint-cert-rand-protocol}, \linebreak $\mathsf{JCertR}^{m,m'}_{n,p'_I(n),p'_O(n), \varepsilon', \delta'_{c}, \delta'_{s}}$ such that $p'_I(n) \le 4m p_I(n) + \mathrm{cost}^{\rm (chall)} + \mathrm{cost}^{\rm (ext)} = O(p_I(n))$ and $4p'_O(n) \ge p_H(n)$ for sufficiently large $n$, $\varepsilon' \le 3\varepsilon_{\rm cf} + 3\varepsilon_{\rm cert}$, $\delta'_{c} = m\delta_{c} + \varepsilon_{\rm cf}$, and $\delta'_{s} = \delta_{s} + \varepsilon_{\rm cf}$. It follows that $p'_I(n) = o(p'_O(n))$ if $p_I(n) = o(p_H(n))$.
\end{lemma}
\begin{proof}
    Let $I \subset [m]$ for $|I| < m'$ index the set of malicious verifiers. We will show that Protocol~\ref{prot:joint_cr} furnishes a construction of $\mathsf{JCertR}^{m,m'}_{n,p'_I(n),p'_O(n), \varepsilon', \delta'_{c}, \delta'_{s}}$. We define $\mathsf{Ver}_{i}$ as the action of party $\mathcal{C}_i$.
    
    Consider first a hybrid of our protocol, in which the output of the coin flipping protocol in Step 2 is replaced by their ideal states, such that after each step, each honest party  $\mathcal{C}_i$ has an external randomness string $R_i$ that is perfectly random and is shared among all honest verifiers, i.e., $R_i = R$ for all $i \in [m]\setminus I$.
    
    We first show the completeness of the hybrid protocol, defining the following honest prover $Q$ that behaves as the honest prover in the completeness definition of Definition~\ref{defn:cert-rand-generation} with the modification that when it receives the same challenge from multiple verifiers, it answers consistently. In this case, $m - m' + 1$ of the transcripts are identical, as at least $m + 1$ verifiers are generating their challenges from randomness $R$. Furthermore each honest party $\mathcal{C}_i$ accepts this transcript with probability at least $\delta_{\rm c}$ by the completeness of the certified entropy generation. By the union bound, \emph{all} honest parties accept a transcript by the end of Step 4, with probability at least $1 - m\delta_{\rm c}$.
    
   We turn now to the soundness in this hybrid setting. First note that if the prover does not reply identically to all honest verifiers, it is always rejected by every honest verifier as otherwise there cannot be $m - m' + 1$ transcripts that are identical upto their last message. In case it does reply identically to all honest provers, soundness of the certified entropy generation guarantees that each party $\mathcal{C}_i$ either rejects with probability at least $1 - \delta_{\rm s}$, or obtains a state in a register $O'_i$ that satisfies  $H_{\min}^{\varepsilon}(O'_i|CE)_{\rho_{O'_i CE\wedge\Omega_i}} \ge p_H(n)$.

    The hybrid protocol, with access to the ideal coin-flipping state, therefore offers the following guarantees:
    \begin{itemize}
        \item \textbf{Hybrid-Completeness:} There exists an honest generator $\mathsf{Gen}$ running in quantum polynomial time and $(m - m' + 1)$ honest verifiers such that the verifiers do not abort with probability at least $1 - m\delta_{\rm c}$ when interacting with any set of fewer than $m'$ malicious adversaries.
        \item \textbf{Hybrid-Soundness:} For any set of malicious adversaries $I \subset [m]$ for $|I| < m'$ and any quantum polynomial time generator $\mathsf{Gen}'$, each honest verifier $\{\mathcal{C}_i\}_{i \in [m]\setminus I}$ either aborts with probability $\delta_{\rm s}$, or the classical register $O'_i$ satisfies $H_{\min}^{\varepsilon_{\rm cert}}(O'_i|CE)_{\rho_{O'_i CE\wedge\Omega_i}} \ge p_H(n)$. Additionally, for each honest verifier that does not abort, $O'_{i}$ holds an identical value.
    \end{itemize}

    We now examine the protocol in the real setting, in which the coin flipping in Step 2 can be biased.
        Let the output of the coin flipping be stored by each party $\mathcal{C}_i$ in a classical register $K_{i}$. Let $E$ be a register with side information correlated with the initial state of the generator and any malicious verifiers. From the guarantees offered by a coin flipping protocol (Definition~\ref{defn:coin-flip}), the output registers after coin-flipping satisfy $\lVert \rho_{\{K_i\}_{i \in [m]\setminus I},E} - \chi_{\{K_i\}_{i \in [m]\setminus I},E}\otimes \rho_E \rVert_{1} \le \varepsilon_{\rm cf}$. Steps 3 and 4 of the protocol (including the simulation of any adversarial verifiers) can be modeled as a quantum operation (or a completely positive trace preserving (CPTP) map) acting on the registers $K_i$ and $E$. In the real setting, the joint state of these registers is within $\varepsilon_{\rm cf}$ trace distance from their state in the hybrid setting. Since CPTP maps do not increase trace distance, we can claim the following:
    \begin{itemize}
        \item Let $\mathrm{acc}_i$ be the registers denoting if party $\mathcal{C}_i$ aborts. The joint distribution of these registers for all honest verifiers in the real case differs from the hybrid case by less than $\varepsilon_{\rm cf}$ in trace distance. Thus the probability that all/any honest verifiers accept/reject differs by no more than $\varepsilon_{\rm cf}$ from the hybrid case.
        \item The output registers in the real case are also identical and the distribution of each $O'_i$ has less than $\varepsilon_{\rm cf}$ trace distance from the hybrid case. By the definition of smooth min-entropy and the triangle inequality, it holds that $H_{\min}^{\varepsilon_{\rm cert} + \varepsilon_{\rm cf}}(O'_i|CE)_{\rho_{O'_i CE\wedge\Omega_i}} \ge p_H(n)$ in the real case whenever $H_{\min}^{\varepsilon_{\rm cert}}(O'_i|CE)_{\rho_{O'_i CE\wedge\Omega_i}} \ge p_H(n)$ in the hybrid case.
    \end{itemize}
    From the previous considerations we have guarantees at the end of Step 3 and 4 that reflect those for Jointly Certifiable Entropy Generation in Definition~\ref{defn:joint-cert-entropy-protocol}. 
    \begin{itemize}
        \item \textbf{Completeness:} There exists an honest generator $\mathsf{Gen}$ running in quantum polynomial time and $(m - m' + 1)$ honest verifiers such that the verifiers do not abort with probability at least $1 - m\delta_{\rm c} - \varepsilon_{\rm cf}$ when interacting with any set of fewer than $m'$ malicious adversaries..
        \item \textbf{Soundness (Entropy):} For any set of malicious adversaries $I \subset [m]$ for $|I| < m'$ and any quantum polynomial time generator $\mathsf{Gen}'$, For all honest verifiers $\mathcal{C}_i$ for $i \in [m]\setminus I$, $\mathcal{C}_i$ either aborts with probability at least $1 -\delta_{\rm s} - \varepsilon_{\rm cf}$, or the classical register $O'_i$ satisfies $H_{\min}^{\varepsilon_{\rm cert} + \varepsilon_{\rm cf}}(O'_i|CE)_{\rho_{O'_i CE\wedge\Omega_i}} \ge p_H(n)$. Additionally, for each honest verifier that does not abort, $O'_{i}$ holds an identical value.
    \end{itemize}

    Finally, for honest verifiers that do not abort, we use a quantum-proof strong randomness extractor to extract a near uniform string from each $O'_{i}$ into register $O_{i}$. After the coin-flipping in Steps 5 and 6, each honest verifier that has not aborted obtains an identical seed which is $\varepsilon_{\rm cf}$ biased from uniformity. It follows from well known results on randomness extraction with near perfect seeds (see for example Appendix~A of~\cite{frauchiger2013true}) that such a biased seed can be used at the cost of an additional bias of $\varepsilon_{\rm cert}$. Since $\mathrm{Ext}$ is a deterministic function of the extractor seed and the contents of $O'_i$ which are both common to all honest verifiers that have not aborted, the contents of $O_{i}$ are also common amongst all such verifiers. To obtain the guarantee $\lVert \rho_{O_i CE\wedge\Omega_i} - \tau_{p_O(n)} \otimes \rho_{CE\wedge\Omega_i} \rVert_{\mathrm{Tr}} \le 3\varepsilon_{\rm cert} + 3\varepsilon_{\rm cf}$ for these verifiers requires a seed length of $n_{\rm seed}$ and the output length $p'_O(n)$ is required to satisfy $p'_O(n) \le p_H(n) - 4 \log_2\left(\frac{1}{\varepsilon_{\rm cert}}\right) - 4 \log_2\left(p'_O(n)\right) - 6$. We choose the largest output length satisfying this condition, which can be easily verified to satisfy $4p'_O(n) \ge p_H(n)$ for sufficiently large $n$. Since $\varepsilon_{\rm cert} \ge \frac{1}{2^{p_I(n)}}$, we have $n_{\rm seed} \le 3p_I(n)$ for sufficiently large $n$. The total randomness required by the protocol is $m \cdot (p^{\mathrm{internal}}_I(n) + n_{\rm seed}) + \mathrm{cost}^{\rm (chall)} + \mathrm{cost}^{\rm (ext)} \le 4m p_I(n) + \mathrm{cost}^{\rm (chall)} + \mathrm{cost}^{\rm (ext)}$. This completes the proof.
\end{proof}

From Lemma~\ref{lem:supermajority-coin-flip}, we can obtain Jointly Certifiable Randomness Expansion (and consequently Jointly Certifiable Entropy Generation) from an underlying Certified Entropy Generation protocol in the case where $m' < m/3$.
\begin{corollary}
    Assume the existence of a certified entropy generation protocol between a classical verifier and a generator $\mathsf{CertE}_{n,p_I,p_O,p_H,\varepsilon_{\rm cert},\delta_{\rm c}, \delta_{\rm s}} = (\mathsf{Ver},\mathsf{Gen})$ as defined in Definition \ref{defn:cert-rand-generation} with associated parameters, such that $\varepsilon_{\rm cert} \ge \frac{1}{2^{p_I(n)}}$.
Then there exists a jointly certifiable randomness expansion scheme, as defined in Definition \ref{defn:joint-cert-rand-protocol}, $\mathsf{JCertR}^{m,m'}_{n,p'_I(n),p'_O(n), \varepsilon', \delta'_{c}, \delta'_{s}}$ such that $p'_I(n) = O(p_I(n))$ and $4p'_O(n) \ge p_H(n)$ for sufficiently large $n$, $\varepsilon' \le 3\varepsilon_{\rm cert}$, $\delta'_{c} = m\delta_{c}$, and $\delta'_{s} = \delta_{s}$. It follows that $p'_I(n) = o(p'_O(n))$ if $p_I(n) = o(p_H(n))$.
\end{corollary}

Finally, we note that our presentation above focuses on a particular case where the correctness of the protocol we construct can be conveniently established. We anticipate at least two generalizations of the setting (in addition to extending the construction to keyed protocols) that we defer to future work. Firstly, coin flipping schemes that achieve negligible bias with polynomial communication have been demonstrated to accommodate more malicious parties (compared to $< m/3$ required by Lemma~\ref{lem:supermajority-coin-flip}). It is known that secure coin flipping schemes can be constructed in the setting without aborts as long as fewer than $m/2$ parties are corrupted, assuming the existence of a one-way function~\cite{prabhanjan2018round}. These protocols can also be compiled into protocols that guarantee output delivery against malicious adversaries, assuming public-key encryption and multi-verifier zero knowledge arguments~\cite[Corollary 1]{prabhanjan2018round}. Indeed, in the security with aborts setting we can also construct coin flipping protocols from one-way functions in the dishonest majority case with two parties \cite{blum1983coin}.  Accommodating these protocols in our construction requires more careful argument, as we do not have an unconditional information theoretic guarantee on the protocol output. Rather, we only have such a guarantee assuming that the underlying functionalities (one-way functions, public-key encryption, multi-verifier zero knowledge arguments) function as intended. This necessitates a more careful formalization and proof that we do not pursue here. Secondly, Protocol~\ref{prot:joint_cr} uses a broadcast channel, even if the coin flipping protocol does not need one. This choice is also made to simplify the proof and we expect that it can be removed via more careful arguments.

The above presents a formal analysis of performing jointly certifiable randomness expansion from an abstract certified entropy generation protocol. One requirement of certified entropy generation is that the protocol output is longer than the consumed randomness input. Since our RCS-based certified randomness protocol generates one sample per circuit and the randomness consumed per circuit is greater than $n$, we cannot achieve randomness expansion.

A straightforward way to obtain randomness expansion is to request multiple samples from a single circuit. One can plausibly extend Theorem \ref{thm:gen_7.10} and Theorem \ref{thm:gen_7.12} to show that entropy grows linearly with the number of requested samples. However, this creates an issue with the classical simulation adversary whose simulation cost does not grow with the number of bitstrings. In this case, instead of using the same circuit, we could use a set of $m$ circuits that are not independently sampled. For example, the first circuit can be sampled with raw randomness, and the rest in the batch of $m$ circuits are deterministically generated using a PRNG. For the quantum adversary, it should be the case that responding to $m$ circuits correlated due to a PRNG cannot be easier than responding to the same circuit $m$ times, although formal analysis may become tricky. For the classical adversary, this completely obviates the simulation advantage of sampling from a single circuit multiple times. In fact, it should not be necessary to use a PRNG and any deterministic transformation that makes the circuits sufficiently different from each should suffice.

If $m$ circuits are submitted per round and the quantum device can only execute one circuit at a time, then we cannot benefit from the latency reduction due to the delayed measurement bases. The $m$ random measurement bases of the round must be deterministically from some seed (e.g. only the first is truly random and the rest are generated using a PRNG). As soon as the quantum server receives the first basis, it can predict the rest immediately, but the adversary would have all the time including the circuit execution time for simulation. If the circuits are all executed in parallel, then the quantum device can measure the quantum states in parallel, and the single measurement time can be set as the latency. Additionally, since RCS-based certified randomness is a multi-round protocol, coin flipping needs to be performed each time a challenge circuit needs to be submitted to the quantum server.

\subsection{Jointly certifiable randomness amplification}

The above section assumes access to nearly perfect randomness as input, which is a different setting from randomness amplification. Nevertheless, we will show that it is possible to perform jointly certifiable entropy generation with imperfect input randomness under suitable assumptions. Specifically, if each of the $m'$ honest participant pre-processes their input weak randomness into something computationally indistinguishable from uniformly random, then the output of the coin flipping protocol must also be computationally indistinguishable from uniformly random by Lemma \ref{lem:comp-dist-ext} and the guarantees of coin flipping. This randomness can then be used by the certified randomness generation protocol.

We note that the resulting randomness from coin flipping is sufficient for validation set selection. However, as discussed in Section \ref{sec:app-pseudorandom-unitary-w-imperfect-rand}, under the assumption that the keyed unitaries are from a random oracle, the generation of the challenge circuits technically requires an actual min-entropy source so that the challenge circuits are QSZK indistinguishable from Haar random. Since the random oracle assumption itself is very strong anyway, the guarantees of using randomness computationally indistinguishable from uniform for challenge circuit generation is sufficient practically speaking.

While this construction allows one to obtain jointly certifiable randomness generation, it does not allow jointly certifiable randomness expansion. To achieve expansion, two-source extraction with another jointly certifiable weak source is required. We leave this possibility for future work.

One benefit of jointly certifiable randomness generation is the reduction in the total validation cost when using the centralized verification scheme in Section \ref{sec:low-budget_client_verification}. The raw bits from the jointly certifiable randomness generation protocol can be certified by pooling the computational resources from all participants, and each participant separately perform low-budget verification of the untrusted central verifier. Once the entropy is verified, each participant can use their own trusted weak source for two-source extraction to obtain private nearly perfectly random bits as discussed in Section \ref{sec:amplify_private}. In this case, the participants would not obtain certified shared randomness, but the gain from this protocol is the ability to reduce the average cost of randomness amplification of their private weak sources.

\subsection{Low-budget client verification}\label{sec:low-budget_client_verification}

Although the bitstrings from the quantum computer can be verified using classical computing, the verification cost is notable. In the case where multiple clients use the same bitstrings for consensus building, each client must individually verify the bitstrings, leading to a large computational cost for the full protocol. Further, individual clients may not be able to afford to devote a substantial amount of computational budget for verification. Additionally, even for clients who can afford validation using cloud computing resources, there is a need to ensure that the communication channel used to send the validation results is not tampered. Therefore, it is better if there is a centralized certified randomness provider model, where the majority of the verification can also be performed by an untrusted third party (possibly the same as the randomness generator), and each client performs some kind of low-budget verification.

The concrete scheme we consider is as follows. A centralized verifier performs regular validation on the validation set, and saves the complex partial amplitudes of all slices for all validation circuits. We shall denote the amplitude of the $j$th slice of the $i$th verification sample as $a_{i,j}$, the full amplitude of the $i$th sample as $A_i$, and the probability of the $i$th sample as $p_i$. Therefore, we have
\begin{equation}
    A_i=\sum_j a_{i,j}, \quad p_i=\vert A_i\vert^2.
\end{equation}
All $a_{i,j}$ are then broadcasted to all low-budget clients. A client then verifies the computational results of a random subset of the slices, checking if they agree with the values provided by the central verifier.

However, it is possible that a \textit{single} dishonest amplitude for a slice of a \textit{single} sample could change the XEB score arbitrarily. Therefore, a malicious central verifier can arbitrarily spoof the XEB score and be almost impossible to detect. As a result, we must put a limit on the maximum allowed amplitude norm, denoted as $a_{\rm max}$, and we have
\begin{equation}
    \forall i, j, \vert a_{i,j}\vert \leq a_{\rm max}.
\end{equation}
For an honest verifier, if $\vert a_{i,j}\vert > a_{\rm max}$ for some $i,j$, we change it into 
\begin{equation}
    a_{i,j}\rightarrow a_{i,j} \cdot a_{\rm max} / \vert a_{i,j}\vert.
\end{equation}

This rescaling of amplitudes may increase or decrease the final XEB score depending on the phase between $a_{i,j}$ and $A_i$, and it affects both the honest and the malicious case. We are therefore interested in upper bounding the increase in XEB due to this rescaling. The worst case scenario is for the unchanged $a_{i,j}$ to have the opposite phase as $A_i$, and rescaling leads to a change in $p_i$ given by
\begin{equation}
    \delta p_i = |A_i|^2 - \left(\vert A_i\vert-\left(\vert a_{i,j} \vert-a_{\rm max}\right)\right)^2\leq 2 \cdot \vert A_i\vert \cdot \left(\vert a_{i,j} \vert-a_{\rm max}\right).
\end{equation}

Unfortunately, there is no theoretical way of bounding this effect, and we have to resort to empirical slice amplitudes. It is easy to check that the probability decreases exponentially as the norm of the slice amplitudes increases.

Let us now consider effect of a single dishonest slice for a single sample. Denote the change in $a_{i,j}$ due to the malicious verifier as $\delta a_{i,j}$, the sample amplitude with the malicious effect as $A_i$. Then, the sample amplitude with the malicious effect is $A_i-\delta a_{i,j}$.
Further, let us make $A_i$ real by changing the phase of $a_{i,j}$ and $A_i$ by some constant, which does not affect $p_i$. In this case, the change in the sample probability $p_i$ is therefore
\begin{equation}
    \delta p_i = A_i^2 - \vert A_i-\delta a_{i,j}\vert^2=2 A_i \cdot \Re\left[\delta a_{i,j}\right]-\vert \delta a_{i,j}\vert^2\leq 2 A_i \cdot \Re\left[\delta a_{i,j}\right].
\end{equation}
For a given malicious $a_{i,j}$, $\Re[\delta a_{i,j}]\leq \Re[a_{i,j}]+a_{\rm max}$. Therefore,
\begin{equation}
    \delta\mathrm{XEB}_i = 2^n \cdot \delta p_i / L_{\rm val} \leq \frac{2^n}{L_{\rm val}}\cdot 2 \sqrt{p_i} \cdot \left(\Re\left[a_{i,j}\right]+a_{\rm max}\right).
\end{equation}
Again, this is only valid for rotated $a_{i,j}$.

For multiple malicious slices for a single sample, let us denote the malicious slice amplitudes as $a^k_{i,j}$, then
\begin{equation}
    \delta p_i = A_i^2 - \Big\vert A_i-\sum_k\delta a^k_{i,j}\Big\vert^2\leq 2 A_i \cdot \Re\left[\sum_k\delta a^k_{i,j}\right]=2 A_i \cdot \sum_k \Re\left[\delta a^k_{i,j}\right],
\end{equation}
and
\begin{equation}
    \delta\mathrm{XEB}_i \leq \frac{2^n}{L_{\rm val}}\cdot 2 \sqrt{p_i} \cdot \sum_k\left(\Re\left[a^k_{i,j}\right]+a_{\rm max}\right).
\end{equation}
We can therefore formally define
\begin{equation}
    \delta\mathrm{XEB}_i\leq\sum_j\delta\mathrm{XEB}_{i,j},
\end{equation}
where
\begin{equation}
    \delta\mathrm{XEB}_{i,j} = 2^n \cdot \delta p_i / L_{\rm val} \leq \frac{2^n}{L_{\rm val}}\cdot 2 \sqrt{p_i} \cdot \left(\Re\left[a_{i,j}\right]+a_{\rm max}\right).
\end{equation}

For each slice with $\delta\mathrm{XEB}_{i,j}$, set the low-budget client verification probability is some function of $\delta\mathrm{XEB}_{i,j}$, $p(\delta\mathrm{XEB}_{i,j})$. Denote the event where the adversarial central verifier is caught cheating as $\Omega$, and the event where it is not caught as $\Omega^{\rm c}$. For an adversarial central verifier that cheats on slices $(i,j)\in S$, the probability of not getting caught is $\Pr[\Omega^{\rm c}]=\prod_{(i,j)\in S}(1-p(\delta\mathrm{XEB}_{i,j}))$. For any given function $p(\delta\mathrm{XEB}_{i,j})$ and target $\Delta\mathrm{XEB}=\sum_{(i,j)\in S}\delta\mathrm{XEB}_{i,j}$, the adversary may choose to optimize how to break the total $\Delta\mathrm{XEB}$ down to individual $\delta\mathrm{XEB}_{i,j}$ to maximize the probability of not getting caught. If $p$ is a constant, then the adversary obviously wants to break $\Delta\mathrm{XEB}$ into as few terms of $\delta\mathrm{XEB}_{i,j}$ as possible. Therefore, the verifier should choose $p$ that increases with $\delta\mathrm{XEB}_{i,j}$.

For an optimal choice of $p(\delta\mathrm{XEB}_{i,j})$, the adversary should not be able to improve upon $\Delta\mathrm{XEB}$ and the probability of not getting caught by changing $\delta\mathrm{XEB}_{i,j}$. Consider the following ansatz:
\begin{equation}
    p(\delta\mathrm{XEB})=1-e^{-c\cdot\delta\mathrm{XEB}}\approx c\delta\mathrm{XEB},
\end{equation}
the probability of not getting caught is
\begin{equation}
    \Pr[\Omega^{\rm c}]=\prod_{(i,j)\in S}(1-p(\delta\mathrm{XEB}_{i,j}))=\prod_{(i,j)\in S}e^{-c\cdot\delta\mathrm{XEB}_{i,j}}=e^{-c\Delta\mathrm{XEB}},\label{eqn:prob_not_get_caught}
\end{equation}
which is independent of how $\Delta\mathrm{XEB}$ is split into contributing $\delta\mathrm{XEB}_{i,j}$. Therefore, the ansatz of $p$ is indeed the optimal form. If the client can expend a fraction $f$ of the total verification cost, we need
\begin{equation}
    fL_{\rm val}C=\frac{C}{n_{\rm slices}}\sum_{\forall (i,j)}p(\delta\mathrm{XEB}_{i,j})\rightarrow \sum_{\forall (i,j)} 1-e^{-c\cdot\delta\mathrm{XEB}_{i,j}} = f L_{\rm val} n_{\rm slices},
\end{equation}
where $C$ is the cost of verifying one sample and $n_{\rm slices}$ is the number of slices per sample. One can then solve for $c$ and obtain the optimal verification probability $p$ as a function of $\delta\mathrm{XEB}_{i,j}$.

One key advantage is that as $n_{\rm slices}$ increases, the cost advantage of our scheme compared to full verification grows. Assume $p(\delta\mathrm{XEB})=c\delta\mathrm{XEB}$, we have $\mathbb{E}[c\delta\mathrm{XEB}_{i,j}]=fL_{\rm val}$. We can reasonably assume that the slice amplitudes are random Gaussian variables on the complex plane. The typical value of single sample amplitudes $\mathrm{Std}[A_i]$ is $\sqrt{n_{\rm slices}}\cdot\mathrm{Std}[\Re[a_{i,j}]]$. To hold $\mathrm{Std}[A_i]$, we need $\mathrm{Std}[\Re[a_{i,j}]]\propto 1/\sqrt{n_{\rm slices}}$. Similarly, $\delta\mathrm{XEB}_{i,j}\propto 1/\sqrt{n_{\rm slices}}$. Therefore, since $\mathbb{E}[c\delta\mathrm{XEB}_{i,j}]=fm$, we must have $c\propto \sqrt{n_{\rm slices}}$. This translates to a reduced $\Pr[\Omega^{\rm c}]$ due to Eq. \ref{eqn:prob_not_get_caught}. If $n_{\rm slices}$ increases by a factor of $F$, we would have $\Pr[\Omega^{\rm c}]\rightarrow\Pr[\Omega^{\rm c}]^{\sqrt{F}}$.

\begin{figure}[!ht]
    \centering
    \includegraphics[width=\textwidth]{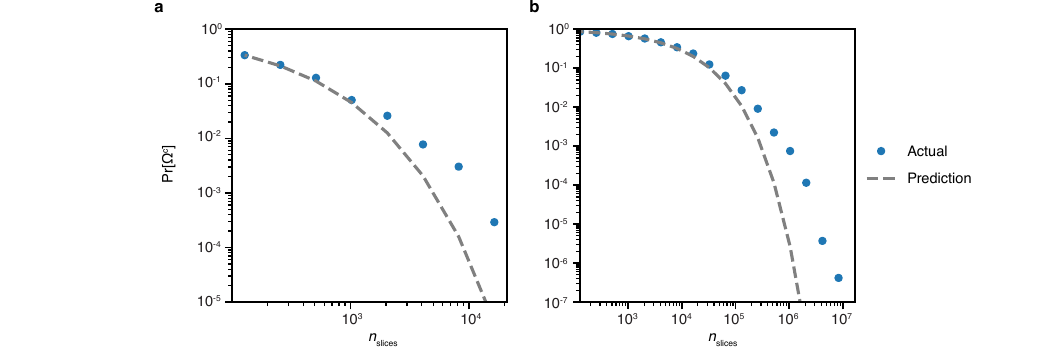}
    \caption{\textbf{$\Pr[\Omega^{\rm c}]$ for different number of slices}. \textbf{a}, Using Frontier results from \cite{jpmc_cr} with 5\% of the total validation budget and $\Delta\mathrm{XEB}=0.02$, \textbf{b}, Using Aurora results from this work with 0.1\% of the total validation budget and $\Delta\mathrm{XEB}=0.01$.}
    \label{fig:abort_probability_scaling}
\end{figure}

In the experimental demonstration of certified randomness expansion using RCS \cite{jpmc_cr}, 351 samples validated on the Frontier supercomputer are published. The contraction tree published has 33,554,432 slices, but every 2,048 slices are grouped together and published as a single data point, resulting in 16,384 chunks. If we treat each chunk as a slice, we have $a_{\rm max}=3.49\times 10^{-10}$, and we can compute $\delta\mathrm{XEB}_{i,j}$ and calculate $\Pr[\Omega^{\rm c}]$ for some fraction of validation budget and tolerable $\Delta\mathrm{XEB}$. Similarly, we can further group the 16,384 chunks into larger chunks. For $5\%$ of the total validation budget and $\Delta\mathrm{XEB}=0.02$, we plot the achievable $\Pr[\Omega^{\rm c}]$ in Fig. \ref{fig:abort_probability_scaling}\textbf{a}. We also assume the distribution of $\delta\mathrm{XEB}_{i,j}$ we obtain from the 351 saved Frontier samples is identical to the full distribution over all 1,522 validated samples in \cite{jpmc_cr}. We compare that against the predicted scaling, taking the not abort probability of $128$ chunks as the baseline and updating it with $\Pr[\Omega^{\rm c}]\rightarrow\Pr[\Omega^{\rm c}]^{\sqrt{2}}$ each time (each data point has $F=2$ times more slices than the previous one). We note that although the predicted scaling is too optimistic, it nevertheless agrees with the general trend that increasing $n_{\rm slices}$ improves the protocol efficiency. If we were to use the original 33,554,432 slices, we can achieve significantly smaller $\Pr[\Omega^{\rm c}]$, validation budget, and $\Delta\mathrm{XEB}$.

In this work, 11,933 circuits are validated with actual tensor network contraction on Aurora. The contraction tree has $2^{23}$ slices. We group every 512 slices into a single chunk and publish the results with 16,384 chunks for all validated samples. We only push 20 samples with all $2^{23}$ slices since the size of the array storing all amplitudes are large. For $0.1\%$ of the total validation budget and $\Delta\mathrm{XEB}=0.01$, we plot the achievable $\Pr[\Omega^{\rm c}]$ in Fig. \ref{fig:abort_probability_scaling}\textbf{b}. To do this, we use 100 samples to estimate the statistics of $\delta\mathrm{XEB}_{i,j}$ and assume it is identical to that of all 11,933 samples. We do use all 11,933 samples to upper bound $a_{\rm max}$.

\section{Other results}\label{sec:other_results}
\subsection{Improving the entropy penalty against quantum side information}\label{sec:improve_quantum}

In this section, we show the deferred proof of the improvement of the entropy penalty factor from $2\log_2 k$ to $\log_2 k$ against quantum side information for the single round. First, we provide the generalization of two technical lemmas.
\begin{lemma}[Generalization of Lemma 3.10 of Ref.~\cite{aaronson2023certifiedArxiv}]\label{lem:sum_state_entropy_upper_bound}
    For a finite-dimensional Hilbert space $A$, let $\{\rho_i\}_{i\in[N]}$ be a set of $N$ normalized quantum states on $A$. For a probability distribution over $N$ outcomes $P$, define state $\rho=\sum_{i\in[N]}P_i\rho_i$. We have
    \begin{equation}
        H(A)_\rho\leq \sum_{i\in[N]}P_i H(A)_{\rho_i}+H(P).
    \end{equation}
\end{lemma}
\begin{proof}
    Define the quantum state
    \begin{equation}
        \psi_{BA}=\sum_{i\in[N]}P_i\vert i\rangle\langle i\vert_B \otimes\rho_{i,A}.
    \end{equation}
    By definition, the von Neumann entropy of $\psi$ is
    \begin{align}
        H(BA)_\psi=-\sum_{i\in[N]}\mathrm{tr}\left(P_i\rho_i\log P_i\rho_i\right)=-\sum_{i\in[N]}\left(P_i\log P_i + P_i\mathrm{tr}(\rho_i\log\rho_i)\right)=\sum_{i\in[N]}P_i H(A)_{\rho_i}+H(P).
    \end{align}
    Since $\mathrm{tr}_B(\psi)=\rho_A$, we have $H(A)_\psi = H(A)_\rho$. Since conditional von Neumann entropy is always non-negative, $H(BA)_\psi\geq H(A)_\psi$, and we conclude the proof.
\end{proof}

\begin{lemma}[Generalization of Lemma 3.12 of Ref.~\cite{aaronson2023certifiedArxiv}]\label{lem:sum_state_mutual_information_bounds}
    For finite-dimensional Hilbert spaces $A$ and $B$, let $\{\rho_{i,AB}\}_{i\in[N]}$ be a set of $N$ normalized bipartite states over $A,B$. For a probability distribution over $N$ outcomes $P$, define state $\rho=\sum_{i\in[N]}P_i\rho_{i,AB}$. We have
    \begin{eqnarray}
        I(A:B)_\rho \geq \sum_{i\in[N]}P_i I(A:B)_{\rho_i}-H(P)\\
        I(A:B)_\rho \leq \sum_{i\in[N]}P_i I(A:B)_{\rho_i}+2H(P).
    \end{eqnarray}
\end{lemma}
\begin{proof}
    By convexity of von Neumann entropy, for $X\in\{A,B\}$,
    \begin{equation}
        H(X)_\rho\geq \sum_{i\in[N]}P_i H(X)_{\rho_i}.
    \end{equation}
    Then, by Lemma \ref{lem:sum_state_entropy_upper_bound},
    \begin{align}
        I(A:B)_\rho&=H(A)_\rho + H(B)_\rho - H(AB)_\rho\geq \sum_{i\in[N]}P_i (H(A)_{\rho_i}+H(B)_{\rho_i}-H(AB)_{\rho_i}-H(P))\\
        &=\sum_{i\in[N]}P_i I(A:B)_{\rho_i}-H(P)\\
        I(A:B)_\rho&=H(A)_\rho + H(B)_\rho - H(AB)_\rho\leq \sum_{i\in[N]}P_i (H(A)_{\rho_i}+H(P)+H(B)_{\rho_i}+H(P)-H(AB)_{\rho_i})\\
        &=\sum_{i\in[N]}P_i I(A:B)_{\rho_i}+2H(P).
    \end{align}
\end{proof}

In the proof of Theorem~7.11 of Ref.~\cite{aaronson2023certifiedArxiv}, it has been shown that the von Neumann entropy conditioned on quantum side information can be bounded by computing the mutual information $I(Z_1\ldots Z_k:Z)_\psi$ for a joint classical state $\psi$ satisfying that $Z\in\{Z_1,\ldots,Z_k\}$ with probability $1-\delta$. Thus it suffices to bound the mutual information.

\begin{theorem}[{Improved Theorem 7.11 of Ref.~\cite{aaronson2023certifiedArxiv}}]\label{thm:improved_7.11}
    Let $\mathcal Z$ be a finite set and $P$ be a distribution over $\mathcal Z$. Let $Z_1,\ldots,Z_k\sim P$ be $k$ i.i.d. random variable sampled according the distribution $P$. Let $Z\in\mathcal Z$ be any random variable that depends on $Z_1,\ldots,Z_k$ and satisfies that $Z\in\{Z_1,\ldots,Z_k\}$ with probability $1-\delta$ and $\psi$ be the joint probability distribution of $Z_1,\ldots,Z_k,Z$. Then $I(Z_1\ldots Z_k : Z)_\psi\geq (1-\delta)(H_{\min}(P)-\log k)-3h(\delta)$, where $h$ is the binary entropy function.
\end{theorem}
\begin{proof}
    To show this, we first introduce a new variable $Z_{k+1}=\bot$ for some special symbol $\bot\notin\mathcal Z$. \ 
    We define another classical state $\phi_{Z_1\ldots Z_{k+1}Z}$ describing the distribution of the following sampling process: \ 
    Sample $Z_1,\ldots,Z_k,Z$ according to $\psi$. \ 
    If $Z\notin\{Z_1,\ldots, Z_k\}$, then set $Z=\bot$ and output $Z_1,\ldots,Z_k,Z_{k+1}=\bot,Z$.

    We first argue that 
    \begin{align}\label{eq:2}
        I(Z_1\ldots Z_k:Z)_\psi \geq I(Z_1\ldots Z_k Z_{k+1}:Z)_\phi - 3 h(\delta).
    \end{align}
    This step ensures that $Z\in\{Z_1,\ldots, Z_k, Z_{k+1}\}$ with probability 1 for independent random variables $Z_1,\ldots,Z_{k+1}$. \ 
    Moreover, $Z$ conveys the information about whether it is in $\{Z_1,\ldots,Z_k\}$. \
    To prove the bound in \eq{2}, define $\psi_1$ be the normalized state of $\psi$ projected onto $\mathcal W=\mathrm{Span}\{\ket{z_1,\ldots,z_k,z}:z_1,\ldots,z_k\in\mathcal Z, z\in\{z_1,\ldots,z_k\}\}$, and $\psi_0$ be $\psi$ projected onto the orthogonal complement $\mathcal W^\bot$. \ 
    By definition, $\psi=(1-\delta)\psi_1+\delta\psi_0$ and $\phi=(1-\delta)\phi_1+\delta\phi_0$. \ 
    By Lemma 3.12 of Ref.~\cite{aaronson2023certified}, 
    \begin{align}\nonumber
        I(Z_1\ldots,Z_k:Z)_{\psi}
        &= I(Z_1\ldots Z_{k}:Z)_\psi \\\nonumber
        &\geq (1-\delta) I(Z_1\ldots,Z_k:Z)_{\psi_1}
        + \delta I(Z_1\ldots,Z_k:Z)_{\psi_0}
        - h(\delta) \\\nonumber
        &\geq (1-\delta) I(Z_1\ldots,Z_k:Z)_{\phi_1}
        + \delta I(Z_1\ldots,Z_k:Z=\bot)_{\phi_0}
        - h(\delta) \\\nonumber
        &\geq I(Z_1\ldots,Z_k:Z)_{\phi} 
        - 3h(\delta) \\
        &\geq I(Z_1\ldots,Z_kZ_{k+1}:Z)_{\phi} 
        - 3h(\delta).
    \end{align}
    The first inequality uses Lemma~3.12 of Ref.~\cite{aaronson2023certified}. \ 
    The second uses the fact that $\psi_1=\phi_1$ and 
    \begin{align}
        I(Z_1\ldots,Z_k:Z)_{\psi_0}
        \geq I(Z_1\ldots,Z_k:Z=\bot)_{\phi_0} = 0. 
    \end{align}
    The third uses Lemma \ref{lem:sum_state_mutual_information_bounds}. \ 
    The last holds since $Z_{k+1}=\bot$ with probability 1. \ 
    We have a state $\phi$ which picks $Z=Z_{k+1}$ with probability $\delta$, and $Z\in\{Z_1,\ldots,Z_k,Z_{k+1}\}$ with probability 1. Next, we bound the mutual information with $\phi$. \
    
    Recall that the marginal is product: $\phi_{Z_1\ldots Z_{k+1}}=\phi_{Z_1}\otimes \ldots \otimes \phi_{Z_{k+1}}$. \
    Now we define a random variable $J\in[k+1]$ which describes the chosen index (i.e., $Z=Z_J$ with probability 1), with ties broken arbitrarily. \ 
    We start with the following inequality:
    \begin{align}\nonumber\label{eq:mutual-phi}
        I(Z_1\ldots Z_{k+1}:Z)_\phi
        &=H(Z_1\ldots Z_{k+1})_\phi
        - H(Z_1\ldots Z_{k+1}:Z)_\phi \\\nonumber
        &= k H(P)
        - H(Z_1\ldots Z_{k+1}|Z)_\phi \\
        &\geq k H(P)
        - H(JZ_1\ldots Z_{k+1}|Z)_\phi,
    \end{align}
    Now we denote $\vec Z=Z_1\ldots Z_{k+1}$ and give an upper bound on $H(J\vec Z|Z)_\phi$. By definition, \ 
    \begin{align}\nonumber
        H(J\vec Z | Z)_\phi 
        &= \sum_{j=1}^{k+1}\sum_{\vec z}\sum_z \Pr[\vec Z=\vec z, J=j,Z=z]\log_2\frac{\Pr[Z=z]}{\Pr[\vec Z=\vec z, J=j]}\\\nonumber
        &= \sum_{j=1}^{k+1}\sum_{\vec z}\Pr[\vec Z=\vec z, J=j,Z=z_j]\log_2\frac{\Pr[Z=z_j]}{\Pr[\vec Z=\vec z, J=j]}\\\nonumber
        &=\sum_{j=1}^{k+1}\sum_{\vec z}\Pr[\vec Z=\vec z, J=j]\log_2\frac{\sum_{j'=1}^{k+1}\Pr[J=j',Z_{j'}=z_j]}{\Pr[\vec Z=\vec z, J=j]}\\\nonumber
        &=\sum_{j=1}^{k}\sum_{\vec z}\Pr[\vec Z=\vec z, J=j]\log_2\frac{\Pr_{Z'\sim P}[Z'=z_j]\sum_{j'=1}^{k}\Pr[J=j'\vert Z_{j'}=z_j]}{\Pr[\vec Z=\vec z, J=j]}\\\nonumber
        &\qquad+\sum_{\vec z}\Pr[\vec Z=\vec z, J=k+1]\log_2\frac{\Pr[J=k+1]}{\Pr[\vec Z=\vec z, J=k+1]}\\\nonumber
        &=\sum_{j=1}^{k}\sum_{\vec z}\Pr[\vec Z=\vec z, J=j]\log_2\frac{\Pr_{Z'\sim P}[Z'=z_j](1-\delta_{z_j})}{\Pr[\vec Z=\vec z, J=j]}+\sum_{\vec z}\Pr[\vec Z=\vec z, J=k+1]\log_2\frac{\delta}{\Pr[\vec Z=\vec z, J=k+1]}\\\nonumber
        &\leq \sum_{j=1}^k \sum_{\vec z} \Pr[\vec Z=\vec z, J=j] \log_2 \Pr_{Z'\sim P}[Z'=z_j]+ 
        \sum_{j=1}^{k+1}\sum_{\vec z} \Pr[\vec Z=\vec z, J=j] \log_2 \frac{1}{\Pr[\vec Z=\vec z, J=j]} \\\nonumber
        &\leq -(1-\delta)H_{\min}(P) + \sum_{j=1}^{k+1} \sum_{\vec z} \Pr[\vec Z=\vec z, J=j] \log_2 \frac{1}{\Pr[\vec Z=\vec z, J=j]} \\\nonumber
        &= - (1-\delta) H_{\min}(P) + H(J\vec Z)_\phi\\\nonumber
        &\leq H(\vec Z)_\phi + H(J)_\phi - (1-\delta) H_{\min}(P)\\\nonumber
        &\leq kH(P) + (1-\delta)\log k + h(\delta) - (1-\delta) H_{\min}(P).
    \end{align}
    The first equality follows from the definition of conditional Shannon entropy. The second equality holds since the probability $\Pr[\vec Z=\vec z, J=j, Z_j=z]=0$ for $z\neq z_j$, and we use the standard convention $0\log(1/0)=0$ in the calculation of Shannon entropy. The third equality also uses the fact that $\Pr[\vec Z=\vec z, J=j, Z_j=z]=0$ for $z\neq z_j$, as well as a decomposition of the probability $P[Z=z_j]$. The forth equality uses the fact that for $j,j'\neq k+1$, the probability $P[Z_{j'}=z_j]$ is independent of $j'$ and only characterized by the distribution $P$, cross terms vanish (i.e. $\Pr[Z_{j'}=z_j]=0$ if $j'=k+1,j\neq k+1$ or $j'\neq k+1,j=k+1$), and $\Pr[Z_{k+1}=\bot]=1$. The fifth equality uses the fact that $\Pr[J=k+1]=\delta$ and defines $1-\delta_{z_{j}}=\sum_{j'=1}^{k}\Pr[J=j'\vert Z_{j'}=z_j]$. The first inequality holds since $\log_2\delta,\log_2(1-\delta_{z_j}\leq0)$. The second inequality follows from the fact that $\log\Pr_{Z'\sim P}[Z'=z_j]\leq \log \left(\max_{z\in\mathcal Z} P(z)\right)\leq - H_{\min}(P)$ for $j\in[k]$ and $\Pr[J\in[k]]=1-\delta$. The third inequality follows from the subadditivity of the Shannon entropy. The fourth inequality holds by the fact that $\Pr[J=k+1]=\delta$, and consequently $H(J)_\phi\leq (1-\delta)\log k + h(\delta)$.
    
    Now, we have shown that $H(J\vec Z|Z_{J})_\phi \leq kH(P) - (1-\delta)\cdot (H_{\min}(P)-\log k)$. \
    It remains to use the bound to conclude the theorem: 
    \begin{align}\nonumber
        I(Z_1\ldots Z_kZ_{k+1}:Z)_\phi 
        &\geq k H(P) - H(J\vec Z|Z_J)_\phi \\\nonumber
        &\geq (1-\delta)\left( H_{\min}(P) - \log k\right)-h(\delta).
    \end{align}
    Finally, by \eq{2}, we have
    \begin{align}\nonumber
        I(Z_1\ldots Z_k:Z)_\phi 
        &\geq I(Z_1\ldots Z_kZ_{k+1}:Z)_\phi - 3h(\delta) \\
        &\geq (1-\delta)\left( H_{\min}(P) - \log k\right) - 4 h(\delta)
    \end{align}
\end{proof}

\begin{remark}
    In the special case of $m=1$, the bound of Theorem \ref{thm:gen_7.12} is more conservative than that of Theorem \ref{thm:improved_7.11} if we change $-3$ to $-4$. Therefore, Theorem \ref{thm:gen_7.12} with $-3$ changed to $-4$ also holds for quantum side information $E$ for $m=1$.
\end{remark}

\subsection{Fidelity of partially contracted state}

One critical assumption is that the quantum state obtained by contracting a fraction $\phi$ of the slices has fidelity $\phi$, which is motivated by prior work \cite{pan2022solving,markov2018quantum}. To give numerical evidence, we construct small quantum circuits and calculate the fidelity of partially contracted states. We do this by first performing exact simulations to obtain the state vector. The small circuit is then converted to a tensor network and sliced into $2048$ slices (even though slicing is not necessary for fitting intermediate tensors on GPU memory), where we forbid slicing of the output indices. We then obtain the approximate state by contracting a fraction $f$ of the slices. Finally, the inner product between the exact state and the approximate state is computed to obtain the fidelity $\phi$.

The circuits geometry we choose is generated using the same method as Ref.~\cite{qntm_rcs}. Specifically, we generate an instance of 24 qubit circuit with depth 8. We choose the same discrete gate set as \cite{morvan2023phase}, which is advantageous for ruling out DMRG-based simulation methods \cite{qntm_rcs, ayral2023density}. We randomly generated a set of single qubit gates using this gate set and obtained one random quantum circuit.

\begin{figure}[!ht]
    \centering
    \includegraphics[width=0.8\textwidth]{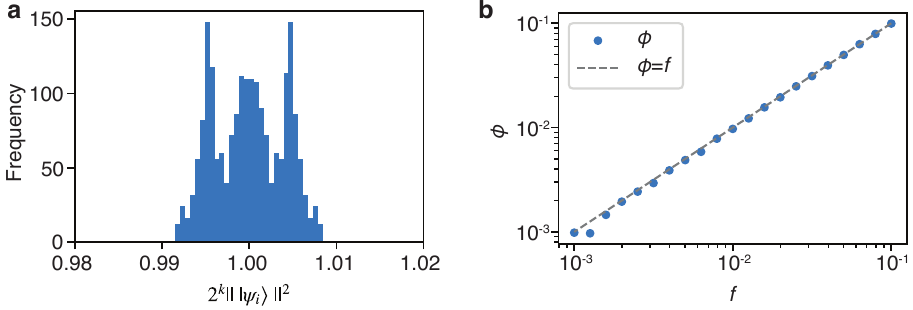}
    \caption{(a) Distribution of statevector norms $||~|\psi_i\rangle~||^2$ normalized by $2^k$ for individual slices. (b) Fidelity $\phi$ of contracting a fraction $f$ of all slices.
    }
    \label{fig:partial_contraction}
\end{figure}

We first observe that the norms of the unnormalized statevector obtained by contracting a single slice are concentrated around $2^{-k}$, where $k$ is the number of sliced indices, and the distribution is shown in Fig. \ref{fig:partial_contraction} a. This is in agreement with Ref.~\cite{kalachev2021classical}. We note that by choosing an $SU(2)$ gate set as in Ref.~\cite{qntm_rcs, jpmc_cr} results in a much wider distribution of the norms. This is not surprising as $SU(2)$ gates may have small rotation angles, and slicing may result in unevenly weighted sectors.

We then proceed to perform partial contraction to obtain the approximate statevectors. The fidelity $\phi$ of the normalized approximate statevectors to the exact state vector is almost $\phi\approx f$ as shown in Fig. \ref{fig:partial_contraction} b.

\subsection{Frugal rejection sampling classical adversary}\label{sec:frugal_rejection_sampling}

We consider the following classical simulation algorithm and show it behaves similar to sampling from a depolarized quantum state. A tensor network is constructed out of the $n$-qubit quantum circuit, with the input indices set to zero, and the collection of $M'$ amplitudes corresponding to $M'$ bitstrings randomly selected from the uniform distribution are computed. Contracting a fraction $\phi$ of all slices of the tensor network thus yields fidelity $\phi$ probability amplitudes of $M'$ bitstrings, and we perform frugal rejection sampling on these bitstrings. We now seek to analyze the distribution of the perfect fidelity probability amplitude of the chosen bitstring.

Assume there are $K$ slices. The amplitude of the $i$th bitstring is given by
\begin{equation}
    A_i = \sum_{j=1}^K a_{ij},
\end{equation}
where $a_{ij}$ is the contribution of the $j$th slice of the tensor network to the amplitude of the $i$th bitstring. We can consider all bitstring amplitudes for slice $j$ as a vector $a_{\cdot j}$ and define a corresponding statevector
\begin{equation}
    \vert\psi_j\rangle=\frac{1}{\sqrt{N_j}}\sum_i a_{ij}\vert i\rangle \quad \text{with}\quad N_j=\sum_i \vert a_{ij}\vert^2,
\end{equation}
where $\vert i\rangle$ is the $i$th computational basis state. Denote the ideal quantum state from the circuit as $\vert\psi\rangle$, we have
\begin{equation}
    \vert\psi\rangle=\sum_j\sqrt{N_j}\vert\psi_j\rangle.
\end{equation}
Note that this holds without requiring $\{\vert\psi_j\rangle\}_{j\in [K]}$ being an orthonormal basis. Now, consider a classical algorithm that only contracts a fraction of $\phi$ slices in the set $S\subset[K]$ such that $\vert S\vert=\phi K$. For a given bitstring, the amplitude of the bitstring estimated with partial contraction is $A^S_i=\sum_{j\in S}a_{ij}$. Define the partial quantum state
\begin{equation}
    \vert\psi_S\rangle=\frac{1}{\sqrt{N_S}}\sum_{j\in S}\sqrt{N_j}\vert\psi_j\rangle,
\end{equation}
where $N_S$ is chosen to normalize $\vert\psi_S\rangle$. It is easy to check that the bitstring amplitude of state $\vert\psi_S\rangle$ is given by
\begin{equation}
    \langle i\vert \psi_S\rangle=\frac{1}{\sqrt{N_S}}A^S_i.
\end{equation}

In frugal rejection sampling, a random set of bitstrings $i$ for $i\in B$ such that $\vert B\vert=M'$ are selected and the corresponding amplitudes $\{A_i^S\}_{i\in B}$ are computed. A bitstring is then chosen from $B$ with weight proportional to $A_i^S$. This is exactly the same as sampling with weight $\langle i\vert\psi_S\rangle$ since $A_i^S\propto\langle i\vert\psi_S\rangle$ for all $i$ with the same proportionality factor $\frac{1}{\sqrt{N_S}}$. If $M'$ is large enough, frugal rejection sampling reproduces sampling from $\vert\psi_S\rangle$ almost exactly.

It is known that contracting a fraction of $\phi$ slices results in a state with fidelity very close to $\phi$ with high probability \cite{markov2018quantum, kalachev2021classical, pan2022solving}, which is to say $\langle\psi_S\vert\psi\rangle\approx\phi$. Therefore,
\begin{equation}
    \vert\psi\rangle \approx \sqrt{\phi}\vert\psi_S\rangle + \sqrt{1-\phi}\vert\psi_S^{\perp}\rangle,
\end{equation}
where $\langle\psi_S\vert\psi_S^{\perp}\rangle=0$. Consider a bitstring $x$ generated by frugal rejection sampling, which faithfully reproduces perfect sampling from the quantum state $\vert\psi_S\rangle$ (up to some small statistical difference \cite{markov2018quantum}). Therefore, the probability with respect to state $\vert\psi\rangle$ is
\begin{align}
    p_{\vert\psi\rangle}(x)&=\vert\langle x\vert\psi\rangle\vert^2\\
    &\approx\phi\vert\langle x\vert\psi_S\rangle\vert^2 + (1-\phi)\vert\langle x\vert\psi_S^{\perp}\rangle\vert^2+2\sqrt{\phi(1-\phi)}\Re[\langle x\vert\psi\rangle\langle \psi_S^{\perp}\vert x\rangle]\\
    &=\phi p_{\vert\psi_S\rangle}(x)+(1-\phi)p_{\vert\psi_S^{\perp}\rangle}(x) +2\sqrt{\phi(1-\phi)}\Re\left[e^{i\theta}\right]\sqrt{p_{\vert\psi_S\rangle}(x)p_{\vert\psi_S^{\perp}\rangle}(x)}.\label{eqn:sampling_process}
\end{align}

The first term $p_{\vert\psi_S\rangle}(x)$ is due to frugal rejection sampling (or equivalently perfect sampling) from $p_{\vert\psi_S\rangle}(x)$, and
\begin{equation}
     \text{PDF}_{p_{\vert\psi_S\rangle}(x)}(t)=N^2 t e^{-Nt}.\label{eqn:frugal_pdf}
\end{equation}
The second term $p_{\vert\psi_S^{\perp}\rangle}(x)$ can be treated as uniform sampling with 
\begin{equation}
    \text{PDF}_{p_{\vert\psi_S^{\perp}\rangle}(x)}(t)=\text{PDF}_U(t)=Ne^{-Nt}.\label{eqn:uniform_pdf}
\end{equation} This is because $\vert\psi_S^{\perp}\rangle$ is approximately independent of $\vert\psi_S\rangle$ if $\vert\psi_S^{\perp}\rangle$ is randomly chosen in the space orthogonal to $\vert\psi_S\rangle$ and $n$ is large. Similarly, this implies $\theta$ is uniformly random between $(-\pi,\pi]$, and we can instead use $R=\cos\theta=\Re[e^{i\theta}]$, and \begin{equation}
    \text{PDF}_R(t)=f_R(t)=\begin{cases}
        \frac{1}{\pi \sqrt{1-t^2}}\quad&\text{if }t^2\leq 1,\\
        0& \text{otherwise}.
    \end{cases}\label{eqn:r_pdf}
\end{equation}

We also consider the distribution of sampling from a depolarized quantum state with fidelity $\phi$. The resulting PDF of $p_{\vert\phi\rangle(x)}$ is given by \cite{jpmc_cr, morvan2023phase}
\begin{equation}
    \mathrm{PDF}_{p_{\vert\psi\rangle(x)}(t)}=\phi N^2 t e^{-Nt} + (1-\phi) N e^{-Nt}.
\end{equation}

\begin{figure}[!ht]
    \centering
    \includegraphics[width=0.8\textwidth]{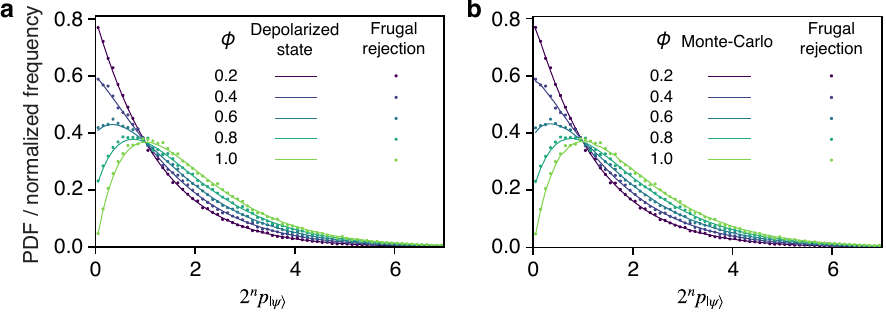}
    \caption{Distribution of the probability $p_{\vert\psi\rangle}(x)$. The number of qubits is $n=20$, and the depth is $d=8$. The Monte-Carlo approach uses 10 million samples, and frugal rejection sampling uses 10 thousand samples.
    }
    \label{fig:frugal_pdf_comparison}
\end{figure}

We now verify that the sampling process described by Eq. \ref{eqn:sampling_process} agrees with classical simulation using frugal rejection sampling, where we perform actual contraction on a fraction of slices and perform frugal rejection sampling based on the calculated probabilities. We also consider a method of sampling based on Eq. \ref{eqn:sampling_process} and term it as Monte-Carlo, where we sample $p_{\vert\psi_S\rangle}(x),p_{\vert\psi_S^\perp\rangle}(x),R$ independently from PDFs given by Eq. \ref{eqn:frugal_pdf}, \ref{eqn:uniform_pdf}, and \ref{eqn:r_pdf}, and compute $p_{\vert\psi_S\rangle}(x)$ with Eq. \ref{eqn:sampling_process}. This is exactly equivalent to sampling from a random fidelity $\phi$ pure state. Finally, we compare everything against the PDF of sampling from the depolarized state. The results are presented in Fig. \ref{fig:frugal_pdf_comparison}, which shows that the Monte-Carlo approach is a faithful representation of frugal rejection sampling since the two traces closely track each other without systematic deviations. This similarly applies to sampling from the depolarized state.

\begin{figure}[!ht]
    \centering
    \includegraphics[width=\textwidth]{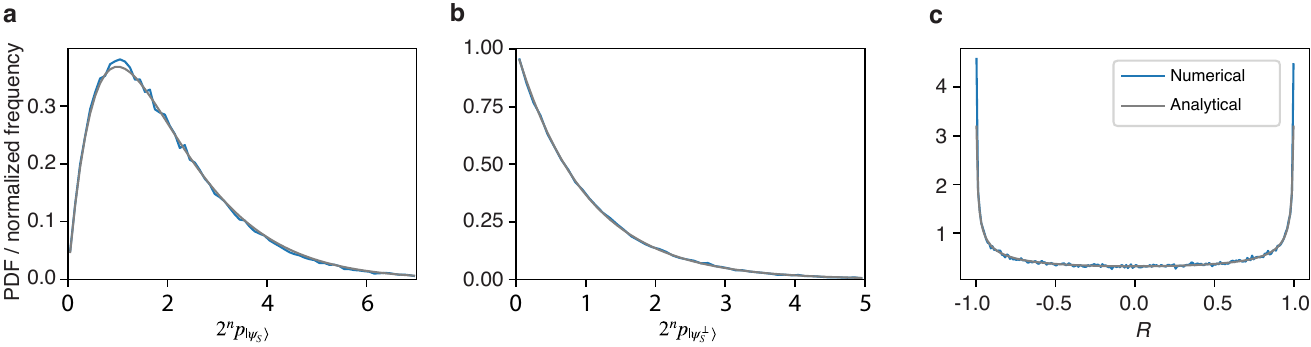}
    \caption{Empirical and analytical distributions of \textbf{a}, $p_{\vert\psi_S\rangle}$, \textbf{b}, $p_{\vert\psi_S^{\perp}\rangle}$, and \textbf{c}, $R$. Configurations are identical to Fig. \ref{fig:frugal_pdf_comparison}.}
    \label{fig:frugal_justification}
\end{figure}

We also provide numerical evidence to justify the assumptions used in the Monte-Carlo approach. We are given $\vert\psi\rangle$. We generate $\vert\psi_S\rangle$ by contracting a random subset of slices, which allows us to compute $\vert\psi_S^{\perp}\rangle=(\vert\psi\rangle-\sqrt{\phi}\vert\psi_S\rangle)/\sqrt{1-\phi}$. Therefore, we can store the values of $\langle x\vert\psi_S\rangle$ and $\langle x\vert\psi_S^{\perp}\rangle$ for each sample, which allows us to compute $p_{\vert\psi_S\rangle}(x), p_{\vert\psi_S^{\perp}\rangle}(x)$, and $R$. We would like to verify that the numerically estimated PDFs of these three quantities follow the analytic results in Eq. \ref{eqn:frugal_pdf}, \ref{eqn:uniform_pdf}, and \ref{eqn:r_pdf}, and the results are shown in Fig. \ref{fig:frugal_justification}. Further, they are independent as one can also verify that the joint distribution agrees with the product of the marginals.

\subsection{Top sampling classical adversary}\label{sec:oversampling_classical_adversary}

We consider the following classical simulation algorithm, which we call top sampling, and show that it behaves similar to sampling from a depolarized quantum state. A tensor network is constructed out of the $n$-qubit quantum circuit, with the input indices set to zero, and the network output is a sparse state of $2^k$ random bitstrings \cite{pan2022solving}. Contracting a fraction $\phi$ of all slices of the tensor network thus yields fidelity $\phi$ probability amplitudes of $2^{k}$ bitstrings, and the bitstring with the maximum probability amplitude is chosen \cite{zhao2024leapfrogging}. We now seek to analyze the distribution of the perfect fidelity probability amplitude of the chosen bitstring.

Denote the ideal quantum state from the circuit as $\vert\psi\rangle$. Just as in discussed in the previous subsection, the quantum state corresponding to contracting a fraction of the slices is close to a fidelity $\phi$ state $\vert\psi'\rangle$. Then,
\begin{equation}
    \vert\psi\rangle = \sqrt{\phi}\vert\psi'\rangle + \sqrt{1-\phi}\vert\psi'^{\bot}\rangle,
\end{equation}
where $\langle\psi'\vert\psi'^{\bot}\rangle=0$. Hence, $\langle\psi'\vert\psi\rangle=\phi$.

Their probability of a bitstring $x$ with respect to state $\vert\psi\rangle$ is
\begin{align}
    p_{\vert\psi\rangle}(x)&=\vert\langle x\vert\psi\rangle\vert^2\\
    &=\phi\vert\langle x\vert\psi'\rangle\vert^2 + (1-\phi)\vert\langle x\vert\psi'^{\bot}\rangle\vert^2+2\sqrt{\phi(1-\phi)}\Re[\langle x\vert\psi\rangle\langle \psi'^{\bot}\vert x\rangle]\\
    &=\phi p_{\vert\psi'\rangle}(x)+(1-\phi)p_{\vert\psi'^{\bot}\rangle}(x) +2\sqrt{\phi(1-\phi)}\Re\left[e^{i\theta}\right]\sqrt{p_{\vert\psi'\rangle}(x)p_{\vert\psi'^{\bot}\rangle}(x)}.
\end{align}

For top sampling, $2^k$ bitstrings $x_i$ for $i\in[2^k]$ are randomly sampled and the bitstring $x$ with the largest $p_{\vert\psi'\rangle}(x)$ is chosen. The probability that the largest $p_{\vert\psi'\rangle}(x_i)$ is less than $t$ is $\Pr[p_{\vert\psi'\rangle}(x_i)\leq t~\forall i\in[2^k]]\approx\mathrm{CDF}_U^{2^k}(t)$ (assuming $k$ is small compared to $n$, we can assume the sampled amplitudes are almost independent), and therefore $\mathrm{PDF}_{p_{\vert\psi'\rangle}(x)}(t)=2^k \cdot \mathrm{CDF}_U^{2^k-1}(t) \cdot\mathrm{PDF}_U(t)$, where the subscript $U$ indicates that it is for uniform sampling. Since $\mathrm{CDF}_U(t)=1-e^{-Nt}$ and $\mathrm{PDF}_U(t)=N e^{-Nt}$, we have

\begin{equation}
     \mathrm{PDF}_{p_{\vert\psi'\rangle}(x)}(t)=2^k\left(1-e^{-Nt}\right)^{2^k-1}N e^{-Nt}.\label{eqn:top_from_uniform_pdf}
\end{equation}

Now consider the second term $p_{\vert\psi'^{\bot}\rangle}(x)$. Bitstring $x$ is sampled with some dependence on $\vert\psi'\rangle$ but independent on the state $\vert\psi'^{\bot}\rangle$. As a result, for Haar random $\vert\psi'^{\bot}\rangle$, $x$ can be treated as being uniformly sampled. Therefore, the distribution of the second term is
\begin{equation}
\mathrm{PDF}_{p_{\vert\psi'^{\bot}\rangle}(x)}(t)=\mathrm{PDF}_U(t)=Ne^{-Nt}.\label{eqn:top_uniform_pdf}
\end{equation}
We justify the approximation $\vert\psi'^{\bot}\rangle$ as Haar random because we assume $\vert\psi'^{\bot}\rangle$ is randomly chosen in the orthogonal space of $\vert\psi'\rangle$ and $n$ is large. Similarly, this implies $\theta$ is uniformly random between $(-\pi,\pi]$, and we can instead use $R=\cos\theta=\Re[e^{i\theta}]$, and \begin{equation}
    \mathrm{PDF}_R(t)=f_R(t)=\begin{cases}
        \frac{1}{\pi \sqrt{1-t^2}}\quad&\text{if }t^2\leq 1,\\
        0& \text{otherwise}.
    \end{cases}\label{eqn:top_r_pdf}
\end{equation}

We can then solve for $R$ in terms of $\phi, p_{\vert\psi\rangle}(x), p_{\vert\psi'\rangle}(x)$, and $p_{\vert\psi'^{\bot}\rangle}(x)$:
\begin{equation}
    R=\frac{p_{\vert\psi\rangle}(x)-\phi p_{\vert\psi'\rangle}(x)-(1-\phi)p_{\vert\psi'^{\bot}\rangle}(x)}{2\sqrt{\phi(1-\phi)p_{\vert\psi'\rangle}(x)p_{\vert\psi'^{\bot}\rangle}(x)}}.
\end{equation}

We can then use the method of transformations to obtain the PDF of $p_{\vert\psi\rangle}(x)$. Setting $X_1,X_2,X_3=R,p_{\vert\psi'\rangle}(x),p_{\vert\psi'^{\bot}\rangle}(x)$ and $Y_1,Y_2,Y_3=p_{\vert\psi\rangle}(x),p_{\vert\psi'\rangle}(x),p_{\vert\psi'^{\bot}\rangle}(x)$. The determinant of the Jacobian is
\begin{equation}
    \vert J\vert=\frac{\partial R}{\partial p_{\vert\psi\rangle}(x)}=\frac{1}{2\sqrt{\phi(1-\phi)p_{\vert\psi'\rangle}(x)p_{\vert\psi'^{\bot}\rangle}(x)}}.
\end{equation}
Therefore, the PDF of $Y_1, Y_2, Y_3$ is
\begin{equation}
    g(y_1, y_2, y_3)=\vert J\vert f(x_1, x_2, x_3),
\end{equation}
and we also have
\begin{align}
    \mathrm{PDF}_{p_{\vert\psi\rangle}(x)}(y_1)&=g(y_1)=\int_0^1 g(y_1, y_2, y_3) dy_2 dy_3=\int_0^1 \vert J\vert f(x_1, x_2, x_3) dx_2 dx_3\\
    &=\int_0^1 \frac{1}{2\sqrt{\phi(1-\phi)x_2 x_3}}\cdot f_R(R(y_1,x_2,x_3)) \cdot \left[2^k\cdot\left(1-e^{-Nx_2}\right)^{2^k-1}N e^{-Nx_2}\right] \cdot Ne^{-Nx_3}dx_2 dx_3,\label{eqn:pdf}
\end{align}
where
\begin{equation}
    R(y_1,x_2,x_3)=\frac{y_1-\phi x_2-(1-\phi)x_3}{2\sqrt{\phi(1-\phi)x_2 x_3}}.
\end{equation}

However, the above analytic expression is not very useful, as it is very hard to integrate numerically. To understand some properties of the distribution without using the integral expression, we provide the first two moments of the distribution analytically. We can estimate the variance of $p_{\vert\psi\rangle}$ described by this sampling process:
\begin{align}
    \text{Var}[p_{\vert\psi\rangle}]&=\mathbb{E}\left[p_{\vert\psi\rangle}^2\right]-\mathbb{E}\left[p_{\vert\psi\rangle}\right]^2.
\end{align}
Since $\mathbb{E}[R]=0$, we have
\begin{align}
\mathbb{E}\left[p_{\vert\psi\rangle}^2\right]=&\mathbb{E}\Big[\phi^2 p_{\vert\psi'\rangle}^2+2\phi(1-\phi) p_{\vert\psi'\rangle} p_{\vert\psi'^{\bot}\rangle}+(1-\phi)^2 p_{\vert\psi'^{\bot}\rangle}^2\\
&+2\left(\phi p_{\vert\psi'\rangle}+(1-\phi) p_{\vert\psi'^{\bot}\rangle}\right)2R\sqrt{\phi(1-\phi)p_{\vert\psi'\rangle} p_{\vert\psi'^{\bot}\rangle}}\\
&+4R^2\phi(1-\phi)p_{\vert\psi'\rangle} p_{\vert\psi'^{\bot}\rangle}\Big]
\\
=&\phi^2\mathbb{E}\left[p_{\vert\psi'\rangle}^2\right]+(2+4\mathbb{E}\left[R^2\right])\phi(1-\phi)\mathbb{E}\left[p_{\vert\psi'\rangle}\right]\mathbb{E}\left[p_{\vert\psi'^{\bot}\rangle}\right]+(1-\phi)^2\mathbb{E}\left[p_{\vert\psi'^{\bot}\rangle}^2\right],\\
\mathbb{E}\left[p_{\vert\psi\rangle}\right]^2=&\phi^2\mathbb{E}\left[p_{\vert\psi'\rangle}\right]^2+2\phi(1-\phi)\mathbb{E}\left[p_{\vert\psi'\rangle}\right]\mathbb{E}\left[p_{\vert\psi'^{\bot}\rangle}\right]+(1-\phi)^2\mathbb{E}\left[p_{\vert\psi'^{\bot}\rangle}\right]^2.
\end{align}
Therefore,
\begin{equation}
    \text{Var}\left[p_{\vert\psi\rangle}\right]=\phi^2 \text{Var}\left[p_{\vert\psi'\rangle}\right]+(1-\phi)^2\text{Var}\left[p_{\vert\psi'^{\bot}\rangle}\right]+4\mathbb{E}\left[R^2\right]\phi(1-\phi)\mathbb{E}\left[p_{\vert\psi'\rangle}\right]\mathbb{E}\left[p_{\vert\psi'^{\bot}\rangle}\right].
\end{equation}
One can integrate $t\mathrm{PDF}_{p_{\vert\psi'\rangle}}(t)$ and $t^2\mathrm{PDF}_{p_{\vert\psi'\rangle}}(t)$ from 0 to infinity to estimate the variance of $p_{\vert\psi'\rangle}$. It can be shown that
\begin{equation}
    \mathbb{E}\left[p_{\vert\psi'\rangle}\right]=\frac{H_{2^k}}{N},
\end{equation}
where $H_{2^k}$ is the $2^k$th Harmonic number, and
\begin{equation}
    \mathbb{E}\left[p_{\vert\psi'\rangle}^2\right]=\frac{1}{N^2}\left(\frac{\pi^2}{6}+H_{2^k}^2-\Psi^{(1)}(2^k+1)\right),
\end{equation}
where $\Psi^{(1)}$ is the trigamma function. Therefore,
\begin{equation}
\text{Var}\left[p_{\vert\psi'\rangle}\right]=\mathbb{E}\left[p_{\vert\psi'\rangle}^2\right]-\mathbb{E}\left[p_{\vert\psi'\rangle}\right]^2=\frac{1}{N^2}\left(\frac{\pi}{6}^2-\Psi^{(1)}(2^k+1)\right).
\end{equation}

Similarly, one can obtain the variance of $p_{\vert\psi'^{\bot}\rangle}$. Since it follows the distribution of probabilities of uniform sampling, the mean and variance are known and are given by
$\mathbb{E}\left[p_{\vert\psi'^{\bot}\rangle}\right]=\frac{1}{N}$ and $\text{Var}\left[p_{\vert\psi'^{\bot}\rangle}\right]=\frac{1}{N^2}$. Finally, $\mathbb{E}[R^2]=\frac{1}{2}$. The final expression of the variance of $p_{\vert\psi\rangle}$ is
\begin{equation}
    \text{Var}\left[p_{\vert\psi\rangle}\right]=\frac{1}{N^2}\left(\frac{\pi^2}{6}\phi^2-\phi^2\Psi^{(1)}(2^k+1)+
    (1-\phi)^2+2\phi(1-\phi)H_{2^k}\right),
\end{equation}
and the expectation value is
\begin{equation}
    \mathbb{E}\left[p_{\vert\psi\rangle}\right]=\frac{1}{N}\left(\phi H_{2^k} + 1-\phi\right).
\end{equation}

\begin{figure}[!ht]
    \centering
    \includegraphics[width=0.6\textwidth]{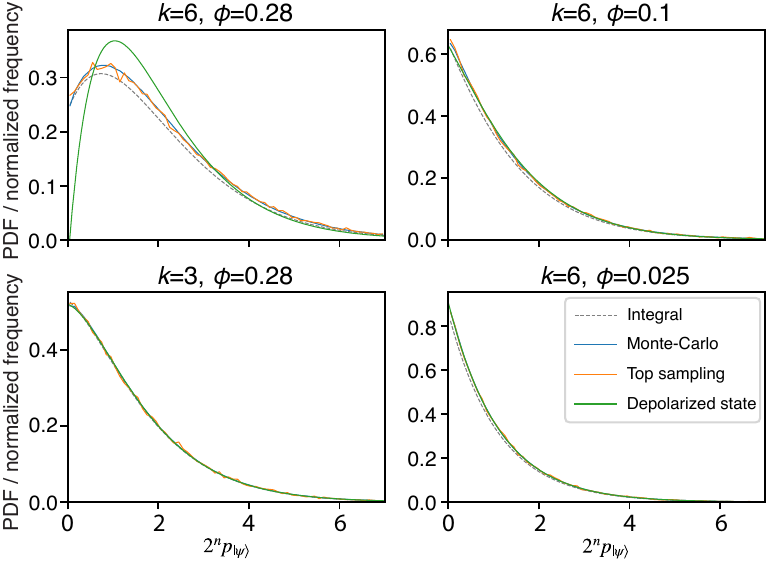}
    \caption{Distribution of the probability $p_{\vert\psi\rangle}(x)$ of bitstring sampled by picking the bitstring $(x)$ that has the largest amplitude norm when a fraction $\phi$ of slices are contracted. The number of qubits is $n=20$, and the depth is $d=8$. The MC approach uses 10 million samples, and top sampling uses 100 thousand samples.
    }
    \label{fig:pdf_comparison}
\end{figure}

\noindent \textbf{Distribution of bitstring probabilities.}
We now verify that the derived probability distribution for the top sampling classical adversary agrees with the experiment. The analytical form of the PDF of the probability of a top sampling adversary is given by Eq. \ref{eqn:pdf}. Therefore, the first method to evaluate the PDF is numerical integration, which we call the integral approach. However, the integral is challenging to evaluate numerically for large $k$ and large $\phi$, and the evaluated value may have some deviation due to limited numerical precision. The second method is to sample $p_{\vert\psi\rangle}$, which we do by independently sampling $p_{\vert\psi'\rangle},p_{\vert\psi'^{\bot}\rangle}$, and $R$ from PDFs given by Eq. \ref{eqn:top_from_uniform_pdf}, \ref{eqn:top_uniform_pdf}, and \ref{eqn:top_r_pdf}. We call this the Monte-Carlo (MC) approach. Both estimates are compared to the results obtained by actually performing top sampling by choosing the bitstring with the largest norm amplitude when a fraction of the slices are contracted. Finally, we compare everything against the PDF of sampling from the depolarized state with the same mean, and the results are shown in Fig. \ref{fig:pdf_comparison}.

We see that numerical integration fails at large $k$ and $\phi$ at small values of $p_{\vert\psi\rangle}$, where the value is slightly lower than those obtained by the MC approach and top sampling. Further, Fig. \ref{fig:pdf_comparison} shows that the MC approach is a faithful representation of top sampling since the two traces closely track each other without systematic deviations. Further, the distribution of sampling from a depolarized state only deviates from the actual distribution for large $k$, $\psi$, and small $p_{\vert\psi\rangle}$.

\begin{figure}[!ht]
    \centering
    \includegraphics[width=\textwidth]{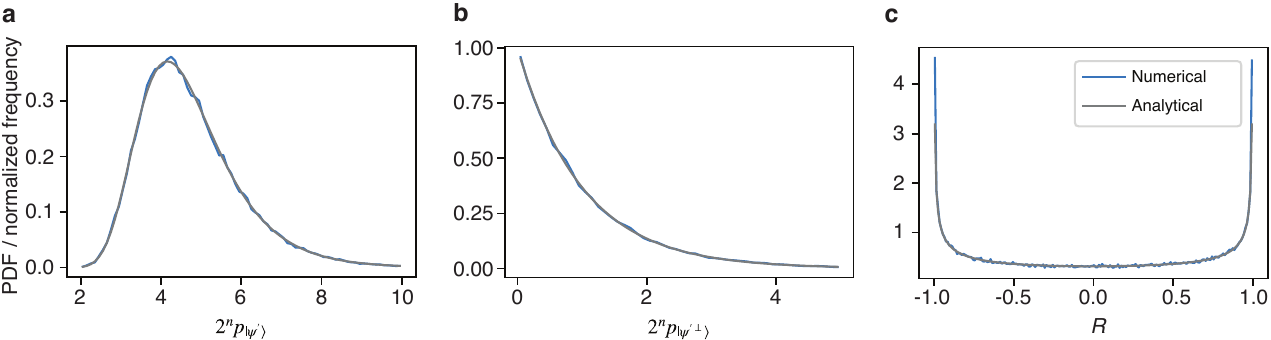}
    \caption{Empirical and analytical distributions of \textbf{a}, $p_{\vert\psi'\rangle}$, \textbf{b}, $p_{\vert\psi'^{\bot}\rangle}$, and \textbf{c}, $R$. Configurations are identical to Fig. \ref{fig:pdf_comparison}.}
    \label{fig:top_justification}
\end{figure}

We also justify the assumptions used in deriving Eq. \ref{eqn:pdf}. In the experiment, we are given $\vert\psi\rangle$. We generate $\vert\psi'\rangle$ by contracting a random subset of slices, which allows us to compute $\vert\psi'^{\bot}\rangle=(\vert\psi\rangle-\sqrt{\phi}\vert\psi'\rangle)/\sqrt{1-\phi}$. Therefore, we can store the values of $\langle x\vert\psi'\rangle$ and $\langle x\vert\psi'^{\bot}\rangle$ for each sample, which allows us to compute $p_{\vert\psi'\rangle}(x), p_{\vert\psi'^{\bot}\rangle}(x)$, and $R$. We would like to verify that the experimentally estimated PDFs of these three quantities follow the analytic results in Eq. \ref{eqn:top_from_uniform_pdf}, \ref{eqn:top_uniform_pdf}, and \ref{eqn:top_r_pdf}, and the results are shown in Fig. \ref{fig:top_justification}. Further, they are independent as one can also verify that the joint distribution agrees with the product of the marginals.

\newpage
\subsection{Matrix product  state simulation feasibility}\label{sec:mps}

Prior work established evidence that approximate simulation methods such as matrix product states (MPSs) cannot efficiently approximate the quantum states corresponding to random quantum circuits of similar scale to those used in our work~\cite{morvan2023phase,qntm_rcs}. This was the key assumption used to justify the focus on exact tensor network contraction style classical simulations in \cite{jpmc_cr}, which we also use here. Since the quantum circuit ensemble is slightly different from prior work due to the use of the initial degree two cycle graph state, we perform the same numerical analysis for the new circuit ensemble.

First, let us summarize the leading MPS approach. For a circuit $C$ that can be partitioned into two parts, i.e. $C=C_2C_1$, the transition amplitude is $\langle x\vert C\vert 0\rangle=\langle x\vert C_2C_1\vert 0\rangle=\langle\psi'\vert\psi\rangle$, where
$\vert\psi\rangle=C_1\vert 0\rangle$ and $\vert\psi'\rangle=C_2^\dagger\vert x\rangle$. Specifically, $\vert\psi\rangle$ and $\vert\psi'\rangle$ are $n$-qubit quantum states whose statevectors can be approximated by two MPSs. To find the transition amplitude, one would take the inner product of the two MPSs by contracting the MPSs.

To reduce the simulation cost of the MPSs, one would find a bipartition of the $n$ qubits into two subsystems such that the entanglement between the two subsystems is minimized. One can heuristically perform this by minimizing the number of two-qubit gates crossing the biparititon. For quantum circuits with random geometries, the optimal bipartitions for $\vert\psi\rangle$ and $\vert\psi'\rangle$ may be different. However, since we need to contract the MPSs, having different bipartitions would introduce swap gates and make the computational cost very large. Therefore, a single bipartition must be chosen such that the entanglement is small for both MPSs, which we optimize heuristically by minimizing the number of two-qubit gates for the whole circuit $C$, instead of just $C_1$ or $C_2$, that cross the bipartition.

To estimate the bond-dimension required to approximate the quantum state with fidelity $F$, we use the method described in \cite{morvan2023phase,qntm_rcs}. We comment that the choice of the single-qubit gate set used the experiment means that we need to analyze the $XY$-Clifford case in \cite{qntm_rcs}. For a depth $d$ quantum circuit, we find the bipartition that minimizes the number of crossing two-qubit gates using the KaHyPar hypergraph partitioning optimizer \cite{schlag2016k,akhremtsev2017engineering}. Then, the depth $d/2$ quantum circuit, corresponding to the first half of the depth $d$ quantum circuit, is simulated using the Clifford tableau formalism to calculate the reduced purity for that bipartition. Having the reduced purity then allows us to estimate the required bond dimension. We term this approach the full bipartition approach, reflecting the fact that the bipartition is optimized over the full circuit.

Alternatively, we can optimize the bipartition for the $d/2$ circuit only. This gives an lower bond dimension but the resulting MPS does not allow easy contraction. Nevertheless, it provides a more conservative bound on the computational resources for MPS simulations. We term this the half bipartition approach.

\begin{figure}[!ht]
    \centering
    \includegraphics[width=\textwidth]{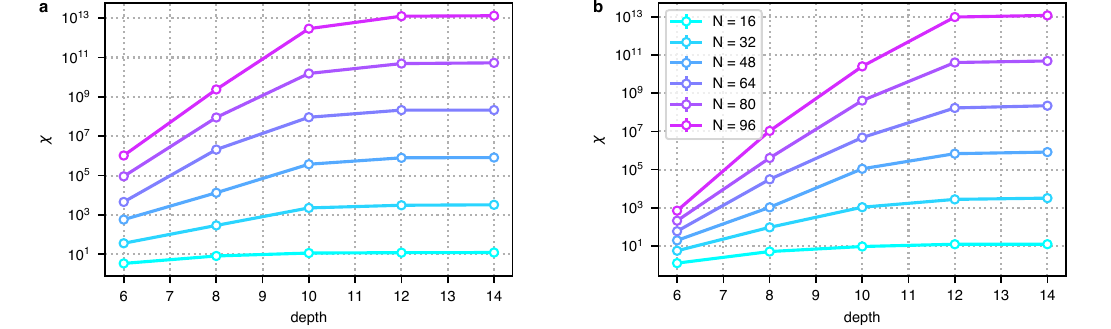}
    \caption{\textbf{MPS bond dimension required to simulate to fidelity or XEB $0.1$}. \textbf{a}, Estimated using the full bipartition scheme. \textbf{b}, Estimated using the half bipartition scheme.}
    \label{fig:mps_bond_dimension}
\end{figure}

Fig. \ref{fig:mps_bond_dimension}\textbf{a} shows that for the $n=64$ quantum circuit that we run, which is essentially depth $9$, the bond dimension required for $\mathrm{XEB}=0.1$ should be around $10^7$. Assuming each complex number requires $8$ bytes (i.e. complex64 format), the memory cost of a bipartition MPS is $2^{n/2}\times\chi\times 2\times 2\times 8=1.37\times10^{18}$ bytes ($2^{n/2}$ is the physical dimension of each MPS tensor, $\chi$ is the bond dimension, there are two tensors in each MPS, and there are two MPSs), which is beyond the storage capability of the largest supercomputers. This ignores the feasibility of actually performing MPS calculations, which requires communication of data. Even for the unrealistic situation where all communication can be performed with inter-GPU communications with the highest efficiency, the peak HBM GPU bandwidth is 208.9PB/s \cite{auroraoverview}. The more realistic number to use is the system interconnect bandwidth which has a peak injection bandwidth of 2.12PB/s \cite{auroraoverview}. This means communication alone would take seconds or hundreds of seconds. Other practical considerations include slow operations such as DMRG sweeps to obtain the approximate quantum state. Additionally, the bond dimension estimate using this method is optimistic due to assuming the uniformity in the singular values.

\subsection{Position verification}\label{sec:position_verification}
While position verification is not the main focus of this work, we briefly discuss the impact of the reduced latency using the delayed measurement bases. In position verification, many verifiers who trust each other but are located far away from each other wish to verify the location of an untrusted prover. The prover is assumed to be far away from all verifiers. It is known that secure protocols using only classical computation is impossible \cite{chandran2014position}, and typical approaches require quantum communication. Recently, it was shown that certified randomness protocols can be compiled into secure classically verifiable position verification protocols that only require classical communication \cite{kaleoglu2024equivalence}. Specifically, RCS-based certified randomness was used to instantiate a position verification protocol. However, the practical latency of the prior experimental demonstration of certified randomness in Ref \cite{jpmc_cr} leads to a position uncertainty that is too large to be useful. As a result, our work that reduces the latency of classical spoofing is a crucial step towards useful position verification protocols. However, additional cryptographic security considerations are needed.

\begin{figure}[h]
    \centering
    \includegraphics[width=0.5\linewidth]{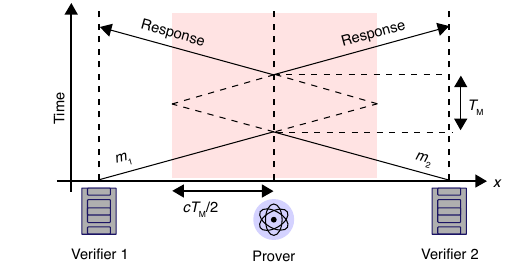}
    \caption{The classical verifiers send, at the speed of light, messages $m_1, m_2$ which together are used to generate the challenge random circuit $C(m_1, m_2)$. The messages are timed such that $m_1$ and $m_2$ arrive at the quantum prover's supposed location at the same time. The prover then responds with the sample drawn from the circuit $C(m_1, m_2)$ to both verifiers at the speed of light. The verifiers require the response to arrive within a short amount of time. This is repeated many times. If the samples meet the timing requirement and pass the XEB test, the verifiers will have ascertained the prover's position up to an uncertainty of  $c\cdot T_{\rm M}/2$. In this context, $T_{\rm M}$ is not the time between the verifier sending the challenge and receiving the response which is large due to the need for the messages to travel a large distance. Instead, it is the time between the prover receiving the challenging and sending the response.}
    \label{fig:position_verification}
\end{figure}

Consider the simple one dimensional case in position verification with classical communication, which is illustrated in Fig.~\ref{fig:position_verification}. We have two verifiers and a quantum prover at the claimed location somewhere between the verifiers in the honest case. The verifiers send half of the classical information necessary to reconstruct the challenge to the verifier such that both halves arrive at the claimed location at the same time assuming the information travels at the speed of light. In the idealized case, the prover retrieves the challenge by combining both pieces of information in some way, performs certified randomness as specified by the challenge, and sends the result back instantaneously. The result travels at the speed of light to the verifiers, and if any of them receives the result later than expected, the protocol aborts. Further, the verifiers check that they received the same result. Finally, the verifiers verifies that the result passes the certified randomness requirement as specified by the challenge.

To understand why this should work, consider the malicious case where there are two colluding provers position to the left and to the right of the claimed location. The provers receive each half of the challenge and forward it to the other prover. The provers will be able to retrieve the challenge, but the timing requirement means that they must immediately produce the result and send it back to the verifiers. The provers must produce the results independently. If the provers want to pass the certified randomness verification of the verifiers, the results must be random and cannot agree. If the provers use deterministic outputs, the results must not pass the certified randomness verification. The positional uncertainty is given by the latency of the quantum prover times the speed of light $c$. This is because if we loosen the time requirement from an instantaneous prover, the two provers could use the time to send the result to each other classically.

In the RCS-based protocol, the challenge is the random quantum circuit, and the time it takes for the quantum circuit to be executed and measured gives the positional uncertainty. In this work, we delay the measurement basis and claim that we can treat the time between receiving the measurement basis and obtaining the measurement result $T_{\rm M}$ as the positional uncertainty under suitable adversary models. Specifically, an adversary can prepare an arbitrary quantum state with polynomial resources shared between the two provers, and this state may depend on the challenge quantum circuit but cannot depend on the measurement basis. This is because if the provers are $c\cdot T_{\rm M}$ apart, the provers can no longer communicate after they receive the measurement basis.

For the restricted adversary considered in \cite{jpmc_cr}, the protocol is secure since there is no mechanism to exploit the additional rounds of communication that is allowed between the two provers after the circuit description is received but before the measurement bases are available. Specifically, the adversary can only perform honest measurement on the quantum state or perform classical simulation. Even though the classical simulation part is deterministic and can be used to allow the two dishonest provers to produce the same output, this effect is already captured in the entropy bound of the protocol. For the quantum measurement part, the restricted adversary is not prescribed a useful strategy and can only prepare the state honestly before measurement. One prover can measure the state in some basis (which they can choose deterministically or randomly but it would not matter since the actual bases are random) and send the result to the other prover, but the randomized bases means that it will fail the XEB test with high probability. Otherwise, the provers must wait for the bases before measurement, which means they cannot produce the same output.

The above argument does not preclude a more general adversary that is not limited to either measuring the quantum state $C\vert 0\rangle$ before or after the basis information is available. It is imaginable that some arbitrary entangled state that depends on $C$ can be prepared between the two provers before the measurement bases are available. Future work is required to understand the security of the protocol against more general adversaries.

\subsection{Validation cost analysis}\label{sec:cost_analysis}

In this section, we analyze the validation cost of the certified randomness protocol in the restricted adversarial model of Ref.~\cite{jpmc_cr} to understand the feasibility of certified randomness. In particular, we analyze the limit where the total number of rounds is infinity but the total validation budget is finite. The adversary has a bounded computational power but can keep running the classical computer instead of bounded total classical computational time. This is to say that the adversary will simulate infinitely many samples. On the other hand, achieving security only requires that a finite number of samples are validated, so the validation cost remains constant. Additionally, we allow $n_{\rm parallel}$ quantum computers to execute multiple circuits at the same time to reduce the average latency. This could be achieved by having separate quantum devices or one large quantum device with significantly more qubits and parallel gate execution capabilities.

We define the verification advantage $\xi$, which is a parameter that is characterized by the capabilities and available resources of the involved parties. 
Specifically,
\begin{equation}
    \xi = \frac{C_{\rm val}T_{\rm val}n_{\rm parallel}}{C_{\rm adv}T_{\rm M}},
\end{equation}
where $n_{\rm parallel}$ is the number of parallel quantum processors, $T_{\rm val}$ is the total time available for verification, $C_{\rm val}$ is the verifier's computational power, $C_{\rm adv}$ is the adversary's computational power. Specifically, the computational power takes into account both the raw computational power of the classical computer as well as the efficiency of the simulation algorithm. $T_{\rm M}$ is the latency of obtaining a quantum sample from when the measurement basis is streamed, or from when the circuit is made available if there is no measurement basis streaming. Note that, $T_{\rm M}/n_{\rm parallel}$ is the effective latency due to parallelization.

The advantage of the $\xi$ parameter is that it alone is sufficient to characterize the performance of the certified randomness protocol, while capturing the effect of timing, resources, computational power, parallelization all at the same time.

Since the total computational budget for the adversary for each round of the protocol is $C_{\rm adv}T_{\rm M}$, the average budget per circuit is therefore $C_{\rm adv}T_{\rm M}/n_{\rm parallel}$. Since the total verification budget is $C_{\rm val}T_{\rm val}$, if a total of $L_{\rm val}$ circuits are verified, then the verification budget per circuit is $C_{\rm val}T_{\rm val}/L_{\rm val}$.

Verification is simulation to fidelity 1, and the fidelity is proportional to the computational budget. Therefore, the adversary can simulate to fidelity
\begin{equation}
    f_{\rm adv}=\frac{C_{\rm adv}T_{\rm M}/n_{\rm parallel}}{C_{\rm val}T_{\rm val}/L_{\rm val}}=\frac{C_{\rm adv}T_{\rm M}L_{\rm val}}{C_{\rm val}T_{\rm val}n_{\rm parallel}}.
\end{equation}
This means
\begin{equation}
    \xi = L_{\rm val} / f_{\rm adv}.
\end{equation}

One can choose the hardness of the quantum circuits such that simulating one circuit costs $C_{\rm val}T_{\rm val}/L_{\rm val}$ given any target $L_{\rm val}$. Specifically, we shall naturally consider the verification advantage $\xi$ as fixed, since it is determined by resources available to the verifier. Then, the verifier would optimize $L_{\rm val}$ such that the experiment performs the best. This is because high $L_{\rm val}$ means more samples and more confidence in the XEB score, but it also leads to higher $f_{\rm adv}$ which requires more samples to differentiate from quantum sampling.

We perform numerical analysis on the protocol performance. In addition, the fidelity of the quantum device also affects the performance. We therefore examine the performance of the protocol with varying quantum computer fidelity $\phi$ and verification advantage $\xi$. In particular, there are two ways to measure the performance, the entropy rate (entropy divided by the number of output bits) and the security parameter. We can therefore calculate the $\xi$ required at different $\phi$ to achieve a certain entropy rate or security parameter.

\begin{figure}[t]
    \centering
    \includegraphics[width=0.8\linewidth]{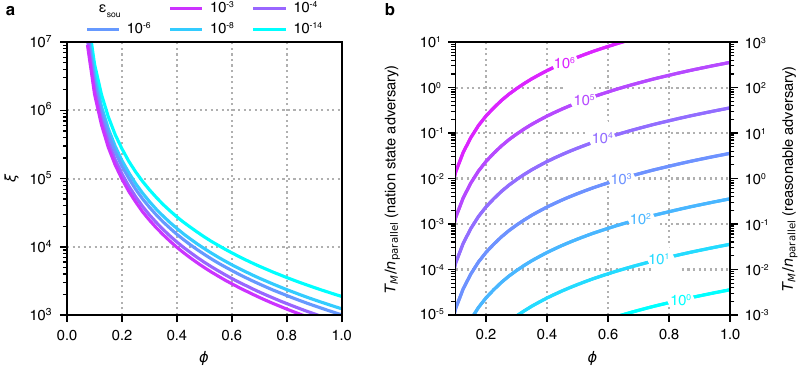}
    \caption{\textbf{Validation cost analysis}. \textbf{a}, Achievable security parameter $\varepsilon$ at varying quantum computer fidelity $\phi$ and verification advantage $\alpha$ for entropy rate $0.04$. \textbf{b}, Required validation budget in GPU hours against nation state ($10^5$ GPUs) and reasonable ($10^3$ GPUs) restricted adversaries at various quantum computer fidelity $\phi$ and measurement latency per parallel sample $T_{\rm M}/n_{\rm parallel}$. Targeted security is $\varepsilon_{\rm sou}=10^{-6}$, and entropy rate is $0.04$.}\label{fig:restricted_adversary_supplement}
\end{figure}

We show the result in Fig. \ref{fig:restricted_adversary_supplement}\textbf{a}, which uses the analysis closely reflecting Supplementary Materials Section V of Ref.~\cite{jpmc_cr}. In particular, the fraction of shape-2 samples, or Porter-Thomas samples, is $R\leq R_{\rm Q}+f_{\rm adv}$, where $R_{\rm Q}$ is the fraction of samples the adversary honestly sample on a quantum computer. In this case, the smooth min-entropy rate is $R_{\rm Q}(n-1)/n$. Normally, the $\varepsilon_{\rm smooth}$-smooth min-entropy has a penalty factor $-\log(1/\varepsilon_{\rm smooth})$, but this factor becomes negligible as the system size and smooth min-entropy goes to infinity.

We also translate the requirement on $\xi$ into more interpretable metrics such as validation budget, adversary power, latencies, and number of parallel samples. We show the result in Fig. \ref{fig:restricted_adversary_supplement}\textbf{b}.

\subsection{Implementation details of classical validation}

For validation, we compute the probability of each bitstring $P_{C_i}(x_i)=\vert\langle x\vert C_i\vert 0\rangle\vert^2$ using the tensor network formalism \cite{Markov2008simulating}. The transition amplitude is a scalar that is the result of a series of tensor contractions, where the tensors corresponds to the unitaries of the quantum gates.

In particular, the order in which the tensors are contracted has a significant effect on the computational cost, and optimizing the contraction order is a significant part of such simulation algorithms \cite{huang2020classical,kalachev2021classical,gray2021hyper,morvan2023phase}. We optimize the contraction order using CoTenGra \cite{gray2021hyper}. The computational time on a GPU depends on the number of floating point operations as well as data movement. The objective function we choose is a weighted sum of the above costs, which corresponds to the combo-256 objective in CoTenGra. We run an ensemble optimization with 50,905 runs on the Crux supercomputer at the Argonne Leadership Computing Facility. Each run consists of 100 optimizations and returns the best result.

The actual performance of the algorithm is highly dependent on complex GPU specific characteristics such as memory bandwidth and software support for specific subroutines. We perform on-device benchmarks using the Intel GPUs on Aurora and choose the contraction order that completes the computation in the shortest amount of time. We perform tensor network contractions using CoTenGra with PyTorch as the backend.

We use the BF16X2 precision for single-precision computations. As a result of the reduced precision, not all contraction orders give the correct result when using the reduced precision, so care is required to ensure that such contraction orders are abandoned or full double precision must be used.

For Intel GPUs, PyTorch does not natively support operations on tensors with greater than 12 dimensions, but tensors with higher dimensions are frequently encountered during contractions in quantum circuit simulations. To address this challenge, we devise a recursive permutation algorithm to limit the maximum dimension of certain tensor operations.

In general, any pair-wise contraction can be expressed as an Einstein summation expression and be deployed using torch.einsum. Any Einstein summation can be deployed using batched matrix multiplication after suitable permutation of tensor dimensions, and this is a standard implementation for many software libraries.

There are four types of indices that we encounter during tensor network contractions for simulating quantum circuits, namely indices that are only in the first input tensor and the output tensor, indices that are only in the second input tensor and the output tensor, contracted indices that are in both input tensors and not the output tensor, and batch indices that are in both input tensors and not in the output tensor. Here, we never encounter indices that are in one of the tensors and not the output tensor. In other words, contraction with itself does not happen. Batch indices may be encountered when there is an index shared by three or more tensors and we are performing contraction between two of the tensors, where the index cannot be eliminated until contraction with the last involved tensor.

To deploy Einstein summation to batched matrix multiplication, we permute all batch indices to the beginning for both input tensors, and the tensors are reshaped such that all these indices become a single dimension. Then, for the first input tensor, what follows are all the uncontracted indices reshaped into a single dimension, and then all the contracted indices reshaped into a single dimension. The second input tensor is manipulated similarly except the uncontracted indices come last. With this, batched matrix multiplication exactly performs the desired computation for Einstein summation. Suitable reshaping and permutation of the output is required to recover the indices of the correct Einstein summation output.

\begin{figure}
    \centering
    \includegraphics[width=\linewidth]{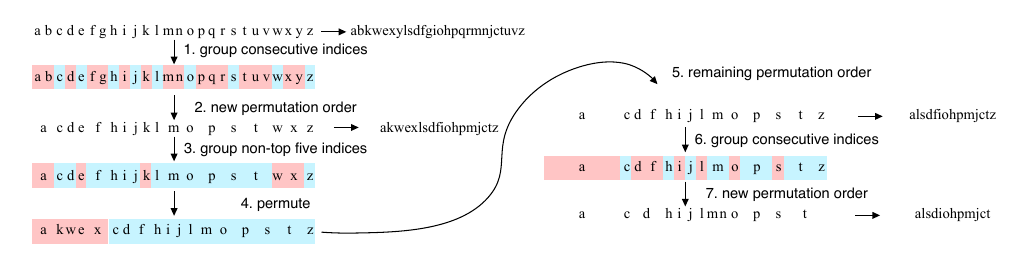}
    \caption{Algorithm for recursive permutation. Step 3 through 7 may need to be repeated many times. In this example, since the tensor at the end of step 7 has 12 dimensions, a native permutation can be applied to end the subroutine.}\label{fig:permutation}
\end{figure}

Here, the step of arbitrary tensor permutation would result in RuntimeError frequently when the tensor dimension is greater than 12. We can break arbitrary index permutation into multiple permutations on a subset of all indices. We first group all indices that remain contiguous in the same order before and after permutation. All grouped indices are shaped into a single dimension. The original permutation order becomes a new permutation order on fewer indices. If the total dimensionality of the tensor is less than or equal to 12 after grouping, we can directly apply permutation with the new order and recover the shape by reshaping. Otherwise, we permute the tensor such that the first 5 indices of the output of the new permutation order are permuted correctly whereas the rest of the indices are untouched. To do this, all contiguous indices that are not the first 5 output indices are reshaped into a single dimension, and the resulting tensor has at most 12 dimensions. This partially permutes the tensor and results in a new permutation order necessary to complete the permutation. We then recursively apply this operation by grouping indices indices moving at most 5 indices to the beginning at a time until all indices are permuted correctly. Fig. \ref{fig:permutation} illustrates the recursive permutation algorithm.

We note that existing implementation of JAX on Intel GPUs does not have the limitation in the tensor dimensionality as we discussed above. However, JAX support for complex number operations on Intel GPUs is not as mature and there is a significant performance gap. Combined with the inability for JAX on Intel to utilize BF16-type precision for complex numbers, PyTorch yields significantly better results in terms of computational throughput.

Another challenge we encounter is the deployment of the validation algorithm on HPC systems in a scalable manner. One complication is the need to save the complex value resulting from all slices. Additionally,
the algorithm must be resilient against node or GPU failures. If we use collective communication to gather the results from all MPI ranks, failed ranks may stall the program. Additionally, since there may be slower GPUs due to fluctuations, when the program scales to tens of thousands of GPUs, such fluctuations can cause significant latency and idling due to blocking collective communication.

As a result, we implement a scheduler to dispatch computational tasks which corresponds to computing a few slices. If a worker fails and does not return results for a long time, the scheduler reroutes the task to another worker. Additionally, there is a separate rank collecting computational results such that the scheduler is not blocked. Each node has 12 GPU tiles that each takes up a rank. To mitigate communication overhead, each node only has one rank responsible for communicating with the scheduler and the results collection rank. Only the collection rank writes the results to the filesystem. This prevents the issue where thousands of ranks try to write to the filesystem simultaneously. We report the efficiency of the overall algorithm at different scales in Table \ref{tab:verification_scaling}.

\begin{table}[h!]
    \centering
        \begin{tabular}{p{2.5cm} p{2.5cm} p{2.5cm} p{2.5cm} p{2.5cm}}
        \toprule
         & Single-tile & 1024 nodes & 2048 nodes & 4096 nodes \\
         \hline
         Aurora & 13.8\% & 13.5\% & 13.7\% & 13.3\%\\
     \bottomrule
\end{tabular}
    \caption{Efficiency of tensor network contraction on different scales. One node has 12 tiles.}
    \label{tab:verification_scaling}
\end{table}

\subsection{Comparison against Ref.~\cite{jpmc_cr}}

\begin{table}
    \centering
        \begin{tabular}{ p{6cm} p{4cm} p{7cm}}
        \hline
          & Ref.~\cite{jpmc_cr} & This work \\
         \hline
         Number of device qubits & 56 & 98 \\
         \hline
         Number of qubits for circuit & 56 & 64 \\
         \hline
         Single-qubit gate set & SU(2) & $Z^p X^{1/2} Z^{-p}$ with $p\in\{-1,-3/4,-1/2,\dots,3/4\}$\\
         \hline
         Number of two-qubit gates & 280 & 276 \\
         \hline
         Full circuit latency & 2.15 seconds & 0.91 seconds \\
         \hline
         Classical spoofing latency & 2.15 seconds & 0.03 seconds \\
         \hline
         Number of submitted circuits & 60,952 & 23,651 \\
         \hline
         Postselected circuits & 30,010 & No postselection \\
         \hline
         Batch size & 15 or 20 & 1 \\
         \hline
         Number of validated circuits & 1,522 & 11,961 (including 28 of the 53 failed rounds) \\
         \hline
         Fidelity (XEB) & 0.32 & 0.59 \\
         \hline
         Classical simulation cost & 90 ExaFLOPs (Frontier) & 3.1 ExaFLOPs (Aurora) \\
         \hline
         Classical simulation time & 93 seconds (Frontier) & 7.7 seconds (Aurora) \\
         \hline
         Classical simulation efficiency & 49\% (Frontier) & 13.8\% (Aurora) \\
         \hline
         Gap between classical and quantum & 13.8 & 151 \\
         \hline
         Security against adaptive adversary & No & Yes \\
         \hline
         Security against postselection & No & Yes \\
         \hline
         Security against general quantum adversary & No & Yes \\
         \hline
         Randomness amplification & No & Yes \\
     \hline
\end{tabular}
    \caption{Comparison of this work and Ref.~\cite{jpmc_cr}.
    }
    \label{tab:prior_work}
\end{table}

Table \ref{tab:prior_work} shows a comparison between this work and Ref.~\cite{jpmc_cr}, showcasing the improvement in the quality of the experiment. Below, we explain various terms used in the table and how certain quantities are calculated.

Full circuit latency is the time between when the first classical bit of information about the circuit is sent and when the bitstring is received. Classical spoofing latency is the same as full circuit latency for \cite{jpmc_cr} since all classical information about the circuit is sent at the beginning of each round. Classical spoofing latency is the time between when the measurement basis is sent and when the bitstring is received in this work, or $T_{\rm M}$ as discussed in the main text.

All floating-point operations we discuss below are single-precision operations. ExaFLOPs stands for $10^{18}$ floating-point operations. The classical simulation time of Ref.~\cite{jpmc_cr} is calculated for the Frontier supercomputer assuming perfect scaling from single-GPU tile to full-machine performance. Frontier has 9,408 nodes, each with 4 AMD MI250X GPUs which has two tiles each. The time it takes to simulate a single circuit used in Ref.~\cite{jpmc_cr} on a single tile is $7\times10^6$ seconds. When calculating the efficiency, we assume a maximum possible single-tile performance of 53 TeraFLOPS (TeraFLOPS stands for $10^{12}$ floating-point operations per second). For this work, the classical simulation time is calculated for the Aurora supercomputer assuming perfect scaling from single-GPU tile to full-machine performance. Aurora has 10,624 nodes, each with 6 Intel Data Center GPU Max Series (codename Ponte Vecchio or PVC) devices which has two tiles each. The time it takes to simulate a single circuit used in this work on a single tile is $9.8\times10^5$ seconds. When calculating the efficiency, we assume a maximum possible single GPU (two tiles) performance of 46 TeraFLOPS. When calculating the gap between classical spoofing and quantum sampling, we use classical simulation time times the fidelity divided by the classical spoofing latency.

We note that the classical simulation efficiencies of Frontier and Aurora should not be directly compared. As Ref.~\cite{jpmc_cr} shows, different tensor network contraction schemes of the same quantum circuit may have very different total compute and data movement costs even if the total simulation time may be similar, and the tradeoff characteristics may vary widely from circuit to circuit and machine to machine. Since the computation efficiency only looks at the compute cost and the theoretical maximum FLOPS, large variations in efficiency are expected.

\subsection{Additional timing information}

\begin{figure}[!ht]
\centering
\includegraphics[width=0.4\linewidth]{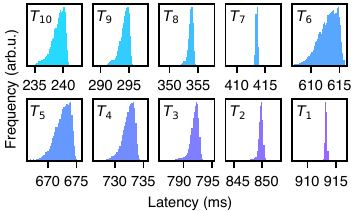}
\caption{Timing results of the streaming protocol. We show the frequency of different latencies ($T_j$ for $j\in[10]$) in arbitrary units.}
\label{fig:timing_results}
\end{figure}

We show the timing information of the streaming protocol in Fig. \ref{fig:timing_results}.

\stoptocentries
\putbib[citations]
\end{bibunit}

\end{document}